\documentclass[final,faculty=fw,department=nat,phddegree=fys]{adsphd}

\title{ Precision tests of the AdS/CFT correspondence }

\author{Xuao}{Zhang}

\supervisor{Prof.~dr.~Nikolay~Bobev}{}
\president{Prof.~dr.~Tjonnie~Li}{}
\jurymember{Prof.~dr.~Thomas~Hertog}{}
\jurymember{Prof.~dr.~Thomas~van~Riet}{}
\externaljurymember{Prof.~dr.~Monica~Guica}{IPhT, Saclay, Ecole Polytechnique, Lausanne, and CERN}
\externaljurymember{Prof.~dr.~Christoph~Uhlemann}{Vrije Universiteit Brussel}

\address{Celestijnenlaan 200D}
\addresscity{B-3001 Leuven} 

\date{August 2024}
\copyyear{2024}

\usepackage{nomencl}   
\renewcommand{\nomname}{List of Symbols}

\makenomenclature%

\usepackage{glossaries} 

\makeglossaries%

\usepackage[utf8]{inputenc}

\usepackage{textcomp} 
\usepackage{pifont}   
\usepackage{booktabs}
\usepackage{amssymb,amsmath,amscd}
\usepackage{amsmath}

\usepackage[T1]{fontenc}
\usepackage{fix-cm}
\usepackage{lmodern}
\usepackage{mathtools}
\usepackage{mathrsfs}
\usepackage{empheq}
\usepackage{physics}
\usepackage{bbm}
\usepackage{bm}
\usepackage{hhline}
\usepackage{tensor}
\usepackage{tikz}
\usepackage{xcolor}
\usepackage{pifont}
\usepackage{latexsym}
\usepackage{slashed}
\usepackage{multirow}
\usepackage{float}
\usepackage{afterpage}
\usepackage{soul}

\usepackage[english]{babel}
\usepackage{titlesec}
\usepackage{slashed}
\usepackage{hyperref}
\usepackage{graphicx,xcolor}
\usepackage{enumerate}
\usepackage{bbm} 						
\usepackage{float}
\usepackage{mathtools}
\usepackage{subfig}

\usepackage{graphicx}

\usepackage{tikz}

\makeatletter
\newcommand*{\rom}[1]{\expandafter\@slowromancap\romannumeral #1@}
\makeatother

\newcommand\dphi{\delta\phi}
\newcommand\mO{\mathcal{O}}
\newcommand\mD{\mathcal{D}}
\newcommand\mP{\mathcal{P}}
\newcommand\mN{\mathcal{N}}
\newcommand\fft[2]{\frac{#1}{#2}}

\newcommand\mQ{\mathcal{Q}}

\newcommand{\cB}{\mathcal{B}}
\newcommand{\cN}{\mathcal{N}}
\newcommand{\cmark}{\ding{51}}
\newcommand{\xmark}{\ding{55}}

\def\cB{{\cal B}}
\def\cD{{\cal D}}
\newcommand{\cE}{\mathcal{E}}
\def\cF{{\cal F}}
\def\cG{{\cal G}}

\def\cI{{\cal I}}

\def\cM{{\cal M}}
\def\cN{{\cal N}}

\def\cQ{{\cal Q}}
\def\cR{{\cal R}}

\def\cZ{{\cal Z}}

\def\Tr{{\rm Tr}\,}

\newcommand{\Dx}{\mathsf{D}}

\def\SU{{\rm SU}}

\newtheorem{conjecture}{Conjecture}

\usepackage{amssymb,amsmath,amscd,extarrows}
\usepackage[english]{babel}
\usepackage{titlesec}
\usepackage{slashed}
\usepackage{hyperref}
\usepackage{graphicx,xcolor}
\usepackage{enumerate}
\usepackage{bbm} 						
\usepackage{float}
\usepackage{mathtools}
\usepackage{subfig}
\usepackage{setspace} 

\newcommand{\bea}{\begin{eqnarray}}
\newcommand{\eea}{\end{eqnarray}}
\newcommand{\be}{\begin{equation}}
\newcommand{\ee}{\end{equation}}

\begin{document}


\makefrontcoverXII

\maketitle

\frontmatter 

\includepreface{preface}
\includeabstract{abstract}
\includeabstractnl{abstractnl}



\tableofcontents


\mainmatter 


%
  \graphicspath{{\chaptersdir/introduction/image/}}%
\chapter{This is introduction}\label{ch:introduction}

\instructionsintroduction


Illustration of how to include citations \cite{Meert2011PhD} and \cite{VandenBroeck2011IJCAI}. Lorem ipsum dolor sit amet, consetetur sadipscing elitr, sed diam nonumy eirmod tempor invidunt ut labore et dolore magna aliquyam erat, sed diam voluptua. At vero eos et accusam et justo duo dolores et ea rebum. Stet clita kasd gubergren, no sea takimata sanctus est Lorem ipsum dolor sit amet.

And yet another citation \cite{FrRo2010Diffusion}.

\nomenclature{$\Theta$}{A nice symbol}
Introducing some symbol: $\Theta$.

\newglossaryentry{md}{name={MD},description={molecular dynamics}}
Introducing an acronym: \gls{md}.

\begin{figure}
  \centering
  \medskip
  \includegraphics[width=.9\textwidth]{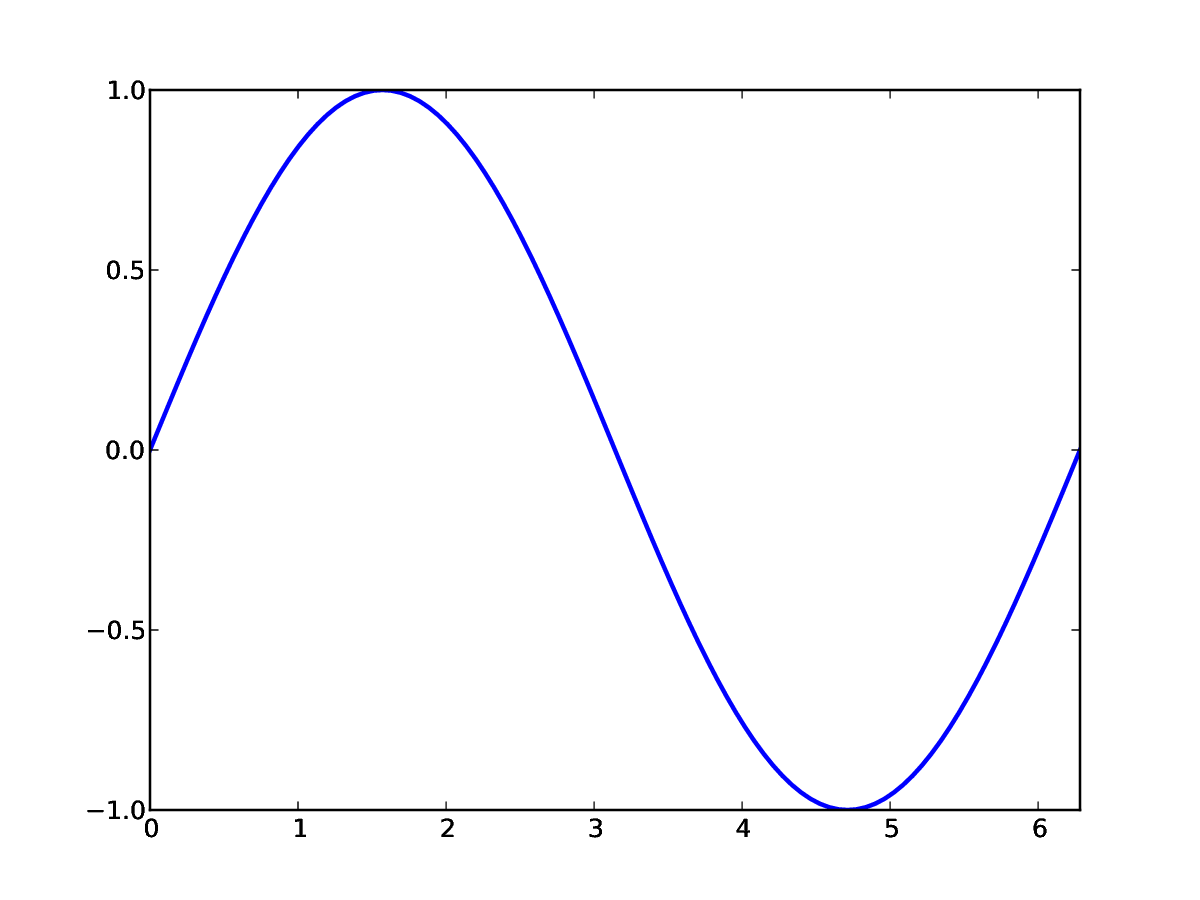}
  \caption[Short caption for Table of Figures]{Illustration of how to
  include a figure (long text, should not go to Table of Figures).}
  \label{fig:sine}
\end{figure}

\cleardoublepage

%



%
  \graphicspath{{\chaptersdir/ads4log/image/}}%
  \chapter{ Log corrections in AdS$_4$ }\label{ch:ads4log}

\section{Introduction}
\label{sec:intro}

The AdS/CFT correspondence is a cornerstone of modern theoretical physics and it is of clear interest to test this duality as precisely as possible. In this chapter, we focus on subleading corrections to the supergravity approximation used in holography, specifically in the context of AdS$_4$/CFT$_3$. The motivation for our analysis comes from the CFT side of the duality and the recent proliferation of techniques to compute QFT observables in the large $N$ limit of supersymmetric holographic CFTs. More concretely, the partition functions $Z_{\mathcal{M}_3}$ of~3d~$\mathcal{N}\geq 2$ holographic SCFTs arising from a stack of $N$ M2-branes on a compact Euclidean manifold $\mathcal{M}_3$ can be calculated in the large $N$ limit using supersymmetric localization. In these theories, the leading term in the free energy $\log Z_{\mathcal{M}_3}$ scales as $N^{3/2}$ and in many examples agrees with the regularized on-shell action of classical asymptotically locally AdS$_4$ Euclidean supergravity solutions which can be thought of as smooth fillings of $\mathcal{M}_3$.\footnote{See \cite{Drukker:2010nc,Jafferis:2011zi,Benini:2015eyy,Hosseini:2016tor,Azzurli:2017kxo} for an incomplete selection of references and \cite{Pestun:2016zxk,Zaffaroni:2019dhb} for a review. } The first subleading term scales as $N^{1/2}$ and it can be accounted for in the bulk by studying the leading four-derivative correction to the classical four-dimensional supergravity action, see \cite{Bobev:2020egg,Bobev:2021oku}. The goal of this chapter is to study how the $\log N$ term that arises at the next order in the large $N$ expansion of the free energy can be computed from the bulk supergravity theory.

An alternative point of view on logarithmic corrections to gravitational path integrals is found in black hole physics. It is expected that the Bekenstein-Hawking formula for the black hole entropy, $S_{\rm BH} = \frac{1}{4}A_{\rm H}$ with $A_{\rm H}$ the area of the horizon in Planck units, receives perturbative quantum corrections. These arise from higher-derivative terms in the gravitational effective action, as well as from quantum effects due to matter fields propagating on a fixed gravitational background. While a lot is known about these corrections to the black hole entropy, they are in general hard to calculate and strongly depend on the details of the UV completion of the effective gravitational theory. It was shown by Ashoke Sen that the logarithmic corrections to black hole entropy are very special in that regard. Notably, the coefficient of the $\log A_{\rm H}$ correction to $S_{\rm BH}$ is determined solely by the one-loop quantum contributions to the gravitational path integral of all fields below the cutoff scale of the effective gravitational theory. This fact represents a powerful ``IR window'' into the UV-complete theory of quantum gravity and can be employed to derive strong consistency condition on the microscopic description of black hole physics. Moreover, in special situations with enough symmetry, it has been shown that the $\log A_{\rm H}$ corrections to black hole entropy can be successfully matched to microscopic results from string theory, see \cite{Mandal:2010cj,Banerjee:2011jp,Sen:2012kpz,Sen:2012cj,Bhattacharyya:2012wz,Sen:2012dw} and references thereof.

These two vantage points on logarithmic corrections to gravitational path integrals lead us to study how such logarithmic terms arise in 4d gravitational theories in AdS,~see \cite{Bhattacharyya:2012ye,Liu:2017vbl,Liu:2017vll,Gang:2019uay,PandoZayas:2020iqr} and \cite{Hristov:2021zai,David:2021eoq,Karan:2022dfy} for previous studies of logarithmic corrections in AdS$_4$/CFT$_3$ in 11d and 4d supergravity, respectively. We follow the approach of Sen and study the quantum effects of matter fields propagating on a fixed background of a 4d gravitational theory with a negative cosmological constant. Focusing on fields with half-integer spin up to 2 and general masses, we employ heat kernel methods to calculate their contributions to the logarithmic term in the gravitational path integral, see \cite{Vassilevich:2003xt} for a review of the heat kernel expansion. In the absence of other scales in the problem, the log correction is of the form $\log L^2/G_N$ where $G_N$ is the 4d Newton constant and $L$ is the length scale set by the cosmological constant.\footnote{The methods we use can be adapted to study situations with a more general cutoff scale not related to the 4d Planck scale. We will comment on such situations further below.} In general the coefficient of $\log L^2/G_N$ can receive contributions from zero modes and non-zero modes of the differential operator that controls the dynamics of the field with a given spin, as well as from boundary terms. For the 4d gravitational backgrounds of interest in this work, we find under reasonable assumptions that the boundary terms have a vanishing contribution to the $\log L^2/G_N$ coefficient. The analysis of the contribution from zero modes is subtle and we can only make precise quantitative statements for very symmetric spaces like Euclidean AdS$_4$ and AdS$_2 \times \Sigma_{\mathfrak{g}}$, where $\Sigma_{\mathfrak{g}}$ is a smooth compact Riemann surface of genus $\mathfrak{g}$. Nevertheless, we argue that even for more general gravitational backgrounds, the zero modes contribute a pure number to the coefficient of $\log L^2/G_N$ which is independent on any continuous parameters that may be present in the 4d gravitational solution.

The results for the contribution of the non-zero modes is more intriguing. We find that general massive fields of spin up to 2 lead to a coefficient of $\log L^2/G_N$ that depends on the continuous parameter of the gravitational background, like squashing deformations of the boundary metric or the angular velocity of a black hole. This seemingly innocuous fact has important repercussions. Focusing on 4d Euclidean supergravity backgrounds, we can compare our results for the coefficients of the $\log L^2/G_N$ corrections to the $\log N$ terms in the path integral of large $N$ 3d holographic SCFTs arising from M2-branes. Using results from supersymmetric localization for a number of explicit examples of such SCFTs, we observe that the coefficient of $\log N$ does not depend on such continuous parameters. We use this to show that the apparent contradiction between the supergravity and SCFT results can only be resolved if the total contribution of the non-zero modes to the $\log L^2/G_N$ term in the gravitational path integral only depends on a specific contribution in the heat kernel expansion that we identify.

We then set out to check this strong ``bootstrap'' constraint on explicit top-down AdS$_4$/CFT$_3$ examples. We start with the familiar ABJM theory at level $k=1$ dual to 11d supergravity on an asymptotically locally AdS$_4 \times S^7$ background. When trying to employ our 4d gravitational approach we are faced with an immediate difficulty, namely that the 4d~$\mathcal{N}=8$ supergravity theory is not a standard EFT with finitely many fields but rather a consistent truncation to 4d $\mathcal{N}=8$ ${\rm SO}(8)$ gauged supergravity coupled to an infinite tower of Kaluza-Klein (KK) modes with masses below the 4d Planck scale. To calculate the non-zero mode contribution to the heat kernel, we thus have to take into account the full KK tower and sum the contributions of the infinitely many KK modes. This calculation leads to a divergent sum which we need to regularize. Using three different regularization methods proposed in the literature, we show that \textit{all} non-zero mode contributions in the heat kernel expansion relevant for the $\log N$ correction vanish.

This result presents a puzzle. Using supersymmetric localization on the round $S^3$, it can be shown that the ABJM free energy has a logarithmic term in its large $N$ expansion given by $\frac{1}{4}\log N$, see \cite{Fuji:2011km,Marino:2011eh}. The factor of $1/4$ was successfully reproduced in 11d supergravity in \cite{Bhattacharyya:2012ye} using a one-loop analysis of the 11d supergravity fields on the dual AdS$_4 \times S^7$ background. In short, one finds that since the heat kernel expansion is used in odd dimensions, only zero modes of the 11d differential operators contribute. The only possible such zero mode in Euclidean AdS$_4$ arises from a 2-form, and it would naively seem that there are no 2-forms among the fields of 11d supergravity. However, the quantization of the 3-form potential of the 11d theory necessitates the introduction of a 2-form ghost, which in turn gives the only non-vanishing contribution to the $\log N$ term. This contribution was calculated in \cite{Bhattacharyya:2012ye} and was shown to agree precisely with the $1/4$ calculated in the holographically dual SCFT. Our 4d analysis of the logarithmic correction yields a different result. Since there are no 2-forms in the KK spectrum of 11d supergravity on AdS$_4 \times S^7$ even after quantization, we find no contribution from any zero modes. As explained above, we also find that the non-zero modes do not contribute. Moreover, one can show that there are no contributions from boundary terms. We are thus led to two possible conclusions: A) The 4d supergravity calculation of the logarithmic correction yields a vanishing result for $\log N$ in clear contradiction with the 11d analysis and with holography; or B) The regularization of the infinite sum over KK modes that contributes to the heat kernel coefficients can be performed in a way that gives a finite result consistent with 11d supergravity and holography. 

While option A) above may appear puzzling, it is not hard to understand how and why it may be possible. While 11d supergravity on AdS$_4 \times S^7$ should be equivalent to 4d $\mathcal{N}=8$ ${\rm SO}(8)$ gauged supergravity coupled to an infinite tower of KK modes at the classical level, there is no guarantee that the two theories produce the same quantum effect. Indeed in this explicit example we can trace the discrepancy to the 2-form ghost needed for the consistent quantization of the 11d 3-form field, which is not present in the 4d supergravity theory or in the full KK spectrum. If indeed option A) is to be taken seriously then it should be viewed as a cautionary tale for any application of lower-dimensional consistent supergravity truncations to holography, especially when studying quantum corrections to the leading supergravity approximation. 

Option B) on the other hand leads to more interesting conclusions. Assuming that the infinite sum over KK modes is regularized in a specific way that yields a non-vanishing heat kernel coefficient, we find that the non-zero modes contribute a factor of $\frac{\chi}{6}$ to the logarithmic correction, where $\chi$ is the Euler number of the four-dimensional background. If we further assume that, apart from the 2-forms mentioned above, there are no zero modes in the asymptotically locally Euclidean AdS$_4$ spaces relevant for holography, we find the following general result for the saddle point approximation to the gravitational path integral:
\begin{equation}
\label{eq:Igenchi}
	I = I_0 + \frac{\chi}{6} \log(L^2/G_N) + \ldots \, .
\end{equation}
Here $I_0$ is the regularized gravitational on-shell action (possibly including higher-derivative corrections) evaluated on the equations of motion, and the $\dots$ include higher order corrections in the semi-classical gravitational expansion. While the result in \eqref{eq:Igenchi} is not derived very rigorously, we show that it is in perfect agreement with results for the large $N$ path integral of the 3d $\mathcal{N}=6$ ABJM SCFT on various compact Euclidean three-manifolds. Moreover, if we assume that \eqref{eq:Igenchi} is valid, we arrive at the following prediction for the logarithmic correction to the entropy of a large AdS-Kerr-Newmann black hole embedded in AdS$_4\times S^7$
\begin{equation}
	S_{\text{AdS-KN}} = \frac{A_{\rm H}}{4 G_N} - \frac{1}{3} \log(A_{\rm H}/G_N) + \ldots \, .
\end{equation}
Here we work in an ensemble with fixed temperature and chemical potentials, the leading term is the standard Bekenstein-Hawking entropy of the black hole, and the $\dots$ are a placeholder for possible further quantum corrections to the black hole entropy. In the supersymmetric limit, our bulk result agrees with the logarithmic term in the large $N$ limit of the superconformal index of the dual SCFT. We consider this agreement to be a strong precision test of the microstate counting for 4d AdS black holes. 

To understand further the interplay between logarithmic corrections in holography, 11d supergravity, and 4d KK supergravity theories, we study four other top-down examples of AdS$_4$ vacua of M-theory with different amounts of supersymmetry and different topology of the internal manifold, and underscore several subtleties in the calculation of the logarithmic terms and the regularization of the infinite sums over KK modes. The conclusions we draw from these other examples are similar to the ones for AdS$_4 \times S^7$, i.e. one either should not use the 4d supergravity theory with the infinite tower of KK modes for the calculation of logarithmic corrections, or one should find a suitable regularization scheme for the infinite sums over KK modes in order to obtain one-loop results consistent with holography.

In addition to being of interest for top-down holographic models with explicit string/M-theory embeddings, our results have important implications for a more agnostic bottom-up approach to holography. Consider an effective theory of gravity coupled to a \textit{finite} number of matter fields of spin less than 2 which is valid up to an energy cutoff scale $\Lambda$ and admits an AdS$_4$ vacuum of scale $L$. Our results imply that the coefficient of the $\log L\Lambda$ term in the semi-classical expansion of the gravitational path integral depends in general on the continuous parameters of an asymptotically locally AdS$_4$ background. If we assume that the gravitational theory has a holographic description, this in turn implies that there exist $\log N$ or $\log \lambda$ terms in various physical observables, such as local correlation functions or the thermal effective action which captures the partition function of the theory on $S^1\times S^2$ in a saddle-point approximation. Note that here we are somewhat abstract and use the integer $N$ to denote the large number of degrees of freedom in the holographic CFTs, and $\lambda$ denotes some notion of a continuous parameter like an exactly marginal coupling. To the best of our knowledge such logarithmic terms do not appear in local correlation functions in any sequence of CFTs and they do not depend on continuous parameters in thermal effective actions. In this work we assume this to be true in general. Equipped with this assumption we are then led to conclude that general effective gravitational theories with a finite number of fields and an AdS$_4$ vacuum cannot be holographic. As we show in detail the only possible exception to this constraint, which somewhat poetically may be referred to as ``the unbearable lightness of the KK scale'' \cite{Kundera:1984}, arises from theories with a very finely tuned spectrum of quadratic fluctuations around the AdS$_4$ vacuum. In particular, we show that the pure 4d $\mathcal{N}=2$ and $\mathcal{N}=4$ gauged supergravity theories do not evade this constraint and are therefore not consistent UV-complete holographic gravitational theories. 

As a final application of our results we can also consider scale-separated vacua of string or M-theory, see \cite{Coudarchet:2023mfs} for a recent review and an extended bibliography. These are putative consistent AdS$_4$ vacua for which the length scale associated with the internal space is much smaller than the scale $L$ that sets the 4d cosmological constant.\footnote{A precise definition of this internal length scale is not crucial for our argument. The scale can be defined through the volume, diameter, or the spectrum of some differential operator on the internal space.} Clearly such a scale-separated AdS$_4$ vacuum defines an effective gravitational theory with finitely many fields valid up to the KK scale set by the internal manifold. If we assume that the scale-separated vacuum is associated with a consistent dual 3d CFT that does not have $\log N$ or $\log \lambda$ terms in local correlations functions, we arrive at very strong consistency constraints. In particular, we find that the mass spectrum of quadratic excitations around the AdS$_4$ vacuum should be such that three of the four heat kernel coefficients that contribute to the bulk logarithmic terms should vanish. These constraints are particularly strong for AdS$_4$ vacua preserving $\mathcal{N}=1$ or more supersymmetry, for which we show that most mass spectra with finitely many fields are inconsistent.

We continue in the next two sections with a summary Kaluza-Klein supergravity and known QFT results for the $\log N$ terms in the large $N$ expansion of partition functions of 3d holographic SCFTs on compact Euclidean manifolds. In Section~\ref{sec:sugra} we review the heat kernel method and apply it to the calculation of logarithmic corrections in the saddle-point approximation of gravitational path integrals in asymptotically AdS$_4$ backgrounds. In Section~\ref{sec:bootstrap} we show how one can use the results for the logarithmic corrections of SCFT partition functions from Section~\ref{sec:SCFT} in conjunction with holography to deduce strong constraints on the Seeley-de Witt coefficients in the heat kernel expansion. Section~\ref{sec:KK} is devoted to a discussion of some top-down examples in AdS$_4$/CFT$_3$ arising from M-theory for which we can compute the $\log N$ corrections by summing over the infinite tower of supergravity KK modes. In Section~\ref{sec:Kundera} we change gears and discuss the important constraints implied by our results for EFTs with finitely many fields coupled to gravity in AdS$_4$, as well as the implications of these constraints for scale-separated AdS$_4$ vacua of string and M-theory. We conclude in Section~\ref{sec:ccl} with a discussion of some open problems and directions for further study. In the five appendices we collect many of the technical details needed for our analysis.

\section{Supergravity and Kaluza-Klein compactification}

	Since we mainly use 4d gauged supergravity in this chapter, we will start with a brief introduction to supergravity in general, followed by explaining the relation between the 4d gauged supergravity with 11d supergravity as an example of the Kaluza-Klein compactification in supergravity.

\subsection{Supergravity and supersymmetry}
In the introduction of the thesis, we have introduced superconformal field theories which play an important role in the AdS/CFT correspondence. Here we will introduce the other part of the correspondence, namely the supergravity theories \cite{Freedman:2012zz, Lauria:2020rhc} which are the low-energy effective theories of string or M-theory. Different from string theory which is considered a unified theory including quantum gravity, the motivation of supergravity is to modify the general relativity with supersymmetry \cite{Golfand:1971iw, Wess:1974tw, Haag:1974qh}, where global supersymmetry is promoted to be local in order to respect the general covariance on a curved space. As the low-energy approximation of string and M-theory, supergravity becomes a highly valuable tool to study the semiclassical dynamics of quantum gravity. There have been developments in this field recently, such as supergravity localization \cite{Dabholkar:2010uh, Dabholkar:2011ec}, the exceptional field theory method \cite{Hohm:2013uia, Malek:2019eaz}, and Euclidean supergravity \cite{deWit:2017cle}. We only consider Euclidean signature in this chapter.

If we were to study top-down holography, we start from the UV complete theories, such as type II string theories or M theory. The corresponding low energy theories are type II supergravities and 11d supergravity, respectively. Studying the phase space of asymptotically AdS backgrounds in these supergravity theories is difficult because the dimension is too high. The first way out comes from supersymmetry. The infinitesimal version of supersymmetry transformation schematically looks like:
\begin{equation}
\delta_\epsilon F = B \epsilon,\quad \delta_\epsilon B = F \epsilon,
\end{equation}
where $B, F$ denote bosonic and fermionic fields, respectively, and $\epsilon$ is a spinor parametrizing the transformation. 
To preserve a certain number of supersymmetry, there must exist at least one $\epsilon$ such that $\delta_\epsilon F = \delta_\epsilon B = 0$ for all the matter fields. This requirement imposes non-trivial relations to the metric and gauge fields, called the ``BPS equations'', which are first order differential equations that are much easier to solve in comparison with the second-order equations of motion. In favorable situations, the BPS equations automatically solve the equations of motion.\footnote{However, it is not true true in general, since there is no reason. But see \cite{Gauntlett:2005ww} for a special case in type IIB supergravity where this is true. } \cite{Pope:2003jp, Gowdigere:2003jf,Nemeschansky:2004yh,Halmagyi:2012ic} This is helpful when looking for supersymmetric backgrounds in 11d or type II supergravities. 

Another method to look for 10d or 11d backgrounds is through compactification. The idea of compactification is that the background should be of a (warped) product form $M_d \times Y_{D-d}$, so all the $D$-dimensional fields can be expanded by products of external fields and internal harmonics. When the internal space $Y_{D-d}$ becomes somehow small, the mass spectrum of fields are dominated by the internal mass spectrum, which is organized into an infinite tower of discrete Kaluza-Klein levels. Thus in the limit where the $Y_{D-d}$ is very small, we can omit the higher KK modes and only look at the massless ones. This is the Kaluza-Klein reduction. However, it turns out to be surprisingly hard to find scale separated backgrounds in 10d/11d supergravity. This will also be the point discussed in section \ref{sec:Kundera}, where we show that our results can be a criterion for the existence of scale-separated background in a given supergravity theory.

Given the lack of scale seperated backgrounds, we need to include infinitely many Kaluza-Klein modes. Because of the conservation of degrees of freedom, the higher-dimensional fields with more Lorentz indices reduce to a large number of fields in the lower dimension. Despite the ramping number of fields, lower dimensional supergravity can still be easier to study compared their 10d/11d counterparts. 

AdS/CFT correspondence identifies the field theories to the full quantum gravity living in 10d/11d. At the classical level, a background ${\rm AdS}_4\times Y_7$ in 11d supergravity is equally described by a 4d gauged supergravity in ${\rm AdS}_4$ with the full tower of Kaluza-Klein modes associated to $Y_7$. In this chapter, we assume that the two descriptions are also equivalent with one-loop quantum corrections, and our results indeed show the precision holography with 4d gauged supergravity. This correspondence is shown by the red dotted arrow in Fig. \ref{fig:AdSCFTKK}. In what follows, we will take the KK reduction from 11d supergravity down to 4d to show how the compactification works.

\tikzstyle{G} = [rectangle, minimum width=3cm, minimum height=1cm, text centered, draw=black, fill=violet!10]
\tikzstyle{arrow} = [thick,->,>=stealth]

\tikzstyle{FT} = [rectangle, minimum width=3cm, minimum height=1cm, text centered, draw=black, fill=cyan!15]
\tikzstyle{arrow} = [thick,->,>=stealth]
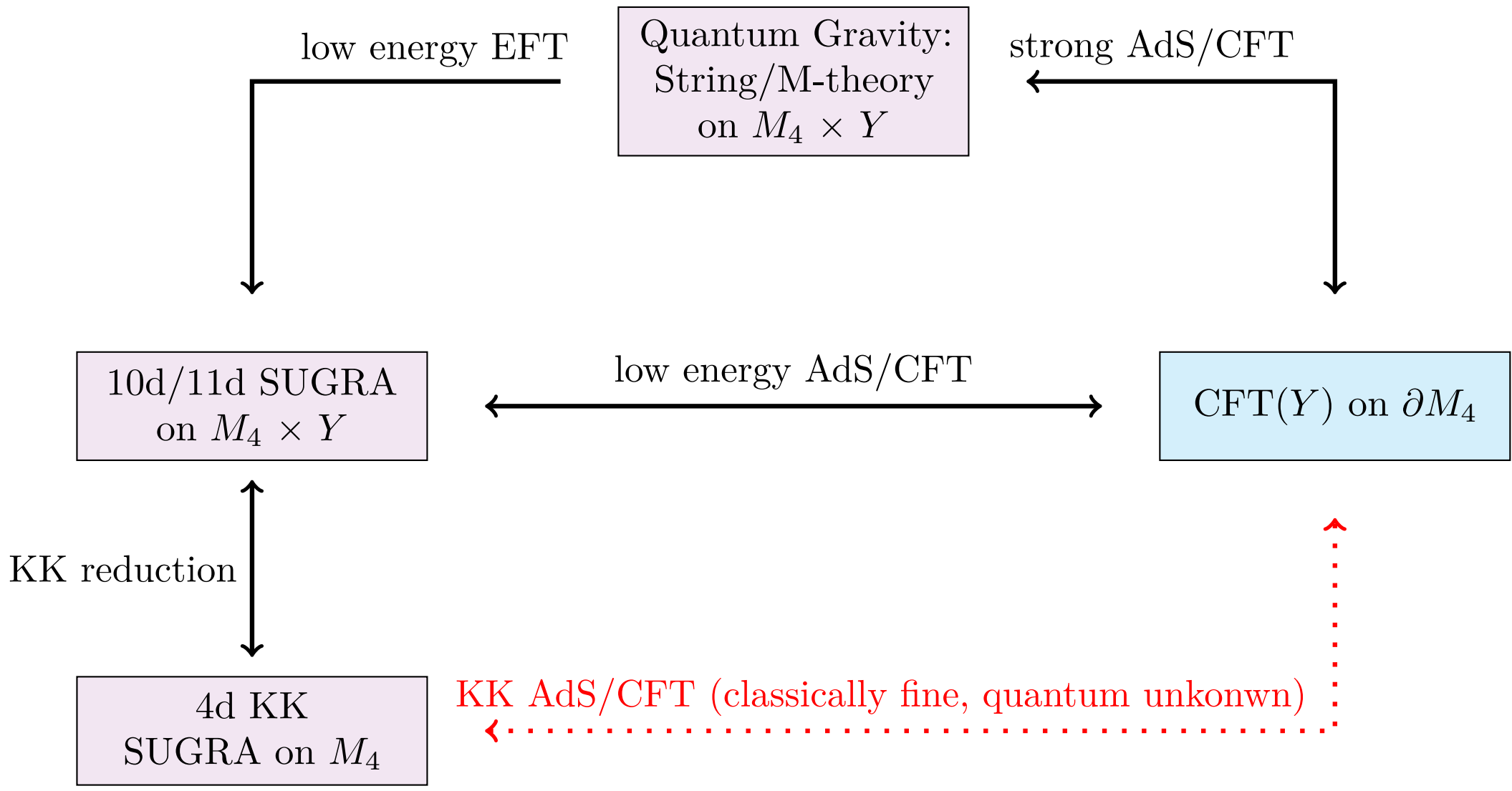
\begin{figure} [!h]
	\centering
	\begin{tikzpicture}[node distance=3cm]
	
	\node (QG) [G, text width=3cm,minimum height=1cm,minimum width=2cm] {Quantum Gravity: \\ String/M-theory on $M_4\times Y$};
	
	\node (CFT) [FT, text width=3cm,minimum height=1cm,minimum width=2cm, below of=QG, xshift=5cm] {CFT($Y$) on $\partial M_{4}$};
	
	\node (DdimSugra) [G, text width=3cm,minimum height=1cm,minimum width=1cm, below of=QG, xshift=-5cm] {10d/11d SUGRA on $M_4\times Y$};
	
	\node (AdS4) [G, text width=3cm,minimum height=1cm,minimum width=1cm, below of=DdimSugra, xshift=0cm] {4d KK SUGRA on $M_4$};
	
	\draw[-latex, very thick, shorten >=15pt,shorten <=15pt, ->]    (QG) -| node[near start,above] {low energy EFT} (DdimSugra);
	
	\draw[-latex, very thick, shorten >=15pt,shorten <=15pt, <->]    (QG) -| node[near start,above] {strong AdS/CFT} (CFT);

	\draw [-latex,very thick,shorten >=5pt,shorten <=5pt, <->] (DdimSugra) -- node[anchor=east] {KK reduction} (AdS4);
	
	
	\draw [-latex,very thick,shorten >=15pt,shorten <=15pt,  <->] (DdimSugra) -- node[anchor=south] {low energy AdS/CFT} (CFT);
	
	\draw[color=red, loosely dotted, -latex, very thick, shorten >=15pt,shorten <=15pt, <->]    (AdS4) -| node[near start,above] {KK AdS/CFT (classically fine, quantum unkonwn)} (CFT);
	\end{tikzpicture}
	\caption{\rm The picture shows different versions of the AdS/CFT correspondence with decreasing stringency. The pink boxes represent gravity theories, while the cyan box denotes the field theory. The dark thick arrows represent procedures that are understood or accepted, while the red dotted arrow is the less understood one. We assume the Kaluza-Klein theory with the full KK tower can be identified with the higher-dimensional counterpart. } \label{fig:AdSCFTKK}
\end{figure}

\subsection{Kaluza-Klein supergravity in 11d}

Among all the supergravity theories, one of them is very special, that is the 11d supergravity \cite{Cremmer:1978km}. Not only because 11 is the maximal number of dimensions permitted by supersymmetry \cite{Nahm:1977tg}, but the theory is uniquely determined, while supergravity theories in lower dimensions are highly constrained but not completely fixed. The field contents is very simple as well: a metric $g_{MN}$ and a three-form potential $A_{MNP}$. The 11d supergravity is regarded as the low energy theory of M theory \cite{Witten:1995ex}, since its reduction on $S^1$ gives the strong-coupling limit of the type IIA supergravity.

To make a connection between this beautiful theory with our real world, one may na\"ively think of putting the 11d supergravity on a background in a direct product\footnote{In fact, one should also consider a warped product background, which is necessary to reproduce some extra theories in 4d guaged supergravity. \cite{Warner:1983vz} } form of $M_4 \times Y_7$ and consider the effective theory on $M_4$, with the hope that $M_4$ gives an effective gravity theory describing the real world. This is roughly the idea of Kaluza-Klein compactification. \cite{Duff:1986hr}

To simplify the discussion further, we shall make an ansatz for the 4-form field strength that singles out the four-dimensional external space: \cite{Freund:1980xh}
\begin{equation} \label{Freund-Rubin}
F_{MNPQ} = \left\{ \begin{aligned}
& 3 m \epsilon_{\mu\nu\rho\sigma}, \quad {\rm along\ }M_4,\\
&0, \quad {\rm otherwise}.
\end{aligned} \right.
\end{equation}
Under this ansatz, the 11d Einstein equations become:
\begin{equation}
R_{\mu\nu} = -12m^2 g_{\mu\nu},\quad R_{mn} = 6m^2 g_{mn},
\end{equation}
where $\mu,\nu,\rho,\sigma$ are coordinates along $M_4$, and $m,n$ are coordinates along $Y_7$. This simple ansatz leads to the requirement that the internal manifold needs to be compact (since it has positive curvature) and Einstein. The external manifold will have negative curvature, and thus asymptotically AdS, happens to be applicable for the AdS/CFT correspondence.

On the product background, one can show that if the ``internal space'' $Y_7$ has an isometry group $G$, the massless states in the effective theory will contain a gauge field with the gauge group $G$. \cite{Salam:1981xd} The maximally symmetric choice of $M_7$ is the seven-sphere with SO(8) isometry, which should correspond to a 4d gauged supergravity with gauge group SO(8). What's more, the ${\rm AdS}_4 \times S^7$ background in 11d supergravity was explicitly found to contain 8 Killing spinors $\epsilon$ \cite{Duff:1983gq}, which will be preserved if we ``spontaneously compactify'' to AdS$_4$, resulting in the 4d $\cN = 8$ SO(8) gauged supergravity \cite{deWit:1982bul}: the Kaluza-Klein modes that descend from the 11d supergravity organize themselves into a tower of 4d $\cN = 8$ supergravity multiplets. 

We can also consider less symmetric internal spaces. The coset spaces can be represented as the group manifold $G/H$ where $H$ is a subgroup of $G$, and they are automatically homogeneous by construction. For example, the seven sphere can be represented as SO(8)/SO(7). The classification of 7d Einstein coset space is done in \cite{Castellani:1983tb}, many of them appear to have Killing spinors and thus correspond to 4d gauged Kaluza-Klein supergravities. These 4d gauged supergravities holographically correspond to 3d superconformal field theories, through the chain of relations in Fig. \ref{fig:AdSCFTKK}. Many of the 3d field theories are identified as supersymmetric gauge theories amenable to supersymmetric localization, as partially listed in Table \ref{tab:CFT-log}. The field theory results motivate us to study the corresponding Kaluza-Klein gauged supergravity obtained by compactifying the 11d backgrounds found in \cite{Castellani:1983tb}.

We can relax the Freund-Rubin ansatz in \eqref{Freund-Rubin} and consider non-vanishing fluxes in the internal space, where things start being complicated. As we discussed above, the straightforward method starts from 11d supergravity, taking advantage of the global symmetry and supersymmetry to get the first-order BPS equations. It may work well when the symmetry is large, for example, in \cite{Englert:1982vs, deWit:1984nz, Corrado:2001nv, Gabella:2012rc}. Another perspective is to look at the 4d $\cN = 8$ SO(8) gauged supergravity, which has 70 scalar fields with a complicated scalar potential. The potential has extrema parametrized by the non-zero VEVs of the scalar fields, they correspond to IR fixed points of RG flows sourced by deformations on the original theory. At such a fixed point, the fields descending from the original OSp(8$|$4) multiplet in the UV reorganize themselves into new multiplets with reduced supersymmetry. Because the 4d $\cN=8$ SO(8) gauged supergravity is a consistent truncation \cite{deWit:1986oxb, Nicolai:2011cy, Varela:2015ywx}, all the 4d theories corresponding to extrema of the scalar potential can be uplifted to 11d.\footnote{This is not true for some other 4d gauged supergravities, for example, the dyonic gauged maximal supergravity \cite{DallAgata:2012mfj} doesn't have an 11d uplift. It is also not true that all 11d supergravity backgrounds can be compactified to 4d gauged supergravity, such as the GMPS solution \cite{Gabella:2012rc, Halmagyi:2012ic}. } Besides, the exceptional field theory method can also be helpful looking for an 11d background when a large global symmetry is preserved, which takes advantage of the full symmetry group of the maximal (or half-maximal) gauged supergravity. \cite{Duboeuf:2022mam}

We will be interested in the quantum aspects of the 4d Kaluza-Klein gauged supergravities. Under the canonical quantization, all the Kaluza-Klein modes will run in loops, thus we need the mass spectrum data for the full tower. In the literature, this can be obtained by doing harmonic analysis of the various differential operators. Nowadays a new method based on the hidden exceptional symmetry \cite{Malek:2019eaz} is more powerful for less symmetric cases, including non-supersymmetric theories. We will study 4d gauged supergravity theories that have uplifts in massive type IIA, type IIB, and 11d supergravity with known holographic duals. Interesting results from the 1-loop quantum effects of the Kaluza-Klein modes will be presented.

\section{Large $N$ partition functions of 3d holographic SCFTs}
\label{sec:SCFT}

Let us begin by briefly reviewing known results about the logarithmic terms in the large $N$ partition functions of 3d holographic SCFTs with $\mathcal{N}\geq2$ supersymmetry placed on compact Euclidean manifolds. Using supersymmetric localization, partition functions of 3d $\mathcal{N}=2$ SCFTs on various compact manifolds can be reduced to matrix models. Of particular interest to us are the squashed\footnote{We will be interested in a particular squashing of $S^3$ \cite{Hama:2011ea} that preserves a U(1)$\times$U(1) isometry and is parametrized by a real positive number $b$, with $b=1$ corresponding to the round sphere. We denote this manifold by $S^3_b$.}  3-sphere partition function \cite{Kapustin:2009kz,Hama:2011ea,Imamura:2011wg}, the Topologically Twisted Index (TTI) obtained by putting the theory on $S^1 \times \Sigma_\mathfrak{g}$ with a supersymmetry-preserving twist along the Riemann surface \cite{Benini:2015noa,Benini:2016hjo,Closset:2016arn,Closset:2017zgf}, and the Superconformal Index (SCI) where the 3d SCFT is put on $S^1 \times_\omega S^2$ after turning on a chemical potential $\omega$ for the angular momentum on the $S^2$ \cite{Bhattacharya:2008zy,Bhattacharya:2008bja,Kim:2009wb,Imamura:2011su}. In what follows, we first introduce the three BPS observables and then discuss their universal behaviors under the large $N$ expansion.

	\subsection{BPS observables in 3d $\cN = 2$ gauge theories }
	
	The classification of three-dimensional BPS manifolds preserving 3d $\cN = 2$ supersymmetry has been done \cite{Closset:2019hyt}. The three geometries we will study, i.e., squashed three-sphere $S_b^3$, $S^1 \times \Sigma_{\mathfrak{g}}$, and $S^1 \times_\omega S^2$, are special cases of the 3d BPS geometries. Application of supersymmetric localization on 3d $\cN=2$ SCFT leads to the precise matrix-integral formulas for the path integral on them. In the last two cases, the partition function evaluates the topologically twisted index (TTI) and the superconformal index (SCI), which are related to the number of black-hole microstates in their AdS duals. \cite{Zaffaroni:2019dhb}
	
	The squashed sphere $S_b^3$ preserves U(1)$\times$U(1) isometry, whose metric can be written as
	\begin{equation}
	ds^2 = b^2 (dx_1^2 + dx_2^2) + 1/b^2 (dx_3^2+dx_4^2),\quad x_1^2+x_2^2+x_3^2+x_4^2=1.
	\end{equation}
	We are interested in 3d superconformal theories with $\cN = 2$ supersymmetry, where the partition function localizes to, schematically: \cite{Hama:2011ea, Imamura:2011wg}
	\begin{equation}
	\begin{aligned}
	Z(R) &= \int \prod_{\rm Cartan} du\, e^{i\pi {\rm tr}\, u^2} \det_{\rm Ad}\left[ \sinh(b \pi u)\sinh(b^{-1} \pi u) \right] \\
	& \times \prod_{{
			\rm chirals\ in\ rep\ }R_i} \det_{R_i} \left[ s_b(\frac{iQ}{2}(1-\Delta_i) + iu) \right],
	\end{aligned}
	\end{equation}
	where $Q = b+1/b$, $\Delta_i$ are the $R$-charges of the chiral multiplets under the $R$-symmetry, and $s_b$ is the double-sine function defined by:
	\begin{equation}
	s_b(x) \equiv \prod_{m,n\ge 0} \frac{mb+nb^{-1} + \frac{Q}{2} - ix}{mb+nb^{-1}+\frac{Q}{2}+ix}.
	\end{equation}
	When the sphere is not squashed, the expressions are dramatically simplified: \cite{Kapustin:2009kz}
	\begin{equation}
	Z(R) = \int \prod_{\rm Cartan} du\, e^{i\pi {\rm tr}\, u^2} \det_{\rm Ad}\left[ \sinh^2(\pi u) \right] \prod_{{
			\rm chirals\ in\ rep\ }R_i} \det_{R_i} \left[ e^{\ell(1-\Delta_i +iu)} \right],
	\end{equation}
	where $\ell(z)$ is a special function:
	\begin{equation}
	\ell(z) \equiv -z\log\left( 1-e^{2\pi i z} \right) + \frac{i}{2} \left( \pi z^2 + \frac{1}{\pi} {\rm Li}_2\left( e^{2\pi i z} \right) \right) - \frac{i\pi}{12}.
	\end{equation}
	For a 3d SCFT with $\cN \ge 3$, the $R$-symmetry SO($\cN$) is non-Abelian, which protects the $R$-charges when going from the UV to the IR fixed point. But for $\cN = 2$ case, the $R$-charge gets quantum corrections and mixes with Abelian non-$R$ flavor symmetries. It was shown that the choice of $R$-charge that maximizes the partition function is the correct superconformal $R$-charge in the infrared \cite{Jafferis:2010un}. This is reminiscent of the $a$-maximization in 4d. \cite{Intriligator:2003jj}
	
	Apart from squashed three-spheres, on some specific backgrounds, the partition functions happen to reproduce the supersymmetric Witten index \cite{Witten:1982df}. The motivation of the superconformal index is to study the Hilbert space of superconformal field theories. We expect the Hilbert space to be a direct sum over irreducible unitary representations of the superconformal algebra, which has been classified in general dimensions. \cite{Dobrev:1985qv, Dobrev:1985vh, Dobrev:1985qz, Minwalla:1997ka, Dobrev:2002dt, Cordova:2016emh} One special set of representations are called ``short'' because they have fewer states compared to the general multiplets, which are called ``long''. Within a continuous family of superconformal field theories, the continuity of the spectrum prevents the short multiplets from becoming long unless a group of short multiplets coincidentally combine into a long one. \cite{Bhattacharya:2008zy} The superconformal index can be defined as special linear combinations of the numbers of the multiplets which are invariant under the multiplet combinations discussed above and are thus a function of the spectrum that remains constant under continuous variation of the spectrum. 
	
	One typical choice of the index is prescribed by Witten \cite{Witten:1982df}, which counts the states annihilated by a particular supercharge $\cQ$ and its conjugate $\bar{\cQ}$. The Witten index is defined as
	\begin{equation}
	\cI = {\rm tr}\, (-1)^F e^{-\beta \{ \cQ, \bar{\cQ}\}} e^{\mu_i F_i} ,
	\end{equation}
	where $F$ is the fermion number operator and $F_i$ are flavor symmetries commuting with $ \cQ$ and $ \bar{\cQ} $. The Witten index receives only contributions from states that are both annihilated by $\cQ$ and $\bar{\cQ}$, and thus by $\{\cQ, \bar{\cQ}\}$, so it is independent of $\beta$ as well as other continuous deformations. The Witten index is the most general index since it can reproduce all the indices that are continuous spectral functions. 
	
	Using supersymmetric localization, the superconformal index of an $\cN \ge 2$ superconformal field theory on $S^1 \times_\omega S^2$ can be reduced to an explicit finite-dimensional matrix integral as well, which involves both the integration over the Cartan generators of the gauge group parametrizing the BPS locus as well as an infinite sum over all the possible magnetic monopole fluxes over the two-sphere valued under the GNO lattice \cite{Goddard:1976qe} of quantized magnetic fluxes. \cite{Bhattacharya:2008bja, Bhattacharya:2008zy, Kim:2009wb, Imamura:2011su}
	
	Preserving supersymmetry on a curved manifold is a non-trivial task. The standard protocol involves coupling the field theory to a supergravity theory and fixing the gravity and gauge flux background. \cite{Festuccia:2011ws} One special scenario is the partial topological twist proposed by Witten \cite{Witten:1988ze, Witten:1991zz}.  The BPS equation schematically reads:
	\begin{equation}
	\nabla_\mu \epsilon = \partial_\mu \epsilon + \omega_\mu^{ab} \Gamma_{ab} \epsilon + A_\mu^R \epsilon.
	\end{equation}
	By specifying the background gauge field $A_\mu^R$ coupled to the $U(1)_R$-symmetry such that the last two terms above cancel, a constant $\epsilon$ automatically solves the BPS equation and thus supersymmetry is preserved. Since the gauge field is along the Riemann surface which is a part of the total space $\Sigma_{\mathfrak{g}}\times S^1$, this operation is called ``partial topological twist'' and its corresponding Witten index is called the topologically twisted index (TTI), which can be expressed explicitly as a finite-dimensional integral using supersymmetric localization. \cite{Benini:2015noa,Closset:2016arn,Benini:2016hjo} The BPS locus is also paramtrized by the GNO lattice and the Cartan algebra of the gauge group. The difference between the twisted index and the superconformal index is whether the gauge field $A_\mu^R$ is turned on or not to preserve the supersymmetry.
	
	Supersymmetric localization makes the exact evaluation for these BPS observables at finite $N$, couplings, and charges possible, which is meaningful in the comparison with gravity in the context of AdS/CFT. In this chapter, we will take advantage of the results in the literature on localizing 3d SCFTs on the three backgrounds, which are summarized in Table \ref{tab:CFT-log}. 

	\subsection{The large $N$ behavior}

Under favorable circumstances related to special values of the 3d background parameters, the matrix model for the $S^3_b$ partition function can be computed in closed form to all orders in the large $N$ expansion \cite{Fuji:2011km,Marino:2011eh,Mezei:2013gqa,Nosaka:2015bhf,Hatsuda:2016uqa}. The matrix models for the other supersymmetric partition functions have not yet been solved analytically, but closed-form expressions were conjectured for various $\mathcal{N}\geq2$ SCFTs based on non-trivial consistency checks with the bulk theory and numerical studies \cite{Bobev:2022jte,Bobev:2022eus,Bobev:2022wem,Bobev:2023lkx}. In all known cases, we have the large $N$ behavior
\begin{equation}
\label{logZ}
	\log Z_{\text{CFT}} = F_0 + \mathcal{C} \log N + \mathcal{O}(N^0) \, ,
\end{equation}
where $F_0$ denotes all terms that dominate over $\log N$ in the large $N$ expansion. In general, $F_0$ is a polynomial in some (possibly fractional) power of $N$ depending on the SCFT of interest, whose coefficients are functions of the parameters of the theory such as the discrete Chern-Simons (CS) level or continuous parameters like squashings, fugacities for flavor symmetries or real masses. In stark contrast, the logarithmic correction in \eqref{logZ} is \emph{universal} and independent of such parameters. We summarize a number of known results for the logarithmic coefficient $\mathcal{C}$ in Table~\ref{tab:CFT-log}, after giving some general explanations of the 3d holographic SCFTs and their observables.\\
\afterpage{
\begin{table}[H]
\centering \small 
\renewcommand*{\arraystretch}{1.3}
\begin{tabular}{|c|c|c|c|c|}
\hline
Theory & $\mathcal{M}_3$ & log coefficient $\mathcal{C}$ & Ref. & 10/11d bulk \\
\hline\hline
\multicolumn{5}{|c|}{M2-brane theories (class I)} \\
\hline
%
%
\multirow{3}{*}{$(S^7/\mathbb{Z}_k)_{\text{free}}$ $(\dagger)$} & $S^3_b$ & $-\frac14$ & \cite{Fuji:2011km,Marino:2011eh,Nosaka:2015iiw,Bobev:2022eus} & \cmark\,\cite{Bhattacharyya:2012ye} (s.c.) \\
\cline{2-5}
& $S^1\times \Sigma_\mathfrak{g}$ & $-\frac12(1 - \mathfrak{g})$ & \cite{Liu:2017vll,Bobev:2022eus} & \cmark\,\cite{Liu:2017vbl} (s.c.) \\
\cline{2-5}
& $S^1\times_\omega S^2$ & $-\frac12$ & \cite{Bobev:2022wem} & \xmark \\
\hline
%
%
\multirow{3}{*}{$(S^7/\mathbb{Z}_{N_f})_{\text{f.p.}}$ $(\dagger)$} & $S^3_b$ & $-\frac14$ & \cite{Mezei:2013gqa,Hatsuda:2016uqa,Minahan:2021pfv,Bobev:2023lkx} & \multirow{3}{*}{\xmark} \\
\cline{2-4}
& $S^1\times \Sigma_\mathfrak{g}$ & $-\frac12(1 - \mathfrak{g})$ & \cite{Bobev:2023lkx} & \\
\cline{2-4}
& $S^1\times_\omega S^2$ & $-\frac12$ & \cite{Bobev:2022wem} & \\
\hline
%
%
\multirow{3}{*}{$N^{010}/\mathbb{Z}_k$} & $S^3_{b=1}$ & $-\frac14$ & \cite{Marino:2011eh,Bobev:2023lkx} & \xmark \\
\cline{2-5}
& $S^1\times \Sigma_\mathfrak{g}$ & $-\frac12(1 - \mathfrak{g})$ & \cite{Bobev:2023lkx} & \cmark\,\cite{PandoZayas:2020iqr} \\
\cline{2-5}
& $S^1\times_\omega S^2$ & $-\frac12$ & \cite{SCI-toappear} & \xmark \\
\hline
%
%
\multirow{2}{*}{$V^{52}/\mathbb{Z}_k$} & $S^1\times \Sigma_\mathfrak{g}$ & $-\frac12(1 - \mathfrak{g})$ & \cite{Bobev:2023lkx} & \cmark\,\cite{PandoZayas:2020iqr} \\
\cline{2-5}
& $S^1\times_\omega S^2$ & $-\frac12$ & \cite{SCI-toappear} & \xmark \\
\hline
%
%
\multirow{2}{*}{$Q^{111}/\mathbb{Z}_k$} & $S^1\times \Sigma_\mathfrak{g}$ & $-\frac12(1 - \mathfrak{g})$ & \cite{Bobev:2023lkx} & \cmark\,\cite{PandoZayas:2020iqr} \\
\cline{2-5}
& $S^1\times_\omega S^2$ & $-\frac12$ & \cite{SCI-toappear} & \xmark \\
\hline\hline
\multicolumn{5}{|c|}{M5-brane theories (class II)} \\
\hline
%
%
\multirow{3}{*}{$A_{N-1}$} & $S^3_b$ & $-\frac12$ & \cite{Gang:2014ema,Bobev:2020zov} & \xmark \\
\cline{2-5}
& $S^1 \times \Sigma_{\mathfrak{g}>1}$ & $(b_1(\mathcal{H}_3)-1)(1-\mathfrak{g})$ & \cite{Gang:2019uay,Benini:2019dyp} & \multirow{2}{*}{\cmark\,\cite{Benini:2019dyp}} \\
\cline{2-4}
& $S^1 \times_\omega S^2$ & $b_1(\mathcal{H}_3) - 1$ & \cite{Benini:2019dyp} & \\
\hline
%
%
\multirow{3}{*}{$D_{N}$} & $S^3_b$ & \multirow{3}{*}{$0$} & \multirow{3}{*}{\cite{Bobev:2020zov}} & \multirow{3}{*}{\xmark} \\
\cline{2-2}
& $S^1 \times \Sigma_{\mathfrak{g}>1}$ & & & \\
\cline{2-2}
& $S^1 \times_\omega S^2$ & & & \\
\hline\hline
\multicolumn{5}{|c|}{IIA theories (class III)} \\
\hline
%
%
\multirow{2}{*}{$\mathbb{CP}^3$ ($\dagger$)} & $S^3_b$ & $-\frac16$ & \cite{Marino:2009dp,Bobev:2022eus} & \multirow{2}{*}{\xmark} \\
\cline{2-4}
& $S^1\times \Sigma_{\mathfrak{g}}$ & $\frac23(1 - \mathfrak{g})$ & \cite{PandoZayas:2019hdb,Bobev:2022eus} & \\
\hline
%
%
$\mathbb{CP}_{\text{def}}^3$ & $S^3_{b=1}$ & $-\frac16$ & \cite{Hong:2021bsb} & \xmark \\
\hline
%
%
\multirow{3}{*}{$S^6_{\text{def}}$} & \multirow{2}{*}{$S^3_{b=1}$} & $-\frac29$ (fixed $k$) & \multirow{2}{*}{\cite{Liu:2019tuk}} & \multirow{3}{*}{\xmark} \\
\cline{3-3}
& & $-\frac16$ (`t Hooft) & & \\
\cline{2-4}
& $S^1\times \Sigma_{\mathfrak{g}}$ & $-\frac7{18}(1 - \mathfrak{g})$ & \cite{Liu:2018bac} & \\
\hline
\end{tabular}
\caption{The logarithmic coefficient in \eqref{logZ} for various 3d $\mathcal{N}\geq 2$ SCFTs. We indicate the compact manifold $\mathcal{M}_3$ on which the theory is placed and the possibility of mass deformations (${}\dagger$), along with relevant references where analytic and/or numerical computations are presented. We refer to the main text for detailed explanations and implicit restrictions of each entry. The last column indicates whether the result has been matched from a one-loop computation in 10d or 11d supergravity with the abbreviation s.c. used to indicate that the supergravity analysis is performed at the superconformal vacuum, i.e. for vanishing squashing and mass deformations.\label{tab:CFT-log}}
\end{table}
\clearpage}

The three broad classes of 3d $\mathcal{N}\geq 2$ SCFTs we consider arise from the worldvolume of M2- and M5-branes in M-theory, and from D2-branes in massive Type IIA string theory. SCFTs in the first class will be distinguished by the Sasaki-Einstein base SE$_7$ of the cone probed by a stack of $N$ M2-branes. Familiar examples in this class are the $\mathcal{N}=6$ ABJM theory~\cite{Aharony:2008ug}, for which SE$_7$ is a freely acting orbifold of the 7-sphere $(S^7/\mathbb{Z}_k)_{\text{free}}$, or the $\mathcal{N}=4$ ADHM theory or ``$N_f$ model''~\cite{Porrati:1996xi,Mezei:2013gqa,Grassi:2014vwa} where the base of the cone is a 7-sphere orbifold $(S^7/\mathbb{Z}_{N_f})_{\text{f.p.}}$ with fixed points. 

SCFTs in the second class are constructed by starting with the $\mathcal{N}=(2,0)$ theory on the worldvolume of $N$ M5-branes and placing it on a compact hyperbolic 3-fold $\mathcal{H}_3$, including a partial twist to preserve $\mathcal{N}=2$ supersymmetry in the three non-compact directions. Taking the limit vol$(\mathcal{H}_3) \rightarrow 0$ produces 3d holographic theories of class $\mathcal{R}$~\cite{Dimofte:2011ju}. In Table~\ref{tab:CFT-log}, we distinguish them by the ADE-type of the ``gauge group'' $G$ in the parent 6d theory. 

The third class of theories consists of 3d holographic theories whose gravity duals uplift to solutions of (massive) Type IIA string theory. They will be distinguished by the internal six-dimensional space. This class includes a particular limit of the ABJM theory where one takes both $N$ and $k$ large while keeping $N/k$ fixed. This implements a dimensional reduction along the Hopf fiber over $\mathbb{CP}^3$ inside $S^7/\mathbb{Z}_k$ and gives access to a regime of the theory dual to massless Type IIA string theory on an asymptotically AdS$_4 \times \mathbb{CP}^3$ background~\cite{Aharony:2008ug}. In addition, this class of theories includes the Gaiotto-Tomasiello (GT) theory~\cite{Gaiotto:2009mv} and the SCFT dual to 4d ISO(7)-gauged maximal supergravity~\cite{Guarino:2015jca} also belong in this class of theories, which uplift to massive Type IIA (mIIA) on a deformed $\mathbb{CP}^3$ and a deformed $S^6$, respectively. In these mIIA SCFTs, the CS level $k$ is controlled by the Romans mass and one can study either the regime of large $N$ and fixed $k$, or the limit of large $N$ and large $k$ with $N/k$ fixed. We call the former the fixed $k$ limit and the latter the `t Hooft limit in Table~\ref{tab:CFT-log}.\\

Once the theory is specified, it can be placed on a compact Euclidean manifold $\mathcal{M}_3$ and the corresponding supersymmetric partition functions can be computed using localization. The resulting matrix models can be studied at large $N$ and, in some cases, the logarithmic term in \eqref{logZ} can be extracted. For M2-brane theories, the squashed 3-sphere partition function ($\mathcal{M}_3 = S^3_b$), the TTI ($\mathcal{M}_3 = S^1 \times \Sigma_\mathfrak{g}$), and the SCI ($\mathcal{M}_3 = S^1 \times_\omega S^2$) have all been studied along these lines. In Table~\ref{tab:CFT-log} we collect the resulting logarithmic coefficients and give the relevant references. We note here that the coefficient of the $\log N$ term in the SCI has only been obtained in the Cardy-like limit where $\omega \rightarrow 0$, so we restrict our summary to this regime.\footnote{Since we expect the logarithmic correction to always be universal, it is likely that higher-order terms in $\omega$ will not change the results. Making this argument precise is however beyond the scope of this work.} Class $\mathcal{R}$ theories can be placed on the same three-manifolds and the logarithmic term in the large $N$ limit can be extracted using the 3d-3d correspondence~\cite{Dimofte:2014ija}. In the squashed sphere case, M5-brane results are available only in the limit $b \rightarrow 0$ and for hyperbolic three-folds with trivial $H^1(\mathcal{H}_3,\mathbb{Z})$. For the SCI, the results have again only been obtained in the Cardy-like limit. These restrictions are implicit in the relevant entries of Table~\ref{tab:CFT-log}. Finally, the logarithmic term in the round 3-sphere partition function of the GT theory has only been obtained for the two-node case, and we again implicitly restrict to this case in our summary.

SCFTs arising from M2-branes can also be deformed away from the superconformal point. For example, the ABJM theory has an SO(4)$\times$U(1) flavor symmetry and one can turn on three $\mathcal{N}=2$ supersymmetry-preserving real masses in the Cartan subgroup of this global symmetry group. The observables introduced above can be studied using localization in the deformed theory, and the logarithmic terms can again be extracted. Theories in which such deformations have been studied are denoted with a $(\dagger)$ in Table~\ref{tab:CFT-log}. We stress that the logarithmic coefficient turns out to be independent of the deformations in all known cases, in line with its universal character. To the best of our knowledge, such mass deformations have not been studied for SCFTs in the second and third class.\\

In Table~\ref{tab:CFT-log}, we also contrast the deluge of SCFT results with the drought of dual supergravity computations. As far as we are aware, there are only few bulk results available in the literature, and they have all been obtained in the eleven-dimensional low-energy effective description of M-theory. There, as mentioned in the Introduction, a one-loop calculation can be performed using the heat kernel method wherein the problem of computing the logarithmic correction reduces to a zero mode counting problem, see e.g.~\cite{Bhattacharyya:2012ye}. While this technique can be used to match with CFT results, it is limited to some simple 11d backgrounds where the spectrum of kinetic operators can be worked out in full. Typically, this requires being at the superconformal point where possible mass or squashing deformations are turned off. To emphasize this point, we include the abbreviation ``s.c.'' in the relevant entries of Table~\ref{tab:CFT-log}. A complementary, although broader, approach to logarithmic corrections in the bulk was put forward in~\cite{Hristov:2021zai} where it was shown how supergravity localization can efficiently compute the coefficient $\mathcal{C}$ in the 4d effective supergravity theory using index theorems. We will discuss this method in more details below.

The rest of the chapter will be devoted to studying logarithmic corrections in the lower-dimensional 4d supergravity theories, with an eye towards replacing some of the ``\xmark'' entries in Table~\ref{tab:CFT-log} with bulk results matching the SCFT predictions. To this end, we set up the 4d Euclidean supergravity path integral and the appropriate heat kernel expansion in the next section.

\section{Logarithms in the 4d Euclidean path integral}
\label{sec:sugra}

The SCFTs we consider in Section \ref{sec:SCFT} are holographically dual in the large $N$ limit to ten- or eleven-dimensional supergravity theories on backgrounds that are asymptotically of the form AdS$_4 \times X$, where $X$ is a 6d or 7d internal manifold. We denote the (common) length scale of these spaces by $L$.\footnote{Here we focus on situations arising from standard top-down AdS/CFT examples in string and M-theory. We will discuss scale-separated AdS$_4$ vacua and more general 4d supergravity theories in Section~\ref{sec:Kundera}.} The KK reduction of the 10d/11d supergravity theory on $X$ produces an infinite tower of massive 4d fields in addition to the massless ones. The Euclidean path integral for the 4d KK supergravity theory involving all these 4d fields is then holographically dual to the 3d SCFT partition function \eqref{logZ}, 
\begin{equation}
	Z_\text{sugra}\approx Z_\text{CFT} \, ,
\end{equation}
where the approximate equality reminds us that we work in the large $N$ limit and are ignoring heavy string and brane states. In the holographic context, specifying a 3d SCFT amounts to choosing a particular KK supergravity theory and the corresponding field content in the bulk. Once the theory is specified, one can study various supersymmetric observables and their bulk incarnations. In particular, the partition function of a given 3d SCFT on a compact Euclidean manifold $\mathcal{M}_3$ is captured by the corresponding KK supergravity Euclidean path integral around a 4d supergravity background whose conformal boundary is $\mathcal{M}_3$. 

In the semi-classical limit where the Newton constant is small in units of the AdS radius, i.e.~$L^2/G_N \gg 1$, the 4d Euclidean path integral can be evaluated in the saddle-point approximation. This yields
\begin{equation}
\label{eq:sugra-saddle}
	\log Z_{\text{sugra}} = - \frac{1}{16\pi G_N}\,S_{\text{cl}}[\mathring{\phi}] + C\log(L/\sqrt{G_N}) + \mO(1)\, ,
\end{equation}
where $\mathring{\phi}$ generically denotes the on-shell values of all fields $\phi$ in the gravity theory for a given background. To leading order (LO) at large $L^2/G_N$, the saddle-point approximation of the supergravity path integral is controlled by the two-derivative on-shell action of the given 4d background. The next-to-leading order (NLO) term in the semi-classical expansion is a four-derivative correction to the action evaluated on the two-derivative solution. Both LO and NLO terms are grouped in the $S_{\text{cl}}[\mathring{\phi}]/G_N$ term in \eqref{eq:sugra-saddle}. This quantity has been computed and shown to match with $F_0$ in \eqref{logZ} in numerous examples, see~\cite{Bobev:2021oku,Bobev:2020zov,Bobev:2022eus,Bobev:2023lkx}, in line with expectations from AdS/CFT away from the strict semi-classical limit.

We now would like to focus on the logarithmic term and ask if the values of the logarithmic coefficient $\mathcal{C}$ in the large $N$ expansion of a given 3d SCFT partition function match the logarithmic coefficient $C$ in the saddle-point approximation \eqref{eq:sugra-saddle}. In this comparison, we will ultimately have to use the AdS$_4$/CFT$_3$ dictionary appropriate for each class of SCFTs introduced in Section~\ref{sec:SCFT} (as explained there, we split class III according to whether the bulk solution admits a IIA or massive IIA uplift):
\begin{equation}
	\fft{L^2}{G_N} \sim N^{\fft32} \;\; (\text{class I}) \, , \qquad \fft{L^2}{G_N} \sim N^{3} \;\; (\text{class II}) \, , \qquad \fft{L^2}{G_N} \sim \begin{cases} N^2 \;\; &(\text{IIA}) \\ N^{\fft53} \;\; &(\text{mIIA}) \end{cases} \, . \label{eq:holo-dict-gen}
\end{equation}
Since we are concerned with a logarithmic term we do not have to worry about the precise numerical factors or possible finite $N$ corrections to the above relations, as these would only contribute $\mathcal{O}(1)$ terms in \eqref{eq:sugra-saddle}. We will use heat kernel techniques to compute $C$. In Section \ref{sec:sugra:log} we first review how the logarithmic coefficient splits into a local and a non-local contribution based on the heat kernel expansion. In Section~\ref{sec:sugra:local} and Section~\ref{sec:sugra:non-local}, we study these contributions in detail.

\subsection{Logarithmic contributions to the Euclidean path integral}
\label{sec:sugra:log}

In this subsection, we review the general structure of the logarithmic coefficient $C$ in the Euclidean path integral for any 4d KK supergravity theory. Note that all supergravity fields with masses below the UV cutoff can run in loops in the Euclidean path integral and thereby contribute a logarithmic term in the bulk partition function. Up to $L$-independent terms, the resulting logarithmic correction generically takes the form \cite{Sen:2012kpz,Sen:2012cj,Sen:2012dw,Bhattacharyya:2012ye}
\begin{equation}
\label{eq:log-sugra}
	C\log(L/\sqrt{G_N}) = \sum_{\phi}\Bigg[-\frac{(-1)^F}{2}\log\text{det}'\mathcal{Q}_\phi + \log\int\mathcal{D}\dphi_0\Bigg] \, ,
\end{equation}
where $\mathcal{Q}_\phi$ is a second-order differential operator that captures the dynamics of a quantum fluctuation $\dphi = \phi - \mathring{\phi}$ around a given background, and the sum is taken over all 4d fields $\phi$ in the KK theory weighted by their fermion number $F$, including ghosts.\footnote{Quantization of a $p$-form requires a tower of $(p-j)$-form ghost fields with $j=1,\ldots, p$ labeling the ghost level~\cite{Siegel:1980jj}. For the purpose of computing the determinant factor, the kinetic terms of these ghost fields can be thought of as differential operators of order $2(j+1)$ and therefore the first term in \eqref{eq:log-sugra} has to be multiplied by a factor of $j+1$, see e.g. \cite{Bhattacharyya:2012ye}. This factor can be understood as arising from ghost number conservation, and the same remark applies for any field with gauge invariance. To  declutter formulas, we will mostly keep the ghost level factor implicit when summing over the spectrum of the theory.\label{foot:ghosts}} The prime on the determinant indicates that it should be computed after removal of the zero modes, which are modes that satisfy
\begin{equation}
\label{eq:ZM-def}
	\mathcal{Q}_\phi\dphi_0 = 0 \, .
\end{equation}
The second term in \eqref{eq:log-sugra} encodes the contribution to the path integral from these zero modes separately. In practice, we expect the spectrum of the $\mathcal{Q}_\phi$ differential operators to be well-behaved, and in particular that the eigenvalues satisfy certain ordering and positivity properties with only a finite number of them being zero. On non-compact spaces with a boundary, such properties depend on the boundary conditions imposed on the fluctuations~$\delta\phi$ and we will discuss some of these aspects in due course. One immediate consequence of this ``good'' spectral behavior is that only massless fluctuations can satisfy~\eqref{eq:ZM-def}, since any positive mass-squared term in~$\mathcal{Q}_\phi$ would lift the $\delta\phi_0$ zero-mode.

To evaluate the non-zero mode contribution to the logarithmic correction \eqref{eq:log-sugra}, it is convenient to introduce the heat kernel associated to the operator $\mathcal{Q}_\phi$ as \cite{Vassilevich:2003xt}
\begin{equation}
\label{eq:K}
	K(x,y;t;\mathcal{Q}_\phi) = \langle x| e^{-t\mathcal{Q}_\phi}|y \rangle \, .
\end{equation}
At coincident space-time points, this can be expanded in the small $t$ limit as
\begin{equation}
	K(x,x;t;\mathcal{Q}_\phi) = \sum_{k=0}^{\infty} a_{2k}(x;\mathcal{Q}_\phi)\,t^{k-2}\,,
\end{equation}
in terms of the Seeley-de Witt (SdW) coefficients $a_{2k}$. Using \eqref{eq:K}, we write the determinant factor in \eqref{eq:log-sugra} as an integral \cite{Sen:2012kpz,Sen:2012cj,Sen:2012dw,Bhattacharyya:2012ye}
\begin{equation}
\label{eq:one-loop-det}
	\log\text{det}'\mathcal{Q}_\phi = -\lim_{\epsilon \to 0}\int_{\epsilon}^\infty \frac{dt}{t}\,\left[\int d^4x \sqrt{g}\,K(x,x;t;\mathcal{Q}_\phi) - n_{\phi_0}\right] \, ,
\end{equation}
where we have explictly removed the $n_{\phi_0}$ zero modes of $\mathcal{Q}_\phi$ and $\epsilon$ is a UV cutoff. By a scaling argument, one can show that the one-loop determinant~\eqref{eq:one-loop-det} produces a logarithmic correction to $\log Z_{\text{sugra}}$ in the large $L^2/G_N$ limit that is entirely controlled by the $k=2$ term in the SdW expansion~\cite{Sen:2012dw,Bhattacharyya:2012ye}:
\begin{equation}
\label{eq:log-sugra:1}
	\log\text{det}'\mathcal{Q}_\phi=-2\Bigg[\int d^4x \sqrt{g}\,a_4(x;\mathcal{Q}_\phi)-n_{\phi_0}\Bigg]\log(L/\sqrt{G_N}) + \mO(1) \, .
\end{equation}

On the other hand, the path integral over the zero modes of the $\mathcal{Q}_\phi$ operator produces a logarithmic correction to \eqref{eq:log-sugra} that can be written as \cite{Sen:2012kpz,Sen:2012cj,Sen:2012dw,Bhattacharyya:2012ye}
\begin{equation}
\label{eq:log-sugra:2}
	\log\int\mathcal{D}\dphi_0 = (-1)^F\beta_{\phi_0}n_{\phi_0}\log(L/\sqrt{G_N}) + \mO(1) \, ,
\end{equation}
where $\beta_{\phi_0}$ is a pure number fixed by demanding locality of the path integral measure $\mathcal{D}\delta\phi_0$. For future reference, we record here the results for fields of spin $3/2$, spin $2$, and for $p$-forms in four dimensions, see \cite{Sen:2012cj,Bhattacharyya:2012ye}:
\begin{equation}
\label{eq:beta}
	\qquad \beta_{3/2} = 3 \, , \qquad \beta_2 = 2 \, , \qquad \beta_{A_p} = \frac{4-2p}{2} \, .
\end{equation}

Together, the non-zero mode contribution \eqref{eq:log-sugra:1} and the zero mode contribution \eqref{eq:log-sugra:2} yield the full logarithmic correction to the 4d Euclidean path integral. Accordingly, we split the resulting coefficient $C$ in \eqref{eq:log-sugra} into two parts,
\begin{equation}
\label{eq:C}
	C = C_{\text{local}} + C_{\text{non-local}} \, ,
\end{equation}
where the local contribution is controlled by the fourth SdW coefficient, 
\begin{equation}
\label{eq:C-local}
	C_{\text{local}} = \sum_{\phi} (-1)^F\int d^4x \sqrt{g}\,a_4(x,\mathcal{Q}_\phi) \, ,
\end{equation}
and the non-local contribution captures the effect of zero modes,
\begin{equation}
\label{eq:C-non-local}
	C_{\text{non-local}} = \sum_{\text{massless }\phi}(-1)^F n_{\phi_0} (\beta_{\phi_0} - j - 1) \, .
\end{equation}
Here the sum is over the massless spectrum, and we have explicitly displayed the ghost level factor discussed in Footnote \ref{foot:ghosts} since it will be important later. In the following, we study these contributions in turn.

\subsection{Local contributions}
\label{sec:sugra:local}

We now review some important features of the SdW coefficient $a_4$ that controls the local contribution (\ref{eq:C-local}). For this discussion, we assume that the second-order differential operator $\mathcal{Q}$ is of Laplace type.\footnote{We temporarily drop the subscript $\phi$ on $\mathcal{Q}_\phi$ since the analysis is valid for generic fields.} This means that it can be represented locally as
\begin{equation}
	\mathcal{Q} = \mD^\mu\mD_\mu + 2\,\omega^\mu\mD_\mu + P \, ,
\end{equation}
where $\mathcal{D}_\mu$ is the covariant derivative with respect to all bosonic local gauge transformations of the theory under consideration\footnote{In particular, $\mathcal{D}_\mu$ includes background gauge fields. We use the symbol $\nabla_\mu$ to denote the spacetime covariant derivative.} and $(\omega_\mu,P)$ are a set of matrices acting in field space. Introducing the differential operator $D_\mu = \mathcal{D}_\mu + \omega_\mu$, we complete the square and write
\begin{equation}
\label{eq:Laplace-op}
	\mathcal{Q} = D^\mu D_\mu + E \, ,
\end{equation}
where $E = P - \omega^\mu\omega_\mu - \mathcal{D}^\mu\omega_\mu$. We stress that the matrix multiplication in field space is implicit in our notation. The curvature associated to $D_\mu$ is denoted by $\Omega_{\mu\nu} = [D_\mu,D_\nu]$.

In general, the SdW coefficients associated to $\mathcal{Q}$ can be written in terms of traces of the matrix-valued quantities $E$ and $\Omega_{\mu\nu}$, together with curvature tensors of the background geometry. If we consider a general 4d Riemannian manifold $(\mathcal{M},g)$ with a non-empty boundary $\partial\mathcal{M}$, the formula for the fourth SdW coefficient of the Laplace operator~\eqref{eq:Laplace-op} integrated over $\mathcal{M}$ reads \cite{Vassilevich:2003xt}\footnote{Note that this reference uses the inward unit vector normal to the boundary. We will use the outward one, and this difference is reflected in various signs in the last term of~\eqref{eq:a4}.} 
\begin{equation}
\label{eq:a4}
	\begin{split}
		\int_{\mathcal{M}} d^4x \sqrt{g}\,a_4(\mathcal{Q}) =&\; \int_{\mathcal{M}} d^4x \sqrt{g}\,a_4^{\text{bulk}}(\mathcal{Q}) + \int_{\mathcal{M}} d^4x \sqrt{g}\,a_4^{\text{tot.der.}}(\mathcal{Q}) \\
		&+ \int_{\partial\mathcal{M}}d^3y\sqrt{\gamma}\,a_4^{\text{bdry}}(\mathcal{Q}) \, ,
	\end{split}
\end{equation}
where $\gamma_{ab}$ is an induced metric on the boundary $\partial\mathcal{M}$ and we define
\begin{align}
	(4\pi)^2a_4^{\text{bulk}}(\mathcal{Q}) =&\; \mathrm{Tr}\bigg[\frac12 E^2 + \frac16 R E + \frac1{12}\Omega^{\mu\nu}\Omega_{\mu\nu} + \frac{1}{360}\bigl(3W^2 - E_4 + 5R^2\bigr)\bigg] \, , \label{eq:a4-details:bulk} \\[1mm]
	(4\pi)^2a_4^{\text{tot.der.}}(\mathcal{Q}) =&\; \mathrm{Tr}\bigg[\frac16\square E + \frac1{30}\square R\bigg] \, , \label{eq:a4-details:tot}
\end{align}
\begin{align}
\label{eq:a4-details:bdry}
	(4\pi)^2a_4^{\text{bdry}}(\mathcal{Q}) =&\; \frac{1}{360}\mathrm{Tr}\bigg[(120\Pi_- - 240\Pi_+)\nabla_n E + (18\Pi_- - 42\Pi_+)\nabla_n R + 24 \widetilde{\nabla}^a\widetilde{\nabla}_aK \nonumber \\
	&\qquad\quad + 120 E K + 20 R K + 4 R_{an}{}^{an} K - 12 R_{anb}{}^n K^{ab} + 4 R_{abc}{}^b K^{ac} \nonumber \\
	&\qquad\quad + \frac1{21}\Big\{(280\Pi_+ + 40\Pi_-)K^3 + (168\Pi_+ - 264\Pi_-)K^{ab}K_{ab}K \nonumber \\
	&\qquad\quad + (224\Pi_+ + 320\Pi_-)K^{ab}K_{bc}K_a{}^c\Big\} \\
	&\qquad\quad - 720 S E - 120 S R - 144 S K^2 - 48 S K^{ab} K_{ab} + 480 S^2 K \nonumber \\[1.5mm]
	&\qquad\quad - 480 S^3 - 120\widetilde{\nabla}^a\widetilde{\nabla}_aS - 60\Omega_{an}\mP\widetilde{\nabla}^a\mP \nonumber \\
	&\qquad\quad + 12(10S - K)\widetilde{\nabla}^a\mP\widetilde{\nabla}_a\mP - 24K^{ab}\widetilde{\nabla}_a\mP\widetilde{\nabla}_b\mP\bigg] \, . \nonumber
\end{align}
Here we have used $i,j,k\in\{1,2,3,4\}$ and $a,b,c\in\{1,2,3\}$ for the local orthonormal frame indices of the tangent bundles on $\mathcal{M}$ and $\partial\mathcal{M}$, respectively. From here on we choose a bulk local orthonormal frame appropriately so that the 4d $i,j,k$ indices become identical to the union of 3d boundary $a,b,c$ indices and the normal vector index $n=4$ on the boundary $\partial\mathcal{M}$. Then the extrinsic curvature of the boundary can be written explicitly as $K_{ab} = -\Gamma^n_{ab}$. The trace of the extrinsic curvature is $K=K^a{}_a$. The boundary covariant derivative $\widetilde{\nabla}$ is distinguished from the bulk one $\nabla$, and the box symbol is defined with respect to the latter as $\square=\nabla^\mu\nabla_\mu$. The quantities $\Pi_\pm$ and $S$ specify the boundary conditions used for the fields on which the differential operator $\mathcal{Q}$ acts, and $\mP=\Pi_+-\Pi_-$. These boundary conditions will be discussed in more detail below. Finally, the trace Tr[\ldots] is taken in field space and over all free indices carried by the $E$ and $\Omega$ matrices, and we have introduced the usual curvature combinations
\begin{equation}
\label{eq:W2}
	\begin{split}
		W^2 =&\; R^{\mu\nu\rho\sigma}R_{\mu\nu\rho\sigma} - 2\,R^{\mu\nu}R_{\mu\nu} + \frac13\,R^2 \, , \\
		E_4 =&\;  R^{\mu\nu\rho\sigma}R_{\mu\nu\rho\sigma} - 4\,R^{\mu\nu}R_{\mu\nu} + R^2 \, .
	\end{split}
\end{equation}
The general structure of~\eqref{eq:a4} is that of a bulk contribution involving four-derivative terms, a total derivative contribution, and an intrinsic boundary contribution. We discuss each of them below.

\subsubsection{Bulk contributions}
\label{sec:sugra:local:bulk}

First we study the bulk contribution \eqref{eq:a4-details:bulk}. For simplicity, from here on we focus on the case where the background $\mathring{\phi}$ in the saddle-point approximation \eqref{eq:sugra-saddle} is a solution of Euclidean $\mN=2$ minimal gauged supergravity with bosonic action
\begin{equation}
\label{eq:EMaction}
	S_\text{minimal}=-\frac{1}{16\pi G_N}\int_{\mathcal{M}} d^4x\,\sqrt{g}\,\bigg[R-2\Lambda-F_{\mu\nu}F^{\mu\nu}\bigg]\,,\qquad \Lambda=-\fft{3}{L^2}\,,
\end{equation}
where the corresponding equations of motion read
\begin{equation}
\label{eom}
	\begin{split}
	R_{\mu\nu}-\fft12g_{\mu\nu}(R-2\Lambda)&=2F_{\mu\rho}F_\nu{}^\rho-\fft12g_{\mu\nu}F_{\rho\sigma}F^{\rho\sigma}\,,\\
	\nabla_\mu F^{\mu\nu}&=0\,.
	\end{split}
\end{equation}
In this case, the bulk contribution \eqref{eq:a4-details:bulk} can be conveniently rewritten using the background equations of motion \eqref{eom} and field redefinitions as \cite{David:2021eoq}
\begin{equation}
\label{eq:a4-bulk}
	(4\pi)^2 a_4^{\text{bulk}}(\mathcal{Q}) = -a_E\,E_4 + c\,W^2 + b_1\,R^2 + b_2\,R\,F^{\mu\nu}F_{\mu\nu} \, ,
\end{equation}
where $(a_E,c,b_1,b_2)$ are a set of coefficients that can be obtained from trace computations. These heat kernel coefficients will be central to our analysis since they govern the bulk part of the SdW contribution to the $C_{\text{local}}$ term. \\

For illustration, consider the quantum fluctuations of a massive neutral scalar (MNS) field around a given background. The relevant second-order operator $\mathcal{Q}_{\text{MNS}} = \square - m^2$ capturing the dynamics of these fluctuations is of Laplace type \eqref{eq:Laplace-op} with
\begin{equation}
	E=-m^2\,,\qquad \Omega_{\mu\nu}=0\,.
\end{equation}
Using this in \eqref{eq:a4-details:bulk} and taking the trace, we obtain the bulk contribution
\begin{equation}
	\text{MNS:} \qquad (4\pi)^2 a_4^{\text{bulk}} = \frac{m^4}{2} - \frac16\,R\,m^2 + \frac{1}{360}\bigl(3W^2 - E_4 + 5R^2\bigr) \, .
\end{equation}
This can be brought into the form~\eqref{eq:a4-bulk} by making use of the trace of the background equation of motion $R=-12/L^2$, which yields the coefficients
\begin{equation}
	\text{MNS:} \qquad a_E = \frac{1}{360} \, , \quad c = \frac{1}{120} \, , \quad b_1 = \frac{1}{288}\bigl((mL)^2 + 2\bigr)^2 \, , \quad b_2 = 0 \, .
\end{equation}
This simple example illustrates how obtaining the bulk contribution to the fourth SdW coefficient for a field $\phi$ amounts to identifying the appropriate second-order $\mathcal{Q}$ operator governing its fluctuations, extracting the $E$ and $\Omega$ matrices, and collecting the $(a_E,c,b_1,b_2)$ coefficients by computing traces and making use of the background equations of motion.

In Appendix~\ref{App:coeffs} we implement this method for both massless and massive fields of spin $0 \leq s \leq 2$. For scalars and fermions, we consider quadratic fluctuations of minimally coupled fields around a generic background satisfying the equations of motion \eqref{eom}. For fields of spin $1 \leq s \leq 2$ we opted to turn off the background Maxwell field for simplicity. In this way, we have obtained the coefficients $(a_E,c,b_1)$ in~\eqref{eq:a4-bulk} for all fields minimally coupled to an Einstein-Maxwell background, while we only have access to the $b_2$ coefficient for scalars and fermions. The results of these lengthy computations are summarized in Table~\ref{tab:coeffs}.
\begin{table}[h]
	\centering
	\renewcommand*{\arraystretch}{1.4}
	\begin{tabular}{|c|c|c|c|c|}
		\hline
		spin & mass & $a_E$ & $c$ & $b_1$ \\
		\hline\hline
		0 & $(mL)^2 = -2$ & $\frac{1}{360}$ & $\frac{1}{120}$ & 0 \\
		\hline
		0 & $m$ & $\frac{1}{360}$ & $\frac{1}{120}$ & $\frac{1}{288}\bigl((mL)^2 + 2\bigr)^2$ \\
		\hline\hline
		1/2 & 0 & $-\frac{11}{720}$ & $-\frac{1}{40}$ & 0 \\
		\hline
		1/2 & $m$ & $-\frac{11}{720}$ & $-\frac{1}{40}$ & $\frac{1}{144}(mL)^2\bigl((mL)^2 - 2\bigr)$ \\
		\hline\hline
		1 & 0 & $\frac{31}{180}$ & $\frac{1}{10}$ & 0 \\
		\hline
		1 & $m$ & $\frac{31}{180} + \frac{1}{360}$ & $\frac{1}{10} + \frac{1}{120}$ & $\frac{1}{288}\bigl(3(mL)^4 - 12(mL)^2 + 4\bigr)$ \\
		\hline\hline
		3/2 & $mL=1$ & $\frac{589}{720}$ & $\frac{137}{120}$ & 0 \\
		\hline
		3/2 & $m$ & $\frac{589}{720} - \frac{11}{720}$ & $\frac{137}{120} - \frac{1}{40}$ & $\frac{1}{72}\bigl((mL)^4 - 8(mL)^2 + 11\bigr)$ \\
		\hline\hline
		2 & 0 & $\frac{571}{180}$ & $\frac{87}{20}$ & 0 \\
		\hline
		2 & $m$ & $\frac{571}{180} + \frac{31}{180} + \frac{1}{360}$ & $\frac{87}{20} + \frac{1}{10} + \frac{1}{120}$ & $\frac{5}{288}\bigl((mL)^4 - 8(mL)^2 + 8\bigr)$ \\
		\hline
	\end{tabular}
	\caption{The quantities controlling the bulk contribution to the SdW coefficient~\eqref{eq:a4-bulk}. For massive fluctuations with gauge invariance, we indicate the effect of adding the appropriate Stückelberg fields as a separate contribution to $a_E$ and $c$. \label{tab:coeffs}}
\end{table}

Let us collect some important remarks on the results. For fields of spins $1\leq s \leq 2$, the operator $\mathcal{Q}$ can be brought to Laplace form \eqref{eq:Laplace-op} only after imposing gauge-fixing conditions inside the path integral. For massless fields, this requires the addition of appropriate ghost fields, as we review in Appendix~\ref{App:coeffs}. Our results for the corresponding $(a_E,c,b_1,b_2)$ coefficients are compatible with previous derivations in the literature, see e.g. \cite{David:2021eoq,Karan:2022dfy}. For massive fields, the mass term typically breaks gauge invariance and to remedy this we must introduce appropriate Stückelberg fields \cite{Stueckelberg:1938zz,Stueckelberg:1938hvi}, see \cite{Ruegg:2003ps} for a modern review. It is important to note that these fields are physical, and while they do not modify the classical equations of motion, they are allowed to run in loops. Therefore, they generically give non-trivial contributions to the heat kernel coefficients. To draw attention to this point, we explicitly separate the contribution from the Stückelberg fields to the $(a_E,c)$ coefficients in Table~\ref{tab:coeffs}. This also makes it clear that the massless case is not obtained as a limit of the massive one, which is a manifestation of the fact that the Stückelberg fields do not decouple at the one-loop level. Introducing the appropriate Stückelberg fields is therefore crucial to obtain the correct coefficients in~\eqref{eq:a4-bulk} for massive fields.

It is also worth mentioning that, because we study fields on an asymptotically AdS$_4$ space, by massless we mean $m^2 = -2/L^2$ for a scalar field and $m = 1/L$ for a gravitino field. This nomenclature takes into account the conformal coupling in the Lagrangian which, in the case of a (pseudo) scalar field for example, reads $\frac16 R \phi^2 = -2L^{-2}\phi^2$ on an asymptotically AdS background. Indeed, such a conformally coupled scalar field sits in the massless supermultiplet of a conserved current.

%
\begin{table}[t]
\centering
\renewcommand*{\arraystretch}{1.4}
\begin{tabular}{|c|c|}
\hline
fields & mass \\
\hline
$s=0,2$ & $(mL)^2 = \Delta(\Delta - 3)$ \\
\hline
$p$-form & $(mL)^2 = (\Delta - p)(\Delta + p - 3)$ \\
\hline
$s=\frac12,\frac32$ & $|mL| = \Delta - \frac32$ \\
\hline
\end{tabular}
\caption{Relation between 4d mass and 3d conformal dimension for various bulk fields.\label{tab:conf-m}}
\end{table}

Finally, the four-derivative quantities $(E_4,W^2,R^2)$ are not all linearly independent on an Einstein background for which $R_{\mu\nu} - \frac12 g_{\mu\nu}(R - 2\Lambda) = 0$. After turning off the background Maxwell field, the linear constraint implied by the Einstein equations \eqref{eom} is
\begin{equation}
	\text{Einstein background:} \qquad E_4-W^2-\fft16R^2 = 0 \, ,\label{Einstein:constraint}
\end{equation}
which translates to an ambiguity in the coefficients $(a_E,c,b_1)$ entering \eqref{eq:a4-bulk}. In Table~\ref{tab:coeffs}, this freedom has been fixed by demanding that $b_1=0$ for massless fields.

\subsubsection{Boundary contributions}
\label{sec:sugra:local:bdry}

For the total derivative contribution \eqref{eq:a4-details:tot}, the fact that $\mathrm{Tr}[E]$ is a two-derivative quantity shows that we can use the background equations of motion~\eqref{eom} together with Stoke's theorem to bring it to the form~\cite{David:2021eoq}
\begin{equation}
\label{eq:a4-tot-der}
	(4\pi)^2 \int_{\mathcal{M}} d^4x\sqrt{g}\,a_4^{\text{tot.der.}}(\mathcal{Q}) = \int_{\partial \mathcal{M}} d^3y\sqrt{\gamma}\,n^\mu \nabla_\mu\Bigl[\alpha_1R + \alpha_2F_{\mu\nu}F^{\mu\nu}\Bigr] \, ,
\end{equation}
where $n^\mu$ denotes the outward unit vector normal to the boundary and $\alpha_{1,2}$ are constants whose values depend on the field under consideration. We can therefore view the total derivative contribution to the integrated SdW coefficient as a surface term.\\

The other surface term in \eqref{eq:a4} depends explicitly on the choice of boundary conditions for each field fluctuation $\delta\phi$. In writing \eqref{eq:a4-details:bdry}, we have implicitly assumed that we can choose so-called ``mixed'' boundary conditions~\cite{Vassilevich:2003xt} for all fields. Such boundary conditions are parametrized as
\begin{equation}
\label{eq:bc-param}
	\Pi_-\delta\phi|_{\partial\mathcal{M}} = 0 \, , \qquad (\nabla_n + S)\Pi_+\delta\phi|_{\partial\mathcal{M}} = 0 \, .
\end{equation}
Within this class, various choices are avalaible for fields in Euclidean gravity. We will further restrict ourselves to boundary conditions that ensure that the differential operator $\mathcal{Q}_\phi$ is elliptic. Roughly speaking, this guarantees that the operator has only finitely-many zero modes so that we can define its determinant in the usual way, as we have indeed assumed we could do in writing~\eqref{eq:log-sugra}. For a nice review of such boundary conditions we refer the reader to~\cite{Witten:2018lgb}, which we follow.

Dirichlet and Neumann boundary conditions on the scalar Laplacian are elliptic. The former amount to choosing $\Pi_- = 1$ and $\Pi_+ = S = 0$ in~\eqref{eq:bc-param}, while the latter are achieved for $\Pi_- = S = 0$ and $\Pi_+ = 1$. To compute the one-loop determinant for a spinor field $\psi$, it turns out to be convenient to use the square of the Dirac operator for the $\mathcal{Q}_\psi$ operator, as we explain in Appendix~\ref{App:coeffs}. A choice of elliptic (and hermitian) boundary conditions for this operator is given by $\Pi_- = \frac12(1 + \mathrm{i}\gamma^n\gamma^5)$, $\Pi_+ = 1 - \Pi_-$ and $S = \frac12 K \Pi_+$, see~\cite{Vassilevich:2003xt}. For Yang-Mills fields, we will use so-called relative boundary conditions where the fluctuations $\delta A_\mu = A_\mu - \mathring{A}_\mu$ satisfy \cite{Vassilevich:2003xt}
\begin{equation}
\label{eq:A-bc-rel}
	(\delta A_a)|_{\partial\mathcal{M}} = 0 \, ,
\end{equation}
where the Latin index $a$ denotes the local orthonormal frame indices of the tangent bundle on the boundary introduced below (\ref{eq:a4-details:bdry}). By BRST invariance, this condition implies that the ghost field $c$ required for quantization vanishes at the boundary, $c|_{\partial\mathcal{M}} = 0$. The same must be true for the anti-ghost field $b$ in order to have a well-defined propagator for the $b$-$c$ system. Invoking BRST again,~\eqref{eq:A-bc-rel} then implies that $\mathcal{G}(\delta A)|_{\partial\mathcal{M}} = 0$ where $\mathcal{G}$ is the gauge-fixing function for the fluctuations. This set of boundary conditions is elliptic, see~\cite{Witten:2018lgb}. Working in the Lorenz gauge as in Appendix~\ref{App:coeffs}, it corresponds to
\begin{equation}
	(\Pi_-)_{ij} = \delta_{ij} - \delta_{in}\delta_{jn}\, , \quad (\Pi_+)_{ij} = \delta_{in}\delta_{jn} \, , \quad S_{ij} = K\delta_{in}\delta_{jn} \, , 
\end{equation} 
in the parameterization~\eqref{eq:bc-param}. Elliptic boundary conditions for spin-3/2 fluctuations can be obtained by using an appropriate combination of elliptic Yang-Mills and spinor boundary projectors $\Pi_\pm$ and $S$, and are also of mixed type.
 
The boundary problem for metric fluctuations is notoriously more involved. While it was shown in~\cite{Anderson:2006lqb} that there exists a set of elliptic boundary conditions, they involve tangential derivatives and therefore cannot be cast in the form~\eqref{eq:bc-param}. Instead, they fall into the more general class of ``oblique'' boundary conditions for which much less is known regarding the general form of $a_4^{\text{bdry}}$ in~\eqref{eq:a4}, see~\cite{Vassilevich:2003xt} for a discussion. To show this explicitly, let us recall that the conformal boundary conditions of~\cite{Anderson:2006lqb} are obtained by fixing the conformal structure of the boundary and requiring that the metric fluctuations preserve the trace of the extrinsic curvature. Writing the perturbed metric as $g_{\mu\nu} = \mathring{g}_{\mu\nu} + h_{\mu\nu}$, the first condition amounts to demanding that the traceless part of $h_{\alpha\beta}|_{\partial\mathcal{M}}$ vanishes, where $\alpha,\beta,\ldots$ denote spacetime indices tangent to the boundary. Just as in the Yang-Mills case, BRST invariance then implies that $\mathcal{G}_\mu(h)|_{\partial\mathcal{M}} = 0$, which gives two additional boundary conditions when split along $\mu =\,\perp$ and $\mu = \alpha$, see also~\cite{Anninos:2023epi}. The extrinsic trace condition gives us one last equation that the fluctuations should satisfy. For concreteness, we will work in the harmonic (de Donder) gauge for the metric perturbations, and use Gaussian normal coordinates for the background so that $\mathring{g}_{\perp\perp} = 1$ and $\mathring{g}_{\perp\alpha} = 0$. Then, the full set of elliptic metric boundary conditions can be summarized as
\begin{equation}
\label{eq:conf-bc}
	\begin{split}
	&\Bigl[h_{\alpha\beta} - \frac13\,\mathring{g}_{\alpha\beta}h^\gamma{}_\gamma\Bigr]\Big\vert_{\partial\mathcal{M}} = 0 \, , \\
	&\Bigl[\partial_\perp h^\gamma{}_\gamma - 2\widetilde{\nabla}^\alpha h_{\alpha\perp} - K h_{\perp\perp}\Bigr]\Big\vert_{\partial\mathcal{M}} = 0 \, , \\
	&\Bigl[\partial_\perp h_{\perp\perp} + K h_{\perp\perp} - 2 K^{\alpha\beta}h_{\alpha\beta}\Bigr]\Big\vert_{\partial\mathcal{M}} = 0 \, , \\
	&\Bigl[\partial_\perp h_{\alpha\perp} + K h_{\alpha\perp} + \widetilde{\nabla}^\beta h_{\alpha\beta} - \frac12\widetilde{\nabla}_\alpha(h_{\perp\perp} + h^{\gamma}{}_\gamma)\Bigr]\Big\vert_{\partial\mathcal{M}} = 0 \, .
	\end{split}
\end{equation}
Since the tangential covariant derivatives $\widetilde{\nabla}$ cannot be eliminated from these equations, we are indeed dealing with oblique boundary conditions. As far as we are aware, the problem of finding good boundary conditions for metric fluctuations in 4d and their explicit contribution to the fourth SdW coefficient has not been solved on a general non-compact manifold with boundary. We believe this is a hard mathematical problem that falls outside the scope of the present work. In order to make progress, we will formally impose the above conformal boundary conditions in order to ensure ellipticity of the $\mathcal{Q}_h$ operator. This will induce a modification of the boundary term~\eqref{eq:a4-details:bdry},
\begin{equation}
\label{eq:a4-oblique}
	a_4^{\text{bdry}} \longrightarrow a_4^{\text{bdry}} + a_4^{\text{oblique}} \, .
\end{equation}
As we will see in Section~\ref{sec:bootstrap}, the contribution from mixed boundary conditions $a_4^{\text{bdry}}$ can always be holographically renormalized away after integrating over a cutoff hypersurface and sending the cutoff to infinity. We will assume that the same is true for the unknown quantity $a_4^{\text{oblique}}$ in the rest of this chapter. This assumption ensures that the oblique nature of the boundary conditions~\eqref{eq:conf-bc} does not affect the local contribution to the logarithmic term in the Euclidean path integral.

\subsection{Non-local contributions}
\label{sec:sugra:non-local}

We now discuss the non-local contribution \eqref{eq:C-non-local}. The first step is to identify whether the Laplace-type operator $\mathcal{Q}_\phi$ admits zero modes. In general, this spectral problem for generic Einstein-Maxwell backgrounds that solve the equations of motion \eqref{eom} is quite involved. We will therefore restrict our attention to two minimal supergravity backgrounds for which we can make explicit statements. These backgrounds are pure Euclidean AdS$_4$ (EAdS$_4$) and Euclidean AdS$_2 \times \Sigma_\mathfrak{g}$ with a non-trivial background gauge field along the Riemann surface. They will serve to illustrate expected general features of the non-local contribution to the log term in the supergravity path integral.

\subsubsection{Zero-modes on EAdS$_4$}
\label{sec:sugra:non-local-AdS4}

For pure EAdS$_4$, an analysis of the zero modes of Laplace-type differential operators acting on $p$-forms, spin-1/2 and metric tensors was conducted in \cite{Camporesi:1994ga,camporesi1994plancherel,Camporesi:1995fb}. There it was shown that there are no square-integrable zero modes of the Laplacian for symmetric tensors on EAdS$_4$. Fermions also do not allow for square-integrable zero modes of the Dirac operator. Lastly for $p$-forms, there are square-integrable zero modes only in even dimension $d$ and for $p = d/2$. Hence, zero modes can only come from 2-forms in a pure EAdS$_4$ background. The number of such zero modes is given by~\cite{camporesi1994plancherel,Bhattacharyya:2012ye} 
\begin{equation}
\label{n:A2}
	n_{A_2} = \fft{3}{4\pi^2L^4}\,\int_{\mathcal M}d^4x \sqrt{g} = 1 \, ,
\end{equation}
for each 2-form field present in the spectrum. In \eqref{n:A2} the divergent volume factor of EAdS$_4$ is renormalized as
\begin{equation}
	\text{vol}(\text{EAdS}_4) = \int_{\mathcal M}d^4x \sqrt{g} = \frac{4\pi^2L^4}{3} \, ,
\end{equation}
by adding appropriate boundary counter-terms. For spin-3/2 fields, one can construct zero modes of the Rarita-Schwinger (RS) operator as follows. We start with an eigenspinor of the Dirac operator on EAdS$_4$,
\begin{equation}
\label{eq:eigenspinors}
	\slashed{\nabla}\Omega^s_{\ell m} = \frac{\mathrm{i}s}{L}\,\lambda\,\Omega^s_{\ell m} \, ,
\end{equation}
indexed by a continuous eigenvalue $\lambda$, two mode numbers $\ell\in\mathbb{N}$ and $m=1,\ldots,d_\ell$ and a sign $s=\pm$ related to chirality. Here $d_\ell$ is the dimension of the spin-$\ell$ representation of Spin(3). Now define the basis spinors
\begin{equation}
\label{eq:build-eigen-RS}
	\Psi_\mu^{(1)s} = \gamma_\mu\Omega^s \, , \qquad \Psi_\mu^{(2)s} = \nabla_\mu\Omega^s \, ,
\end{equation} 
where we momentarily suppress the mode indices. Consider the RS operator for a spin-3/2 field of mass $M$ on EAdS$_4$,
\begin{equation}
\label{eq:RS-op}
	(D_\pm)^{\mu\nu} = \gamma^{\mu\nu\rho}\nabla_\rho \pm M\gamma^{\mu\nu} \, ,
\end{equation}
where the sign factor encodes chirality. This operator acts on~\eqref{eq:build-eigen-RS} as
\begin{equation}
	\begin{split}
	(D_\pm)_\mu{}^\nu\Psi_\nu^{(1)s} =&\; \bigg(\pm3M - \frac{2\mathrm{i}s}{L}\,\lambda\bigg)\Psi_\mu^{(1)s} + 2\,\Psi_\mu^{(2)s} \, , \\
	(D_\pm)_\mu{}^\nu\Psi_\nu^{(2)s} =&\; \bigg(\frac{R}{8} \pm \frac{\mathrm{i}s}{L}\,M\,\lambda\bigg)\Psi_\mu^{(1)s} \mp M\,\Psi_\mu^{(2)s} \, ,
	\end{split}
\end{equation}
where we have used the defining relation~\eqref{eq:eigenspinors}. This shows that appropriate linear combinations of the spinors $\Psi^{(1)s}_\mu$ and $\Psi^{(2)s}_\mu$ generate a (non-orthonormal) basis for the eingenfunctions of the RS operator. In particular, consider the following linear combination, 
\begin{equation}
	\Psi^s_{\mu\ell m}(\lambda) = \mathcal{N}^s_\ell\bigg(\nabla_\mu + \frac{s}{2}\,M\gamma_\mu\bigg)\Omega^s_{\ell m} \, .
\end{equation}
Here, the normalization constant $\mathcal{N}^s_\ell$ should be fixed by demanding 
\begin{equation}
	\langle\Psi^s_{\mu\ell m}(\lambda),\Psi^{\mu s'}_{\ell'm'}(\lambda')\rangle = \delta_{ss'}\delta_{\ell\ell'}\delta_{mm'}\delta(\lambda - \lambda') \, ,
\end{equation}
where we define the Euclidean inner product as
\begin{equation}
	\langle f,g\rangle = \int d^4 x \sqrt{g}\,f^\dagger g \, .
\end{equation}
For a massless gravitino field in EAdS$_4$ with $M=1/L$, it is easy to see that $\Psi^s_{\mu\ell m}$ is in fact a zero-mode of the RS operator:
\begin{equation}
(D_s)_\mu{}^\nu \Psi^s_{\nu\ell m}(\lambda) = 0 \, ,
\end{equation}
where the sign factors are now correlated. Thus, it appears that quantum fluctuations of a spin-3/2 field can include zero modes of the RS operator on EAdS$_4$. We should now ask whether such modes are square-integrable with
\begin{equation}
	\langle\Psi^s_{\mu\ell m}(\lambda),\Psi^{\mu s}_{\ell m}(\lambda)\rangle < \infty\, .
\end{equation}
To answer this question, we can use the known expression for the Dirac eigenspinors $\Omega^s_{\ell m}$ in~\cite{Camporesi:1995fb}. We give details in Appendix~\ref{App:RS-non-local}, where we show that the zero modes $\Psi^s_{\mu\ell m}$ either have vanishing norm or are \emph{not} square-integrable and must therefore be discarded.\\

The upshot of this analysis is that, on EAdS$_4$, the non-local contribution \eqref{eq:C-non-local} comes only from 2-forms as
\begin{equation}
\label{eq:Cnon-local-AdS4}
	C_\text{non-local} = \sum_{\text{massless 2-forms}}(-1)^F(-j-1) \, .
\end{equation}
A common feature of all 4d KK supergravity theories we will consider is that the spectrum does not include $p$-forms with $p > 2$. As a result, any 2-form is necessarily bosonic and at ghost level $j=0$ (see Footnote~\ref{foot:ghosts}), in which case the above simplifies to 
\begin{equation}
	C_\text{non-local}(\text{EAdS}_4) = -N_{A_2} \, ,
\end{equation} 
with $N_{A_2}$ the total number of massless 2-form fields. We stress that the non-local contribution~\eqref{eq:Cnon-local-AdS4} to the EAdS$_4$ logarithmic correction is always a pure number, regardless of the field content of the gravitational theory we consider.

\subsubsection{Zero-modes on EAdS$_2 \times \Sigma_\mathfrak{g}$}
\label{sec:sugra:non-local-AdS2xS}

On a product space like EAdS$_2 \times \Sigma_\mathfrak{g}$, the quadratic operator $\mathcal{Q}$ acting on arbitrary field fluctuations can be split into
\begin{equation}
	\mathcal{Q}_{4\text{d}} = \mathcal{Q}_{\text{EAdS}_2} + \mathcal{Q}_{\Sigma_{\mathfrak{g}}} \, .
\end{equation}
We will always consider compact Riemann surfaces, which implies that the operator $\mathcal{Q}_{\Sigma_{\mathfrak{g}}}$ has real non-negative eigenvalues. We will see that, as originally shown in \cite{Camporesi:1994ga,camporesi1994plancherel,Camporesi:1995fb}, normalizable modes on EAdS$_2$ also have non-negative eigenvalues. Therefore, we can study the modes that are in the kernel of both two-dimensional operators separately to obtain the zero-modes on the 4d space.

The EAdS$_2 \times \Sigma_{\mathfrak{g}>1}$ solution to \eqref{eom} with the corresponding graviphoton reads 
\begin{equation}
\label{eq:NHG-Romans-ZM}
\begin{split}
	ds^2 & = \frac{L^2}{4}\bigg[(\rho^2-1)\,d\tau^2 + \frac{d\rho^2}{\rho^2-1} + 2\,\frac{dx^2 + dy^2}{y^2}\bigg] \, ,\\
	dA & = \frac{L}{2y^2}\,dx \wedge dy \, .
\end{split}
\end{equation}
We restrict ourselves to a Riemann surface with genus $\mathfrak{g} > 1$ since this is the situation arising from the near-horizon limit of supersymmetric extremal black holes in AdS$_4$, see Section~\ref{sec:bootstrap:Romans}. On this background, a 4d field can be dimensionally reduced down to the EAdS$_2$ factor with coordinates $(\tau,\rho)$. In this two-dimensional space, the relevant Laplace-type operators for scalar and spinor fields are the usual 2d scalar Laplacian and the square of the Dirac operator. Their eigenvalues are well-known,
\begin{equation}
\begin{split}
	-\square_{\text{EAdS}_2}\phi =&\; \Bigl(\lambda^2 + \frac14\Bigr)\phi \, , \qquad \lambda \in \mathbb{R} \, , \\
	-\slashed{\nabla}^2_{\text{EAdS}_2}\psi =&\; \lambda^2\,\psi \, , \qquad\qquad\;\;\; \lambda > 0 \, ,
\end{split}
\end{equation}
which shows that there are no zero modes for these 2d fields.  In contrast, 2d fields with spin $1\leq s \leq 2$ have normalizable zero modes whose explicit expressions can be found in~\cite{Sen:2012cj}. These zero modes are indexed by a non-zero integer $\ell$ whose range depends on the spin, and they satisfy
\begin{equation}
	\sum_{|\ell|\geq1} |a^{(\ell)}_\alpha|^2 = \frac{2}{\pi L^2} \, , \qquad \sum_{\ell \geq 1} |\xi^{(\ell)}_\alpha|^2 = \frac{4}{\pi L^2} \, , \qquad \sum_{|\ell|\geq2} |w_{\alpha\beta}^{(\ell)}|^2 = \frac{6}{\pi L^2} \, ,\label{sq:sum}
\end{equation}
for a vector, gravitino and metric zero mode, respectively. Here $\alpha$ denotes a space-time EAdS$_2$ index and the norm is taken with appropriate 2d metric contractions. Multiplying \eqref{sq:sum} by the regularized volume $\text{vol}(\text{EAdS}_2) = -\pi L^2/2$, we obtain the number of zero-modes for each two-dimensional field,
\begin{equation}
\label{eq:n0-H2}
	n^{\text{EAdS}_2}_0 = n^{\text{EAdS}_2}_{1/2} = 0 \, , \qquad n^{\text{EAdS}_2}_1 = -1 \, , \qquad n^{\text{EAdS}_2}_{3/2} = -2 \, , \qquad n^{\text{EAdS}_2}_2 = -3 \, ,
\end{equation}
%

On the Riemann surface, the only zero-mode of the scalar Laplacian is the constant function.\footnote{This can be shown, for instance, by studying the Laplacian on the upper half-plane and quotienting by a discrete subgroup of $PSL(2,\mathbb{R})$ to obtain the spectrum on the compact Riemann surface.} For spin-$1/2$, we must take into account the fact that there is a non-trivial gauge field along $\Sigma_\mathfrak{g}$ as in~\eqref{eq:NHG-Romans-ZM}. This implies that the relevant differential operator acting on spinor fluctuations is the Dirac operator on the compact Riemann surface twisted by the line bundle $\mathcal{L}$ defined by the gauge connection. Denoting this operator by $\slashed{\nabla}_A$, we can make use of the result of~\cite{ALMOROX20062069} for $\mathfrak{g} > 1$, 
\begin{equation}
	\text{dim}\,\text{Ker}\slashed{\nabla}_A = \text{deg}(\mathcal{L}) = 2(\mathfrak{g}-1) \, ,
\end{equation}
Thus, the number of zero modes for scalar and spinor fields on the Riemann surface are
\begin{equation}
\label{eq:n0-S}
	n^{\Sigma}_0 = 1 \, , \qquad n^{\Sigma}_{1/2} = 2(\mathfrak{g} - 1) \, .
\end{equation}
We will not need the number of zero modes for fields of spin $1 \leq s \leq 2$ since they always come tensored with scalar or spinor zero modes on the EAdS$_2$ factor, which as we saw above do not exist.

We can now decompose all 4d fields according to their 2d spins and use the above results to compute the number of 4d zero modes. The 4d scalar and spinor fields decompose into combinations of scalars and fermions on EAdS$_2$ and the Riemann surface, so they have no zero modes by~\eqref{eq:n0-H2}. A vector field in 4d decomposes into a spin-1 field on EAdS$_2$ and a scalar on $\Sigma_{\mathfrak{g}}$, so we have $-1$ such zero-modes. A 4d gravitino field decomposes into a combination of spin-3/2 and spin-1/2 two-dimensional fields, producing a total of $-4(\mathfrak{g}-1)$ gravitino zero-modes in 4d. For the metric, the absence of globally defined Killing vectors on the Riemann surface of genus $\mathfrak{g}>1$ implies that the 4d mode decomposes into a combination of metric and scalar modes on the 2d spaces. Thus,
\begin{equation}
	n^{4\text{d}}_0 = n^{4\text{d}}_{1/2} = 0 \, , \qquad n^{4\text{d}}_1 = -1 \, , \qquad n^{4\text{d}}_{3/2} = -4(\mathfrak{g} - 1) \, , \qquad n^{4\text{d}}_2 = -3 \, .\label{n:EAdS2}
\end{equation}
Observe that the dependence on the genus comes from the fact that the gravitino is charged under the background gauge field that defines the line bundle $\mathcal{L}$.\\

The non-local contribution to the logarithmic correction from the above zero modes is obtained by substituting \eqref{eq:beta} and \eqref{n:EAdS2} into \eqref{eq:C-non-local}. The result is
\begin{equation}
\label{eq:C:non-local-AdS2xS-gen}
	C_{\text{non-local}} = \Bigl[\sum_{s=1}(-1)^F j\Bigr] + 4(1-\mathfrak{g})\Bigl[\sum_{s=3/2}(-1)^F(2-j)\Bigr] - 3\Bigl[\sum_{s=2}(-1)^F(1-j)\Bigr] \, ,
\end{equation}
where the sums run over the massless fields of spin $s$. The 4d KK supergravity theories we want to consider do not contain ghost fields of spin $s>1$, so that the last two sums are restricted to the $j=0$ sector and have definite fermion numbers. Gauge-fixing in the spin-2 fluctuation sector requires the introduction of a pair of vector ghosts (see also Appendix~\ref{App:coeffs-graviton}). If we assume that this gauge-fixing is the only source of $s=1$ ghosts, the above simplifies to
\begin{equation}
\label{eq:C:non-local-AdS2xS}
	C_\text{non-local}(\text{EAdS}_2\times\Sigma_{\mathfrak{g}>1}) = 8N_{3/2}(\mathfrak{g}-1) - 4N_{2} \, ,
\end{equation}
where $N_{3/2}$ and $N_2$ denote the number of massless gravitino and metric fluctuations in the spectrum. Note that the contribution~\eqref{eq:C:non-local-AdS2xS-gen} to the logarithmic correction is always integer, regardless of the KK theory under consideration. Moreover, it depends on the parameters of the background geometry only through the genus of the Riemann surface, which arises from the non-trivial charge of the gravitino fluctuations under the background gauge field.

\subsubsection{Generic zero-modes}
\label{sec:sugra:non-local-gen}

The results~\eqref{eq:Cnon-local-AdS4} and~\eqref{eq:C:non-local-AdS2xS} illustrate how the spectral problem of finding the zero modes of the $\mathcal{Q}$ operators depends on the supergravity background. In particular, when the latter includes a factor of $d$-dimensional Euclidean AdS, one can leverage known results in the literature to study the zero modes. Unfortunately, the task is much more arduous on generic backgrounds. Nevertheless, our examples make it clear that the non-local contributions to the logarithmic corrections on a general $\mathcal{M}_4$ will always take the form of pure numbers related to the counting of zero modes and possibly a simple dependence on discrete parameters of topological origin like the rank of some fiber bundles. In particular, we do not expect that $C_{\text{non-local}}$ can depend on continuous parameters specifying the 4d supergravity background.

Before closing this section, we would like to point out that the discussion above was focused on the calculation of the Euclidean gravitational path integral in the so-called grand canonical ensemble of fixed temperature and chemical potentials. If one is interested in a different thermodynamic ensemble, for example in order to calculate the entropy of a black hole, an appropriate Laplace transform according to the standard rules of thermodynamics should be performed. This change of ensemble can lead to additional contribution to the logarithmic corrections of the partition function, see \cite{Sen:2012dw}. Notably, these additional terms in the coefficient of the logarithmic correction are always pure numbers independent of continuous parameters in the supergravity background of interest.

\section{Bootstrapping the local contributions}
\label{sec:bootstrap}

Following the procedure described in Section \ref{sec:sugra}, one can in principle compute the logarithmic coefficient $C$ in the 4d Euclidean path integral \eqref{eq:sugra-saddle} around any asymptotically EAdS$_4$ Einstein-Maxwell background in a given 4d KK supergravity theory. As we have pointed out, there are a few technical hurdles:
\begin{enumerate}
	\item Computing the local contribution $C_\text{local}$ for generic Einstein-Maxwell backgrounds with a non-vanishing graviphoton remains out of reach, since we have restricted our results in Table~\ref{tab:coeffs} to pure Einstein backgrounds, and we therefore do not have access to the $b_2$ coefficient in~\eqref{eq:a4-bulk} for every KK supergravity field.

	\item While imposing conformal boundary conditions on the metric fluctuations ensures that the differential operator $\mathcal{Q}_h$ is elliptic, this choice falls outside of the class for which explicit surface contributions to the SdW coefficient are known.
	
	\item Computing the non-local contribution $C_\text{non-local}$ involves a spectral problem that we cannot solve in general for arbitrary Einstein-Maxwell backgrounds.
\end{enumerate}
In view of these difficulties, it would seem that a first-principle computation of $C$ for various 4d supergravity backgrounds still eludes us. However, we will now leverage holography and explain how the various logarithmic coefficients $\mathcal{C}$ on the field theory side (see Table~\ref{tab:CFT-log}) can be used to ``bootstrap'' the heat kernel coefficients $(a_E,c,b_1,b_2)$ in the bulk. This will allow us to elegantly circumvent the first two issues in the above list. The main idea is to study the logarithmic corrections as a function of various backgrounds supporting arbitrary quadratic field fluctuations, rather than studying individual fields in the KK spectrum evolving on a fixed background. Before we illustrate how this procedure works in detail on several examples, we note that from the expressions for the SdW coefficient in \eqref{eq:a4} and \eqref{eq:a4-bulk}, it is clear that the contribution to $C_\text{local}$ from the $a_E$ coefficient comes accompanied by the integrated Euler density $\int E_4$. This integral yields a topological invariant of the 4d Euclidean manifold $\mathcal{M}$ used as a supergravity background, i.e. its Euler characteristic~$\chi(\mathcal{M})$, and thus the contribution of $a_E$ to $C_\text{local}$ is always a pure number independent of any continuous parameters that may be present in the supergravity background. As we show below this general expectation is realized in all examples we study.

\subsection{EAdS$_4$}
\label{sec:bootstrap:EAdS4}

We start with the pure EAdS$_4$ solution to \eqref{eom} given by
\begin{equation}
\label{sol:EAdS4}
	ds^2 = \frac{L^2}{L^2 + r^2}\,dr^2 + r^2\,d\Omega_3^2 \, , \qquad A = 0 \, ,
\end{equation}
where $d\Omega_3^2$ is the metric on the round $S^3$ with unit radius. This space is conformally flat and so has vanishing Weyl tensor. Introducing appropriate counter-terms, the regularized Euler characteristic and $R^2$ invariant are given by
\begin{equation}
\label{eq:chi-AdS}
	\chi = \frac{1}{32\pi^2}\int d^4 x \sqrt{g}\,E_4 = 1 \, , \qquad \frac{1}{32\pi^2} \int d^4 x \sqrt{g}\,R^2 = 6 \, .
\end{equation}
Thus, we find that an arbitrary fluctuation of a field $\phi$ (of general mass and spin) with corresponding kinetic operator $\mathcal{Q}_\phi$ propagating in this background produces a bulk contribution to the integrated SdW coefficient~\eqref{eq:a4-bulk} given by
\begin{equation}
\label{eq:a4-bulk-round}
	\int d^4x \sqrt{g}\,a_4^{\text{bulk}}(\mathcal{Q}_\phi) = 2\bigl[6\,b_1(\phi) - a_E(\phi)\bigr] \, .
\end{equation}
Observe that the coefficient of $a_E$ equals $-2\,\chi$ with $\chi$ given in~\eqref{eq:chi-AdS}. In addition, the total derivative contribution~\eqref{eq:a4-tot-der} vanishes trivially since the bulk Ricci scalar is constant and $A=0$. It remains to analyze the boundary contributions \eqref{eq:a4-details:bdry}. The extrinsic curvature of EAdS$_4$ in the coordinate system \eqref{sol:EAdS4} reads 
\begin{equation}
	K_{cd} = \frac{\sqrt{L^2 + r_b^2}}{L\,r_b}\,\delta_{cd}  = \frac{1}{L}\,\delta_{cd} +\mathcal{O}(r_b^{-2}) \, ,
\end{equation}
where $r_b$ is the location of the boundary in the radial direction. This in turn determines the projectors $\Pi_\pm$ and $S$ specifying the boundary conditions. Note that following the discussion in Section~\ref{sec:sugra:local:bdry}, the latter can be parameterized as $S = S' K$ where the $S'$ field space matrix is constant. Working in a radial expansion with a large cutoff $r_b$, we find the following surface contribution to the integrated SdW coefficient:
\begin{equation}
\label{eq:a4-bdry-AdS4}
	\begin{split}
		\int_{\partial\mathcal M}&d^3y\,\sqrt{\gamma}\,a_4^{\text{bdry}}(\mQ_\phi) \\
		=&\; \frac{1}{360}\Tr\Bigl[45L^2 E(1 - 6 S') - 2(\Pi_- - 29\Pi_+) + 1620S'^2(1 - S') - 93\Bigr]\,\frac{r_b^3}{L^3} \\[1mm]
		&\;+ \frac{1}{240}\Tr\Bigl[15L^2 E(1 - 6 S') - 2(\Pi_- - 29\Pi_+) - 180S'(2 - 9S' + 9S'^2) - 31\Bigr]\,\frac{r_b}{L} \\[1mm]
		&\;+ \mathcal{O}(r_b^{-1}) \, ,
	\end{split}
\end{equation}
where $\Pi_\pm$, $S'$ and $E$ are all independent of the radial cutoff. This boundary contribution can be holographically renormalized upon introducing the counter-term
\begin{equation}
\label{eq:CT}
	\int_{\partial \mathcal{M}} d^3y \,\sqrt{\gamma}\,\bigl(\mathfrak{c}_1 + \mathfrak{c}_2 \mathcal{R}\bigr) \, ,
\end{equation}
where $\mathcal{R}$ denotes the Ricci scalar of the boundary. The constant $\mathfrak{c}_1$ can be chosen so as to cancel the cubic divergence in~\eqref{eq:a4-bdry-AdS4} while $\mathfrak{c}_2$ can be tuned to cancel the linear term. Importantly, this renormalization scheme does not introduce any constant term on EAdS$_4$. In this way, the boundary contribution to the SdW coefficient vanishes by holographic renormalization. With this prescription, the total local contribution \eqref{eq:C-local} is
\begin{equation}
	C_{\text{local}} = 2\sum_\phi (-1)^F\,\bigl[6\,b_1(\phi) - a_E(\phi)\bigr] \, .\label{eq:C-local:EAdS4}
\end{equation}
Here we have taken the sum of the contributions of all fields propagating in the background weighted by their fermion number $F$.\\

Let us make the following comment on our result. Using the heat kernel coefficients $(a_E,b_1)$ given in Table~\ref{tab:coeffs} and expressing the masses of the bulk fields in terms of conformal dimensions of the dual operators using Table~\ref{tab:conf-m}, we find that the summand in~\eqref{eq:C-local:EAdS4} can be written as
\begin{equation}
6\,b_1(\phi) - a_E(\phi) = \frac{2s+1}{48}\,\Bigl[\Bigl(\Delta - \frac32\Bigr)^4 - \Bigl(s + \frac12\Bigr)^2\Bigl(2\Bigl(\Delta - \frac32\Bigr)^2 + \frac16\Bigr) - \frac7{240}\Bigr] \, ,
\end{equation}
for any bosonic field of spin $s \in \{0,1,2\}$, and as
\begin{equation}
6\,b_1(\phi) - a_E(\phi) = \frac{2s+1}{48}\,\Bigl[\Bigl(\Delta - \frac32\Bigr)^4 - \Bigl(s + \frac12\Bigr)^2\Bigl(2\Bigl(\Delta - \frac32\Bigr)^2 - \frac13\Bigr) + \frac1{30}\Bigr] \, ,
\end{equation}
for any fermionic field of spin $s \in \{1/2,3/2\}$. In other words, we have
\begin{equation}
\label{eq:Cloc-S3-G}
	C_{\text{local}} = \sum_\phi\,G_s\Bigl(\Delta - \frac32\Bigr) \, , 
\end{equation}
where the function $G_s$ is defined by
\begin{equation}
	G_s(x) = \begin{cases} \frac{2s+1}{24}\,\bigl(x^4 - \bigl(s + \frac12\bigr)^2\bigl(2x^2 + \frac16\bigr) - \frac7{240}\bigr) \quad &s \in \{0,1,2\} \\[2mm] -\frac{2s+1}{24}\,\bigl(x^4 - \bigl(s + \frac12)^2\bigl(2x^2 - \frac13\bigr) + \frac1{30}\bigr) \quad &s \in \{1/2,3/2\} \end{cases} \, .
\end{equation}
The appearance of this function is not an accident. Rather, as shown in~\cite{Camporesi:1993mz,Liu:2016dau,Binder:2021euo}, the function $G_s$ is precisely the one that controls the logarithmic divergence in the free energy $F_{\Delta,s}$ of a free field in EAdS$_4$ dual to an operator of spin $s$ and conformal dimension $\Delta$,
\begin{equation}
\label{eq:3-sphere-log}
F_{\Delta,s} = G_s\Bigl(\Delta - \frac32\Bigr)\log(L/\epsilon) + \text{finite} \, ,
\end{equation}
where $\epsilon$ is a UV cutoff. The result~\eqref{eq:3-sphere-log} has been obtained by a direct spectral analysis on pure Euclidean AdS$_4$, and what we have shown is that our heat kernel coefficients precisely reproduce the expected logarithmic behavior. Note that to achieve this agreement, it was crucial to use the precise values of the $a_E$ and $b_1$ coefficients given in Table~\ref{tab:coeffs}, which in particular take into account the contribution from St\"uckelberg fields when $1 \leq s \leq 2$. Therefore, we can view~\eqref{eq:Cloc-S3-G} as a non-trivial check of our computations in Appendix~\ref{App:coeffs}. Moreover, we can now argue on the basis of~\eqref{eq:3-sphere-log} that there cannot be local contributions to the EAdS$_4$ logarithmic correction due to the oblique nature of boundary conditions on the metric fluctuations, in agreement with our assumption at the end of Section~\ref{sec:sugra:local:bdry}.\footnote{Recall from~\eqref{eq:Cnon-local-AdS4} that the non-local contribution vanishes in the absence of bulk massless 2-forms in the spectrum, which is assumed in deriving~\eqref{eq:3-sphere-log}.} In the simple EAdS$_4$ background, we can therefore address all three points raised at the beginning of the present section and conclude that the logarithmic correction to the path integral is entirely controlled by $C=C_{\text{local}}$ given in~\eqref{eq:C-local:EAdS4}. To compare with the CFT coefficient $\mathcal{C}$, we still need to explicitly sum over the field content of the relevant KK supergravity theory and use a suitable regularization. We will discuss this in Section~\ref{sec:KK}, but for the time being we move on to other minimal supergravity backgrounds.

\subsection{Euclidean Romans}
\label{sec:bootstrap:Romans}

Next we consider the Euclidean Romans solution to~\eqref{eom} given by \cite{Romans:1991nq,Bobev:2020pjk}
\begin{equation}
	\label{eq:romans}
	\begin{split}
		ds^2 =&\; U(r)\,d\tau^2 + \frac{dr^2}{U(r)} + r^2 ds^2_{\Sigma} \, , \qquad U(r) = \bigg(\frac{r}{L} + \frac{\kappa L}{2r}\bigg)^2 - \frac{Q^2}{4r^2} \, , \\
		dA =&\;  \frac{Q}{2r^2}\,d\tau\wedge dr + \frac{\kappa L}{2}\,V_{\Sigma} \, ,
	\end{split}
\end{equation}
where $\kappa\in\{0,\pm1\}$ is the normalized curvature of the Riemann surface $\Sigma$ and $Q$ is a free, continuous parameter. Evaluating the relevant curvature-squared terms with appropriate counter-terms, we find that the bulk contribution \eqref{eq:a4-bulk} for a field $\phi$ of general spin and mass is given by
\begin{equation}
\label{eq:a4-bulk:Romans}
	\begin{split}
	\int d^4x \sqrt{g}\,a_4^{\text{bulk}}(\mathcal{Q}_\phi) =&\; -4(1 - \mathfrak{g})a_E(\phi) + \frac{3(|Q| + \kappa L)^2}{10\pi L|Q|}\text{vol}(\Sigma)\,c(\phi) \\
	&\;- \frac{3(|Q| - \kappa L)^2}{2\pi L |Q|}\text{vol}(\Sigma)\,b_1(\phi) - \frac{3(Q^2 + \kappa^2 L^2)}{4\pi L |Q|}\text{vol}(\Sigma)\,b_2(\phi) \, .
	\end{split}
\end{equation}
The coefficient of $a_E$ equals $-2\,\chi$ where the regularized Euler characteristic is $\chi = 2(1 - \mathfrak{g})$. We stress that these results are valid for quadratic fluctuations of an arbitrary field $\phi$ around the Romans background. Using the same holographic renormalization prescription as in the case of pure EAdS$_4$, the total derivative and boundary contributions to $C_{\text{local}}$ can be shown to vanish. This implies that the only contribution to \eqref{eq:C-local} reads
\begin{equation}
\label{eq:C-local:Romans}
	C_\text{local} = \sum_\phi (-1)^F\int d^4x\,\sqrt{g}\,a_4^\text{bulk}(\mQ_\phi) \, ,
\end{equation}
with the integrated $a_4^\text{bulk}(\mQ_\phi)$ given in \eqref{eq:a4-bulk:Romans} and the sum is taken over all fields propagating in the background weighted by their fermion number $F$.

The 4d Euclidean path integral around the Romans background is holographically dual to the TTI of the boundary SCFT on $S^1 \times \Sigma_\mathfrak{g}$. The logarithmic coefficient $\mathcal{C}$ extracted from the large $N$ limit of the TTI is given in Table~\ref{tab:CFT-log} for various SCFTs, and one readily checks that this coefficient is always a pure number. On the bulk side, we have argued in Section~\ref{sec:sugra:non-local-gen} that the non-local contribution to $C$ is a pure number for arbitrary backgrounds. Therefore, holography dictates that $C_{\text{local}}$ given in~\eqref{eq:C-local:Romans} should also be a pure number, and in particular should be independent of the continuous parameter $Q$ of the background. This requirement then translates into a linear constraint on the heat kernel coefficients summed over the spectrum,
\begin{equation}
\label{constraints:1}
\begin{split}
	\sum_{\phi}(-1)^F\bigl[2\,c(\phi) - 10\,b_1(\phi) - 5\,b_2(\phi)\bigr] = 0 \, .
\end{split}
\end{equation}
We stress that this should be true in any four-dimensional KK supergravity theory whose SCFT dual falls in one of the classes discussed in Table~\ref{tab:CFT-log}. Indeed, this constraint must be statisfied in order to fit the expectations of the AdS/CFT correspondence at order $\mathcal{O}(\log N)$ in the large $N$ expansion.

The background~\eqref{eq:romans} admits a supersymmetric limit as $Q \rightarrow 0$. For hyperbolic Riemann surfaces, i.e. for $\mathfrak{g}>1$, the resulting geometry has a Lorentzian interpretation as a magnetically charged Reissner-Nordström (RN) black hole in AdS$_4$. In Section~\ref{sec:log-conj} we will comment on the implications of our results for the thermodynamics of AdS black holes.

\subsection{U(1)$\times$U(1) squashing}
\label{sec:bootstrap:U(1)-squash}

Another solution to \eqref{eom} we can consider is one where the conformal boundary is a squashed 3-sphere. A particular squashing that preserves supersymmetry and a U(1)$\times$U(1) subgroup of the ${\rm SO}(4)$ isometry group can be obtained in the bulk by turning on an anti-self-dual graviphoton compared to the pure EAdS$_4$ case. In a convenient coordinate system, this squashed background takes the form~\cite{Martelli:2011fu}
\begin{align}
	ds^2 =&\; f_1(x,y)^2\,dx^2 + f_2(x,y)^2\,dy^2 + L^2\,\frac{(x^2 - 1)(y^2 - 1)}{\mathrm{s}^2 - 1}\,d\phi_1^2 + L^2\,\frac{(\mathrm{s}^2 - x^2)(y^2 - \mathrm{s}^2)}{\mathrm{s}^2(\mathrm{s}^2-1)}\,d\phi_2^2 \, , \nonumber \\
	A =&\; \frac{L}{2(x+y)}\bigg(\frac{\mathrm{s}^2 + xy}{\mathrm{s}}\,d\phi_2 - (1 + xy)\,d\phi_1\bigg) \, ,
\end{align}
where the squashing parameter is $\mathrm{s}$ (not to be confused with the spin $s$ of a field), and 
\begin{equation}
	f_1(x,y)^2 = L^2\frac{y^2 - x^2}{(x^2 - 1)(\mathrm{s}^2 - x^2)} \, , \qquad f_2(x,y)^2 = L^2\frac{y^2 - x^2}{(y^2 - 1)(y^2 - \mathrm{s}^2)} \, .
\end{equation}
Computing the relevant curvature invariants, we obtain the bulk contribution
\begin{equation}
\label{eq:a4-bulk-U1-squash}
	\int d^4x \sqrt{g}\,a_4^\text{bulk}(\mQ_\phi) = 2\bigl[6\,b_1(\phi) - a_E(\phi)\bigr] - \frac{3(\mathrm{s}-1)^2}{2\,\mathrm{s}}\,b_2(\phi) \, ,
\end{equation}
for any field $\phi$ of general mass and spin propagating on this background. Again, the coefficient of $a_E$ in this expression is minus twice the Euler characteristic of the squashed background, which is topologically indistinguishable from the pure EAdS$_4$ geometry.

When $\mathrm{s}=1$, we recover the result for pure EAdS$_4$ with a round 3-sphere boundary. In addition, our renormalization prescription ensures that there are no other contributions to $C_{\text{local}}$ besides~\eqref{eq:a4-bulk-U1-squash}. Following the logic described in the previous subsection, AdS/CFT requires that the above integrated SdW coefficient be independent of the squashing parameter after summing over all fields in the spectrum.\footnote{The universal CFT coefficient $\mathcal{C}$ in Table~\ref{tab:CFT-log} has been obtained analytically in the large $N$ limit of the squashed 3-sphere partition function for $\mathrm{s}=1$ \cite{Fuji:2011km,Marino:2011eh} and $\mathrm{s}=3$ \cite{Hatsuda:2016uqa}, and conjectured for a generic squashing parameter in \cite{Bobev:2022eus,Bobev:2023lkx}. One-loop calculations in the dual 11d backgrounds, however, make the universality of the logarithmic coefficient clear for any value of $\mathrm{s}$, since in this approach the logarithmic coefficient is entirely controlled by the number of 11d zero-modes \cite{Bhattacharyya:2012ye} that should not depend on any continuous parameter.} This immediately imposes
\begin{equation}
	\sum_{\phi}(-1)^F b_2(\phi) = 0 \, ,
\end{equation}
which shows that while we do not have access to the $b_2$ coefficient for a given field in general, universality of the logarithmic correction and holography imply a strong constraint on the total contribution after summing over the spectrum. This addresses the first point raised at the beginning of this section, at least partially. We can combine the above constraint with~\eqref{constraints:1} to obtain a more stringent relation among the remaining heat kernel coefficients,
\begin{equation}
\label{constraints:2}
	\sum_{\phi}(-1)^F\bigl[c(\phi) - 5\,b_1(\phi)\bigr] = 0 \, .
\end{equation}

\subsection{SU(2)$\times$U(1) squashing }
\label{sec:bootstrap:SU(2)-squash}

We now consider another supersymmetric bulk solution whose conformal boundary is a squashed 3-sphere, with the squashing preserving an SU(2)$\times$U(1) isometry. The minimal supergravity background is given by \cite{Martelli:2011fw}
\begin{equation}
	\begin{split}
		ds^2 =&\; \frac{r^2 - \mathrm{s}^2}{\Omega(r)}\,dr^2 + (r^2 - \mathrm{s}^2)(\sigma_1^2 + \sigma_2^2) + \frac{4\mathrm{s}^2\Omega(r)}{r^2 - \mathrm{s}^2}\,\sigma_3^2 \, ,\\
		A =&\; \mathrm{s}\,\sigma_3\,\frac{r-\mathrm{s}}{r+\mathrm{s}}\sqrt{\frac{4\mathrm{s}^2}{L^2} - 1} \, ,
	\end{split}
\end{equation}
where
\begin{equation}
	\Omega(r) = (r - \mathrm{s})^2\bigg(1 + \frac{(r-\mathrm{s})(r+3\mathrm{s})}{L^2}\bigg) \, ,
\end{equation}
$\sigma_i$ are the SU(2) left-invariant one-forms, and $\mathrm{s}$ is the squashing parameter. We can proceed as before and compute the renormalized bulk contribution \eqref{eq:a4-bulk} which now takes the form 
\begin{equation}
\begin{split}
	\int d^4x \sqrt{g}\,a_4^\text{bulk}(\mQ_\phi) =&\; -2\,a_E(\phi) + \frac{2(L^2 - 4\mathrm{s}^2)^2}{L^4}\,c(\phi) \\
&\;+ \frac{96\mathrm{s}^2(L^2 - 2\mathrm{s}^2)}{L^4}\,b_1(\phi) + \frac{24\mathrm{s}^2(L^2 - 4\mathrm{s}^2)}{L^4}\,b_2(\phi) \, ,
\end{split}
\end{equation}
and the surface terms again vanish by holographic renormalization. Since the logarithmic term in the dual squashed sphere partition function is independent of the squashing parameter, we can apply the same logic as before and deduce that AdS/CFT imposes a new constraint on heat kernel coefficients summed over the spectrum. Together with the previous constraints \eqref{constraints:1} and \eqref{constraints:2}, we arrive at the final conclusion that
\begin{equation}
	\label{eq:no-Cloc}
	\sum_{\phi}(-1)^Fc(\phi) = \sum_{\phi}(-1)^Fb_1(\phi) = \sum_{\phi}(-1)^Fb_2(\phi) = 0 \, .
\end{equation}
In other words, the universal character of the logarithmic corrections to the large $N$ limit of SCFT partition functions imposes, via holography, that the sum of the heat kernel coefficients $(c,b_1,b_2)$ over the spectrum of the theory should vanish. This must be true \emph{regardless of the particular field content of the 4d KK supergravity theory under consideration}. As we will see below, these constraints have important consequences. Recall also that the~$a_E$ coefficient is not subject to any constraint since the Euler characteristic $\chi(\mathcal{M})$ cannot depend on continuous parameters.

\subsection{AdS-Taub-NUT}
\label{sec:AdSTNsol}

The bootstrap result (\ref{eq:no-Cloc}) is obtained by studying the Euclidean path integrals around three distinct supersymmetric backgrounds holographically dual to 3d $\mN=2$ SCFT partition functions whose logarithmic coefficients have been argued to be pure numbers by supersymmetric localization and the index theorem. As we discuss in Section~\ref{sec:Kundera}, one can derive similar constraints on heat kernel coefficients without the need to assume $\mN=2$ supersymmetry. To illustrate this point we now consider two different non-BPS supergravity backgrounds.

As a first example, we study the following Euclidean AdS-Taub-NUT background~\cite{Emparan:1999pm}
\begin{equation}
ds^2 = 4n^2 V(r)\,\sigma_3^2 + V(r)^{-1}dr^2 + (r^2 - n^2)(\sigma_1^2 + \sigma_2^2) \, ,
\end{equation}
where
\begin{equation}
V(r) = \frac{r^2 + n^2 - 2mr + L^{-2}(r^4 - 6n^2r^2 - 3n^4)}{r^2 - n^2} \, .
\end{equation}
The above metric solves the Einstein equations in \eqref{eom} with a vanishing Maxwell field, and the solution fully breaks supersymmetry. The mass parameter $m$ is related to the NUT charge $n$ as
\begin{equation}
m = n - \frac{4n^3}{L^2} \, ,
\end{equation}
and the boundary at $r\rightarrow \infty$ is a squashed 3-sphere with metric
\begin{equation}
ds^2_\infty = r^2\Bigl(\sigma_1^2 + \sigma_2^2 + \frac{1}{1+\mathrm{s}}\sigma_3^2\Bigr) \, .
\end{equation}
The squashing parameter is related to the NUT charge as
\begin{equation}
n = \frac{L}{2\sqrt{1+\mathrm{s}}} \, .
\end{equation}
With this at hand, we compute the bulk part of the integrated fourth SdW coefficient:
\begin{align}
\int d^4 x \sqrt{g}\,a_4^\text{bulk}(\mathcal{Q}_\phi) =&\; \frac{8}{\sqrt{1+\mathrm{s}}}\bigl[6\,b_1(\phi) - a_E(\phi)\bigr]\,\frac{r_b^3}{L^3} - \frac{6}{(1+\mathrm{s})^{3/2}}\bigl[6\,b_1(\phi) - a_E(\phi)\bigr]\,\frac{r_b}{L} \nonumber \\
&\;+ \frac{2}{(1+\mathrm{s})^2}\Bigl[6\,b_1(\phi) - (1 + \mathrm{s}^2)\,a_E(\phi) + \mathrm{s}^2\,c(\phi)\Bigr] \\[1mm]
&\;+ \mathcal{O}(r_b^{-1}) \, , \nonumber
\end{align}
where we have introduced a boundary cut-off $r_b$. Using the counter-term~\eqref{eq:CT}, the divergent terms above and the surface part of the integrated SdW coefficient can both be holographically renormalized away, and we find that the local contribution to the logarithmic correction fo the AdS-Taub-NUT free energy $F(\mathrm{s})$ is given by
\begin{equation}
\label{eq:Cloc-TN}
	C_{\text{local}}(\mathrm{s}) = -2\sum_{\phi}(-1)^F\Bigl[a_E(\phi) - \frac{1}{(1+\mathrm{s})^2}\Bigl(\mathrm{s}^2 c(\phi) + 6(1 + 2\mathrm{s})\,b_1(\phi)\Bigr)\Bigr] \, ,
\end{equation}
where as usual we need to add up the heat kernel coefficients of all bulk fields in the gravitational effective theory. In addition, we do not expect the non-local contribution to depend on the squashing parameter since they arise from zero modes. The correction~\eqref{eq:Cloc-TN} translates into a logarithmic correction to the stress tensor two-point function coefficient $C_T$ in the boundary CFT via the relation \cite{Bobev:2017asb}
\begin{equation}
\label{eq:CT-def}
	C_T = -\frac{48}{\pi^2}\,\partial_{\mathrm{s}}^2 F(\mathrm{s})\big\vert_{\mathrm{s} = 0} \, .
\end{equation}
Using the above, we thus find that $C_T$ generically contains a term
\begin{equation}
\label{eq:CT-gen}
	C_T \ni \frac{192}{\pi^2}\sum_{\phi}(-1)^F\bigl[6\,b_1(\phi) - c(\phi)\bigr] \log N \, .
\end{equation}
If we now assume that the stress tensor two-point function coefficient in any holographic CFT does not contain a logarithmic term in its large $N$ expansion, we arrive at the bootstrap constraint 
\begin{equation}
	\sum_{\phi}(-1)^F\bigl[6\,b_1(\phi) - c(\phi)\bigr] = 0 \, .
\end{equation}
This is compatible with our previous results using supersymmetric backgrounds and SCFT observables. We will discuss our assumption about the absence of log terms in $C_T$ and further implications of this result in Section~\ref{sec:Kundera}.

\subsection{Kerr-Newman}
\label{sec:KNsol}

Lastly, we come to the Euclidean incarnation of the 4d electric Kerr-Newman (KN) black hole solution in AdS. The metric and Maxwell field take the form~\cite{Bobev:2019zmz,Caldarelli:1998hg,Cassani:2019mms}
\begin{align}
	\label{eq:KN-AdS}
            ds^2 =&\; \frac{\Delta_r}{V}\left( d\tau + \frac{\alpha}{\Xi}\sin^2\theta d\phi \right)^2 + V \left( \frac{dr^2}{\Delta_r} + \frac{d\theta^2}{\Delta_\theta}\right) + \frac{\Delta_\theta\sin^2\theta}{V}\left( \alpha d\tau - \frac{\tilde{r}^2 - \alpha^2}{\Xi} d\phi \right)^2\, ,\nonumber\\
            A =&\; \mathrm{i}\,m \sinh(2\delta) \frac{\tilde{r}}{V} \left( d\tau + \frac{\alpha}{\Xi} \sin^2\theta d\phi \right)\, , 
\end{align}
where
\begin{equation}
\label{eq:KN-param}
\begin{split}
            \tilde{r} =&\; r + 2m \sinh^2 \delta \, ,\quad \Xi = 1 + \frac{\alpha^2}{L^2}\, ,\quad V = \tilde{r}^2 - \alpha^2 \cos^2\theta\, ,\\
            \Delta_r =&\; r^2 - \alpha^2 - 2mr + \frac{\tilde{r}^2}{L^2}(\tilde{r}^2 - \alpha^2)\, ,\quad \Delta_\theta = 1 + \frac{\alpha^2}{L^2}\cos^2\theta\, .
\end{split}
\end{equation}
The parameters $(m,\alpha,\delta)$ are related to the energy, angular momentum and electric charge of the solution. After a lengthy computation, we find that the bulk contribution to the integrated SdW coefficient can be holographically renormalized to give
\begin{equation}
\begin{split}
\int d^4 x \sqrt{g}\,a_4^{\text{bulk}}(\mathcal{Q}_\phi) =&\; - 4\,a_E(\phi) + f_1(m,\alpha,\delta)\,c(\phi) \\
&\; + f_2 (m,\alpha, \delta)\,b_1(\phi) + f_3 (m,\alpha, \delta)\,b_2(\phi)\, ,
\end{split}
\end{equation}
where the various functions on the right-hand side are given by
\begin{align}
\label{eq:fi-KN-AdS}
	f_1 =&\; \frac{\beta  m^2}{2\pi \alpha ^5 \Xi}\Biggl\{\frac{3 m^2 c^4 s^4}{\tilde{r}_{+}^4}(\alpha ^4-\tilde{r}_{+}^4) \log\Bigl(\frac{\tilde{r}_{+}+\alpha}{\tilde{r}_{+}-\alpha }\Bigr) -\frac{16 \alpha^5 m c^2 s^2 (c^2+s^2) \tilde{r}_{+}^3 (3 \tilde{r}_{+}^2+\alpha ^2)}{\tilde{r}_{+}^3(\tilde{r}_{+}^2-\alpha ^2)^3} \nonumber \\
&\;\; +\frac{2 \alpha(m^2 c^4 s^4(3 \tilde{r}_{+}^8-8 \alpha ^6 \tilde{r}_{+}^2+42 \alpha ^4 \tilde{r}_{+}^4-8 \alpha ^2 \tilde{r}_{+}^6+3 \alpha ^8)+4 \alpha ^4 (c^2+s^2)^2 \tilde{r}_{+}^4 (\tilde{r}_{+}^2+\alpha ^2))}{\tilde{r}_{+}^3(\tilde{r}_{+}^2-\alpha ^2)^3}\Biggr\}\, , \nonumber\\[1mm]
	f_2 =&\; \frac{12 \beta}{\pi L^4 \Xi}  \left(\alpha ^2 \tilde{r}_{+}-\tilde{r}_{+}^3+mL^2(c^2 + s^2)\right) \, , \nonumber \\[3mm]
	f_3 =&\; \frac{24 \beta m^2  c^2 s^2 \tilde{r}_+}{\pi \Xi (\tilde{r}_{+}^2-\alpha^2)} \, . 
\end{align}
In these expressions we have used the shorthand $s=\sinh\delta$ and $c = \cosh\delta$. The quantity $\tilde{r}_+ = r_+ + 2m\sinh^2\delta$ is given in~\eqref{eq:KN-param} with $r_+$ being the largest real root of $\Delta_{r}(r_+) = 0$, and $\beta$ is the periodicity of the Euclidean time circle. Regularity of the solution at $r=r_+$ fixes the value of $\beta$ in terms of the solution parameters as
\begin{equation}
\beta = 4\pi(\tilde{r}^2 - \alpha^2)\Bigl[\frac{d\Delta_r}{dr}\Bigr]^{-1}\Big\vert_{r=r_+} \, .
\end{equation}
The mass parameter $m$ can also be related to the value of the radial coordinate $r_+$ at which the $\tau$ circle shrinks to zero size, see e.g.~\cite{Cassani:2019mms}.

Some interesting special cases can be studied from the above general expressions. First, if we set
\begin{equation}
\label{eq:KN-BPS}
\alpha = \mathrm{i}L\bigl(\coth(2\delta) - 1\bigr) \, ,
\end{equation}
the solution~\eqref{eq:KN-AdS} preserves two real supercharges. In this supersymmetric limit, it is not hard to see from~\eqref{eq:fi-KN-AdS} that the bulk contribution to the integrated SdW coefficient will in general depend on the value of the continuous parameter $\delta$. The holographically dual description of this gravitational solution is in terms of the superconformal index of the dual SCFT, i.e. the supersymmetric partition function on $S^1\times_\omega S^2$ with the angular fugacity $\omega$ related to $\delta$. As summarized in Table~\ref{tab:CFT-log}, the logarithmic contribution to the superconformal index is independent of $\omega$ and therefore $C_{\text{local}}$ should not depend on $\delta$. This in turn leads to the following constraints for the total values of the heat kernel coefficients for fields propagating on the AdS-KN background
\begin{equation}
	\label{eq:no-ClocKN}
\sum_{\phi}(-1)^Fc(\phi) = \sum_{\phi}(-1)^Fb_1(\phi) = \sum_{\phi}(-1)^Fb_2(\phi) = 0 \, .
\end{equation}
This confirms the result we have obtained previously in~\eqref{eq:no-Cloc}. We can also consider the non-BPS limit of vanishing electric charge $\delta \rightarrow 0$ in which case the above solution reduces to the Euclidean version of the AdS-Kerr black hole. In this limit,
\begin{equation}
\label{eq:fi-K-AdS}
\begin{split}
	f_1(m,\alpha,0) =&\; \frac{\beta (r_{+}^2+L^2)^2 (r_{+}^2+\alpha ^2)}{\pi  L^4 \Xi \,r_{+} (r_{+}^2-\alpha^2)}\, , \\[1mm]
	f_2(m,\alpha,0) =&\; -\frac{6 \beta (r_{+}^2-L^2) (r_{+}^2-\alpha^2)}{\pi  L^4 \Xi  \,r_{+}} \, , \\[1mm]
	f_3(m,\alpha,0) =&\; 0 \, .
\end{split}
\end{equation}
The AdS-Kerr solution is holographically dual to the grand canonical thermal partition function of a CFT placed on $S_\beta^1 \times S^2$, where the circle has size $\beta$ which sets the (inverse) temperature. This QFT observable can be studied in a saddle-point approximation which defines the so-called thermal effective action of the theory, see~\cite{Benjamin:2023qsc}. The thermal effective action of a given holographic CFT may contain a logarithmic term in the large $N$ limit. If we assume that the coefficient of this $\log N$ term does not depend on the temperature or the spin fugacity, then by our general reasoning the dependence of $C_{\text{local}}$ on the dual continuous parameters arising from~\eqref{eq:fi-K-AdS} can only be suppressed if we impose 
\begin{equation}
\sum_{\phi}(-1)^Fc(\phi) = \sum_{\phi}(-1)^Fb_1(\phi) = 0 \, . 
\end{equation}
This example illustrates how the strong constraints on the spectrum of the gravitational theory we have obtained are not necessarily a consequence of supersymmetry. Rather, they can be implied more generally by the topological nature of logarithmic terms in the large $N$ limit of holographic CFT observables. We are not aware of a general proof that the $\log N$ term in the thermal effective action of any holographic CFT cannot depend on chemical potentials, so our statements in the non-supersymmetric setting are weaker than the ones we made using supersymmetric observables. It would be very interesting to sharpen these statements using QFT arguments.

\section{Explicit KK supergravity examples}
\label{sec:KK}

We have just learned that AdS/CFT dictates that the sum over the spectrum of individual heat kernel coefficients $(c,b_1,b_2)$ must vanish as in~\eqref{eq:no-Cloc}. In this section, we will explicitly check this property by studying the logarithmic coefficient $C$ in the semi-classical expansion of the 4d Euclidean path integral \eqref{eq:sugra-saddle} in certain supergravity theories for which the KK spectrum is known. We will focus in the present section on supergravity theories with at least $\mathcal{N}=2$ supersymmetry. Generically, supersymmetry requires the presence of non-minimal couplings that enter the quadratic expansion of the action around an arbitrary background. Thus, it would seem that we cannot make use of the results we have obtained in Table~\ref{tab:coeffs} for minimally coupled fields. However, we can adapt a useful trick put forward in~\cite{Sen:2012kpz} in the context of asymptotically flat space-times to the AdS setting. Rather than repeating the trace computations in Appendix~\ref{App:coeffs} with non-minimal couplings, we will instead supersymmetrize the $C_{\text{local}}$ term~\eqref{eq:C-local} directly. To do so, we use the results of~\cite{Bobev:2021oku} to relate the four-derivative invariants controlling the fourth SdW coefficient. On any background of $\mathcal{N}=2$ minimal supergravity, this relation reads
\begin{equation}
\label{eq:tshirt}
W^2\vert_{\mathcal{N}=2} = E_4 + \frac{4}{L^2}\,\Bigl[R + \frac{6}{L^2} - F_{\mu\nu}F^{\mu\nu}\Bigr] \, ,
\end{equation}
modulo terms that vanish using the equations of motion~\eqref{eom}. Here the subscript on the left-hand side denotes the supersymmetrization of the Weyl-squared density~\eqref{eq:W2}, see~\cite{Bobev:2021oku} for the explicit expression. The relation~\eqref{eq:tshirt} implies that the bulk part of the fourth SdW coefficient~\eqref{eq:a4-bulk} in all theories that can be expressed in an $\mathcal{N}=2$ language is straightforwardly supersymmetrized:
\begin{equation}
(4\pi)^2a_4^{\text{bulk}}\vert_{\mathcal{N}\geq 2} = (c - a_E)\,E_4 + \Bigl(b_1 - \frac{c}{6}\Bigr)\,R^2 + \Bigl(b_2 + \frac{c}{3}\Bigr)\,RF_{\mu\nu}F^{\mu\nu} \, .
\end{equation}
Since all non-minimal couplings required by supersymmetry have been taken into account in this formula, we can use the minimally coupled results of Table~\ref{tab:coeffs} to infer the value of the heat kernel coefficients $(a_E,c,b_1)\vert_{\mathcal{N}\geq2}$ in supersymmetric theories,
\begin{equation}
\label{eq:hk-non-min}
a_E\vert_{\mathcal{N}\geq2} = a_E - c \, , \qquad c\vert_{\mathcal{N}\geq2} = 0 \, , \qquad b_1\vert_{\mathcal{N}\geq2} = b_1 - \frac{c}{6} \, .
\end{equation}
In the rest of this chapter, we will systematically compute heat kernel coefficients as if all fields were minimally coupled according to Table~\ref{tab:coeffs}, and only restore the effect of non-minimal couplings using~\eqref{eq:hk-non-min} when necessary. We also note that if a particular spectrum produces a vanishing total $\sum_\phi (-1)^F c(\phi) = 0$, the relation~\eqref{eq:hk-non-min} immediately shows that $\mathcal{N}\geq2$ non-minimal couplings have no bearing on the other heat kernel coefficients. We will make extensive use of this fact  below when studying specific KK spectra.

\subsection{KK supergravity on $S^7$}
\label{sec:KK:S7}

As a first example, we consider the simple case of eleven-dimensional supergravity compactified on the round $S^7$. Since the length scales of the internal and external spaces are of the same order, this compactification gives rise to maximal gauged supergravity in 4d along with an infinite KK tower of massive fields. All fields can be arranged into 4d $\mathcal{N}=8$ supermultiplets, and the complete spectrum was obtained long ago in~\cite{Freedman:1983na,Sezgin:1983ik,Biran:1983iy,Casher:1984ym,Duff:1986hr}. In what follows, we will denote the SO(8) $R$-symmetry representations by their Dynkin labels $(\alpha_1,\alpha_2,\alpha_3,\alpha_4)$ and follow the convention of Appendix A in~\cite{Bobev:2021wir} for the SO(8) triality frame. The massless and massive field content of the KK supergravity theory is indexed by a single positive integer $k$ that labels the KK level and also controls the mass of the field. The spectrum is summarized in Table~\ref{tab:S7} and Table~\ref{tab:S7-m}. \\
\begin{table}[t]
	\centering
	\renewcommand*{\arraystretch}{1.4}
	\begin{tabular}{|c|c|c|c|}
		\hline
		spin & SO(8) irrep & Dynkin label & $\Delta$ \\
		\hline
		2 & $\mathbf{1}$ & $(0,0,0,0)$ & 3 \\
		\hline
		3/2 & $\mathbf{8}_s$ & $(0,0,0,1)$ & $\frac52$ \\
		\hline
		1 & $\mathbf{28}$ & $(0,1,0,0)$ & 2 \\
		\hline
		1/2 & $\mathbf{56}_s$ & $(1,0,1,0)$ & $\frac32$ \\
		\hline
		$0_+$ & $\mathbf{35}_v$ & $(2,0,0,0)$ & 1 \\
		\hline
		$0_-$ & $\mathbf{35}_c$ & $(0,0,2,0)$ & 2 \\
		\hline
	\end{tabular}
	\caption{The massless $\mathcal{N}=8$ supermultiplet.\label{tab:S7}}
\end{table}
\begin{table}[t]
	\centering
	\renewcommand*{\arraystretch}{1.4}
	\begin{tabular}{|c|c|c|}
		\hline
		spin & Dynkin label & $\Delta$ \\
		\hline
		2 & $(k,0,0,0)_{k\geq0}$ & $3 + \frac{k}{2}$ \\
		\hline
		\multirow{2}{*}{3/2} & $(k,0,0,1)_{k\geq0}$ & $\frac52 + \frac{k}{2}$ \\[-1mm] 
		& $(k-1,0,1,0)_{k\geq1}$ & $\frac72 + \frac{k}{2}$ \\
		\hline
		\multirow{3}{*}{1} & $(k,1,0,0)_{k\geq0}$ & $2 + \frac{k}{2}$ \\[-1mm] 
		& $(k-1,0,1,1)_{k\geq1}$ & $3 + \frac{k}{2}$ \\[-1mm]
		& $(k-2,1,0,0)_{k\geq2}$ & $4 + \frac{k}{2}$ \\
		\hline
		\multirow{4}{*}{1/2} & $(k+1,0,1,0)_{k\geq0}$ & $\frac32 + \frac{k}{2}$ \\[-1mm] 
		& $(k-1,1,1,0)_{k\geq1}$ & $\frac52 + \frac{k}{2}$ \\[-1mm]
		& $(k-2,1,0,1)_{k\geq2}$ & $\frac72 + \frac{k}{2}$ \\[-1mm]
		& $(k-2,0,0,1)_{k\geq2}$ & $\frac92 + \frac{k}{2}$ \\
		\hline
		\multirow{3}{*}{$0_+$} & $(k+2,0,0,0)_{k\geq0}$ & $1 + \frac{k}{2}$ \\[-1mm] 
		& $(k-2,2,0,0)_{k\geq2}$ & $3 + \frac{k}{2}$ \\[-1mm]
		& $(k-2,0,0,0)_{k\geq2}$ & $5 + \frac{k}{2}$ \\
		\hline
		\multirow{2}{*}{$0_-$} & $(k,0,2,0)_{k\geq0}$ & $2 + \frac{k}{2}$ \\[-1mm] 
		& $(k-2,0,0,2)_{k\geq2}$ & $4 + \frac{k}{2}$ \\
		\hline
	\end{tabular}
	\caption{Massive $\mathcal{N}=8$ supermultiplets at KK level $k$. For $k=0$ we recover Table~\ref{tab:S7}.\label{tab:S7-m}}
\end{table}

To compute the logarithmic correction to the path integral around a given 4d background $\mathcal{M}$ in this KK theory, we use the general formulae of Section~\ref{sec:sugra}. The local contribution \eqref{eq:C-local} is controlled by the SdW coefficient, which receives a bulk contribution of the form~\eqref{eq:a4-bulk}. In Appendix~\ref{App:KK:S7}, we explain how to compute the heat kernel coefficients $(a_E,c,b_1)$ by combining the results summarized in Table~\ref{tab:coeffs} together with the KK spectrum given in Tables~\ref{tab:S7} and~\ref{tab:S7-m}. For the $c$ and $b_1$ coefficients, we find a non-trivial cancellation ``level-by-level'',\footnote{This result is reminiscent of the level-by-level cancellation in the one-loop beta function of $\mN=8$ gauged supergravity observed in~\cite{Gibbons:1984dg}.}
\begin{equation}
	\label{eq:c-b1-S7}
	c(k) = 0 \, , \qquad b_1(k) = 0 \, , \qquad \forall \; k \geq 0 \, .
\end{equation}
Since the sum over the spectrum reduces in this simple case to a sum over the KK level $k$, we obtain
\begin{equation}
	c^{\text{tot}} \equiv \sum_{k\geq 0} (-1)^F c(k) = 0 \, , \quad \text{and} \quad b_1^{\text{tot}} \equiv \sum_{k\geq 0} (-1)^F b_1(k) = 0 \, .
\end{equation}
Since $c^{\text{tot}}$ vanishes,~\eqref{eq:hk-non-min} implies that the vanishing of $b_1^{\text{tot}}$ also holds for non-minimally coupled fields around any background $\mathcal{M}$. Likewise, the vanishing of $\sum_k (-1)^F c(k)\vert_{\mathcal{N}=8}$ automatically follows from~\eqref{eq:hk-non-min}. These results are in perfect agreement with the bootstrap constraints in~\eqref{eq:no-Cloc} applied to KK supergravity on the 7-sphere. Having knowledge of the full spectrum also allows us to compute the $a_E$ coefficient, for which no such level-by-level cancellation occurs. Instead we find that its contribution to $C_{\text{local}}$ is given by
\begin{equation}
\label{eq:aE-S7}
	-\frac{1}{72}\,\chi(\mathcal{M})\sum_{k\geq 0}(k+1)(k+2)(k+3)^2(k+4)(k+5) \, ,
\end{equation}
where $\chi(\mathcal{M})$ is the regularized Euler characteristic of the 4d background. For a generic background $\mathcal{M}$, another contribution to $C_{\text{local}}$ is of the form
\begin{equation} 
\label{eq:b2-S7}
	\frac{1}{16\pi^2}\Bigl(\int_\mathcal{M} d^4x \sqrt{g}\,R\,F_{\mu\nu}F^{\mu\nu}\Bigr)\,b_2^{\text{tot}} \, ,
\end{equation}
where $b_2^{\text{tot}} = \sum_k (-1)^F b_2(k)$, but this term vanishes according to~\eqref{eq:no-Cloc}. Lastly, the local part of the logarithmic correction also receives contributions from surface terms in the expression of the SdW coefficient~\eqref{eq:a4}. Since these cannot be written compactly for generic backgrounds and boundary conditions, we generically denote them by $S(\mathcal{M})$ after summing over the spectrum. With this notation, the full local contribution is
\begin{equation}
\label{eq:Cloc-S7-M}
	C_{\text{local}}(\mathcal{M}) = S(\mathcal{M}) - \frac{1}{72}\,\chi(\mathcal{M})\sum_{k\geq 0}(k+1)(k+2)(k+3)^2(k+4)(k+5) \, .
\end{equation}
Clearly this expression contains divergent sums that need to be appropriately regularized. Before doing so, we recall that yet another contribution to the logarithmic term is given by $C_{\text{non-local}}$ in~\eqref{eq:C-non-local}. As discussed in Section~\ref{sec:sugra:non-local}, the corresponding spectral problem for generic backgrounds $\mathcal{M}$ remains out of reach at present. So from now on, we will focus on specific backgrounds of the $S^7$ KK supergravity theory where we can make progress.\\

An interesting (and tractable) case is to come back to the situation where $\mathcal{M}$ is pure EAdS$_4$. There, we argued in Section~\ref{sec:bootstrap} that the surface terms could be holographically renormalized away prior to summing over the spectrum of the KK theory. As a result, we have $S(\text{EAdS}_4) = 0$. In addition, the KK spectrum does not include any 2-form fields, which by~\eqref{eq:Cnon-local-AdS4} implies that the non-local contribution vanishes. In total, we arrive at the logarithmic correction to the path integral
\begin{equation}
\label{eq:C-tot-AdS4}
C(\text{EAdS}_4) = -\frac{1}{72}\,\sum_{k\geq 0}(k+1)(k+2)(k+3)^2(k+4)(k+5) \, ,
\end{equation}
where we have used that $\chi(\text{EAdS}_4) = 1$. Note that we could have arrived at this result starting from~\eqref{eq:Cloc-S3-G} and explicitly implementing the sum over the spectrum by keeping track of the dimensions of the SO(8) representations and the conformal dimensions of the dual operators at a given KK level. Indeed,~\eqref{eq:C-tot-AdS4} is compatible with~\cite{Camporesi:1993mz,Liu:2016dau,Binder:2021euo}. 

What remains is to discuss a possible regularization of the infinite series. For instance, the method used in~\cite{Liu:2016dau} consists of introducing a $z^k$ regulator to the summand with $|z| < 1$ and defining the regularized series as the finite term in the expansion around $z=1$. This attaches the value
\begin{equation}
\label{eq:C-tot-AdS4-reg-Liu}
	C(\text{EAdS}_4) \stackrel{?}{=} 0 \, ,
\end{equation} 
to the logarithmic coefficient in the one-loop EAdS$_4$ free energy. In Appendix~\ref{app:reg}, we present two other methods based on zeta-function regularization that yield the same result. However,~\eqref{eq:C-tot-AdS4-reg-Liu} should give us pause. Indeed, if correct, we are then forced to conclude that the bulk computation of the logarithmic coefficient cannot possibly agree with the holographically dual ABJM result
\begin{equation}
\label{eq:C-M2-round}
\mathcal{C}(S^3_{b=1}) = -\frac14 \, ,
\end{equation}
given in Table~\ref{tab:CFT-log}. At this stage, it seems that there are a couple of possible resolutions for this discrepancy. Recall that in 11d supergravity on EAdS$_4 \times S^7$~\cite{Bhattacharyya:2012ye}, the logarithmic term comes entirely from the zero mode of the 2-form ghost required to quantize the 3-form potential of the theory. After dimensional reduction, we have insisted that the KK supergravity fields arrange themselves into $\mathcal{N}=8$ multiplets that do not accommodate 2-form fields. As a result, the non-local contribution in the 4d theory vanishes on EAdS$_4$ according to~\eqref{eq:Cnon-local-AdS4}. To remedy this, we could opt to dualize some (or all) of the massless scalars in the KK theory to 2-forms and hope to obtain a non-zero contribution to the $C_{\text{non-local}}$ part of the logarithmic correction. While this change of duality frame explicitly breaks the $\mathcal{N}=8$ multiplet structure, it may be necessary to obtain a match between bulk and boundary contributions at the one-loop level. However, a short computation shows that this cannot resolve the discrepancy. The heat kernel coefficient $a_E$ for fluctuations of a massless 2-form can be obtained by the same trace computations that led to the results in Table~\ref{tab:coeffs}. It was shown in~\cite{Larsen:2015aia} that there is a simple relation between the $a_E$ coefficient for a massless scalar field and a massless 2-form,
\begin{equation}
	a_E(\text{2-form}) = a_E(\text{scalar}) - \frac12 \, .
\end{equation}
This shows that dualizing $q$ scalar fields into 2-forms shifts the local contribution to the logarithmic coefficient as
\begin{equation}
	C_{\text{local}} \longrightarrow C_{\text{local}} + q \, .
\end{equation}
On the other hand, the non-local contribution also acquire a shift after dualizing,
\begin{equation}
C_{\text{non-local}} \longrightarrow C_{\text{non-local}} - q \, ,
\end{equation}
where we have used~\eqref{eq:Cnon-local-AdS4} with $j=0$. In this case, the two shifts precisely cancel and we see that the total logarithmic coefficient $C(\text{EAdS}_4)$ does not depend on the duality frame. This shows that dualizing massless fields cannot help in resolving the discrepancy between bulk and boundary computations. A second possible resolution of the tension between~\eqref{eq:C-tot-AdS4-reg-Liu} and~\eqref{eq:C-M2-round} would be to find a regularization scheme that yields a non-zero result for the total $\sum_k (-1)^F a_E(k)$ contribution to $C_{\text{local}}$. Although we have discussed three distinct methods to regularize the series and arrive at a vanishing result, it is interesting to note that the regularization
\begin{equation}
\label{eq:good-reg}
\sum_{k\geq 0}(k+1)(k+2)(k+3)^2(k+4)(k+5) = 24 \, ,
\end{equation}
is compatible with AdS/CFT expectations. Indeed, this would imply 
\begin{equation}
C(\text{EAdS}_4) = -\frac13 \, , 
\end{equation}
which together with the holographic dictionary for class I SCFTs~\eqref{eq:holo-dict-gen}, $L^2/G_N \sim N^{\frac32}$, precisely matches~\eqref{eq:C-M2-round}. Unfortunately, we are not aware of a rigorous way to arrive at the result~\eqref{eq:good-reg}. If it does exist, it will be interesting to understand why holography seems to favor this putative regularization method over the ones regularizing the infinite sum to zero.\\

We can provide arguments in favor of the regularization~\eqref{eq:good-reg} by considering supergravity backgrounds $\mathcal{M}$ that are not pure EAdS$_4$. For the 4d background whose conformal boundary is the U(1)$\times$U(1) squashed 3-sphere, we have also explained previously how the surface terms can be renormalized away. The Euler characteristic of this background is the same as pure EAdS$_4$ since $\chi$ cannot depend on the squashing parameter, and therefore~\eqref{eq:good-reg} implies that the Euclidean path integral around the squashed background contains a logarithmic term $-\frac14\log N$, again in agreement with the ABJM results in Table~\ref{tab:CFT-log}.

Let us also consider the EAdS$_2 \times \Sigma_{\mathfrak{g}}$ near-horizon region of the Romans solution with genus $\mathfrak{g} > 1$. As we have explained in Section~\ref{sec:bootstrap:Romans} the surface term can be holographically renormalized to zero. In addition, we have $\chi(\text{EAdS}_2 \times \Sigma_{\mathfrak{g}}) = 2(1 - \mathfrak{g})$ and putting this together with the non-local contribution in~\eqref{eq:C:non-local-AdS2xS}, we obtain
\begin{equation}
C(\text{EAdS}_2 \times \Sigma_{\mathfrak{g}}) = (\mathfrak{g}-1)\Bigl[64 + \frac{1}{36}\sum_{k\geq 0}(k+1)(k+2)(k+3)^2(k+4)(k+5)\Bigr] - 4 \, .
\end{equation}
Regardless of the regularization chosen for the infinite series, this cannot match the logarithmic coefficient in the TTI of the boundary theory given in Table~\ref{tab:CFT-log},
\begin{equation}
\label{eq:C-M2-TTI}
\mathcal{C}(S^1\times\Sigma_{\mathfrak{g}}) = \frac12(\mathfrak{g} - 1) \, .
\end{equation}
However, the comparison in this case is more subtle since there could be degrees of freedom localized outside the near-horizon region of the Romans solution that contribute to the logarithmic term seen from the asymptotic boundary. In fact, it was argued in~\cite{Liu:2017vbl} that taking into account the full Romans geometry modifies the zero-mode counting in a drastic way compared to Section~\ref{sec:sugra:non-local-AdS2xS}. In particular, they argued that on the full geometry,
\begin{equation}
\label{eq:Cnon-local-Romans}
	C_{\text{non-local}}(\text{Romans}) = 2(1 - \mathfrak{g})\sum_{\text{massless 2-form}}(-1)^F(-j-1) \, .
\end{equation}
In the absence of two-forms in the spectrum, we therefore find
\begin{equation}
C(\text{Romans}) = \frac{2}{3}(\mathfrak{g} - 1) \, ,
\end{equation}
after using~\eqref{eq:good-reg}. With the holographic dictionary relevant for M2-branes, this again perfectly matches~\eqref{eq:C-M2-TTI} found in the ABJM TTI.

One last example we can consider is the supersymmetric AdS-KN background holographically dual to the SCI of the boundary theory. Because we can again set $S(\text{KN}) = 0$ using holographic renormalization, combining $\chi(\text{KN}) = 2$ together with the regularization~\eqref{eq:good-reg} yields
\begin{equation}
C(\text{KN}) = -\frac23 + C_{\text{non-local}}(\text{KN}) \, .
\end{equation}
For this to match the ABJM result
\begin{equation}
\mathcal{C}(S^1 \times_\omega S^2) = -\frac12 \, ,
\end{equation} 
upon using $L^2/G_N \sim N^{\frac32}$, the zero-modes of various Laplacian operators on the AdS-KN background must be such that the non-local contribution~\eqref{eq:C-non-local} vanishes. This is a highly non-trivial prediction for the spectral problem of Laplace-type differential operators on this complicated background.

\subsection{A conjecture for $C$ and black hole entropy}
\label{sec:log-conj}

At this stage, we assume that the regularization~\eqref{eq:good-reg} can be made rigorous and we will use it to predict the logarithmic correction of general 4d backgrounds. For this, we observe that in all bulk geometries holographically dual to the squashed 3-sphere, TTI and SCI observables of the ABJM theory, the non-local contribution vanishes and the logarithmic term comes entirely from the heat kernel coefficient $a_E^{\text{tot}}$ after summing over the spectrum. The situation should be contrasted with the 11d computations in~\cite{Bhattacharyya:2012ye, Liu:2017vbl} where the coefficient of the log correction can only come from zero modes since the heat kernel vanishes in odd dimensions. We now would like to conjecture that, in all minimal gauged supergravity background that are relevant for AdS$_4$/CFT$_3$, a non-local contribution to $C$ can only  arise from the presence of massless 2-forms in the spectrum of the theory. In KK supergravity on $S^7$, maximal supersymmetry in 4d does not accommodate such 2-forms (in a duality frame where we keep all 70 massless scalars), which then becomes the reason behind the fact that $C_{\text{non-local}}$ vanishes. This conjecture is certainly compatible with~\eqref{eq:Cnon-local-AdS4} and~\eqref{eq:Cnon-local-Romans}, and it would be most interesting to derive similar formulae for other asymptotically locally AdS$_4$ backgrounds. 

Assuming that the general picture holds, we can go ahead and compute the logarithmic correction in the Euclidean path integral around any $S^7$ KK supergravity background $\mathcal{M}$. The result for $C$ takes a very simple form:
\begin{equation}
\label{eq:C-S7-gen}
	C(\mathcal{M}) = -\frac13\,\chi(\mathcal{M}) \, ,
\end{equation}
where we have also assumed that the boundary contribution $S(\mathcal{M})$ can be set to zero using holographic renormalization on $\mathcal{M}$ with the counter-term~\eqref{eq:CT}. Recall that this was checked explicitly for all backgrounds of Section~\ref{sec:bootstrap}. We note that~\eqref{eq:C-S7-gen} takes the same functional form as the one derived in~\cite{Hristov:2021zai} based on 4d $\mathcal{N}=2$ gauged supergravity localization. In that work, the specific field content of the theory was left arbitrary and the dependence of the logarithmic correction on the Euler characteristic of $\mathcal{M}$ was derived using an index theorem. In this framework, each multiplet of the 4d $\mathcal{N}=2$ theory contributes a fixed amount to $C$, and that amount was computed for vector and hypermultiplets in~\cite{Hristov:2021zai}. However, the contribution from the graviton and KK multiplets was left unspecified. Our heat kernel result~\eqref{eq:C-S7-gen} effectively resums these contributions in the case of KK supergravity on $S^7$. Moreover, it predicts a $\log N$ term in the large $N$ limit of the dual ABJM free energy on $\partial\mathcal{M}$ with a coefficient
\begin{equation}
\label{eq:C-M2-gen}
	\mathcal{C}(\partial\mathcal{M}) = -\frac14\,\chi(\mathcal{M}) \, .
\end{equation}
This holographic formula agrees with the SCFT results in Table~\ref{tab:CFT-log} but is also valid regardless of the amount of supersymmetry, so that we can use it to make predictions for non-BPS backgrounds with interesting CFT duals. One such example is the AdS-Taub-NUT solution discussed in Section~\ref{sec:AdSTNsol}. We have already shown in~\eqref{eq:c-b1-S7} that $c^{\text{tot}} = b_1^{\text{tot}} = 0$ in KK supergravity on $S^7$. According to~\eqref{eq:CT-gen}, this implies that there is no logarithmic term in the large $N$ expansion of the stress tensor two-point function coefficient $C_T$ in ABJM theory. This is in agreement with our general formula~\eqref{eq:C-M2-gen}, since the Euler characteristic of AdS-Taub-NUT is a constant equal to one, and taking derivatives with respect to the squashing parameter or the NUT charge as in~\eqref{eq:CT-def} trivially gives zero.

Another non-BPS background we can consider is the so-called AdS soliton~\cite{Horowitz:1998ha}. The Euclidean metric reads
\begin{equation}
\label{eq:AdS-soliton}
ds^2 = \frac{r^2}{L^2}\Bigl[\Bigl(1 - \frac{r_0^3}{r^3}\Bigr)\,d\tau^2 + dx^2 + dy^2\Bigr] + \frac{L^2}{r^2}\Bigl(1 - \frac{r_0^3}{r^3}\Bigr)^{-1}dr^2 \, ,
\end{equation}
and the asymptotic boundary at $r\rightarrow \infty$ is the three-manifold $S^1_\beta \times \mathbb{R}^2$. Here the inverse temperature $\beta$ is related to the parameter $r_0$ by demanding absence of conical singularity in the bulk, which leads to $r_0 = 4\pi L^2/3\beta$. When the ABJM theory is put on this thermal background, the free energy computes the one-point function of the stress tensor which is non-vanishing since conformal invariance is broken by the finite temperature.\footnote{For a review of thermal observables in CFT see~\cite{Iliesiu:2018fao}, and for a discussion of higher-derivative effects in the stress tensor one-point function for general $\mathcal{N}=2$ holographic CFTs see~\cite{Bobev:2023ggk}.} Using~\eqref{eq:AdS-soliton}, it is straightforward to show that the Euler characteristic of the AdS soliton vanishes. Thus, by~\eqref{eq:C-M2-gen}, we predict that the large $N$ expansion of the stress tensor one-point function coefficient does not contain a $\log N$ term,
\begin{equation}
\mathcal{C}(S^1_\beta \times \mathbb{R}^2) = 0 \, .
\end{equation}
To derive this result from a QFT analysis seems arduous, because it pertains to strongly-coupled theories at finite temperature. Our holographic formula~\eqref{eq:C-M2-gen} elegantly sidesteps this difficulty and allows us to make a strong prediction.\\

In light of~\eqref{eq:C-S7-gen}, it is also instructive to consider non-BPS asymptotically AdS black hole backgrounds embedded in 11d supergravity on $S^7$. The simplest such example is the Euclidean AdS-Schwarzschild solution
\begin{equation}
ds^2 = \Bigl(\frac{r^2}{L^2} + 1 - \frac{m}{r}\Bigr)\,d\tau^2 + \Bigl(\frac{r^2}{L^2} + 1 - \frac{m}{r}\Bigr)^{-1}dr^2 + r^2d\Omega_2^2 \, .
\end{equation}
The location of the outer horizon $r_+$ is related to the mass parameter as $m = r_+(r_+^2/L^2 + 1)$. The regularized Euler characteristic is given by $\chi(\text{AdS-Sch}) = 2$ which, by~\eqref{eq:C-S7-gen}, implies that the logarithm of the Euclidean path integral in the semi-classical expansion contains a $\log L$ term:
\begin{equation}
\label{eq:IonshelSch}
I = I_0 + \frac13\log(L^2/G_N) + \mathcal{O}(1) \, .
\end{equation}
Here $I_0$ is the classical on-shell action of the solution. We can now make use of the quantum statistical relation~\cite{Gibbons:1976ue,Gibbons:2004ai}
\begin{equation}
I = \beta M - S \, ,
\end{equation}
to find that the AdS-Schwarzschild black hole entropy receives a logarithmic correction at one-loop
\begin{equation}
S_{\text{AdS-Sch}} = \frac{A_{\rm H}}{4G_N} - \frac13 \log(A_{\rm H}/G_N) + \mathcal{O}(1) \, ,
\end{equation}
with $A_{\rm H} = 4\pi r_+^2$ the horizon area. To obtain this result, we have assumed that we have a large black hole in AdS, i.e. the length scale $r_+$ that sets the size of the black hole horizon is comparable to the AdS scale $L$. In the logarithmic term, this means we can trade $L^2 \sim r_+^2$ up to $\mathcal{O}(1)$ terms.

Using the same approach we can also calculate the correction to the entropy and on-shell action of the general Kerr-Newmann black hole in AdS$_4$ presented in Section~\ref{sec:KNsol}. The quantum statistical relation in this case reads
\begin{equation}
I = \beta M - S - \beta\Phi Q-\beta\omega J \, ,
\end{equation}
where $Q$ and $J$ are the electric charge and angular momentum and $\Phi$ and $\omega$ are the corresponding electric chemical potential and angular velocity. The logarithmic correction to the on-shell action takes the same form as in \eqref{eq:IonshelSch} and since the black hole background is not changed when computing this correction, we conclude that the Bekenstein-Hawking entropy of the AdS-KN black hole is corrected as follows
\begin{equation}\label{eq:SKNlogcorr}
S_{\text{AdS-KN}} = \frac{A_{\rm H}}{4G_N} - \frac13 \log(A_{\rm H}/G_N) + \mathcal{O}(1) \, .
\end{equation}
The explicit expressions for the regularized on-shell action $I_0$ and the horizon area $A_{\rm H}$, as well as those for the black hole charges and fugacities, are quite lengthy and can be found in \cite{Cassani:2019mms,Bobev:2021oku}. Importantly this result for the logarithmic correction to the AdS-KN entropy is valid in a thermodynamic ensemble of fixed $(T,\Phi,\omega)$ since this is the natural ensemble used in the evaluation of the on-shell action. One is of course free to change to a different thermodynamic ensemble by employing a suitable Legendre transformation. However, as emphasized by Sen \cite{Sen:2012dw}, upon such a change in ensemble one should carefully track how the black hole charges and fugacities scale with the large parameter in the theory that controls the logarithmic corrections, i.e. the rank $N$ of the gauge group in our M-theory examples, since this could lead to additional (constant) contributions to the logarithmic coefficient.  

Finally, we note that the AdS-KN black hole admits a supersymmetric limit as discussed in Section~\ref{sec:KNsol}. By taking the $Q\rightarrow 0$ limit of the Romans solution presented in Section~\ref{sec:bootstrap:Romans} with $\mathfrak{g}>1$, we can also study the supersymmetric AdS-Reissner-Nordström (AdS-RN) black hole. In both of these examples we find that the supersymmetric black hole entropy receives the same type of logarithmic correction as in \eqref{eq:SKNlogcorr}, i.e. $- \frac{\chi}{6} \log(A_{\rm H}/G_N)$. For theories arising from M2-branes this can be compared to the logarithmic corrections in the large $N$ limit of the superconformal index (for AdS-KN) or the topologically twisted index (for AdS-RN). Comparing the supergravity logarithmic corrections to the corresponding $\log N$ entries for the $S^1 \times_{\omega} S^2$ and $S^1\times\Sigma_{\mathfrak{g}}$ path integrals in Table~\ref{tab:CFT-log}, we indeed find perfect agreement. This constitutes an important precision test of the microscopic counting of supersymmetric black hole entropy in AdS$_4$ using our 4d supergravity approach.

\subsection{Other KK supergravity examples}
\label{sec:KK:other}

We now proceed with examples of KK compactifications of 11d supergravity to four dimensions with less supersymmetry. We focus on three examples for which the KK spectrum is known in detail and comment on a fourth example that illustrates some challenges and subtleties.

\subsubsection{KK supergravity on $S^7/\mathbb{Z}_k$}
\label{sec:KK-S7-k}

The KK supergravity spectrum of the AdS$_4\times S^7/\mathbb{Z}_k$ 11d supergravity solution\footnote{We caution the reader that here $k$ refers to the order of the orbifold and should not be confused with the label of the KK level in the previous section.} can be obtained by branching the $\mN=8$ KK spectrum labeled by $\mathfrak{so}(8)$ representations into the $\mN=6$ KK spectrum labeled by the $\mathfrak{so}(6)\oplus\mathfrak{u}(1)$ representations and keeping the supermultiplets with vanishing U(1) charges modulo $k$ only \cite{Liu:2016dau}. In Appendix~\ref{App:KK:S7modk} we present the resulting $\mN=6$ KK spectrum following this procedure, and then evaluate the heat kernel coefficients. The KK level is now labelled by two integers $h$ and $r$ that satisfy $h - |r| \geq 0$ and $k\,|\,2r$. We again find a non-trivial cancellation level-by-level for the $c$ and $b_1$ coefficients, which immediately leads to 
\begin{equation}
	c^\text{tot}=b_1^\text{tot}=0 \, .
\end{equation}
This provides another non-trivial test of our bootstrap results in Section~\ref{sec:bootstrap}. The total $a_E$ coefficient is obtained as the following sum over the spectrum,
\begin{equation}
		a_E^{\text{tot}}=\sum_{h=1}^\infty\fft{1+2h}{24}\sum_{\ell=-\left\lfloor h/k'\right\rfloor}^{\left\lfloor h/k'\right\rfloor}\Bigl[h(1+h)(-4+5h+5h^2)+(7-10h-10h^2)(\ell k')^2+5(\ell k')^4\Bigr] \, ,
\end{equation}
where we have defined $k'$ as
\begin{equation}
	k'=\begin{cases}
		k/2 &\text{for} \;\; k\in2\mathbb Z \\
		k &\text{for} \;\; k\in2\mathbb Z+1
	\end{cases} \, ,
\end{equation}
and imposed the constraints $h - |r| \geq 0$ and $k\,|\,2r$ to translate the sum over $r$ to a finite sum over an auxiliary label $\ell$. Computing this finite sum explicitly, we obtain:
\begin{align}
\label{N=6:aEtot:k-finite}
		a_E^{\text{tot}} =&\; \sum_{h=1}^\infty\fft{(1+2h)(1+2\left\lfloor h/k'\right\rfloor)}{72}\bigg[3h(1+h)(-4+5h+5h^2) \\
		&\quad+\left\lfloor h/k'\right\rfloor\left(1+\left\lfloor h/k'\right\rfloor\right)k^2(7-10h+10h^2-k^2)+3\left\lfloor h/k'\right\rfloor^2\left(1+\left\lfloor h/k'\right\rfloor\right)^2k^4\bigg]\,. \nonumber
\end{align}
In addition,~\eqref{eq:no-Cloc} imposes that $b_2^{\text{tot}} = 0$ and we arrive at the conclusion that
\begin{equation}
\label{eq:Cloc-S7-modk}
	C_{\text{local}}(\mathcal{M}) = -2\,\chi(\mathcal{M})\,a_E^{\text{tot}} \, ,
\end{equation}
modulo surface terms $S(\mathcal{M})$ that can be renormalized to zero on all 4d backgrounds of interest. According to our zero mode conjecture in Section~\ref{sec:log-conj},~\eqref{eq:Cloc-S7-modk} is in fact the complete contribution to the logarithmic coefficient in the Euclidean path integral. AdS/CFT then demands that the regularized value of the infinite series~\eqref{N=6:aEtot:k-finite} must be independent of $k$ and equal to $1/6$. Once again, we are not aware of a regularization method that would yield such a simple result. In fact, the situation in KK supergravity on $S^7/\mathbb{Z}_k$ is richer than for $k=1$, since we can now consider the limit where $k\rightarrow \infty$ and study logarithmic corrections to supersymmetric observables in the IIA limit of the ABJM theory. In this limit, we have
\begin{equation}
\label{N=6:aEtot:k-infinite}
	\lim_{k\to\infty}a_E^\text{tot}=\sum_{h=1}^\infty\fft{h(1+h)(1+2h)(-4+5h+5h^2)}{24} \, ,
\end{equation}
and the corresponding supergravity logarithmic coefficient is
\begin{equation}
C(\mathcal{M}) = -2\,\chi(\mathcal{M})\sum_{h=1}^\infty\fft{h(1+h)(1+2h)(-4+5h+5h^2)}{24} \, .
\end{equation}
For this to be compatible with the dual $S^3$ free energy, we must require that the series be regularized to $1/12$ so that $\mathcal{C}(S^3) = -1/6$, after using the dictionary~\eqref{eq:holo-dict-gen} for SCFTs whose dual uplifts to massless Type IIA. However, this regularization then fails to give the correct log coefficient for the IIA limit of the ABJM TTI given in Table~\ref{tab:CFT-log}. We will not dwell on this further, but simply view it as a clear indication that a holographic match heavily depends on the choice of regularization when summing over the infinitely many fields of the dual KK spectra. Moreover, one should keep in mind that in the Type IIA limit there could also be $\log\lambda$ corrections to the free energy, where $\lambda = N/k$ is the 't Hooft coupling. Such terms could compete with the $\log N$ term we study above and disentangling these two contributions could lead to additional subtleties in the bulk supergravity calculation and the holographic comparison.

\subsubsection{mABJM}
\label{subsubsec:mABJM}

We now consider the gravity dual of a 3d $\mathcal{N}=2$ SCFT sometimes referred to as the mABJM theory. The theory can be obtained by focusing on the $k=1,2$ ABJM model and adding a particular superpotential mass term that involves also the light monopole operators responsible for the enhancement to  $\mathcal{N}=8$ supersymmetry in the ABJM theory. The RG flow that ensues ends at an IR fixed point which has been studied with various QFT methods, see \cite{Benna:2008zy,Jafferis:2011zi,Bobev:2018uxk} for more details. Notably the known $\log N$ corrections to various partition functions in the mABJM theory coincide with the ones in the parent ABJM model. 

The holographic studies of the mABJM theory are facilitated by the presence of an explicitly known AdS$_4$ dual background. This AdS solution was first found as a supersymmetric critical point of the potential in the 4d $\mathcal{N}=8$ ${\rm SO}(8)$ gauged supergravity in \cite{Warner:1983vz}. The 4d solution was then uplifted to a background of 11d supergravity in \cite{Corrado:2001nv} which we will refer to as the CPW solution. Importantly the full KK spectrum of the CPW solution was explicitly presented in \cite{Klebanov:2008vq,Malek:2020yue}. This allows us to use the results for the heat kernel coefficients in Table~\ref{tab:coeffs} and calculate the total contributions from all KK modes. The calculation proceeds in an analogous way to the one for the AdS$_4\times S^7$ solution presented in Section~\ref{sec:KK:S7}. There are a number of technical differences due to the lower amount of supersymmetry and the treatment of long multiplets. We discuss all this in Appendix~\ref{App:KK:mABJM} where we present the details of this analysis. While the calculation is somewhat arduous, the final result is easy to state and simple. The $c$ and $b_1$ coefficients once again vanish level-by-level,  and therefore we have
\begin{equation}
c^{\text{tot}} = b_1^{\text{tot}} =0 \, ,
\end{equation}
in agreement with~\eqref{eq:no-Cloc}. The $a_E$ coefficient is non-vanishing and one finds that the sum over KK levels is given by
\begin{equation}
\label{eq:aE-S7}
a_E^{\text{tot}} = \frac{1}{144}\sum_{n\geq 0}(n+1)(n+2)(n+3)^2(n+4)(n+5) \, .
\end{equation}
Notably, this is the same sum we encountered in \eqref{eq:C-tot-AdS4} for the $a^{\text{tot}}_E$ contribution from all KK modes on the round $S^7$. We thus arrive at an interesting observation, namely that the total $(a_E^{\text{tot}},c^{\text{tot}},b_1^{\text{tot}})$ coefficients for the KK spectrum on AdS$_4\times S^7$ are the same as those for the CPW solution. While we do not have a complete explanation for this result, it is reasonable to suspect that this is due to the fact that the two AdS$_4$ solutions are connected by a smooth gravitational domain wall, i.e. a holographic RG flow. Along this flow the spin of all KK modes does not change. The mass of these modes does change, but since they are organized into multiplets of the $\mathcal{N}=2$ supersymmetry preserved by the flow, the results in Table~\ref{HeatKernelCoefficientsMultiplets} dictate that the heat kernel coefficients for these multiplets are independent of the mass. Nevertheless, this line of reasoning does not immediately lead to the equivalence of the two sets of coefficients since the $(a_E,c,b_1)$ coefficients of the KK multiplets of the CPW background depend on the R-charges. Perhaps a more conceptual explanation of the equivalence between the UV and IR heat kernel coefficients may be given by some type of anomaly matching à la 't Hooft, but we were unable to make this precise at this stage. It will certainly be most interesting to understand this phenomenon better and to uncover whether it is a general feature of AdS vacua connected by smooth domain walls.

Under the same assumptions as in the $S^7$ case, we then arrive at the same logarithmic correction in the KK supergravity theory dual to mABJM, namely
\begin{equation}
	C(\mathcal{M}) = -\frac13\chi(\mathcal{M}) \, .
\end{equation}
This once again matches the results obtained in the dual CFT. Indeed, the logarithmic term in the large $N$ limit of the mABJM free energy on $S^3_b$, $S^1 \times \Sigma_\mathfrak{g}$ and $S^1 \times_\omega S^2$ can be obtained using supersymmetric localization by starting from the same UV quiver as that of the ABJM theory. Supersymmetry ensures that the log term in these observables is protected along the flow to the IR, and therefore the boundary logarithmic coefficient $\mathcal{C}(\partial\mathcal{M})$ is expected to be the same as that of the ABJM theory given in Table~\ref{tab:CFT-log}. We refer to~\cite{Bobev:2018wbt,Bobev:2023lkx} for additional details on mABJM supersymmetric observables.

\subsubsection{$N^{0,1,0}$}
\label{subsubsec:N010}

Another example we can study is the Freund-Rubin AdS$_4 \times N^{0,1,0}$ solution of 11d supergravity, see \cite{Castellani:1983yg}. The internal 7d space $N^{0,1,0}$ is Einstein and Tri-Sasakian and thus the dual 3d SCFT preserves $\mathcal{N}=3$ supersymmetry. The KK spectrum for this background has been computed in \cite{Fre:1999gok,Termonia:1999cs} where it was also organized in multiplets of the 3d $\mathcal{N}=3$ superconformal algebra and representations of the additional ${\rm SU}(3)$ isometry of $N^{0,1,0}$, which plays the role of a flavor symmetry in the dual SCFT. To perform this arduous calculation the authors of \cite{Fre:1999gok,Termonia:1999cs} used the fact that $N^{0,1,0}$ is a coset space and therefore the spectrum of various operators on it can be calculated using group theory techniques. The 3d $\mathcal{N}=3$ SCFT dual to the AdS$_4 \times N^{0,1,0}$ solution has been studied in \cite{Gaiotto:2009tk,Hohenegger:2009as,Hikida:2009tp,Cheon:2011th}.

The details of this explicit example of an AdS/CFT dual pair of theories will not be of great importance for our discussion. What is crucial is that on the field theory side one can employ supersymmetric localization to compute the partition function of the SCFT on various compact manifolds, including the omnipresent $\log N$ term of interest to us. We refer to Table~\ref{tab:CFT-log} for more details and a list of relevant references. To access the $\log N$ terms we can follow the logic outlined in Section~\ref{sec:KK:S7} for the AdS$_4\times S^7$ solution of 11d supergravity. To this end we use the explicit results on the KK spectrum from \cite{Fre:1999gok,Termonia:1999cs} to find the contributions of each KK mode to the heat kernel coefficients. The details of this calculation are important but not very illuminating and we present them in Appendix~\ref{App:KK:N010}. Here we will only summarize the results. 

Using the organization of the KK spectrum into multiplets of the $\mathcal{N}=3$ superconformal algebra, we find that each multiplet has vanishing $b_1$ coefficient. Thus, the total contribution from all KK modes is 
\begin{equation}
\label{eq:b1-N3}
b_1^{\rm tot} = 0 \, . 
\end{equation}
The situation is more interesting for the total $c^{\rm tot}$ coefficient. The long superconformal multiplets have vanishing $c$ coefficients, while the various short multiplets have a non-zero value for $c$ that depends on their R-charge. Since we have an infinite number of short multiplets in the KK spectrum, we arrive at a divergent sum indexed by a single integer parametrizing the R-charge. Interestingly, in Appendix~\ref{App:KK:N010}, we find a regularization method based on~\cite{friedman2012special} that yields a vanishing $c^{\text{tot}}$. Combining this with~\eqref{eq:b1-N3} and the discussion around~\eqref{eq:hk-non-min}, we come to the conclusion that $c^{\text{tot}}\vert_{\mathcal{N}=3} = b_1^{\text{tot}}\vert_{\mathcal{N}=3} = 0$ for the $N^{0,1,0}$ spectrum, in accordance with the bootstrap constraints~\eqref{eq:no-Cloc}.\footnote{In contrast, we have checked that the regularization method of~\cite{Liu:2016dau} for $c^{\text{tot}}$ yields results for $b_1^{\text{tot}}\vert_{\mathcal{N}=3}$ that are not compatible with~\eqref{eq:no-Cloc}.} To arrive at this result, we use a ``spin-by-spin'' prescription where the contributions from short graviton, short gravitini and short vector multiplets are regularized individually using the method in Appendix~\ref{app:reg} before assembling them together. We hasten to stress that the method we use to regularize the infinite sums in each spin sector fails to produce the result~\eqref{eq:good-reg} for $a_E^{\text{tot}}$ in KK supergravity on $S^7$, as we have mentionned in Section~\ref{sec:KK:S7}. It is therefore clear that we currently lack a good understanding of the prescriptions needed to obtain results that are compatible with the AdS/CFT correspondence in all situations described so far. Nevertheless, it is encouraging to find some prescription ensuring that the bootstrap constraints~\eqref{eq:no-Cloc} are compatible with a somewhat standard way of regularizing divergent series.

The total $a_E^{\rm tot}$ coefficient is even more involved, since we need to take into account the contribution of all long and short superconformal multiplets in the KK spectrum. Once the dust settles, we find various contributions that are spelled out in detail in Appendix~\ref{App:KK:N010}. In particular, there are double series indexed by two integers coming from the long sector, and we do not know how to systematically regularize such expressions. What we can do is once again combine the various series contributing to $a_E^{\text{tot}}$ with our zero mode conjecture to eventually attach a finite value to all the divergent sums. That value must be compatible with the dual~$N^{0,1,0}$ CFT results presented in Table~\ref{tab:CFT-log}. However, this exercise is not very illuminating and we will not go through it explicitly here.

\subsubsection{$Q^{1,1,1}$}
\label{subsubsec:Q111}

The last example we will study is the AdS$_4 \times Q^{1,1,1}$ solution of 11d supergravity. The dual 3d SCFT preserves $\mathcal{N}=2$ supersymmetry and has SU(2)$^3$ flavor group. The spectrum of the KK theory has been organized into $\mathcal{N}=2$ multiplets in~\cite{Merlatti:2000ed} by leveraging the coset structure of the $Q^{1,1,1}$ internal Sasaki-Einstein space. In general there are short and long multiplets, and they are indexed by the three quantum numbers of SU(2)$^3$ and by the R-charge. We denote the tuple of quantum numbers by $(j_1,j_2,j_3,k)$. In the short sector, the sum over the spectrum reduces to a sum over $k \in \mathbb{N}$ and using the tables provided in~\cite{Merlatti:2000ed} we find
\begin{equation}
\label{eq:HK-Q111-short}
\begin{split}
a_E^\text{short} =&\; \frac18\sum_{k\geq0}(k+1)(13k^2+26k-31) \, , \\
c^\text{short} =&\;  -\frac14\sum_{k\geq0}(k+1)(17k^2 + 58k +109) \, , \\ 
b_1^\text{short} =&\; \frac{1}{24}\sum_{k\geq0}(k+1)(3k^4 + 16k^3 + 84k^2 + 204k + 109) \, .
\end{split}
\end{equation}
We assume a regularization scheme that assigns a finite value to these infinite sums. The long sector is more involved and depends on the SU(2)$^3$ quantum numbers. However, using the spectrum in~\cite{Merlatti:2000ed}, we find that the contributions to the $(a_E, c, b_1)$ coefficients from the long multiplets take a simple form:
\begin{equation}
\label{eq:Q111-long}
(a_E^\text{long},c^\text{long},b_1^\text{long}) = \Bigl(\frac94,-\frac14,-\frac{1}{24}\Bigr) \sum_{k,j_1,j_2,j_3} \text{dim}(j_1,j_2,j_3) \, ,
\end{equation}
where $\text{dim}(j_1,j_2,_3)$ is the dimension of the SU(2)$^3$ representation. We do not know of a rigorous way to regularize the infinite sums on the right-hand side. However, we can turn the logic around and see what can be learned from the bootstrap constraints~\eqref{eq:no-Cloc}. In particular, we see from~\eqref{eq:hk-non-min} and~\eqref{eq:Q111-long} that $b_1^{\text{long}}\vert_{\mathcal{N}=2} = b_1^{\text{long}} - c^{\text{long}}/6 = 0$. Thus, the bootstrap constraints imply that the spectrum in the short sector must be such that
\begin{equation}
\label{eq:HK-Q111-long}
c^\text{short} = 6\,b_1^\text{short}
\end{equation}
to achieve $b_1^\text{short}\vert_{\mathcal{N}=2}=0$. The regularization scheme used for~\eqref{eq:HK-Q111-short} should satisfy the above constraint. It is straightforward to check that using the method in Appendix~\ref{app:reg} to regularize $c^{\text{short}}$ and $b_1^{\text{short}}$, we do not find finite values that satisfy~\eqref{eq:HK-Q111-long}. Since our computation is highly sensitive to the full spectrum of the KK theory, this contradiction could point to issues in the analysis of~\cite{Merlatti:2000ed}. We note that this possibility has been raised independently in~\cite{Cheon:2011th} and further emphasized in \cite{Eager:2013mua}. Of course, we cannot exclude that there exists some renormalization scheme for the sums in the short sector that is compatible with the bootstrap constraints~\eqref{eq:HK-Q111-long}. At present, we are unable to make a definite statement, but we feel that this example highlights how sensitive our log computations are to the intricate details of the KK spectrum in a given supergravity compactification.

\subsection{Heat-kernel coefficients and RG flows}
	
	\tikzstyle{G} = [rectangle, minimum width=3cm, minimum height=1cm, text centered, draw=black, fill=violet!10]
	\tikzstyle{arrow} = [thick,->,>=stealth]
	\begin{figure}[!h]
		\centering
		\small 
		\begin{tikzpicture}[node distance=3cm]
		
		\node (N8SO8) [G, text width=3cm,minimum height=1cm,minimum width=1cm, xshift=0cm] {${\cal N} = 8$, SO(8)};
		
		\node (N1G2) [G, text width=3cm,minimum height=1cm,minimum width=2cm, below of=N8SO8, xshift=0cm] {${\cal N} = 1$, $G_2$};
		
		\node (N1U1xU1) [G, text width=3cm,minimum height=1cm,minimum width=2cm, below of=N8SO8, xshift=-5cm] {${\cal N} = 1$, $U(1)^2$};
		
		\node (N1SO3) [G, text width=3cm,minimum height=1cm,minimum width=2cm, below of=N8SO8, xshift=5cm] {${\cal N} = 1$, $SO(3)$};
		
		\node (N2SU3U1) [G, text width=3cm,minimum height=1cm,minimum width=1cm, below of=N1G2, xshift=0cm] {${\cal N} = 2$, $SU(3)\times U(1)$};

		\draw[-latex, very thick, shorten >=5pt,shorten <=5pt, ->]    (N8SO8) to (N1G2);
		
		\draw[-latex, very thick, shorten >=5pt,shorten <=5pt, ->]    (N1G2) to (N2SU3U1);
		
		\draw[-latex, very thick, shorten >=15pt,shorten <=15pt, ->]    (N8SO8) to (N1U1xU1);
		
		\draw[-latex, very thick, shorten >=15pt,shorten <=15pt, ->]    (N8SO8) to (N1SO3);
		
		\end{tikzpicture}
		\caption{\rm The web of RG flows between holographic CFTs starting from the ABJM theory. } \label{fig:ABJMRGWebs}
	\end{figure}
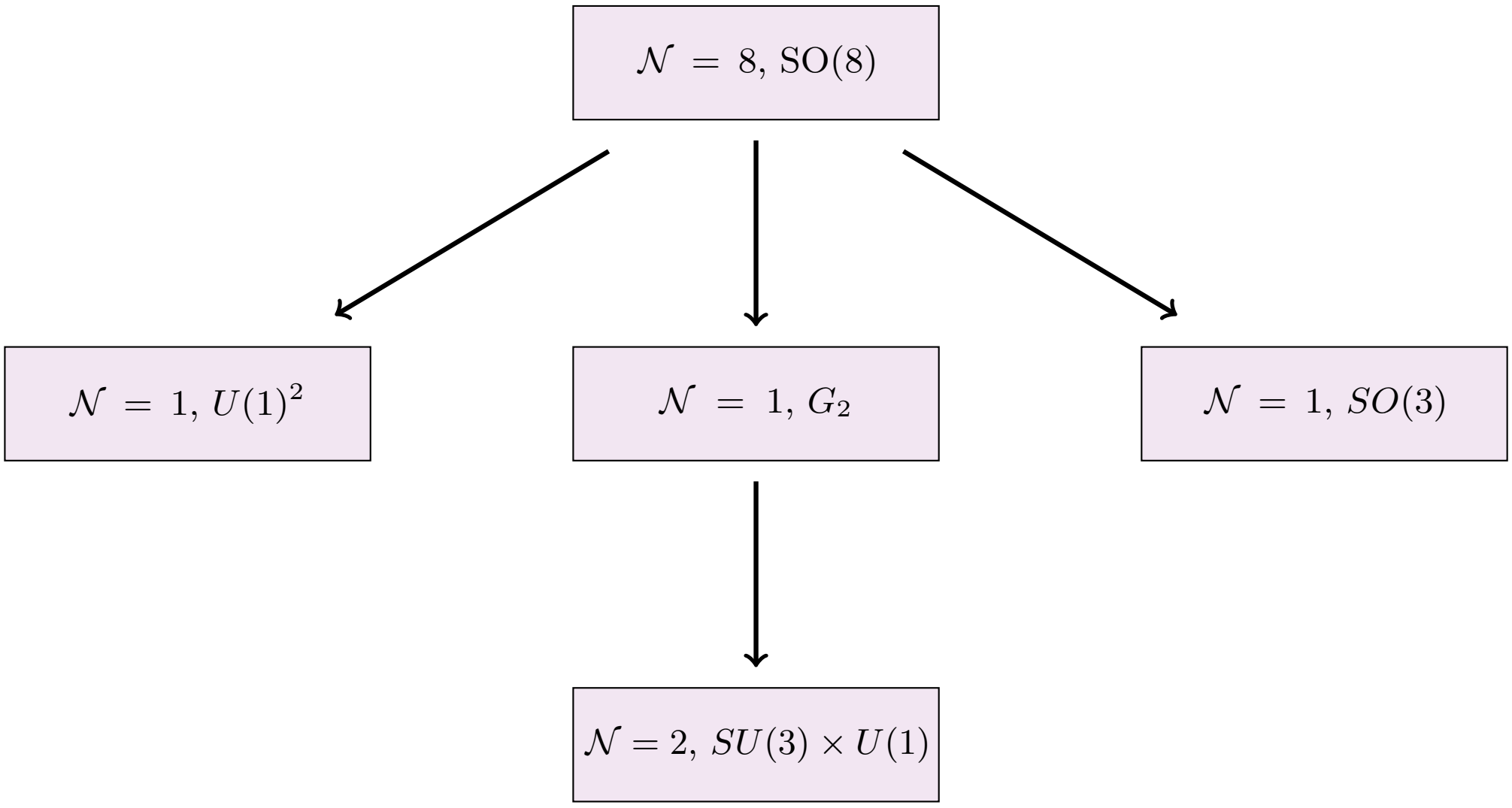
	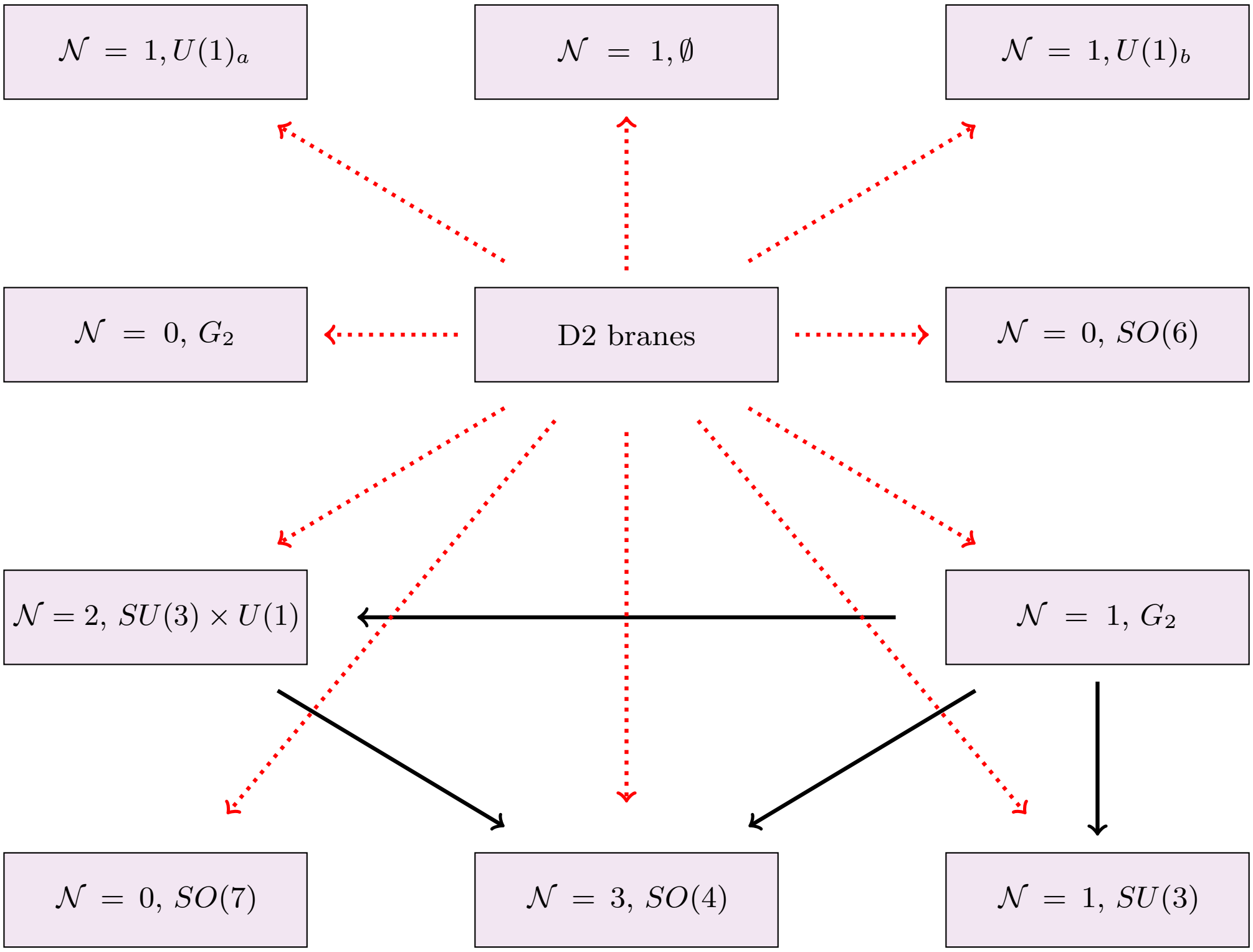
\begin{figure}[!h]
		\small 
	\centering
	\begin{tikzpicture}[node distance=3cm]
	
	\node (1U1a) [G, text width=3cm,minimum height=1cm,minimum width=1cm, xshift=-5cm] {${\cal N}=1, U(1)_a$};
	
	\node (1U1b) [G, text width=3cm,minimum height=1cm,minimum width=1cm, xshift=5cm] {${\cal N}=1, U(1)_b$};
	
	\node (1Phi) [G, text width=3cm,minimum height=1cm,minimum width=1cm, xshift=0cm] {${\cal N}=1, \emptyset $};
	
	
	\node (N0G2) [G, text width=3cm,minimum height=1cm,minimum width=2cm, below of=1Phi, xshift=-5cm] {${\cal N} = 0$, $G_2$};
	
	\node (D2) [G, text width=3cm,minimum height=1cm,minimum width=1cm, below of=1Phi, xshift=0cm] {D2 branes};
	
	\node (N0SO6) [G, text width=3cm,minimum height=1cm,minimum width=2cm, below of=1Phi, xshift=5cm] {${\cal N} = 0$, $SO(6)$};
	
	
	\node (N2SU3U1) [G, text width=3cm,minimum height=1cm,minimum width=1cm, below of=D2, xshift=-5cm] {${\cal N} = 2$, $SU(3)\times U(1)$};
	
	\node (N1G2) [G, text width=3cm,minimum height=1cm,minimum width=2cm, below of=D2, xshift=5cm] {${\cal N} = 1$, $G_2$};
	
	
	\node (N0SO7) [G, text width=3cm,minimum height=1cm,minimum width=2cm, below of=N2SU3U1, xshift=0cm] {${\cal N} = 0$, $SO(7)$};
	
	\node (N3SO4) [G, text width=3cm,minimum height=1cm,minimum width=2cm, below of=N2SU3U1, xshift=5cm] {${\cal N} = 3$, $SO(4)$};
	
	\node (N1SU3) [G, text width=3cm,minimum height=1cm,minimum width=2cm, below of=N1G2, xshift=0cm] {${\cal N} = 1$, $SU(3)$};
	
	\draw[-latex, very thick, shorten >=15pt,shorten <=15pt, ->]    (N1G2) to (N2SU3U1);
	
	\draw[-latex, very thick, shorten >=15pt,shorten <=15pt, ->]    (N2SU3U1) to (N3SO4);
	
	\draw[-latex, very thick, shorten >=15pt,shorten <=15pt, ->]    (N1G2) to (N3SO4);
	
	\draw[-latex, very thick, shorten >=5pt,shorten <=5pt, ->]    (N1G2) to (N1SU3);
	
	\draw[color=red, dotted, -latex, very thick, shorten >=15pt,shorten <=15pt, ->]    (D2) to (1U1a);
	
	\draw[color=red, dotted, -latex, very thick, shorten >=15pt,shorten <=15pt, ->]    (D2) to (1U1b);
	
	\draw[color=red, dotted, -latex, very thick, shorten >=5pt,shorten <=5pt, ->]    (D2) to (1Phi);
	
	\draw[color=red, dotted, -latex, very thick, shorten >=5pt,shorten <=5pt, ->]    (D2) to (N0G2);
	
	\draw[color=red, dotted, -latex, very thick, shorten >=15pt,shorten <=15pt, ->]    (D2) to (N1SU3);
	
	\draw[color=red, dotted, -latex, very thick, shorten >=15pt,shorten <=15pt, ->]    (D2) to (N2SU3U1);
	
	\draw[color=red, dotted, -latex, very thick, shorten >=15pt,shorten <=15pt, ->]    (D2) to (N1G2);
	
	\draw[color=red, dotted, -latex, very thick, shorten >=15pt,shorten <=15pt, ->]    (D2) to (N3SO4);
	
	\draw[color=red, dotted, -latex, very thick, shorten >=15pt,shorten <=15pt, ->]    (D2) to (N0SO7);
	
	\draw[color=red, dotted, -latex, very thick, shorten >=5pt,shorten <=5pt, ->]    (D2) to (N0SO6);
	
	\end{tikzpicture}
	\caption{\rm The web of RG flows between holographic CFTs from the holographic dual of the ISO(7) gauged supergravity uplifting to massive type IIA theory. The thick lines denotes the known RG flows between fixed points reviewed in \cite{Guarino:2019snw}, and the red dashed lines are those coming from the field theory dual of the $\cN=8$ ISO(7) gauged supergravity, which is not conformal. } \label{fig:ISO(7)gaugedSugraRGWebs}
\end{figure}

	In our study of the mABJM theory above, we realize the interesting property that the heat-kernal coefficients $(a_E, c, b_1)$ are identical to those of the ABJM theory for any given Kaluza-Klein level $k$. To make sure that this identity is not a coincidence, we evaluate the coefficients in other gauged supergravity theories whose corresponding field theories are connected by an RG flow to the others. The theories we study as well as the RG flow webs connecting them are shown in Fig. \ref{fig:ABJMRGWebs} and \ref{fig:ISO(7)gaugedSugraRGWebs}. The theories we study include theories connected to the ABJM theory, dual to the $\cN = 8$ SO(8) gauged supergravity; theories descending from the D2-brane theory which is dual to the $\cN = 8$ ISO(7) gauged supergravity\footnote{The field theory dual of this supergravity is not conformal, thus not a valid UV fixed point. } with uplifts in massive type IIA; and $J_n$ theories that are dual to the Janus backgrounds in type IIB supergravity. The details are presented in appendix \ref{App:KK}, where all of the results conducted so far point to the following universal behavior: the coefficients $(a_E, c, b_1)$ are identical among all theories connected by an RG flow. 

	This unexpected coincidence might have something to do with protected quantities such as the 't Hooft anomalies, as discussed in section \ref{sec:KK:other} when we study the mABJM theory. Technically, one could try to explain this for $a_E, c$ coefficients. Because of their independence of the masses, for fields with a given spin, their contribution to the total coefficient is proportional to the dimension of their representations under the global symmetries. As long as the representations are obtained by the branching rules for theories connected by an RG flow, the contributions of each particle to these HK coefficients are always preserved, since the branching rules don't change the total dimension. However, the coefficient $b_1$ does depend on the masses. Since the masses are not preserved along the RG flow, it seems non-trivial that they finally match. However, if both the IR and UV fixed points preserve at least $\cN = 2$ supersymmetry,\footnote{I am not requiring the flow itself to preserve $\cN = 2$ supersymmetry.} and the branching rule doesn't break apart the $\cN = 2$ multiplets\footnote{Such as the RG flow from ABJM to the SU(3)$\times$U(1) massive ABJM.}, since $b_1$ for the long multiplets as a whole are also constants (see Table \ref{HeatKernelCoefficientsMultiplets}), the matching can also be regarded as trivial, except for the short multiplets whose contributions are subleading for very high KK levels. However, for RG flows where the $\cN = 2$ multiplets are broken apart and recombined, the matching is still non-trivial.
	
	The calculations we have done, as shown in the appendix, give strong evidence for the following conjecture:
	\begin{conjecture} \label{LogNRGFlowConjecture}
		All the heat-kernel coefficients ($a_E, c, b_1, b_2$) are identical level-by-level between two SCFTs connected by an RG flow.
	\end{conjecture}
	And a stronger version of it:
	\begin{conjecture} \label{LogNRGFlowConjecture2}
		All the heat-kernel coefficients ($a_E, c, b_1, b_2$) are identical level-by-level and spin-by-spin between two SCFTs connected by an RG flow, if we take into account the effect where a massless field with spin $s+1$ eats a field with spin $s$ and becomes massive.
	\end{conjecture}
	This conjecture appears to be consistent with the swampland conjecture \eqref{HKE:Einstein-Maxwell} proposed in the next section: within many gauge supergravity theories which correspond to different vacua of the scalar potential, as long as one theory satisfies \eqref{HKE:Einstein-Maxwell}, so do all of the rest. This conjecture should also be applied on RG flow across dimensions, which will involve supergravity theories in different dimensions whose heat-kernel coefficients are not fully available.
	
\section{The unbearable lightness of the KK scale}
\label{sec:Kundera}

So far we have discussed how to compute logarithmic corrections to the gravitational path integral on asymptotically AdS$_4$ backgrounds. Our focus was on 4d KK supergravity theories that preserve some amount of supersymmetry and arise from consistent truncations of 11d supergravity on a 7d manifold. In these explicit top-down constructions the size of the internal manifold is comparable to the size of the AdS$_4$ vacuum. This in turn implies that the four-dimensional gravitational theory does not really fall into the framework of standard EFT with finitely many fields, i.e. there is an infinite tower of KK modes that have masses of the same order that is parametrically smaller than the Planck scale. We now turn our attention to a more agnostic, or bottom-up, approach to EFTs coupled to gravity in AdS and their holographic physics in the context of the logarithmic corrections of interest in this work.

In order to be as general as possible we will work with a 4d gravitational theory (with at least one AdS$_4$ vacuum) coupled to a finite number of massive matter fields with spin up to 2 that obey some form of positive energy condition. We will assume that this theory has a consistent UV completion into a quantum gravitational theory, or to use modern parlance, that it is not in the Swampland. We will also assume that the 4d gravitational EFT is valid up to some energy scale which for concreteness we take to be the 4d Planck scale, although our arguments are general and can be adapted to other UV cutoff scales of interest. Lastly, we assume that the UV completion of the theory is such that it admits an equivalent holographic description in terms of a consistent unitary 3d CFT. As usual in AdS/CFT, the semi-classical gravity approximation is controlled by a large parameter which we call $N$. We emphasize that, at this somewhat abstract level of discourse, $N$ is just a label for a quantity much bigger than~1 that measures the number of degrees of freedom in the CFT, or rather the sequence of CFTs. For concreteness we can think of $N$ as being defined by the free energy of the CFT on $S^3$. Of course, in concrete and familiar AdS/CFT examples, $N$ can often be thought of as the rank of a gauge group. By evaluating the regularized on-shell action of the Euclidean AdS$_4$ vacuum of the gravitational theory we can find a map between the 4d Newton constant $G_N$, the scale $L$ of AdS$_4$, and the large parameter $N$. By dimensional analysis, this map will take the form
\begin{equation}
\frac{L^2}{G_N} = \beta N^{\gamma}\,,
\end{equation}
where the equality is to be understood to leading order in the large $N$ limit, the parameters $\beta$ and $\gamma$ are positive real numbers, and in all known AdS/CFT examples $\gamma$ is rational.

The path integral of the CFT on a compact Euclidean manifold should admit a large $N$ expansion of the form we already discussed around \eqref{logZ}, i.e.
\begin{equation}
	\log Z_{\text{CFT}} = F_0 + \mathcal{C} \log N + \mathcal{O}(N^0) \,.
\end{equation}
Here $F_0$ contains all positive powers of $N$ while the terms in $\mathcal{O}(N^0)$ are constant, have negative powers of $N$, or are exponentially suppressed at large $N$. These terms may have a complicated dependence on various continuous parameters of the theory including geometric deformations of the Euclidean manifold and dependence on relevant or exactly marginal couplings. Importantly, we will assume that the quantity $\mathcal{C}$ does not depend on any continuous parameters, i.e. it is a fixed real number for any given large $N$ CFT. Since this is a crucial assumption in our reasoning it is important to scrutinize it. We are not aware of any example of a sequence of CFTs (holographic or not) parametrized by $N$ for which, in the large $N$ limit, one finds that $\mathcal{C}$ depends on continuous parameters. One way to understand this may be to look at the 3d CFT placed on a squashed $S^3$ background and study the implications of any potential dependence in $\mathcal{C}$ on the continuous squashing parameter(s). If there is such a dependence one can expand the path integral in the limit of small squashing and deduce that there is a $\log N$ term in the large $N$ expansion of integrated correlators of the CFT on the round $S^3$. In particular, there will be a $\log N$ term in the $C_T$ coefficient of the two-point function of the energy momentum tensor.\footnote{See \eqref{eq:CT-def} for an explicit example of such a relation between $C_T$ and a squashed sphere partition function.} We are not aware of an example of a sequence of 3d CFTs where such a term appears, and it does not seem compatible with the usual large $N$ diagrammatics of gauge theories à la 't Hooft. The absence of any dependence of $\mathcal{C}$ on continuous parameters is also manifest in various explicit examples of partition functions of 3d $\mathcal{N}=2$ SCFTs (holographic or not) that can be studied by supersymmetric localization. It will be very interesting to construct a more robust argument or a proof that $\mathcal{C}$ does not depend on continuous parameters. One possible avenue to do this is to follow the approach in \cite{Cassani:2021fyv} and study the supersymmetric effective action of a 3d $\mathcal{N}=2$ SCFT on $S^1\times_{\omega} S^2$ with supersymmetric boundary conditions on $S^1$ and with angular momentum fugacity $\omega$, in the limit where the radius of the $S^1$ is small. It was shown in \cite{Cassani:2021fyv} that for 4d SCFTs the logarithmic term in the analogous effective action does not depend on the continuous parameter $\omega$ and has a topological nature. Furthermore, one can try to break supersymmetry and study the question more generally in the context of the thermal path integral of the 3d CFT on $S^1\times S^2$ which is controlled by the thermal effective action, see \cite{Benjamin:2023qsc} for a recent discussion. It may be possible to use this formalism to prove that no logarithmic terms can appear in $n$-point functions. In the absence of a solid proof, from now on we will assume that $\mathcal{C}$ is independent of continuous parameters and will proceed to study the implications of this assumption for the holographically dual 4d gravitational theory. Before doing so, we would like to emphasize that the essence of the discussion above is not specific to the $\log N$ terms in CFT path integrals and local correlation functions. If the CFT at hand has any dimensionless parameter, like an exactly marginal coupling $\lambda$, we also do not expect the coefficient of a $\log \lambda $ term in the path integral to depend on continuous parameters like squashing deformations or chemical potentials. 

We can now proceed in a similar manner as in Section~\ref{sec:bootstrap}. Studying the logarithmic corrections to the gravitational path integral on various explicit gravitational backgrounds,\footnote{We expect that the path integral of a standard 4d gravitational EFT allows for a standard diagrammatic expansion with respect to the large parameter $L^2/G_N$. The coefficient of the logarithmic correction $\log\relax(L^2/G_N)$ that is dual to $\mathcal{C}$ in the field theory side is then anticipated to be insensitive to the details of the UV cutoff and thereby determined unambiguously in the diagrammatic expansion. In particular, this requires the absence of exotic contributions like $[\relax(L^2/G_N)]^p[\log\relax(L^2/G_N)]^q$ with $p\geq 0$, $q>1$ in the expansion. In the discussion below we assume that such contributions indeed do not arise.} we find that the coefficient of $\log N$ (or $\log \lambda$) will depend on continuous parameters unless the total contribution to the $(c,b_1,b_2)$ heat kernel coefficients from the finite number of matter fields in the effective gravitational theory vanishes, i.e. we find the constraint
\begin{equation}
c^\text{tot}=b_1^\text{tot}=b_2^\text{tot}=0 \, . \label{HKE:Einstein-Maxwell}
\end{equation}
We emphasize that this conclusion is independent of the supersymmetry preserved by the gravitational theory or the background. This result imposes very strong constraint on any candidate consistent effective theory of gravity in AdS with finitely many fields. We now illustrate this with several examples.

Consider the 4d $\mathcal{N}=8$ ${\rm SO}(8)$ gauged supergravity theory constructed in \cite{deWit:1982bul}, i.e. the nonlinear theory for the fields in the massless $\mathcal{N}=8$ gravity multiplet presented in Table~\ref{tab:S7}. Using \eqref{eq:hk-non-min} and the results summarized in Section~\ref{sec:KK:S7} we find that the heat kernel coefficients for the maximally supersymmetric AdS$_4$ vacuum in this theory read\footnote{We note that if we choose to break $\mathcal{N}=8$ supersymmetry we can dualize some of the 35 pseudoscalars in the $\mathcal{N}=8$ gravity multiplet into 2-forms which will in turn yield different values for the $a_E$ coefficient. However, the $c$ and $b_1$ coefficients remain invariant under dualization~\cite{Larsen:2015aia}.}
\begin{equation}\label{eq:N=8sugraabc}
a_E^{\mathcal{N}=8} = \frac{5}{2}\,, \qquad c^{\mathcal{N}=8}=b_1^{\mathcal{N}=8}=0\,.
\end{equation}
Following the arguments above we therefore cannot conclude that this gauged supergravity is not a consistent holographic theory of quantum gravity, see also \cite{Alday:2022ldo,Montero:2022ghl} for recent discussion on the viability of the 4d $\mathcal{N}=8$ ${\rm SO}(8)$ gauged supergravity as a consistent quantum gravity theory. Arguments, similar in spirit to ours, have been used to exclude 4d $\mathcal{N}=8$ ungauged supergravity as a consistent quantum gravitational theory \cite{Green:2007zzb}.

We can proceed in a similar manner to study 4d $\mathcal{N}=4$ and $\mathcal{N}=2$ minimal gauged supergravity, i.e. the gauged supergravity theory of the gravity multiplet with the respective amount of supersymmetry. We find the following results:
\begin{equation}
\begin{alignedat}{3}
a_E^{\mathcal{N}=4} &= \frac12\,,& \qquad c^{\mathcal{N}=4} &= 0\,,& \qquad b_1^{\mathcal{N}=4}&=-\frac1{12}\,,\\[1mm]
a_E^{\mathcal{N}=2} &= -\frac{11}{24}\,,& \qquad c^{\mathcal{N}=2} &= 0 \,,& \qquad b_1^{\mathcal{N}=2}&= - \frac{13}{36}\,.
\end{alignedat}
\end{equation}
We therefore conclude that our logarithmic constraint excludes these two models as consistent quantum gravitational holographic theories. At this point one may be emboldened by the relative simplicity of our arguments and proceed to analyze all matter coupled (super)gravity theories and check whether they stand a chance of acting as consistent effective theories of gravity. This analysis is however complicated by the fact that we are interested in gravitational theories, like gauged supergravity, in which the scalar fields generically have highly non-trivial potentials. In such models it is a very involved task to study the space of AdS$_4$ vacua and the mass spectrum of linear fluctuations around them. As an illustration, consider the 4d $\mathcal{N}=8$ gauged supergravity with an $\rm{ISO} (7)$ gauging. This theory has the same matter content as the maximal ${\rm SO}(8)$ gauged supergravity, i.e. the fields in Table~\ref{tab:S7}, but a very different scalar potential. In particular, none of the 219 known AdS$_4$ vacua in this theory, see \cite{Bobev:2020qev}, is located at the origin of the scalar manifold where the full gauge group and maximal supersymmetry is preserved. This means that the heat kernel coefficients for any of the AdS vacua of this theory are not simply given by \eqref{eq:N=8sugraabc}. To find the correct logarithmic corrections to the gravitational path integral, one first needs to choose one of the vacua of the gauged supergravity theory, compute the mass spectrum of excitations around it (as was done in \cite{Bobev:2020qev}) and only then use the results in Table~\ref{tab:coeffs} above to calculate the total $(a_E,c,b_1)$ coefficients. Given these considerations we are left with the conclusion that the validity of a given effective theory of gravity in AdS with finitely many fields, supersymmetric or not, has to be studied on a cases-by-case basis. This can be done systematically for any given theory with fields with spin up to 2 and a known mass spectrum by using the results in Table~\ref{tab:coeffs} and checking whether the total heat kernel coefficients obey the constraints in \eqref{HKE:Einstein-Maxwell}.\\

In several corners of string theory the notion of a scale-separated AdS$_4$ vacuum appears, see for instance \cite{Kachru:2003aw,Balasubramanian:2005zx,DeWolfe:2005uu,Polchinski:2009ch} for several well-known constructions of this kind and \cite{Coudarchet:2023mfs} for a recent review. These are proposed consistent backgrounds of string or M-theory of the schematic form AdS$_4\times Y$ where there may be a warp factor in front of the AdS$_4$ metric and the internal space $Y$ is either completely geometric, i.e. a smooth manifold with properly quantized R-R and NS-NS fluxes, or has some mild singularities allowed by string theory such as orbifolds and orientifolds. The defining feature of such backgrounds is that the length scale $\ell_Y$ associated to $Y$ is parametrically smaller than the scale $L$ of AdS$_4$,
\begin{equation}
\ell_Y\ll L\,.
\end{equation}
We hasten to add that there could be some ambiguity in the definition of the scale $\ell_Y$. It can be associated to the volume or the diameter of $Y$, or be defined in terms of the eigenvalues of some appropriate differential operator on $Y$ used in the determination of the KK spectrum. What is important for our discussion is that below the scale set by $\ell_Y$ one should have a consistent effective gravitational theory in AdS$_4$ with finitely many matter fields of spin up to 2. In this setup, we can employ our results for the logarithmic corrections to the gravitational path integral to shed new light on the consistency of such scale-separated AdS$_4$ vacua. More specifically, we conclude that the spectrum of low-lying excitations around a given scale-separated AdS$_4$ vacuum, i.e. all fields with masses up to the scale set by $\ell_Y$, should be such that their total contribution to the $(c,b_1,b_2)$ coefficients vanishes as in \eqref{HKE:Einstein-Maxwell}. We note that these strong constraints should be viewed as an addition to the constraints and obstructions for the existence of scale-separated vacua discussed previously in the literature, see \cite{Gautason:2015tig,Alday:2019qrf,Lust:2019zwm,DeLuca:2021mcj,DeLuca:2021ojx,Collins:2022nux} and references thereof.\\

Since many AdS$_4$ vacua discussed in the literature, including scale-separated constructions, preserve $\mathcal{N}=1$ supersymmetry, we finish this section with a short discussion on the calculation of the heat kernel coefficients for 4d $\mathcal{N}=1$ matter multiplets. This should facilitate the exploration of the consistency condition in \eqref{HKE:Einstein-Maxwell} for any given AdS$_4$ vacuum with a known spectrum of light modes. To organize our results we first present the possible 4d $\mathcal{N}=1$ multiplets. We use notation similar to the one used for 4d $\mathcal{N}=2$ multiplets in Appendix~\ref{App:KK:mABJM} and find the following list of multiplets
   \begin{itemize}
        \item LGRAV = $L_1\left[E_0, \frac{3}{2}\right],\quad $ SGRAV = $A_1\left[ \frac{5}{2}, \frac{3}{2}\right],\quad $ 
        
        \item LGINO = $L_1\left[E_0, 1 \right],\quad $ SGINO = $A_1\left[ 2, 1 \right],\quad $

        \item LVEC = $L_1\left[E_0, \frac{1}{2}\right],\quad $ SVEC = $A_1\left[ \frac{3}{2}, \frac{1}{2}\right],\quad $

        \item LSCA = $L_2\left[E_0, 0 \right],\quad $ SSCA = $A_2\left[ \frac{1}{2}, 0 \right].\quad $
    \end{itemize}
On the right-hand side we have indicated how each supergravity multiplet can be related to the 3d $\mathcal{N}=1$ SCFT multiplets tabulated in Section 2.1.1 of \cite{Cordova:2016emh}. This facilitates the comparison with the holographically dual CFT and also allows one to consult the explicit field content of each multiplet using the detailed presentation in \cite{Cordova:2016emh}. We note that in the list above we have not included the identity multiplet denoted by $B_1[0,0]$ in that reference. The list above contains long multiplets (LGRAV, LGINO, LVEC, LSCA), for which all fields with $s\ge 1$ are massive, as well as short multiplets (SGRAV, SGINO, SVEC, SSCA), for which the fields with $s\ge 1$ are massless. Finally, $E_0$ is equal to the dimension $\Delta$ of the dual CFT primary operator and the mass of the field in AdS$_4$ can be found by using the relations between $\Delta$ and $m$ in Table~\ref{tab:conf-m}. The heat kernel coefficients $(a_E, c, b_1)$ of each 4d $\cN = 1$ multiplets are presented in Table~\ref{HeatKernelCoefficientsN=1Multiplets}.
%
        \begin{table}[h]
    \renewcommand{\arraystretch}{1.5}
        \centering
        \begin{tabular}{|c|c|c|c|}\hline
             & $a_E$ & $c$ & $b_1$  \\\hline 
            LGRAV & $\frac{23}{12}$ & $\frac{7}{3}$ & $\frac{5}{16}-\frac{1}{12}(E_0 - 1)^2$\\\hline
            SGRAV & $\frac{113}{48} $ & $\frac{77}{24}$ & $ 0 $ \\\hline
            LGINO & $-\frac{7}{16}$ & $-\frac{7}{8}$ & $-\frac{1}{8}+\frac{1}{16}(E_0 - 1)^2$\\\hline
            SGINO & $- \frac{31}{48}$ & $ - \frac{25}{24}$ & $ 0 $\\\hline
            LVEC & $\frac{5}{24}$ & $\frac{1}{6}$ & $\frac{1}{32}-\frac{1}{24}(E_0 - 1)^2$\\\hline
            SVEC & $\frac{3}{16}$ & $\frac{1}{8}$ & $0 $\\\hline
            LSCA  & $\frac{1}{48}$ & $\frac{1}{24}$ & $ \frac{1}{48}(E_0 - 1)^2 $\\\hline
             SSCA & $\frac{13}{720}$ & $\frac{1}{30}$ & $\frac{23}{4608}$\\\hline
         \end{tabular}
        \caption{\rm The heat kernel coefficients for 4d $\cN = 1$ multiplets. }
        \label{HeatKernelCoefficientsN=1Multiplets}
    \end{table}

We stress that for all entries in Table~\ref{HeatKernelCoefficientsN=1Multiplets} we have already incorporated the $(-1)^F$ that accounts for the fermion number of a given field in sums like the one in \eqref{eq:C-local:EAdS4}. Therefore, for a given AdS$_4$ vacuum with a known mass spectrum organized into 4d $\cN = 1$ multiplets, one simply needs to add the contributions from the table above to check whether the constraint in \eqref{HKE:Einstein-Maxwell} is obeyed. Since most multiplets in Table~\ref{HeatKernelCoefficientsN=1Multiplets} have positive $c$ coefficients, it is clear that imposing \eqref{HKE:Einstein-Maxwell} will result in non-trivial constraints. It will be very interesting to explore these constraints in detail for many of the well-known AdS$_4$ vacua in string and M-theory.

\section{Outlook}
\label{sec:ccl}

In this work we studied several aspects of logarithmic corrections to gravitational path integrals in AdS$_4$ and their relation to holography. In addition to a compendium of new technical results on this subject, we can extract at least two general lessons from our analysis. First, for a given explicit AdS$_4$ vacuum of string or M-theory, it is crucial to study the logarithmic corrections in the proper setup, i.e. in the higher-dimensional supergravity theory or in the 4d consistent truncation accompanied by a proper regularization scheme for the infinite tower of KK modes that contribute to the heat kernel. Second, we find that the form of the logarithmic corrections to the path integral on asymptotically locally AdS$_4$ backgrounds impose non-trivial constraints on the possible consistent 4d effective gravitational theories. In particular, assuming the validity of holography and the 4d gravitational EFT up to some cutoff scale, we conclude that either the total $(c,b_1,b_2)$ coefficients of the matter fields in the theory need to vanish, or that the dual 3d CFTs necessarily contain certain exotic logarithmic terms in their local correlation functions and thermal observables.

Our analysis points to several interesting open questions and directions for future work which we now briefly discuss.

\begin{itemize}

\item As emphasized in Section~\ref{sec:KK:S7}, the 4d KK supergravity calculation of the $\log N$ correction to the free energy on AdS$_4 \times S^7$ does not agree with the field theory result or the 11d supergravity analysis in \cite{Bhattacharyya:2012ye} if one chooses a regularization scheme in which $a_E=0$. It may be possible to resolve this discrepancy by using that the non-trivial contribution to $\log N$ in the 11d calculation comes from a zero mode of a 2-form ghost field needed for the proper quantization of the 3-form in 11d supergravity. To this end, one may contemplate a KK reduction of all ghost fields used in the quantization of the 11d graviton, gravitino, and 3-form around AdS$_4 \times S^7$ which can perhaps be organized into 4d $\mathcal{N}=8$ ``ghost KK multiplets''. With this at hand one should repeat our 4d KK supergravity analysis to account for the contribution of these ``KK ghosts'' to the coefficient of the $\log N$ term. Even if this calculation ends up successfully accounting for the correct $\frac14\log N$ contribution to the free energy, it is still unclear to us why such an analysis is justified. It will be interesting to understand this point better and draw lessons for the analysis of similar logarithmic corrections in more general gravitational backgrounds.

\item As discussed in Section~\ref{sec:log-conj}, if we adopt a particular regularization scheme for the infinite tower of KK modes on $S^7$ we can find an expression for the total $a_E$ heat kernel coefficient that agrees with the $\frac14\log N$ corrections to the AdS$_4\times S^7$ path integral. If in addition we assume the absence of zero modes other than for 2-forms in asymptotically locally AdS$_4$ spaces, we find results for the logarithmic corrections to the gravitational path integral that are consistent with the supersymmetric localization results in the ABJM theory summarized in Table~\ref{tab:CFT-log}. Moreover, this allows us to calculate the logarithmic corrections to the Bekenstein-Hawking entropy of non-supersymmetric black holes in AdS$_4$, generalizing the work of Sen \cite{Sen:2012dw} for asymptotically flat black holes. It is clearly important to put these calculations and assumptions on a more solid footing and find efficient methods to treat the contributions of zero modes to the logarithmic corrections in asymptotically AdS$_4$ backgrounds. We should also emphasize that we carried the calculation of the $b_2$ heat kernel coefficient in Appendix~\ref{App:coeffs} for minimally coupled charged scalars and spin-$1/2$ fermions and relied on the bootstrap argument in Section~\ref{sec:bootstrap} in most of our analysis. It will be nice to fill this gap and compute the $b_2$ coefficient for matter fields of arbitrary spin, mass, and charge.

\item An interesting generalization of the topologically twisted index of 3d $\mathcal{N}=2$ SCFTs on $S^1 \times \Sigma_{\mathfrak{g}}$ is the supersymmetric partition function on the Seifert manifold $\mathcal{M}_{\mathfrak{g},p}$ given by a degree $p$ fibration of the $S^1$ over $\Sigma_{\mathfrak{g}}$. This path integral was discussed in detail in~\cite{Closset:2017zgf} for general SCFTs where it was shown how it can be computed using supersymmetric localization. The holographic description of this path integral is given by the supersymmetric Euclidean AdS-Taub-Bolt solution presented in \cite{Alonso-Alberca:2000zeh,Toldo:2017qsh} and was studied for theories arising from M2- and M5-branes in \cite{Toldo:2017qsh} and \cite{Gang:2019uay,Benini:2019dyp,Bobev:2020zov}, respectively. It is interesting to apply the conjecture in Section~\ref{sec:log-conj} to the AdS-Taub-Bolt background. As shown in \cite{Bobev:2020zov}, the Euler number of this 4d supergravity background is independent of $p$ and is given by $\chi = 2(1-\mathfrak{g})$. This implies that the logarithmic correction to the SCFT free energy on $\mathcal{M}_{\mathfrak{g},p}$ does not depend on $p$. Indeed, as shown in \cite{Gang:2019uay,Bobev:2020zov}, this is true for class $\mathcal{R}$ theories arising from M5-branes. This is yet another piece of evidence that supports the conjectural results in Section~\ref{sec:log-conj}. Using this we can then conclude that the free energy on $\mathcal{M}_{\mathfrak{g},p}$ for the ABJM theory, discussed to leading $N^{3/2}$ and subleading $N^{1/2}$ order in \cite{Toldo:2017qsh} and \cite{Bobev:2020egg,Bobev:2021oku}, respectively, should receive a $\frac{1}{2}(1 - \mathfrak{g})\log N$ correction. Confirming this prediction by an explicit matrix model calculation based on the results of \cite{Toldo:2017qsh} will amount to a very non-trivial test of our results. We hope to report on this in the near future.

\item A central assumption underlying the discussion in Section~\ref{sec:Kundera} is the independence of the coefficients of the $\log N$ and $\log\lambda$ terms in CFT partition functions on continuous parameters like squashing deformations or relevant operator couplings. This can also be phrased as the absence of $\log N$ and $\log\lambda$ terms in correlation functions of local operators. Given the strong constraint this assumption imposes on gravitational theories in AdS, it is important to scrutinize it and understand whether it can be put on a more rigorous footing.

\item In many of the discussions above we focused on the $\log N$ terms in the path integrals of 3d $\mathcal{N}=2$ holographic SCFTs in M-theory. There can of course be $\log \lambda$ corrections to holographic SCFTs path integrals if a suitable notion of a 't Hooft coupling exists in the theory. These corrections should then correspond to $\log L/\ell_s$ corrections to the AdS$_4$ path integral where $\ell_s$ is the string length. Studying these corrections in 10d supergravity is in principle possible with the heat kernel methods discussed here and will be very interesting to pursue. However, a technical complication arises since one will first need to calculate the 10d heat kernel coefficients of the various supergravity fields in order to compute the local contribution to the coefficient of $\log L/\ell_s$. It will be very interesting to pursue this analysis, contrast the result with a 4d KK supergravity approach along the lines presented in this work, and ultimately compare the outcome of these calculations with the available 3d SCFT results for the relevant path integrals. More generally, even in the absence of supersymmetry, when there are dimensionful parameters like temperature, electric charges, and angular momenta present in the gravitational path integral, one can combine them with the AdS scale to form dimensionless parameters that can be scaled in various ways. It will be interesting to explore these scaling limits systematically and understand whether one can compare the coefficient of the logarithmic terms with dual 3d CFT results. Moreover, it will be nice to understand whether this analysis can lead to any interesting constraints on effective gravitational theories in AdS$_4$ along the lines of Section~\ref{sec:Kundera}.

\end{itemize}

\cleardoublepage%

  \graphicspath{{\chaptersdir/etheory/image/}}%

\chapter{ The planar limit of the $\mathcal{N} = 2$ $\mathbf{E}$-theory: numerical calculations and the large $\lambda$ expansion }\label{ch:etheory}

\section{Introduction and summary of results}

Since the seminal work of 't Hooft \cite{tHooft:1973alw} the planar limit of four-dimensional gauge theories has served as an important approximation that provides new insights into their dynamics. In subsequent developments a plethora of additional tools, amongst which integrability, supersymmetric localization, and AdS/CFT, were applied with great efficacy in conjunction to the planar limit to improve our understanding of gauge and string theory. A central example which often serves as a testbed for new calculational techniques and conceptual advances is the $\mathcal{N}=4$ SYM theory. Our modest goal in this work is to build on recent results from supersymmetric localization for four-dimensional $\mathcal{N}=2$ SCFTs, see \cite{Pestun:2016zxk} for a review, and derive some new results for physical observables in a particular planar gauge theory that shares many properties with the $\mathcal{N}=4$ SYM theory.

The theory of interest here is an $\mathcal{N}=2$ $\SU(N)$ gauge theory with one hypermultiplet in the symmetric representation of the gauge group and one hypermultiplet in the antisymmetric representation. This was referred to as the $\mathbf{E}$ theory in \cite{Billo1} and we will also adopt this monicker. As discussed in \cite{Billo1} the $\mathbf{E}$ theory is closely related to $\SU(N)$ $\mathcal{N}=4$ SYM theory. For instance, the conformal anomaly coefficients of the two theories are
\begin{equation}\label{eq:acEN4}
a^{\mathcal{N}=4}=c^{\mathcal{N}=4}=\frac{1}{4}(N^2-1)\,, \qquad a^{\mathbf{E}} = a^{\mathcal{N}=4} + \frac{1}{24}\,, \qquad c^{\mathbf{E}} = c^{\mathcal{N}=4} + \frac{1}{12}\,.
\end{equation}
Moreover, similarly to $\mathcal{N}=4$ SYM, the $\mathbf{E}$ theory theory has a holographic description in terms of type IIB string theory on a $\mathbb{Z}_2$ orientifold of AdS$_5\times S^5$, see \cite{Park:1999ep,Ennes:2000fu}. 

Certain physical observables in both the $\mathbf{E}$ theory and $\mathcal{N}=4$ SYM can be calculated by placing the theory on $S^4$ and employing supersymmetric localization as in \cite{Pestun:2007rz}. Importantly, in the planar limit the instanton contributions to the supersymmetric localization matrix model trivialize and the calculation of some physical observables as a function of the 't Hooft coupling $\lambda$ becomes feasible. This in turn provides the interesting possibility for explicit gauge theory calculations in the strong coupling limit, $\lambda \gg 1$, which should lead to insights into the $\alpha'$ corrections of type IIB string theory around the AdS$_5\times S^5/ \mathbb{Z}_2$ orientifold. Obtaining such strong coupling results in the planar limit of the $\mathbf{E}$ theory is the main objective of this work. 

In particular, we will be mainly interested in the two-point correlation functions of single trace operators in the $\mathbf{E}$ theory built out of the complex scalar field in the $\mathcal{N}=2$ vector multiplet. Conformal covariance and the R-symmetry of the theory dictate that these two-point functions take the following form
\begin{equation}
	\langle {\rm tr}\varphi^m(x) {\rm tr}\bar{\varphi}^n(0) \rangle = \frac{G_m(\lambda,N)\delta_{m,n}}{(4\pi^2 x^2)^{2m}}\,,
\end{equation}
where $G_m(\lambda,N)$ is a non-trivial function that can in principle be computed by supersymmetric localization, however explicit calculations for general values of $N$ are difficult. To organize the calculation it is useful to take the ratio of this function to the corresponding two-point function in $\mathcal{N}=4$ SYM, $G_m^{(0)}$. Moreover, it is important to distinguish the operators with $m$ even and odd in the planar limit. The operators with odd $m$ correspond to twisted sector modes in the AdS$_5\times S^5/ \mathbb{Z}_2$ orientifold and have distinct correlators from those of $\mathcal{N}=4$ SYM already at leading order in the planar limit. The untwisted sector modes correspond to operators with even $m$ and their two-point functions are identical with those in $\mathcal{N}=4$ SYM to leading order at large $N$ but differ at subleading orders. This information can be conveniently organized into the following large $N$ expansions 
\begin{equation}
	\frac{G_{2k+1}}{G_{2k+1}^{(0)}} = 1 + \Delta_k(\lambda) + O(N^{-2})\,,\qquad 	\frac{G_{2k}}{G_{2k}^{(0)}} = 1 + \frac{\delta_k(\lambda)}{N^2} + O(N^{-4})\,.
	\label{DeltaAnddelta}
\end{equation}
As explained in \cite{Billo1}, $\Delta_k(\lambda)$ and $\delta_k(\lambda)$ can be calculated by using supersymmetric localization and matrix model techniques. It was shown in \cite{Billo1} that a central role in this calculation is played by an infinite matrix $\mathsf{X}$ with matrix elements:
\begin{align} \label{Xx}
	\mathsf{X}_{kl} = (-1)^{k+l+1}8 \sqrt{(2k+1)(2l+1)} \int_0^\infty \! dt\ W(t) 
	J_{2k+1}\Big(\frac{t\sqrt{\lambda}}{2\pi}\Big)\, 
	J_{2l+1}\Big(\frac{t\sqrt{\lambda}}{2\pi}\Big)\,,	
\end{align}
with $k,l = 1 , 2 , \ldots$, where $J_n$ are Bessel functions, and 
\begin{equation} \label{defW}
	W(t) =\frac{e^t}{t (e^t-1)^2} \,.
\end{equation}
The calculation of $\Delta_k(\lambda)$ then amounts to inverting $1-\mathsf{X}$ and calculating the following ratio of determinants\footnote{Here the symbol $\Dx_{(k)}$ is used to denote the upper left $k \times k$ block of the matrix $\Dx$. }
\begin{align}
	\label{defDeltaINTRO}
	1 + \Delta_k(\lambda) = 
	\frac{\det \Dx_{(k)}}{\det \Dx_{(k-1)}},\qquad \Dx_{k,l}\equiv \Big(\frac{1}{1 - \mathsf{X}}\Big)_{k,l}\,.
\end{align} 		
It is relatively easy to calculate the small $\lambda$ expansion of $\Delta_k(\lambda)$ using~\eqref{defDeltaINTRO}. Indeed, for each order in the small $\lambda$ expansion, the matrix $\mathsf{X}$ is essentially a finite matrix, and the calculation of $\Delta_k(\lambda)$ order by order in the small $\lambda$ limit amounts to straightforward calculations with finite matrices. However, it is more challenging to calculate the large $\lambda$ expansion of $\Delta_k(\lambda)$. To tackle this problem we develop numerical techniques based on solving an integral equation that allow for the efficient calculation of $\Delta_k(\lambda)$ for general finite values of $\lambda$. Using this method for large values of $\lambda$ and performing a precise numerical fitting procedure allows us to propose a conjecture for the analytic form of the first six terms in the large $\lambda$ expansion of $\Delta_k(\lambda)$.  The first three terms in this expansion read
\begin{equation}\label{eq:NikolayINTRO}
1+\Delta_k(\lambda) = 8 \pi^2 k (2 k+1) \left[ \frac{1}{\lambda}-\frac{16 k \log 2}{\lambda^{3/2}}
+\frac{32 k (4 k -1)\log^2 2}{\lambda^{2}}\right]+O(\lambda^{-5/2})\,,
\end{equation}
while three more subleading terms are presented in Section~\ref{sec:Deltamoreterms}. The leading $1/\lambda$ term in \eqref{eq:NikolayINTRO} agrees with the analysis in \cite{Billo1} where it was calculated using analytic methods. We have also attempted to provide an analytic derivation of \eqref{eq:NikolayINTRO} but we were not successful due to certain subtleties we encountered when working with infinite matrices. This attempt is summarized in Appendix~\ref{sec:expansionDeltak}.

It was argued in \cite{Billo1} that $\delta_k(\lambda)$ in \eqref{DeltaAnddelta} can be calculated by taking derivatives with respect to $\lambda$ of the difference between the $S^4$ free energy of the $\mathbf{E}$ theory and $\mathcal{N}=4$ SYM. It was shown in \cite{Tseytlin} that this difference in free energies is related to the determinant of the matrix $\mathsf{X}$ as follows: 
\begin{equation}
	{\cal F} \equiv F^{\rm E-theory} - F^{\cN=4} = \frac{1}{2} \log \det ( 1 - \mathsf{X})\,.
\end{equation}
Again, it is relatively straightforward to calculate the small $\lambda$ expansion of ${\cal F}$, but it is much more difficult to calculate its large $\lambda$ expansion. To this end we develop a numerical method to compute the determinant of $1 - \mathsf{X}$ for general values of $\lambda$. Detailed numerical analysis in the large $\lambda$ regime leads us to conjecture the following analytic form of the first four terms in the large $\lambda$ expansion of $\mathcal{F}$
\begin{equation}\label{eq:Flargelambdaintro}
{\cal F} = \frac{1}{8} \sqrt{\lambda}- \frac{3}{8} \log \left(\frac{\lambda}{\lambda_0}\right) - \frac{3}{4} \frac{\log 2}{\sqrt{\lambda}} 
- \frac{3}{2} \frac{\log^2 2}{\lambda} + O( \lambda^{-3/2})\,,
\end{equation}
where $\lambda_0 \approx 7.72390117$ is a numerical constant.\footnote{ An analytical form has been found by other methods, see the discussion around \eqref{FF}. } We note that the $1/8$ coefficient of the leading $\sqrt{\lambda}$ term in \eqref{eq:Flargelambdaintro} differs from the result in \cite{Tseytlin} where analytical methods were used to find the value $1/2\pi$ for this coefficient. We have no reason to doubt our very precise numerical analysis, which also allows us to find three more subleading terms in the large $\lambda$ expansion as compared to \cite{Tseytlin}, and believe that the discrepancy may be due to subtleties with the analytic method used in \cite{Tseytlin} when applied to matrices of infinite rank.
The result for $\mathcal{F}$ in \eqref{eq:Flargelambdaintro} can be combined with the arguments in \cite{Billo1} to find the large $\lambda$ expansion of the two-point function $\delta_{k}(\lambda)$. We find the following explicit result for the leading four terms in the large $\lambda$ expansion
\begin{multline}
\delta_k(\lambda) =  -\frac{k(4k^2-1) }{16}\sqrt{\lambda} + \frac{3}{8}k(4k^2-2)\\- \frac{3}{8}\frac{k(4k^2-3)\log 2}{\sqrt{\lambda}} -\frac{3}{2}  \frac{k(4k^2-4) \log^22}{\lambda}+O(\lambda^{-3/2})\,.
\end{multline}

Knowing the difference of free energies $\mathcal{F}$ allows us also to calculate the vacuum expectation value of a circular supersymmetric Wilson loop wrapping the equator of $S^4$. For $\mathcal{N}=4$ SYM in the planar limit this Wilson loop vev is given in terms of a Bessel function, see \cite{Drukker:2000rr}, 
\begin{equation}
\langle W^{\mathcal{N}=4}\rangle_0 = \frac{2N}{\sqrt{\lambda}} I_1(\sqrt{\lambda})\,.
\end{equation}
The deviations from this leading planar result are captured by a function $q(\lambda)$, see \cite{Tseytlin}
\begin{equation}\label{eq:Deltaqdef}
\frac{\langle W^{\mathcal{N}=4}\rangle}{\langle W^{\mathcal{N}=4}\rangle_0} = 1 + \frac{q^{\mathcal{N}=4}(\lambda)}{N^2} + O(N^{-4})\,, ~~ \text{where} ~~ q^{\mathcal{N}=4}(\lambda) = \frac{\lambda^{3/2}}{96} \frac{I_2(\sqrt{\lambda})}{I_1(\sqrt{\lambda})} - \frac{\lambda}{8}\,.
\end{equation}
As discussed in \cite{Tseytlin} for the $\mathbf{E}$ theory one finds that the leading order result for the Wilson loop vev in the planar limit is the same as that for $\mathcal{N}=4$ SYM, $\langle W^{\mathbf{E}}\rangle_0  = \langle W^{\mathcal{N}=4}\rangle_0 $. There are differences however at order $N^{-2}$ which can be captured by the function~$\Delta q(\lambda) = q^{\mathbf{E}}(\lambda)-q^{\mathcal{N}=4}(\lambda)$. It was furthermore shown in \cite{Tseytlin} that $\Delta q(\lambda) = -\frac{\lambda^2}{4}\partial_\lambda \mathcal{F}$. Using this relation and the result in \eqref{eq:Flargelambdaintro} we find the following large $\lambda$ behavior
\begin{equation}\label{eq:Deltaqres}
\Delta q (\lambda) = - \frac{1}{64} \lambda^{3/2}+\frac{3}{32} \lambda - \frac{3\log2}{32} \lambda^{1/2} - \frac{3\log^22}{8} +O(\lambda^{-1/2})\,.
\end{equation}

Our results for the large $\lambda$ behavior of $\mathcal{F}$ and $\Dx_{k,l}$ can be used in conjunction with the recent studies in \cite{Billo:2022xas} to derive the strong coupling behavior of the planar limit of some three-point extremal correlators of single trace operators in the $\mathbf{E}$ theory. We are able to compute these correlations functions up to order $\lambda^{-2}$ in the strong coupling expansion, which improves on the results of \cite{Billo:2022xas} by providing three additional terms in the large $\lambda$ expansion. These results are presented in Section~\ref{sec:Threeptfc}.

We stress that our numerical methods to calculate  $\Delta_k(\lambda)$ and  ${\cal F}$ are useful independently of the large $\lambda$ results presented above. The numerical method we propose is fast and efficient and provides accurate results for a wide range of values for the coupling  $\lambda$, $0 \le  \lambda \lesssim 10^7 $, for both $\Delta_k(\lambda)$ and  ${\cal F}$. The numerical techniques we use are explained in some detail in Sections~\ref{sec:Deltanum} and \ref{sec:Fnum}, respectively, and we hope they may also find some use in other similar problems.

The structure of the chapter is as follows. In Section~\ref{sec:N2summary} we summarize some relevant facts about correlation functions in the $\mathbf{E}$ theory and their calculations by supersymmetric localization techniques. In Section~\ref{sec:Deltakinteq}, we develop a method involving integral equations to calculate $\Delta_k(\lambda)$. We solve the resulting integral equation numerically to calculate $\Delta_k(\lambda)$ accurately for many values of $\lambda$ and we use these results to conjecture analytic expressions for the large $\lambda$ behaviour of $\Delta_k(\lambda)$. In Section~\ref{sec:FreeEnergy}, we rewrite the free energy $\cal{F}$ as a Fredholm determinant and explain how to calculate it numerically for many values of $\lambda$. We again use these numerical results to conjecture the large $\lambda$ behaviour of $\cal{F}$ and therefore for $\delta_k(\lambda)$ and $\Delta q(\lambda)$. In Section~\ref{sec:Threeptfc}, we show how to use these results to determine the large $\lambda$ behaviour of three point extremal correlators. We conclude in Section~\ref{sec:Discussion} with a short discussion on some open questions for future research. The four appendices contain some details on our numerical algorithm as well as an illustrative example of an analytic approach to the calculation of $\Delta_k$ which we believe is wrong for subtle reasons.

\section{$\cN = 2$ superconformal field theories and matrix models}
\label{sec:N2summary}

	In this section, we start with a briefly recall on some background material on supersymmetric localization and matrix model results for correlation functions in 4d $\mathcal{N}=2$ SCFTs. More details on this material can be found in \cite{Billo1, Billo:2022xas,Billo:2019st,Gerchkovitz:2016bo,Beccaria:2007ln,Baggio:2014sna}. Then we introduce our main objects: the chiral primary operators and extremal correlators.

	One can use supersymmetric localization techniques\footnote{There are two localization schemes in the literature; in this chapter we use the full Lie algebra localization \cite{Billo1, Billo:2022xas, Billo:2019st, Beccaria:2007ln, Billo:2021rdb}.} to study ${\cal N} = 2$ SCFTs, see \cite{Pestun:2016zxk} for a review. After suitable regularization of the superdeterminant on $S^4$, the partition function can be localized to a matrix integral:
	\begin{equation} \label{eq:PestunScheme}
	Z_{S^4} = \int d^N a \Delta(a) Z_{\rm cl}(a) Z_{\rm 1-loop}(a),
	\end{equation}
	where $N$ is the rank of the gauge group, $\Delta$ denotes the Vandermonde determinant, $Z_{\rm cl}(a)$ and $Z_{\rm 1-loop}(a)$ correspond to $e^{-S[X_0]}$ and the superdeterminant terms in \eqref{eq:LocalizedPathIntegral}, both of which can be written explicitly as functions of the Cartan of the gauge group:
	\begin{equation} \label{eq:PestunLoc}
	\begin{aligned}
	Z_{\rm cl}(a) &= e^{-\frac{8\pi^2}{g^2} {\rm tr}\, a^2},\quad Z_{\rm 1-loop}(a) = \frac{\prod_{w\in W({\rm adj})} H(i {\bf w} \cdot {\bf a}) }{ \prod_{w\in W(\cR)} H(i {\bf w} \cdot {\bf a}) },\\
	&  \quad H(x) \equiv G(1+x)G(1-x),
	\end{aligned}
	\end{equation}
	where $G(x)$ is the Barnes $G$-function, $\cR$ is the representation under which the hypermultiplet transfers under the gauge group. Thus everything is known explicitly if we ignore the instanton contributions \cite{Nekrasov:2002qd} which are suppressed exponentially at large $N$. This localisation scheme \eqref{eq:PestunScheme} is the original one obtained by Pestun \cite{Pestun:2007rz}, where we integrate over the Cartan subalgebra. 
	
	In this chapter, alternatively, we use the scheme where the partition function on $S^4$ can be written as the following matrix integral, which uses the fact that $d^Na\Delta(a)$ is the Haar measure of the SU($N$) group:
		\begin{equation}\label{PartitionFunction}
		\cZ_{S^4} = \int da\,  e^{- {\rm tr}\, a^2 - S_{\rm int}(a) }\ ,
	\end{equation}
	where we slightly changed the measure over the BPS locus, where $a = a^bT_b$ with $T_b$ the generators of the full gauge algebra instead of the Cartan algebra as in \eqref{eq:PestunLoc}, and the new measure is given by
	\begin{equation}
		da = \prod_{b=1}^{N^2-1} \frac{da^b}{\sqrt{2\pi}}\,.
	\end{equation}
	One can show that $S_{\rm int}(a)=0$ for the ${\cal N} = 4$ SYM theory and the matrix model is therefore Gaussian in that case. There is a closed form expression for the interaction term $S_{\rm int}(a)$ in general. The general form of this expression is not presented here, but can be found for the {\bf E} theory in equation~\eqref{EtheorySint} below. 

	\subsection{Extremal correlators on $\mathbb{R}^4$}
	An interesting set of objects in 4d $\cN = 2$ SCFTs are single trace chiral primary operators built out of the complex scalar $\varphi$ in the vector multiplet:
	\begin{equation}\label{CPO}
	O_n(x) \equiv {\rm tr}\,\varphi^n(x)\,.
	\end{equation}
	These operators are chiral because they are annihilated by half of the supersymmetries, and they are automatically normal-ordered because of $R$-charge conservation. The anti-chiral primary operators are denoted by $\bar{O}_n(x)$ and are constructed from $\bar{\varphi}(x)$ in the same way. The properties of the OPE for $4d\ \cN = 2$ theories leads to the chiral-ring relation:
	\begin{equation}\label{ChiralRing}
	O_m(x) O_n(0) = O_{m,n}(x) + \cdots,\quad O_{m,n}(x) \equiv {\rm tr}\, \varphi^m(x) {\rm tr}\, \varphi^n(x),
	\end{equation}
	where ``$\cdots$'' denotes $\bar{Q}$-exact terms. In this chapter, we are interested in the two-point  and three-point functions involving chiral and anti-chiral operators in $\mathbb{R}^4$. These correlators are constrained by the chiral-ring relation and R-symmetry selection rules and take the following form
	\begin{equation}\label{FTCorrelators}
	\langle O_m(0) \bar{O}_n(x)\rangle = \frac{G_m(N, \lambda) \delta _{m,n}}{(4\pi^2 x^2)^{2m}},\qquad 	\langle O_m(0)O_n(x) \bar{O}_p(y)\rangle = \frac{G_{m,n}(N, \lambda)  \delta _{m+n,p }}{(4\pi^2 x^2)^{m}(4\pi^2 y^2)^{n}  }.
	\end{equation}
	More generally, the extremal correlators in $\mathbb{R}^4$ involve chiral primaries with one anti-chiral primary. The superconformal symmetry constrains their forms up to a constant: \cite{Papadodimas:2009eu}
	\begin{equation}
	\langle O_{i_1}(x_1) O_{i_2}(x_2) \cdots O_{i_n}(x_n) \bar{O}_j(y)\rangle_{\rm \mathbb{R}^4} = \cG_{i_1,\cdots, i_n; j}(\tau, \bar{\tau}) \prod_{k=1}^n \frac{1}{(y-x_k)^{2\Delta_{O_{i_k}}}},
	\end{equation}
	where $\cG_{i_1,\cdots, i_n; j}(\tau, \bar{\tau})$ is a function of the exactly marginal coupling $\tau = \frac{\theta}{2\pi} + \frac{4\pi i}{g_{\rm YM}^2}$. We push the anti-chiral operator to the spatial infinity using the conformal invariance through $\bar{O}_{j}(\infty) \equiv \lim_{y\rightarrow\infty} y^{2\Delta_j} \bar{O}_j(y)$. Using supersymmetry Ward identity \cite{Baggio:2012rr}, one can prove that the extremal correlator with $\bar{O}$ located at infinity doesn't depend on any of $
	\{x_i\}$. Thus, we can put all the chiral operators on top of each other at the origin, and thus:
	\begin{equation}
	\cG_{i_1,\cdots, i_n; j} = \langle O_{i_1}(0) O_{i_2}(0) \cdots O_{i_n}(0) \bar{O}_j(\infty)\rangle_{\rm \mathbb{R}^4}.
	\end{equation}
	Now we put the theory on $S^4$, it follows that supersymmetric localization on $S^4$ reduces $O_n$ to $\Omega_n \equiv {\rm tr}\, a^n$.\footnote{By definition, $a$ are Hermitian matrices, thus the anti-chiral operator is localised to $\bar{\Omega}_k = {\rm tr}\,\bar{a}^k = {\rm tr}\, a^k = \Omega_k$. } To make a connection with the 4d $\cN = 2$ theory on $\mathbb{R}^4$, one na\"ively expects:\footnote{In the last equality, the differentiations of the partition function should give an integrated correlator, which has UV divergences to be regularized. The regularization we used, where we put the anti-chiral operator to the north pole and the chiral operators to the south pole, is the one consistent with the Ward Identity. }
	\begin{equation}
	\begin{aligned}
	& \langle O_{i_1}(0) \cdots O_{i_n} (0) \bar{O}_j(\infty)\rangle_{\rm \mathbb{R}^4} \overset{?}{=}  \langle O_{i_1}(S) \cdots O_{i_n} (S) \bar{O}_j(N)\rangle_{\rm S^4} = \\
	& = \langle \Omega_{i_1} \cdots \Omega_{i_n} \Omega_j(N)\rangle =  Z_{S^4}^{-1} \partial_\tau^{i_1}\cdots \partial_\tau^{i_n} \partial_{\bar{\tau}}^j Z_{S^4}(\tau, \bar{\tau}),
	\end{aligned}
	\end{equation}
	where $N, S$ denotes the north and south poles of the sphere. However, an important complication happens because of the conformal anomalies on $S^4$: an operator $O_n(x)$ on $S^4$ will mix with the full tower of lower-dimensional operators below it, \cite{Gerchkovitz:2016gxx}
	\begin{equation}
	O_n^{\mathbb{R}^4} \rightarrow \Omega_n + \alpha_1(\tau, \bar{\tau}) \Omega_{n-2} R^{-2} + \alpha_2 (\tau, \bar{\tau}) \Omega_{n-4} R^{-4} + \cdots,
	\end{equation}
	where $R$ is the radius of the round $S^4$. Diagonalizing the operator mixing matrix on $S^4$ \`a la Gram-Schmidt leads to the representation of the extremal correlators on $\mathbb{R}^4$:
	\begin{equation}
	\cG_{n,n} = \langle O_n(0) \bar{O}_n(\infty)\rangle_{\mathbb{R}^4} = \frac{\det \mathsf{D}_{(n,n)}}{\det \mathsf{D}_{(n-1,n-1)}},\quad \mathsf{D}_{nm} \equiv Z_{S^4}^{-1} \partial_{\tau}^n \partial_{\bar{\tau}}^m Z_{S^4}(\tau, \bar{\tau}),
	\end{equation}
	where the suffix $(n,n)$ means the $n\times n$ minor in the upper-left corner. The matrix elements $\mathsf{D}_{mn}$ are two-point correlators $\langle \Omega_n \Omega_n\rangle$ on the $S^4$ which contain the mixing.
	
	\subsection{The superconformal $\bf E$-theory}

For 4d $\mathcal{N}=2$ $\mathfrak{su}(N)$ gauge theories with $N_F$ hypermultiplets in the fundamental representation, $N_A$ hypermultiplets in the anti-symmetric representation and $N_S$ in the symmetric representation, the gauge coupling admits corrections only at 1-loop order, and the coefficient of its $\beta$-function is
	\begin{equation}
		\beta_0 = 2N - N_F - (N+2)N_S - (N-2)N_A \,.
	\end{equation}
	There exist five families of theories with vanishing $\beta_0$ \cite{Koh:1984}, which are called {\bf ABCDE} theories respectively. Their matter content is listed below:
	$$
	\begin{array}{|c||c|c|c|c|c|}
		\hline
		{\rm theory}  & {\rm\bf A} & {\rm\bf B} & {\rm\bf C} & {\rm\bf D} & {\rm\bf E} \\
		\hline 
		\hline
		N_F & 2N & N-2 & N+2 & 4 & 0\\
		N_S & 0 & 1 & 0 & 0 & 1\\
		N_A & 0 & 0 & 1 & 2 & 1\\
		\hline
	\end{array}
	$$
	We focus on the {\bf E}-theory in this work. The main reason for this is that the numerical method we employ below relies on the results of \cite{Billo1} which have been derived so far for this model only. It should be possible to generalize our discussion also to the {\bf D}-theory which shares many similarities with the {\bf E}-theory but we refrain from discussing this here since, among other differences, the corrections to the untwisted correlation functions in the {\bf D}-theory have a different scaling with $N$. 
	The {\bf A}, {\bf B}, and {\bf C}- theories have more qualitative differences with $\mathcal{N}=4$ SYM and their study may require different techniques. 
	
	One can calculate that for {\bf E} theory, the interaction term in the matrix model~\eqref{PartitionFunction} is given by the following expression
	\begin{equation}\label{EtheorySint}
		S_{\rm int}^{\rm \bf E} (a)= -4 \sum_{l,m = 1}^\infty \left(- \frac{g^2 }{8\pi^2 }  \right)^{l+m+1} \frac{(2l+2m+1)!}{(2l+1)!(2m+1)!}  \zeta(2l+2m+1) {\rm tr}\, a^{2l+1}\, {\rm tr}\, a^{2m+1},
	\end{equation}
where $\zeta(s)$ is the Riemann zeta function. 
	Expectation values in the matrix model are given by the usual expression
		\begin{equation}\label{EMM}
				\langle f(a) \rangle = \frac{\int da\, f(a)\, e^{-{\rm tr}\, a^2 - S_{\rm int}(a)} }{ \int da\, e^{-{\rm tr}\, a^2 - S_{\rm int}(a)}  } \,.
			\end{equation}
		This can also be written as
		\begin{equation}
			\langle f(a) \rangle = \frac{\left\langle f(a) e^{-S_{\rm int}} \right\rangle_{(0)} }{  \left\langle e^{-S_{\rm int}} \right\rangle_{(0)} },
		\end{equation}
		with the expectation value in the Gaussian model defined as
		\begin{equation}
			\langle f(a) \rangle_{(0)} =\int da e^{-{\rm tr}\, a^2} f(a) \,.
		\end{equation}
	One can also use the matrix model to calculate the correlation functions~\eqref{FTCorrelators}. Namely, in the planar limit, one finds 
	\begin{equation}\label{Def3ptFunctions}
		G_n =  \langle O_n(a) O_n(a) \rangle\,, \quad\text{and}\quad G_{m,n} =  \langle O_m(a) O_n(a) O_{m+n}(a) \rangle\,.
	\end{equation}
	The expectation values on the right hand sides are given by~\eqref{EMM}. The operators $O_n(a)$ are defined as \cite{Billo:2022xas, Gerchkovitz:2016bo}
	\begin{equation}\label{O2Omega}
	O_n(a) = \Omega_n(a) - \sum_{m < n} C_{n,m}O_m(a)\,,
\end{equation} 
	with $\Omega_n(a) = {\rm tr}\, a^{n}$ and the mixing coefficients $C_{n,m}$ given by
	\begin{equation}\label{mixingCmn}
		C_{n,m} =\frac{\langle \Omega_n(a) \Omega_m(a) \rangle }{\langle \Omega_m(a) \Omega_m(a) \rangle }\,.
	\end{equation} 
Equation~\eqref{O2Omega} can be viewed as resulting from applying Gram-Schmidt orthogonalization. 
	
From now on, we will denote the quantities related to the zero coupling limit with a suffix $^{(0)}$, which is identical to $\cN = 4$ YM. Then we define the normalized normal-ordered operators in the matrix model, focusing on the twisted sector:
	\begin{equation}\label{Defomega}
	\omega_k(a) \equiv O_{2k+1}^{(0)}(a) \Big/ \sqrt{G_{2k+1}^{(0)}},\quad \langle \omega_k(a) \omega_l(a) \rangle_{(0)} = \delta_{kl} ,
	\end{equation}
	where $G_{2k+1}^{(0)} = (2k+1)(N/2)^{2k+1}$ in the large $N$ limit.	As is pointed out in \cite{Beccaria:2007ln}, the correlators of $\omega_k(a)$ can be evaluated using Wick's theorem. We can regard the matrix operators $\omega_k(a)$ as a set of normally distributed real variables $\omega_k$ and write
	\begin{equation}\label{FreePFInomega}
		\langle \omega_{k_1}(a)  \omega_{k_2}(a) \cdots  \omega_{k_n}(a) \rangle = \int [\cD \omega]\omega_{k_1} \omega_{k_2}\cdots \omega_{k_n} e^{-\frac{1}{2}\mathbf{\omega}^T\mathbf{\omega}} ,\quad [\cD\omega]\equiv  \prod_{i=1}^\infty \frac{d\omega_i}{\sqrt{2\pi}}\,,
	\end{equation}
	where we denote $\omega$ as an infinite vector whose components are $\omega_k$. It was pointed out in \cite{Beccaria:2007ln} that one can re-express $S_{\rm int}$ for {\bf E} theory \eqref{EtheorySint} in terms of $\omega$ as
	\begin{equation}
		S^{\rm \bf E}_{\rm int} = - \frac{1}{2} \omega^T \mathsf{X}\,  \omega\,,
	\end{equation}
	where the infinite matrix $\mathsf{X}$ is given in \eqref{Xx}.
	
	We will study several observables using this formalism. As an example, the partition function of the matrix model \eqref{PartitionFunction}, ignoring the normalization factor, is given by:\footnote{The fact that $S_{\rm int}$ of {\bf E} theory only contains odd double-traces is important here and is the main reason we restrict ourselves to the {\bf E} theory. }
	\begin{equation}
		\cZ = \left\langle e^{-S_{\rm int}} \right\rangle_{(0)} =  \left\langle e^{\frac{1}{2} \omega^T \mathsf{X} \omega } \,\right\rangle_{(0)} .
	\end{equation}
	Using \eqref{FreePFInomega}, one obtains:
	\begin{equation}
	\cZ = \int [\cD\omega] e^{-\frac{1}{2}\omega^T (\mathbf{1} - \mathsf{X}) \omega} = {\rm det}^{-1/2}\left(\mathbf{1} - \mathsf{X} \right) \,.
	\end{equation}
	The corresponding free energy, which is actually the free energy of the ``difference theory'', is given by \cite{Tseytlin}
	\begin{equation}
		{\cal F} \equiv F^{\rm E-theory} - F^{\cN=4} = -\log\cZ = \frac{1}{2} \log \det ( 1 - \mathsf{X})\,.
		\label{FreeEnergyDifference}
	\end{equation} 

\section{Twisted correlators}\label{sec:Deltakinteq}

		Following the last section, the expectation value of any operator $f(\omega(a))$ containing only odd traces can be written in terms of the free model quantities as
		\begin{equation}
		\langle f(\omega(a)) \rangle = \frac{1}{\cZ } \int [\cD\omega] f(\omega) e^{-\frac{1}{2} \omega^T (\mathbf{1} - \mathsf{X}) \omega} \,.
		\end{equation}
		Thus the propagators of $\omega_k(a)$ in the interacting theory are given by 
		\begin{equation}\label{2ptFunctionomega}
		\langle \omega_k (a) \omega_l(a) \rangle =  \Big(\frac{1}{1 - \mathsf{X}}\Big)_{k,l } \equiv	 \Dx_{k,l} \,.
		\end{equation}
		Multiple correlators can be obtained by Wick contraction with propagator $\mathsf{D}$. 
		
		What we are really interested in are the operators $O_n(a)$. Thus we define the analogue of $\omega_k(a)$, which are by definition diagonal and suitably normalized:
		\begin{equation}
			\widehat{\omega}_k(a) \equiv \frac{O_{2k+1}(a) } {\sqrt{G_{2k+1}^{(0)}} },\quad \langle \widehat{\omega}_k(a) \widehat{\omega}_l(a) \rangle = \frac{\langle {\rm tr}\, \varphi^{2k+1}\, {\rm tr}\, \bar{\varphi}^{2l+1} \rangle }{ \langle {\rm tr}\, \varphi^{2k+1}\, {\rm tr}\, \bar{\varphi}^{2l+1} \rangle_{(0)}  } = \delta_{kl} \left( 1 + \Delta_k(\lambda) + O(1/N)^2\right)\,.
		\end{equation}
		Comparing with \eqref{Defomega}, we find that $\Delta_k(0) = 0$. As pointed out in \cite{Billo1}, from the Gram-Schmidt procedure, one finds from \eqref{2ptFunctionomega} that:\footnote{Here we use $A_{(k)}$ to denote the upper left $k \times k$ block of the matrix $A$. }
	\begin{align}\label{defDelta}
		1 + \Delta_k(\lambda) = 
		\frac{\det \Dx_{(k)}}{\det \Dx_{(k-1)}}\,.
	\end{align}
		In the remainder of this section, we will introduce a novel method to evaluate the matrix elements $\mathsf{D}_{kl}$ and the quantity $\Delta_k$ both numerically and analytically.

\subsection{Fredholm integral equations}
\label{sec:FredhomeIntegralEqn}

To calculate  $\Delta_k(\lambda)$ numerically one could truncate the matrix \eqref{Xx} to size $M\times M$ with $M$ large. The calculation of \eqref{defDelta} then only involves linear algebra with finite matrices. However, for large $M$, there are many integrals to calculate numerically which is slow. We therefore proceed as follows.
If $V$ is an $n \times m$ matrix, it is easy to check that
\begin{equation}\label{eq:eq1}
	(1 + V V^T)^{-1} = 1 - V(1 + V^T V)^{-1} V^{T}\,.
\end{equation}
Notice that the inverse on the left is of an $n\times n$ matrix, whereas the inverse on the right is of an $m \times m$ matrix.
This can also be written as
\begin{equation}\label{eq:eq2}
	(1 + V V^T)^{-1} = 1 - V Z\,,
\end{equation}
where $Z$ is the solution of the equation 
$ (1 + V^T V) Z = V^{T}$.

A limiting procedure for $m \to \infty$ then gives the following result. If $V_k(t)$ are functions with $k = 1 , \ldots, n$ and $t \in [ 0, + \infty[ $ and if we define a matrix $V V^T$ as
\begin{equation}
	( V V^T)_{kl} = \int_0^{+\infty}\!\!\!\! dt\ V_k(t)\ V_l(t)\,, 
	\label{VVT}
\end{equation}
then
\begin{equation}\label{eq:eq3}
	(1 + V V^T)^{-1}_{kl} = \delta_{kl} -  \int_0^{+\infty}\!\!\!\! dt\ V_k(t)\ Z_l(t)\,,
\end{equation}
where $Z_k(t)$ is the solution of the integral equation
\begin{equation}\label{eq:eq4}
	Z_k(t) +  \int_0^{+\infty}\!\!\!\! ds\  M(t,s) Z_k(s)= V_k(t)\,,
\end{equation}
with $ M(t,s) = \sum_{k=1}^n V_k(t) V_k(s)$. The matrix $\mathsf{X}$ defined in \eqref{Xx} has the form \eqref{VVT}, thus its inverse matrix $\mathsf{D}$ can be written using \eqref{eq:eq3} as
\begin{equation}\label{eq:eq6}
	\Dx_{kl} = \delta_{kl} - (-1)^{k+l} \sqrt{2k+1} \sqrt{2l+1}  
	\int_0^{+\infty}\!\!\!\! dt\ v(t) J_{2 k+1} (t) \Psi_{2 l+1} (t)\,,
\end{equation}
where  $\Psi_{2 l+1} (t)$ is the solution of the integral equation
\begin{equation}
		\Psi_{2 k+1} (t) + \int_0^{+\infty}\!\!\!\! ds\ K(t,s) \Psi_{2 k+1} (s) = v(t) J_{2 k+1} (t)\,, \label{eq:eq7}
\end{equation}
with
\begin{equation}
		\qquad K(t,s) = v(t) K_B(t,s) v(s)\,,\qquad\text{and}\qquad 	v(t)^2 = \dfrac{16 \pi}{\sqrt{\lambda}}\ W\left(\dfrac{2 \pi t}{\sqrt{\lambda}}\right)\,.\nonumber
\end{equation}
	Here $K_B(t,s)$ is the Bessel kernel \cite{TracyWidom}, which can be evaluated in closed form as \cite{Billo1}\footnote{A short proof of this identity can be found in \cite{Tygert}.}: 
\begin{equation}\label{eq:kernelK2}
	K_B(t,s) \equiv \sum_{k=1}^{+\infty} (2k+1)J_{2 k+1}(t) J_{2 k +1}(s) =  - \frac{1}{2} \frac{t s}{t^2 - s^2} \left(t J_1(t) J_2(s) - s J_2(t)J_1(s)\right)\,.
\end{equation}
All in all, we have converted the original definition \eqref{2ptFunctionomega} of $\Dx_{kl}$, which uses the inverse of an infinite matrix,
to an expression which uses instead the solution of an integral equation. In Section~\ref{sec:smallDeltaAN} we will perform a check of  the equations derived above by comparing against the expansion in small $\lambda$. 
We are not able to solve the integral equation \eqref{eq:eq7} analytically, therefore we resort to a numerical method which we discuss in Section~\ref{sec:Deltanum}.

\subsection{Analytical comparison against small $\lambda$ expansion}
\label{sec:smallDeltaAN}

In this section we calculate the small $\lambda$ expansion of $\Dx_{k,l}$ using equations \eqref{eq:eq6} and \eqref{eq:eq7}. For convenience, in this section, we use the notation $\mu = \dfrac{\sqrt{\lambda}}{2 \pi}$. We also change variables $t / \mu = x$ and  
$s/ \mu = y$ and define $ \chi_{2 k+1}(x) = \Psi_{2 k+1}(\mu x)$.
Equations \eqref{eq:eq6} and \eqref{eq:eq7} then become
\begin{equation}\label{small:D}
	\Dx_{kl} = \delta_{kl} - (-1)^{k+l} \sqrt{2k+1} \sqrt{2l+1}\ ( 8 \mu)^{1/2}  
	\int_0^{+\infty}\!\!\!\! dx\ W(x)^{1/2} J_{2 k+1} (\mu x) \chi_{2 l+1} (x)\,,
\end{equation}
and
\begin{equation}\label{small:inteq}
	\chi_{2 k+1} (x) + 8 \int_0^{+\infty}\!\!\!\! dy\ W(x)^{1/2}\ K_B(\mu x, \mu y) W(y)^{1/2}\chi_{2 k+1} (y) 
	= \left(\frac{8}{\mu} \right)^{1/2} W(x)^{1/2}\ J_{2 k+1} (\mu x) \,.
\end{equation}
Using the Taylor series expansion of Bessel functions, equation \eqref{eq:kernelK2} leads to the following small $\lambda$ (or equivalently small $\mu$) expansion
$$
K_B(\mu x, \mu y) = \frac{1}{768} x^3 y^3 \mu^6 - \frac{1}{12288} ( x^5 y^3 + x^3 y^5  ) \mu^8 +O(\mu)^{10}\,.$$
Therefore, up to this order, the kernel of the integral operator in \eqref{small:inteq} is degenerate, and the integral equation 
can be solved exactly.\footnote{See Appendix~\ref{sec:degkernels} for more details on how to implement this and for the notation we use.}
We can thus write
\begin{equation}
K(x,y) = \sum_{i=1}^3 a_i(x) b_i(y) + O(\mu)^{10}\,,
\end{equation}
with
\begin{align*}
	a_1(x) & = \frac{8}{768} x^3 W(x)^{1/2}\,, & b_1(y) & = y^3 W(y)^{1/2}\mu^6\,, \\
	a_2(x) & = - \frac{8}{12288}  x^5 W(x)^{1/2}\,, & b_2(y) & = y^3 W(y)^{1/2}\mu^8\,, \\
	a_3(x) & = - \frac{8}{12288}  x^3 W(x)^{1/2}\,, & b_3(y) & = y^5 W(y)^{1/2}\mu^8\,.
\end{align*}
The $3 \times 3 $ matrix
\begin{equation}
	A_{ki} = \int_0^{+\infty}\!\!\!\! dx\ b_k(x) a_i(x)\,,
\end{equation}
can be calculated analytically by employing the Mellin transform
\begin{equation}\label{small:Mellin}
	\int_0^{+\infty}\!\! \frac{dx}{x} x^p\ W(x) = \Gamma(p-1) \zeta(p-2)\,,
\end{equation}
to find
\begin{equation}\label{small:A}
	A = 
	\left(
	\begin{array}{ccc}
		\dfrac{5  }{4} \zeta (5)\mu ^6& -\dfrac{105}{32} \zeta (7) \mu ^6 & -\dfrac{5}{64} \zeta (5) \mu ^6 \\[4mm]
		\dfrac{5 }{4} \zeta (5) \mu ^8 & -\dfrac{105}{32}  \zeta (7) \mu ^8 & -\dfrac{5}{64} \zeta (5) \mu ^8 \\[4mm]
		\dfrac{105 }{2} \zeta (7) \mu ^8 & -\dfrac{945}{4}  \zeta (9) \mu ^8 & -\dfrac{105}{32} \zeta (7) \mu ^8 \\
	\end{array}
	\right)+O(\mu)^{10}\,.
\end{equation}
The inverse is given by
$$
(1 + A)^{-1} =  
\left(
\begin{array}{ccc}
	1-\dfrac{5 }{4} \zeta (5)\mu ^6  & \dfrac{105 }{32}\zeta (7)\mu ^6  & \dfrac{5}{64} \zeta (5)\mu ^6  \\[4mm]
	-\dfrac{5 }{4}\zeta (5)\mu ^8  & 1+\dfrac{105 }{32}\zeta (7)\mu ^8  & \dfrac{5 }{64}\zeta (5)\mu ^8  \\[4mm]
	-\dfrac{105 }{2}\zeta (7)\mu ^8  & \dfrac{945 }{4}\zeta (9)\mu ^8  & 1+\dfrac{105 }{32}\zeta (7)\mu ^8
\end{array}
\right)+O(\mu)^{10}\,.
$$
Equation \eqref{eq:deg8} then gives
\begin{multline*}
\chi_{2 k +1}(x) =  \left(\frac{8}{\mu} \right)^{1/2}W(x)^{1/2} J_{2 k+1} (\mu x) \\
- \sum_{i,j=1}^3 \int_0^{+\infty}\!\!\!\! dy\ a_i(x) ( 1 + A)^{-1}_{ij} b_j(y)  \left(\frac{8}{\mu} \right)^{1/2} W(y)^{1/2} J_{2 k+1} (\mu y)\,.
\end{multline*}
 
One can now expand the Bessel functions again as a series in small $\mu$ and calculate the integral with \eqref{small:Mellin}. This gives an expansion for $\chi_{2 k +1}(x)$ at small $\mu$. The result of this procedure is fairly complicated and we do not present it explicitly here. Finally, inserting $\chi_{2 k +1}(x)$ into equation \eqref{small:D} gives $\Dx_{kl}$ as a series in $\mu$:
\begin{align*}
	D_{11} & = 1-\frac{5 }{4}\zeta (5)\mu ^6 +\frac{105}{16} \zeta (7)\mu ^8 -\frac{1701}{64} \zeta (9)\mu ^{10}
	+ \left(\frac{25 }{16}\zeta (5)^2+\frac{12705}{128} \zeta (11)\right)\mu ^{12}+O\left(\mu\right)^{13}\,,\\[4mm]
	D_{12}&=D_{21} = \frac{7}{32} \sqrt{15}\ \zeta (7)\mu ^8 -\frac{105}{64} \sqrt{15}\ \zeta (9)\mu ^{10} +\frac{1089}{128} \sqrt{15}\  \zeta (11) \mu ^{12}+O\left(\mu\right)^{13}\,,\\[4mm]
	D_{22}&= 1-\frac{63}{64} \zeta (9)\mu ^{10} +\frac{1155 }{128}\zeta (11)\mu ^{12} +O\left(\mu\right)^{13}\,.
\end{align*}
This then finally leads to 
\begin{align*}
	\Delta_1(\lambda) & = -\frac{5 }{4}\zeta (5)\mu ^6 +\frac{105}{16} \zeta (7)\mu ^8 -\frac{1701}{64} \zeta (9)\mu ^{10}
	+ \left(\frac{25 }{16}\zeta (5)^2+\frac{12705}{128} \zeta (11)\right)\mu ^{12}+O\left(\mu\right)^{13}\,,\\[4mm]
	\Delta_2(\lambda) & =-\frac{63}{64} \zeta (9)\mu ^{10} +\frac{1155 }{128}\zeta (11)\mu ^{12} -\frac{27885 }{512} \zeta (13)\mu ^{14}+O\left(\mu\right)^{15}\,.
\end{align*}
This agrees with equations (3.37) and (3.38) in \cite{Billo1}. Since the results in \cite{Billo1} were obtained by a different method we view this as a consistency check of our approach.

\subsection{Numerical method for the calculation of $\mathsf{D}_{kl}$ and $\Delta_k(\lambda)$}\label{sec:Deltanum}

To solve the integral equation \eqref{eq:eq7} we employ numerics and use the Nystr\"om method. 
This method is well known and is based on discretizing the integral appearing in the integral equation in the schematic form
\begin{equation}\label{eq:eq11}
	\int_0^{+\infty}\!\!\!\! dt\ f(t) \approx \sum_{a=1}^m w_a f(t_a)\quad\text{with $m$ large}\,.
\end{equation}
Here, $w_a \ge 0$ are weights and $t_a$ are the discretization points. The upper left $p \times p$ block of the matrix $\Dx$ is then equal to
\begin{equation}\label{eq:eq12}
	\Dx_{(p)} = \mathbf{1}_{ p \times p} - \mathbb{V} \left( \mathbf{1}_{m \times m} + \mathbb{K}\right)^{-1} \mathbb{V}^T\,,
\end{equation}
where $\mathbb{V}$ is a $p \times m$ matrix and $\mathbb{K}$ is an $m \times m$ matrix given by
\begin{equation}
\begin{split}
	\mathbb{V}_{ka} &= \sqrt{w_a}\ (-1)^k \sqrt{2 k+1}\ J_{2 k+1}(t_a)\ v(t_a)\,,\\
\mathbb{K}_{ab} &= \sqrt{w_a}\ K(t_a, t_b)\ \sqrt{w_b}\,.	
\end{split}
\end{equation}
If $m \to \infty$, \eqref{eq:eq12} should converge to the correct result. 
All in all, what we have done is instead of truncating the original matrix \eqref{Xx} to size $M \times M$ and treating the integrals over the Bessel functions exactly, we have taken $M \to \infty$ and discretized the integrals. 

\begin{figure}[ht]
	\centering
	\includegraphics[scale = .7]{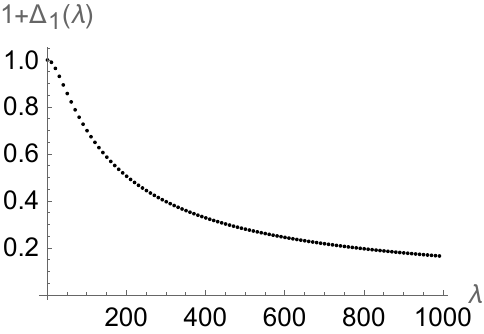} \includegraphics[scale = .7]{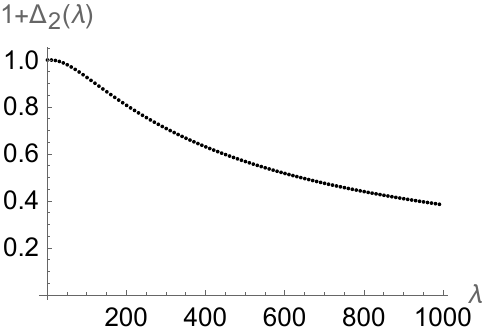}
	\caption{ $1 + \Delta_k(\lambda)$ as function of $\lambda$ with $k=1$ (left) and $k=2$ (right).}\label{fig1}
\end{figure}

\begin{figure}[H]
	\centering
	\includegraphics[scale = .9]{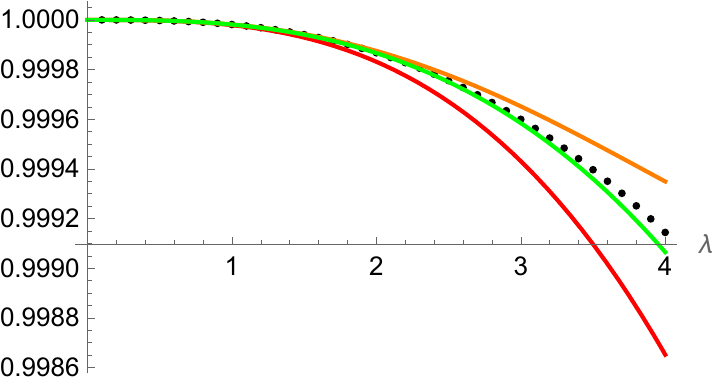}
	\caption{Black dots: numerical results for $1+\Delta_1(\lambda)$.	Red line: $ 1-\frac{5\,\zeta (5)}{256\pi^6}\lambda^3$;
	Orange line: $ 1-\frac{5\,\zeta (5)}{256\pi^6}\lambda^3+\frac{105\, \zeta (7)}{4096\pi^8}\lambda^4$;
	Green line: $1-\frac{5\,\zeta (5)}{256\pi^6}\lambda^3+\frac{105\, \zeta (7)}{4096\pi^8}\lambda^4-\frac{1701\, \zeta (9)}{65536\pi^{10}} \lambda^5$.}\label{Figure0630a}
\end{figure}

There are many discretization schemes (also known as quadrature rules) that one can use in~\eqref{eq:eq11}. Some of these are discussed in Appendix \ref{app:discretisation}. We chose the Fej\'er type 1 quadrature rule. In this rule,  we also have to truncate the integral
$$\int_0^{+\infty}\!\!\! dt\ \ f(t) \approx \int_0^L \!\!\! dt\ \ f(t) \approx \sum_{a=1}^m w_a f(x_a)\,.$$
So we have to take $L$ and $m$ both large to get accurate results. More details on Fej\'er type 1 and the values of $L$ and $m$ can be found in
Appendix \ref{app:discretisation}.
As an illustration, the two graphs in Figure \ref{fig1} were made using equation~\eqref{eq:eq12} with appropriate settings for $L$ and $m$.
As a check on the  accuracy of the numerical algorithm we can compare our numerical result against the analytic calculations of the small $\lambda$ expansion of $\Delta_k(\lambda)$ discussed in \cite{Billo1}. The first few terms in the small $\lambda$ expansion for $\Delta_1$ read 
\begin{equation}\label{Delta1small}
	\Delta_1(\lambda)  =-\frac{5\,\zeta (5)}{256\pi^6}\lambda^3+\frac{105\, \zeta (7)}{4096\pi^8}\lambda^4-\frac{1701\, \zeta (9)}{65536\pi^{10}} \lambda^5+\cdots\,.
\end{equation}
In Figure~\ref{Figure0630a} we compare this analytic result against our numerics. It is clear that there is very good agreement between the two calculations for small values of $\lambda$.\footnote{Since the numerical coefficients multiplying the powers of $\lambda$ in \eqref{Delta1small} are small ($10^{-5}$ or smaller) we can treat order 1 values for $\lambda$ as ``small''.} 

As discussed in \cite{Beccaria:2007ln}, the radius of convergence of the series for $\Delta_1$ is $\lambda_c = \pi^2$ because of the branch point located at $\lambda = -\pi^2$, see also \cite{Russo:2013sba}.

In Table~\ref{tab:smallDelta1} we provide additional numerical evidence for the agreement between the two methods. 
\begin{table}[H]
	\centering
	$$
	\begin{array}{|l|lllll|}
		\hline
		n \backslash \lambda & 1 & 2 & 3 & 4 & 20 \\
		\hline
		1 & -0.0000210659 & -0.000168527 & -0.000568778 & -0.00134822 & -0.168527 \\
		2 & -0.0000183417 & -0.000124939 & -0.000348117 & -0.000650816 & 0.267347 \\
		3 & -0.0000186194 & -0.000133826 & -0.000415601 & -0.000935195 & -0.621336\\
		4 & -0.0000185927 & -0.000132119 & -0.000396155 & -0.000825934 & 1.08587 \\
		5 & -0.0000185952 & -0.000132442 & -0.000401667 & -0.000867224 & -2.1399 \\
		\hline
		\text{numerics} & -0.000018595 & -0.000132391 & -0.000400453 & -0.000855928 &
		-0.0326525 \\
		\hline
	\end{array}
	$$
	\caption{Comparison of the small $\lambda$ expansion against numerical calculations of $\Delta_1(\lambda)$ for different values of $\lambda$. The last row gives the value which is calculated with the numerical method explained in Section~\ref{sec:Deltanum}. The first row with $n=1$ gives the result using 1 term in the expansion~\eqref{Delta1small}, the second row with $n=2$ gives the result using 2 terms in the expansion~\eqref{Delta1small} and so on. It is clear that for small $\lambda < \lambda_c $ ($\lambda = 1,2,3,4$ in the table) the series~\eqref{Delta1small} converges to the numerical result as $n$ is increased. For large $\lambda > \lambda_c $ ($\lambda = 20$ in the table), the series~\eqref{Delta1small} does not converge, but the numerical method of Section~\ref{sec:Deltanum} still works.}\label{tab:smallDelta1}
\end{table}

We can proceed similarly and test our method for $k=2$, see Table~\ref{tab:smallDelta2}. To do this we use the result for the small $\lambda$ expansion of $\Delta_2$ from \cite{Billo1} 
\begin{equation}\label{Delta2small}
	\Delta_2(\lambda)  =-\frac{63\,\zeta (9)}{65536\pi^{10}}\lambda^5+\frac{1155\, \zeta (11)}{524288\pi^{12}}\lambda^6-\frac{27885\, \zeta (13)}{8388608\pi^{14}} \lambda^7+\cdots \,.
\end{equation}

\begin{table}[H]
	\centering 
	$$
	{\tiny \begin{array}{|l|lllll|}
		\hline
		n \backslash \lambda & 1 & 2 & 3 & 4 & 20 \\
		\hline
		1 & -1.02857\times 10^{-8} & -3.29142\times 10^{-7} & -2.49942\times 10^{-6}
		& -1.05325\times 10^{-5} & -3.29142\times 10^{-2} \\
		2 & -7.90102\times 10^{-9} & -1.76523\times 10^{-7} & -7.60998\times 10^{-7}
		& -7.64937\times 10^{-7} & 1.19705\times 10^{-1} \\
		3 & -8.26546\times 10^{-9} & -2.23172\times 10^{-7} & -1.55805\times 10^{-6}
		& -6.73606\times 10^{-6} & -3.46789\times 10^{-1} \\
		4 & -8.21887\times 10^{-9} & -2.11244\times 10^{-7} & -1.25232\times 10^{-6}
		& -3.68225\times 10^{-6} & 8.46105\times 10^{-1} \\
		5 & -8.22428\times 10^{-9} & -2.14016\times 10^{-7} & -1.35889\times 10^{-6}
		& -5.1016\times 10^{-6} & -1.92606 \\
		10 & -8.22374\times 10^{-9} & -2.13514\times 10^{-7} & -1.33223\times
		10^{-6} & -4.65977\times 10^{-6} & 7.26826\times 10^1 \\
		\hline
		\text{num.} & -8.22374\times 10^{-9} & -2.13514\times 10^{-7} &
		-1.3323\times 10^{-6} & -4.66487\times 10^{-6} & -1.79469\times 10^{-3} \\
		\hline
	\end{array} }
	$$
	\caption{Comparison of the small $\lambda$ expansion against numerical calculations of $\Delta_2(\lambda)$ for different values of $\lambda$. The notation is the same as in Table~\ref{tab:smallDelta1} and the conclusions are similar.}\label{tab:smallDelta2}
\end{table}

\subsection{Large $\lambda$ expansion of $\mathsf{D}_{kl}$ and $\Delta_k$: a conjecture}
	
As outlined above and discussed in detail in \cite{Billo1} it is relatively easy to calculate the small $\lambda$ expansion of $\mathsf{D}_{kl}$. The large $\lambda$ expansion appears to be much more challenging. The first term was calculated in \cite{Billo1} and reads (assuming $k \le l$), 
		\begin{equation}
		\mathsf{D}_{kl} =  \frac{4\pi^2}{\lambda}  \sqrt{(2k+1)(2l+1)}\ k(k+1)+O(1/\lambda^{3/2})\,.
		\end{equation}
The matrix elements for $k \ge l$ follow from the symmetry $\mathsf{D}_{kl} = \mathsf{D}_{lk}$. We are not able to rigorously calculate the subleading terms in the expansion analytically. However, based on numerical investigations, we make the following conjecture for $\mathsf{D}_{kl}$ with $k \le l$,
		\begin{equation}\label{Dkl}
	\begin{aligned}
		\mathsf{D}_{kl}  = 4\pi^2 &\sqrt{(2k+1)(2l+1)} \frac{k(k+1)}{\lambda} \left\{1 - 8l(l+1) \frac{\log 2}{\lambda^{1/2}} \right.\\
		& 	\left. +\frac{2 \pi^2 }{9 \lambda} \left[ 3 l(l+1)(l-k)(l+k+1)+2(k-1)k(k+1)(k+2)  \right] \right.\\
		& \left. + \frac{16\log^22}{\lambda} l(l+1)[k(k+1)+l(l+1)-1]  \right.\\
	 	& \left. - \frac{16\pi^2\log 2}{9\lambda^{3/2}} L [K ( K -2) +L(L-2)] \right.\\ 
		& \left. - \frac{64\log^32}{9\lambda^{3/2}} L [2 L^2 + 6 K L - 7 L - 7 K + 2 K^2 +6] \right.\\
		& \left. - \frac{\zeta(3)}{3 \lambda^{3/2} } L  [3 + 4 ( 4 K^2 + 4 L^2 - 5 K - 5 L - 6 K L)] + O(\lambda)^{-2} \right\} 
	\end{aligned}
\end{equation}
We have used the notation $K = k (k+1)$ and $L = l ( l+1)$ to shorten some of the coefficients.	The expression \eqref{Dkl} was obtained as follows. First, we used the numerical method outlined in Section~\ref{sec:Deltanum} to calculate $\mathsf{D}_{kl}$ for many values of $k,l$ and many $\lambda$ ranging\footnote{For $\lambda$ greater than $e^{16}$, our algorithm converges too slowly, and the precision obtained is not good enough. } from $e^8$ to $e^{16}$. We then used this numerical data and the function {\tt LinearModelFit[]} of {\tt Mathematica} to estimate numerically the coefficients in the large $\lambda$ expansion of $\mathsf{D}_{kl}$. Finally, the precision in these fitted coefficients was sufficiently high so that we could guess\footnote{We were able to guess a closed form expression of this additional term only after the publication of \cite{Beccaria:2022ypy}. In \cite{Beccaria:2022ypy} an analytical expression for the {\it determinant} of $D_{kl}$ is obtained, this helped us to guess closed from expressions of the components $D_{kl}$ themselves. } the closed form expression \eqref{Dkl}. 
	
%
The analytical form \eqref{Dkl} for the  strong coupling behavior of $\mathsf{D}_{kl}$ is used as ingredient in the calculation of the three point function of single trace chiral/anti-chiral correlators discussed in Section \ref{sec:Threeptfc}. It can also be directly used in \eqref{defDelta} to find the quantity $\Delta_k(\lambda)$ that controls the two-point function of twisted operators. Using \eqref{defDelta} and \eqref{Dkl} we find
	\begin{equation}
		1+\Delta_k(\lambda) = \frac{8 \pi^2  k ( 2 k+1)}{\lambda} \left[1 
		-\frac{16 k\log 2}{\lambda^{1/2}}
		+\frac{32 k( 4 k -1)\log^2 2}{\lambda} +O(\lambda^{-3/2})
		\right]\,.
		\label{NikolayConjecture}
	\end{equation}
A notable fact is that the $\pi^4$ terms appearing at $O(1/\lambda^{2})$ order in $\mathsf{D}_{kl}$, see \eqref{Dkl}, cancel with each other in $\Delta_k$, leading to a relatively simple result that improves on the calculations in \cite{Billo1} by providing two more terms in the large $\lambda$ expansion. Some observations can be made based on the leading three terms. First, the powers series is in inverse powers of $\sqrt{\lambda}$, which is compatible with the string theory perturbation series in $\alpha'$ after using the identification $\lambda^{-1} \sim \alpha'^2$. Second, the coefficients of the three terms are polynomials of $k$, whose order increases by 1 with each term in the series and which exhibit a factorized form. Third, we find an increasing power of $\log 2$ in each term in the perturbative series. This factor increases the degree of transcedentality of each perturbative coefficient and perhaps suggests a renormalization of the coupling $\lambda$.

\begin{figure}[H]
	\centering
	\subfloat[][ 
	Red dots: $\log\left(1+\Delta_1(\lambda)\right)$.\\ Red line: $\log\left(\Delta_1^{(1)}(\lambda)\right)$. ]{
		\includegraphics[width=0.5\textwidth]{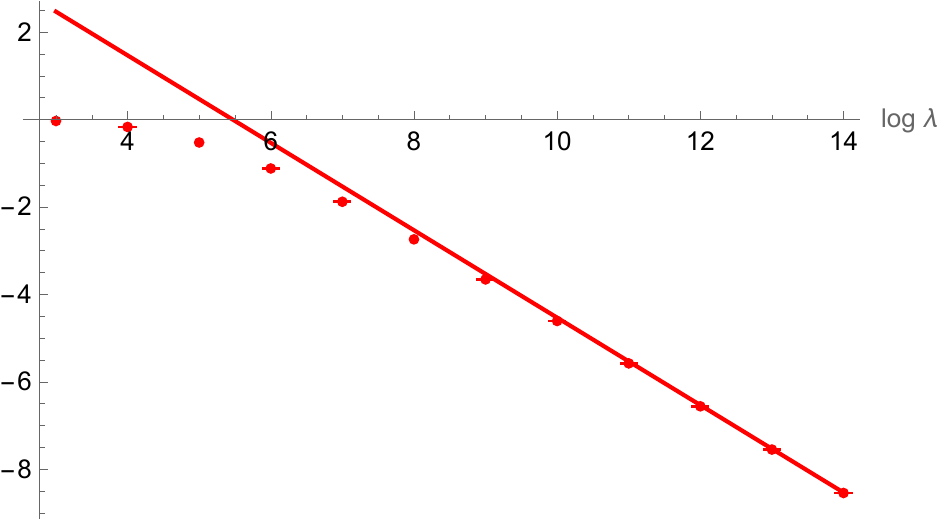}
				\label{Figure0611a}}
	\subfloat[][Orange dots: $\log\left( - (1+\Delta_1(\lambda) - \Delta_1^{(1)}(\lambda) \right) $. 
	Orange line: $\log\left( - \Delta_1^{(2)}(\lambda)\right) $. ]{
		\includegraphics[width=0.5\textwidth]{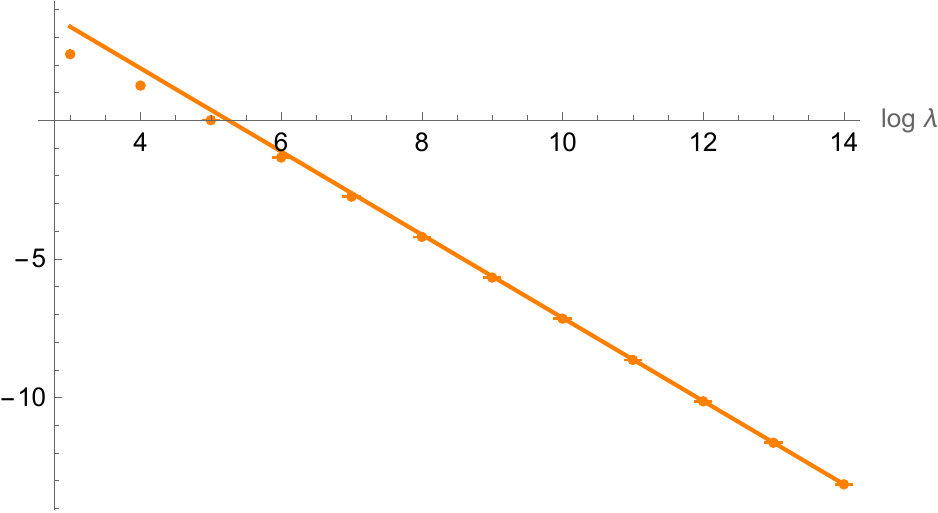}
				\label{Figure0611b}}\\
	\subfloat[][Green dots: $\log\left(1+\Delta_1(\lambda) - \Delta_1^{(1)}(\lambda) - \Delta_1^{(2)}(\lambda) \right) $. 
	Green line: $\log\left(\Delta_1^{(3)}(\lambda)\right) $.]{
		\includegraphics[width=0.5\textwidth]{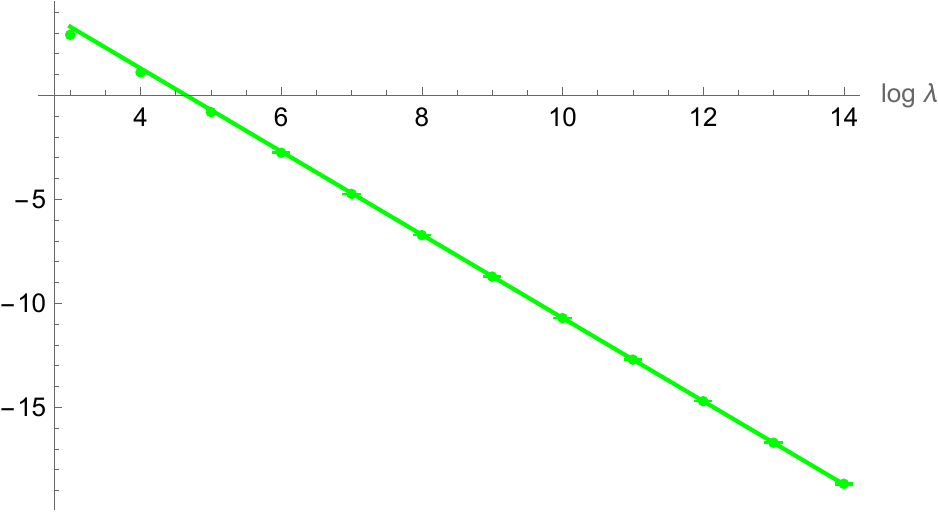}
				\label{Figure0611c}}
	\caption{A loglog plot of $1+\Delta_1(\lambda)$ and its asymptotic expansion. For large $\lambda$, the numerical data agrees very well with the asymptotic expansion. This indicates that all terms in the asymptotic expansion \eqref{NikolayConjecture} are correct.}
	\label{fig:globfig}
\end{figure}

\begin{figure}[H]
	\centering
	\includegraphics[scale = .5]{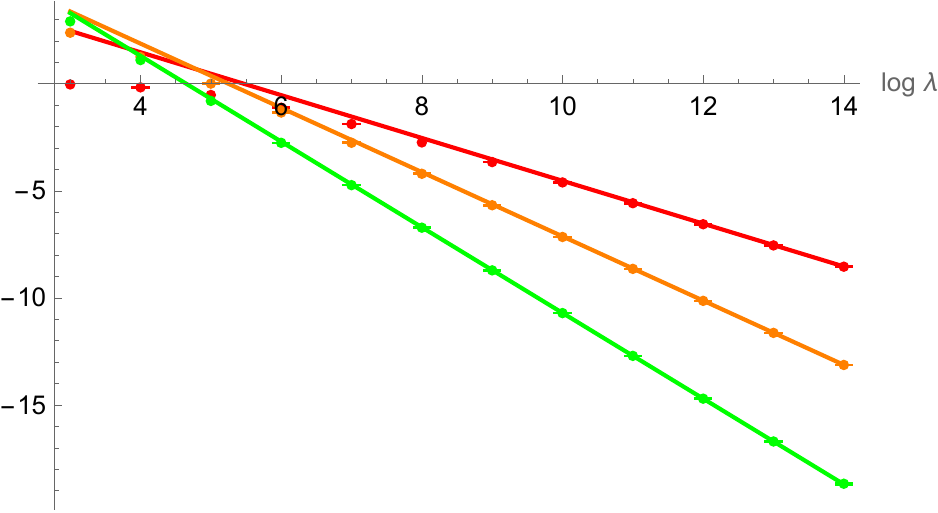}
	\caption{A combination of Figures \ref{Figure0611a}-\ref{Figure0611c}. The slope of the lines decrease if more terms in the asymptotic expansion are included. This illustrates that the coefficients in the asymptotic expansion \eqref{NikolayConjecture} are correct.  
	}\label{Figure0611d}
\end{figure}

\begin{figure}[H]
	\centering
	\includegraphics[width=0.45\textwidth]{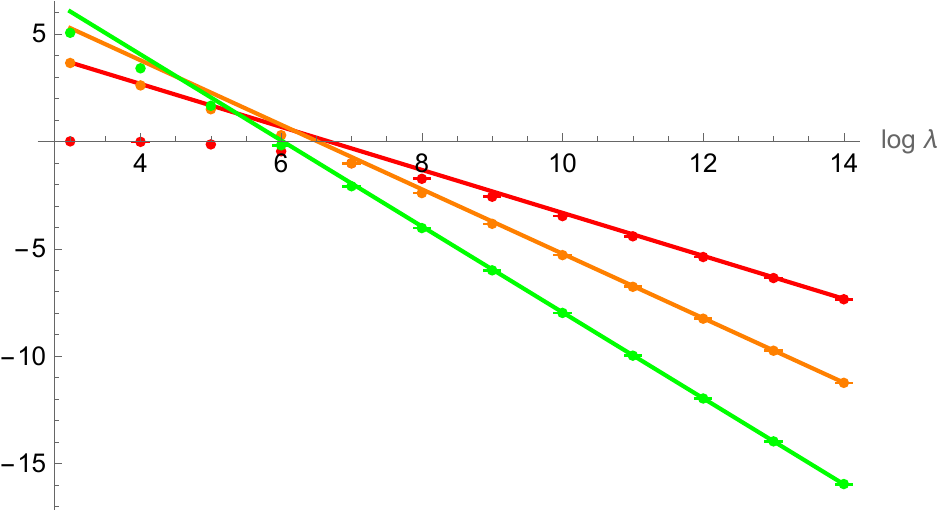}
	\includegraphics[width=0.45\textwidth]{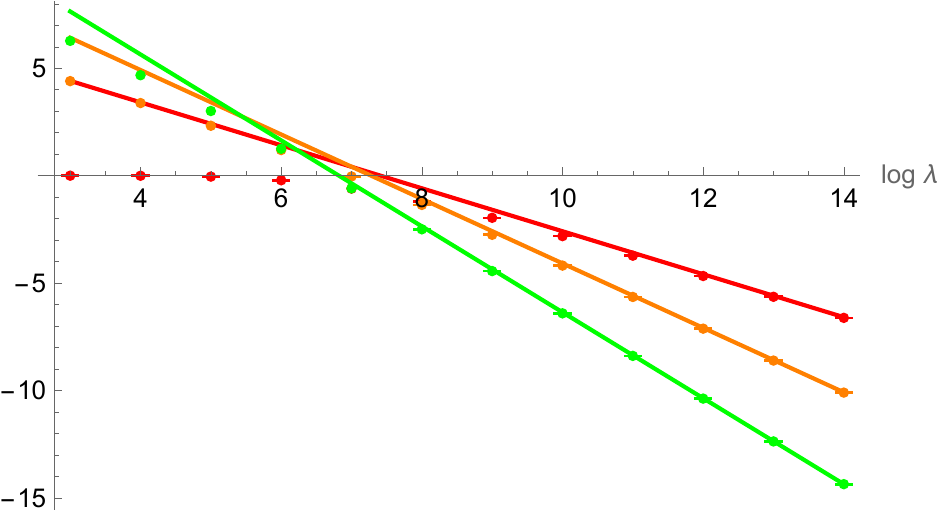}
	\caption{Similar to Figure \ref{Figure0611d}, but with $k=2$ (left) and $k=3$ (right) respectively.
	}\label{Figure0611ef}
\end{figure}

To illustrate the agreement between the analytic conjecture~\eqref{NikolayConjecture} and our numerical results we present a few plots. To facilitate the comparison we define the following three quantities that represent the three terms on the right hand side of the expansion~\eqref{NikolayConjecture}
\begin{equation}
\begin{split}
	\Delta_k^{(1)}(\lambda) & = 8 \pi^2 \left[ \frac{k ( 2 k+1)}{\lambda}
	\right]\,,\\
	\Delta_k^{(2)}(\lambda) & = 8 \pi^2 \left[-\frac{16 k^2 ( 2 k +1) \log 2}{\lambda^{3/2}}\right]\,,\\
	\Delta_k^{(3)}(\lambda) & = 8 \pi^2 \left[\frac{32 k^2 ( 2 k +1) (4 k -1)\log^2 2}{\lambda^{2}}
	\right]\,.
\end{split}
\end{equation}
In the figures below, we have added error bars on the numerical data of $1+\Delta_k(\lambda)$ with an estimation of accuracy and we have set $k=1$ in Figures~\ref{fig:globfig} and \ref{Figure0611d}. Similar results can be obtained for other values of $k$ which we illustrate for $k=2,3$ in Figure~\ref{Figure0611ef}. 

Similarly, Figure \ref{Figure0611ef} serves as evidence that the asymptotic expansion \eqref{NikolayConjecture} is valid for $k=2$ and $k=3$.

\subsection{Large $\lambda$ expansion of $\Delta_k(\lambda)$: more terms}
\label{sec:Deltamoreterms}
The matrix elements $\mathsf{D}_{kl}$ are harder to evaluate than the single-index quantity $\Delta_k$ that controls the two-point functions of physical interest.  Utilizing the same numerical method described in Section~\ref{sec:Deltanum}, we can evaluate $\Delta_k(\lambda)$ directly for $\lambda$ going up to $e^{16}$ and $k$ up to 25. A selection of the numerical data we have obtained can be found in Appendix~\ref{app:tables}. This numerical data agrees very well with the three terms in the expansion in \eqref{NikolayConjecture} as expected and allows us to find an expression for more subleading terms in the large $\lambda$ expansion of $\Delta_k(\lambda)$.
	
For future convenience, let us denote the coefficients in the large $\lambda$ expansion with $C_i$ such that $1 + \Delta_k(\lambda) = \sum_i C_i\left( \frac{1}{\lambda}\right)^i$. At $O(1/\lambda^{5/2})$, the structure of the coefficient is more or less the same as in \eqref{NikolayConjecture}, but there is a curious shift in the polynomial factor $(4k-3)$, namely we find:
\begin{equation}
	\label{C5p2}
	C_{5/2} = -8\pi^2 \cdot 32(\log^32)\cdot k^2(2k+1)(4k-1)\left[(4+4\epsilon_{5/2})k - (3-\epsilon_{5/2}) \right]\,,
\end{equation}
where $\epsilon_{5/2} \approx 0.107738$, which does not  appear to be any simple ``closed form'' number. Proceeding to higher order we find that at $O(1/\lambda^{3})$, the factor $(4k-3)$ which is shifted in $C_{5/2}$ is however unshifted, but there is a new factor $(4k-7)$ that is shifted:
\begin{equation}
	\label{C3}
	C_3 = 8\pi^2 \cdot 16 (\log^4 2) \cdot k^2  (2k+1)(4k-1)(4k-3)[(4+4\epsilon_3)k - (7-\epsilon_3)],\quad \epsilon_3 \approx -0.1381\,.
\end{equation}
	We have also observed a numerical coincidence between the two shifts. Namely, we find the following relation to be correct up to four digits:
\begin{equation}
	\epsilon_3 \approx 8 \epsilon_{5/2} - 1 \approx -0.13809664\,.
\end{equation}
The fact that we do not have closed form fully analytic expressions for $C_{5/2}$ and $C_3$ but rather have these coefficient	depend on the numerical constants $\epsilon_{5/2} $ and $\epsilon_3$ limits the accuracy of our numerical investigations due to the limited accuracy in the {\tt LinearModelFit[]} function in {\tt Mathematica}. The highest order in the large $\lambda$ expansion we could reliably evaluate is at $O(1/\lambda^{7/2})$ for which we find the coefficient 
\begin{equation}
	\label{C7p2}
	C_{7/2} = 8\pi^2 \cdot (\log^52)\cdot 2^8 k^2(2k+1)(4k-1)(4k-3)\cdot [(1+\epsilon_{7/2})k^2 - (2-2\epsilon_{7/2})]\,,
\end{equation}
where $\epsilon_{7/2} \approx 0.023$. The numerical data we used to obtain the expressions for $C_{5/2}$, $C_{3}$, and $C_{7/2}$ above is presented in Table~\ref{DataDeltakLargeLambda}.
	
	\begin{table}[h]
		\centering
		\begin{tabular}{|c||c|c|c|c|}
			\hline
			$k$ & $C_{5/2}$  & $C_{3}$ & $C_{7/2}$ & $L$ \\\hline\hline 
			1  & $-11652.$ & $-9.6\times 10^3$ & $3\times 10^4$  & 5000\\
			2 & $-70322(2\pm 2).$ & $-5.0 \times 10^{4}$ & $5 \times 10^{6}$ & 6000\\
			3 & $-6.06468 \times 10^6$& $5.83 \times 10^6$& $1.46 \times 10^8$ & 6000\\
			4 & $-2.695(60\pm 1) \times 10^7$& $5.447 \times 10^7$& $1.31 \times 10^9$ & 6000\\
			5 & $-8.4686(7/8) \times 10^7$ & $2.616 \times 10^8$&$6.8 \times 10^{9}$ & 8000\\
			6 & $-2.145944 \times 10^8$ & $8.930 \times 10^8$ & $2.55 \times 10^{10}$ & 8000\\
			7 & $-4.69624 \times 10^8$ & $2.459 \times 10^9$ & $7.7(4\pm 2)\times 10^{10}$ & 8000\\
			8 & $-9.239(10\pm 1) \times 10^8$ & $5.831 \times 10^9$ & $2.01\times 10^{11}$ & 8000\\
			9 & $-1.67634 \times 10^9$ & $1.2384 \times 10^{10}$ & $4.65\times 10^{11}$ & 8000\\
			10 & $-2.85418 \times 10^9$ & $2.416 \times 10^{10}$ & $9.8(3\pm 2)\times 10^{11}$ & 8000\\
			15 & $-2.20129\times 10^{10}$ & $3.0492 \times 10^{11}$ & $1.731 \times 10^{13}$ & 8000\\
			20 & $-9.34553 \times 10^{10} $ & $1.7983 \times 10^{12}$ & $1.313 \times 10^{14}$ & 8000\\
			25 & $-2.8645\times 10^{11}$ & $7.056 \times 10^{12}$ & $6.31\times 10^{14}$ & 8000\\\hline 
		\end{tabular}
	\caption{The numerical values for the coefficients $C_i$ with different $k$. The range of our discretization points is $[0,L]$. }
		\label{DataDeltakLargeLambda}
	\end{table}

\textit{Note added:}
	While we were finalizing version 1 of this manuscript, \cite{Beccaria:2022ypy} appeared in which the large $\lambda$ expansion of $\Delta_k(\lambda)$ is calculated analytically. The result in \cite{Beccaria:2022ypy} agrees with our conjecture~\eqref{NikolayConjecture}. Their expansion also agrees with the expressions in~\eqref{C5p2} and~\eqref{C3}. Additionally, they obtained the closed form expressions
\begin{equation}
\epsilon_{5/2}  = \frac{1}{3} - \frac{\zeta(3)}{ 16 \log^3 2 }
\qquad \text{and} \qquad
\epsilon_{3}  = \frac{5}{3} - \frac{\zeta(3)}{ 2 \log^3 2 }\, ,
\end{equation}
which agree with our numerical estimates discussed above. It turns out that our numerics was not accurate enough to determine the full structure of the coefficient $C_{7/2}$ in~\eqref{C7p2}. Namely, instead of the factor in the square brackets $[(1+\epsilon_{7/2})k^2 - (2-2\epsilon_{7/2})]$, the analytical calculation in \cite{Beccaria:2022ypy} yields
\begin{equation}
	\label{neweqC72}
	\begin{aligned} 
		& \left(-\frac{9 \zeta_5 }{128 \log^52}+\frac{\zeta_3 }{\log ^32}-\frac{32}{15}\right) k^2  +
		\left(-\frac{9 \zeta_5 }{128 \log^52}-\frac{3 \zeta_3 }{4 \log
			^32}+\frac{16}{5}\right) k \\
		& \qquad + \left( -\frac{27 \zeta_5 }{2048 \log^52}-\frac{\zeta_3 }{4 \log
			^32}-\frac{16}{15} \right) \\
		& \approx [ 1.02051 k^2 +0.0371916 k-2.05448]\,.
	\end{aligned}
\end{equation}
The rest of the expression for $C_{7/2}$ in~\eqref{C7p2} agrees with the analytic result in \cite{Beccaria:2022ypy}.

\section{Free energy ${\cal F}$ and untwisted correlators}
\label{sec:FreeEnergy}

Our goal in this section is to compute the free energy of the matrix model at hand. To this end we employ the Weinstein-Aronszajn determinant identity for an $m \times n$ matrix $A$ and an $n \times m$ matrix $B$ which reads
\begin{equation} 
	\det(\mathbf{1}_{m\times m} + AB) = \det(\mathbf{1}_{n\times n} + BA)\,.
	\label{TaoMagicalFormula}
\end{equation} 
We can identify the matrix $\mathsf{X}$ defined in \eqref{Xx} as the product of two $A_k(t)$ which can be regarded as a generalized matrix with one discrete index $k\in \mathbb{N_+}$ and one continuous index $t\in (0,\infty)$. After the change of variables $t \rightarrow \frac{2\pi t}{\sqrt{\lambda}}$ we can rewrite \eqref{Xx} as
\begin{equation}
	\mathsf{X}_{kl} = -\int dt A_k(t)A_l(t),\quad A_k(t)\equiv (-1)^k \sqrt{(2k+1)\ \frac{16 \pi}{\sqrt{\lambda}}\ W\left( \frac{2\pi t}{\sqrt{\lambda}} \right)} J_{2k+1}(t)\,.
\end{equation}
Using a continuous version of formula (\ref{TaoMagicalFormula}), one can transform the free energy difference ${\cal F}$ defined in \eqref{FreeEnergyDifference} as follows:
\begin{equation}
	\begin{aligned}
		{\cal F} &= \frac{1}{2} \log \det (\mathbf{1}-\mathsf{X}) =  \frac{1}{2} \log \det \left( \mathbf{1} + \int dt A_k(t)A_l(t) \right)\\
		&\qquad  =  \frac{1}{2} \log \det \left( \mathbf{1} +\sum_k A_k(t) A_k(s) \right) \,.
	\end{aligned}
\end{equation}
We therefore conclude that ${\cal F}$ is the logarithm of a Fredholm determinant: 
\begin{equation}	\label{FredholmDeterminantF}
	{\cal F}= \frac{1}{2} \log \det( \mathbf{1} + K)\,,
\end{equation}
where $\mathbf{1}+ K$ is the functional operator that we encountered in \eqref{eq:eq7}. This will be our starting point for the calculation of the free energy. 

As a consistency check on the manipulations performed above, we can compare our results derived from \eqref{FredholmDeterminantF} against the small  $\lambda$ expansion derived in \cite{Tseytlin} by other methods.  In Section~\ref{sec:smallDeltaAN}, we showed that when performing an expansion for small $\lambda$ the kernel is degenerate. Therefore, we can calculate the Fredholm determinant in \eqref{FredholmDeterminantF} using Equation \eqref{eq:deg10}. To this end we employ the results and notation of Section~\ref{sec:smallDeltaAN}. For the free energy up to order $\mu^{10}$ we find
\begin{equation}
{\cal F} = \dfrac{1}{2} \log\det(1 + A)+O\left(\mu\right)^{10}\,,
\end{equation}
where the $3 \times 3$ matrix $A$ is given in equation \eqref{small:A}. This in turn leads to
\begin{equation}
{\cal F } = \frac{5 }{8}\zeta (5)\mu ^6 -\frac{105 }{32}\zeta (7)\mu ^8 +O\left(\mu\right)^{10}\,.
\end{equation}
This result agrees with the first two terms in Equation (3.4) in \cite{Tseytlin}. It is straightforward but tedious to extend our method to higher orders in $\mu$. In particular we have successfully compared our results to the terms up to order $\mu^{20}$ in Equation (3.4) of \cite{Tseytlin}.\footnote{All the terms agree, apart from one. In \cite{Tseytlin} there is a term $\frac{3.6355}{8} \zeta(13) \hat{\lambda}^7$, but we find instead $\frac{212355}{8} \zeta(13) \hat{\lambda}^7$. Given that $\frac{3.6355}{8} \zeta(13) \hat{\lambda}^7$ is the only term with irrational coefficient multiplying a $\zeta$ function in Equation (3.4) in \cite{Tseytlin}, this is probably a typo.}

\subsection{Numerical method for the calculation of ${\cal F}$}\label{sec:Fnum}

\begin{figure}[H]
	\centering
	\includegraphics[scale = 1]{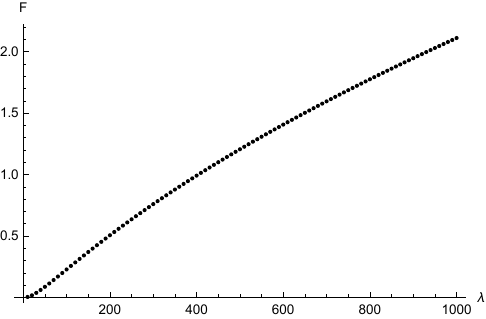}
	\caption{\rm $\cF$ as a function of $\lambda$ computed with the Bornemann method.}
	\label{FLambda}
\end{figure}

\begin{figure}[H]
	\centering
	\includegraphics[scale = .9]{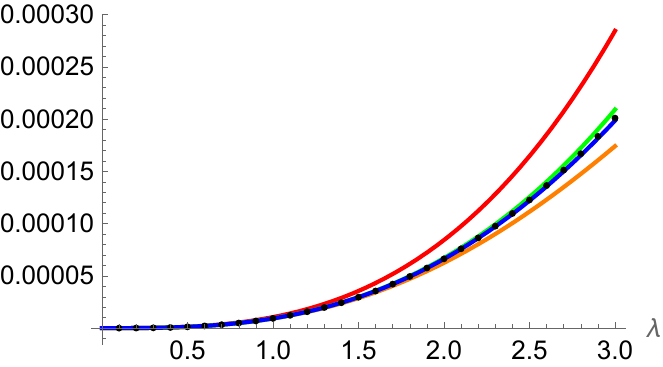}
	\caption{Black dots: numerical calculation of ${\cal F}$ with the Bornemann method. Red line: $ F_3 \lambda^3$; orange line: $ F_3 \lambda^3 + F_4 \lambda^4$; green line: $F_3 \lambda^3 + F_4 \lambda^4 + F_5 \lambda^5$; blue line: $F_3 \lambda^3 + F_4 \lambda^4 + F_5 \lambda^5 + F_6 \lambda^6 $. Clearly the numerical and analytic results agree very well.}\label{Figure0630b}
\end{figure}

Unfortunately we are not able to analytically calculate the Fredholm determinant in \eqref{FredholmDeterminantF} for all values of $\lambda$. To proceed we calculate ${\cal F}$ numerically with the Bornemann method~\cite{Bornemann}. This numerical method is based on the Nystr\"om method for solving integral equations. Namely, if $w_a \ge 0$ are weights and $t_a$ are discretization points of a quadrature method, with $a = 1, \ldots, m$, then
\begin{equation}\label{Bornemann}
\det( \mathbf{1} + K) \approx \det_{a,b=1}^m \Big( \delta_{ab} + \sqrt{w_a} K(t_a,t_b) \sqrt{w_b} \Big)\,.
\end{equation}
This algorithm is simple but very efficient for smooth kernels. There are many discretization schemes one can use in \eqref{Bornemann}. As in Section~\ref{sec:Deltanum} we have chosen Fej\'er type 1. 
As an illustration of the results obtained via this method, in Figure~\ref{FLambda} we present the numerical results for the free energy with some appropriate settings for the quadrature parameters $L$ and $m$.
A table with numerical values of  ${\cal F}$ can be found in Appendix~\ref{app:tables}. As a consistency check on the numerical implementation of the Fredholm determinant algorithm we can compare the numerical results against the analytic small $\lambda$ expansion. We use the small $\lambda$ expansion of Equation (3.4) in \cite{Tseytlin} which we have independently reproduced:
\begin{equation}\label{eq:Fsmall}
	{\cal F} = F_3 \lambda^3 + F_4 \lambda^4 + F_5 \lambda^5 + F_6 \lambda^6 + \cdots
\end{equation}
with  $F_3 = \dfrac{5}{512 \pi ^6}\zeta (5)$, $F_4 = -\dfrac{105 }{8192 \pi ^8}\zeta (7)$ etc. The results are illustrated in figure \ref{Figure0630b}.
In Table~\ref{tab:smallFlambda} we provide more evidence for the validity of our numerical method. As discussed in \cite{Tseytlin}, the radius of convergence of $\lambda$ is $\lambda_c = \pi^2$, which is the same as for $\cN = 4$ SYM.
\begin{table}[H]
	\centering
	$$
	\begin{array}{|l|lllll|}
		\hline
		n \backslash \lambda & 1 & 2 & 3 & 4 & 20 \\
		\hline
		1 & 0.0000105329 & 0.0000842635 & 0.000284389 & 0.000674108 & 0.0842635 \\
		2 & 0.00000917083 & 0.0000624697 & 0.000174058 & 0.000325408 & -0.133674 \\
		3 & 0.00000931482 & 0.0000670777 & 0.00020905 & 0.000472864 & 0.327125 \\
		4 & 0.00000930041 & 0.0000661549 & 0.000198539 & 0.000413804 & -0.595688 \\
		5 & 0.00000930182 & 0.0000663362 & 0.000201637 & 0.00043701 & 1.2173 \\
		10 & 0.0000093017 & 0.0000663066 & 0.000200935 & 0.000430498 & 249.84 \\
		15 & 0.0000093017 & 0.0000663066 & 0.000200935 & 0.000430495 & 960.633 \\
		\hline
		\text{numerics} & 0.0000093017 & 0.0000663066 & 0.000200935 & 0.000430494 &
		0.0175565 \\
		\hline
	\end{array}
	$$
	\caption{Comparison of small $\lambda$ expansion against numerical calculation of the free energy for different values of $\lambda$. The last row gives the value calculated with the Bornemann method. The first row with $n=1$ gives the result using 1 term in the expansion \eqref{eq:Fsmall}, the second row with $n=2$ gives the result using two terms in the expansion \eqref{eq:Fsmall} and so on. One can see that for small $\lambda < \lambda_c$ ($\lambda = 1,2,3,4$ in the table) the series \eqref{eq:Fsmall} converges to the numerical result as $n$ increases. For large $\lambda > \lambda_c $ ($\lambda = 20$ in the table), the series \eqref{eq:Fsmall} does not seem to converge, but the numerical method still yields sensible results.}\label{tab:smallFlambda}
\end{table}

\subsection{Large $\lambda$ expansion of ${\cal F}$: a conjecture}

Using the numerical method detailed above we calculated ${\cal F}$ for $\lambda$ going up to $e^{18}$. Fitting the numerical data listed in table \ref{TblLambdaF}, we observed that the following asymptotic expansion with coefficients in closed form agrees very accurately with the data
\begin{equation} 
	{\cal F} = \frac{1}{8} \sqrt{\lambda}- \frac{3}{8} \log \left(\frac{\lambda}{\lambda_0}\right) - \frac{3}{4} \frac{\log 2}{\sqrt{\lambda}} 
	- \frac{3}{2} \frac{\log^2 2}{\lambda} + O( \lambda^{-3/2})\,.
	\label{FF}
\end{equation}
The constant $\lambda_0 \approx 7.723901172$ can be determined with high precision by our algorithm whose closed-form formula was found later on by  the authors of \cite{Beccaria:2022ypy}:
\begin{equation}\label{eq:lambda0v2}
\lambda_0 = 2^{ 14/9} e^{ 8 \zeta'(-1)} \pi^2\, ,
\end{equation}
which agrees with the numerical value we found. It is tempting to speculate that the coefficient of the $\log\lambda$ term above is related to the difference in conformal anomalies between the $\mathbf{E}$ theory and $\mathcal{N}=4$ SYM, see \eqref{eq:acEN4}, but we were not able to make this statement quantitatively precise. 

\begin{table}[h]
	\centering
	\begin{tabular}{|c||c|c|c|c|}
		\hline
		$\lambda$ & $\cF$  \\\hline\hline 
		$e^8$ & $4.5816147173   (9\! \pm \! 6 )$\\
		$e^{8.5}$ & $ 6.3347294273322   (8\! \pm \! 8 ) $\\
		$e^{9}$ & $ 8.637894890881   (0\! \pm \! 9 ) $ \\
		$e^{9.5}$ & $ 11.64760271190   (0\! \pm \! 6 ) $ \\
		$e^{10}$ & $ 15.5647287113   (94\! \pm \! 21 ) $\\
		$e^{10.5}$ & $ 20.6471553368   (9\! \pm \! 4 )$\\
		$e^{11}$ & $ 27.225974672885   (68\! \pm \! 15 )$\\
		$e^{11.5}$ & $ 35.726290419555   (6\! \pm \! 8 )$\\
		$e^{12}$ & $ 46.69392596735   (85\! \pm \! 24 )$ \\
		$e^{12.5}$ & $ 60.8297166581   (16\! \pm \! 11 )$\\
		$e^{13}$ & $ 79.033540756198   (14\! \pm \! 11 )$\\
		$e^{13.5}$ & $ 102.460855469831   (90\! \pm \! 14 )$\\
		$e^{14}$ & $ 132.595289990373   (00\! \pm \! 26 )$\\
		$e^{14.5}$ & $ 171.34185630862   (66\! \pm \! 11 )$\\
		$e^{15}$ & $ 221.1466338982   (52\! \pm \! 31 )$\\
		$e^{15.5}$ & $ 285.1504476051873   (00\! \pm \! 31 )$\\
		$e^{16}$ & $ 367.3861937408442   (07\! \pm \! 28 )$\\\hline 
	\end{tabular}
	\caption{The numerical values of $\cF$ that we used for numerical fitting. The range of our sampling points is $(0,L)$, where in practice we take $L = 12000$. }
	\label{TblLambdaF}
\end{table}

\begin{figure}
	\centering
	\includegraphics[width=9cm]{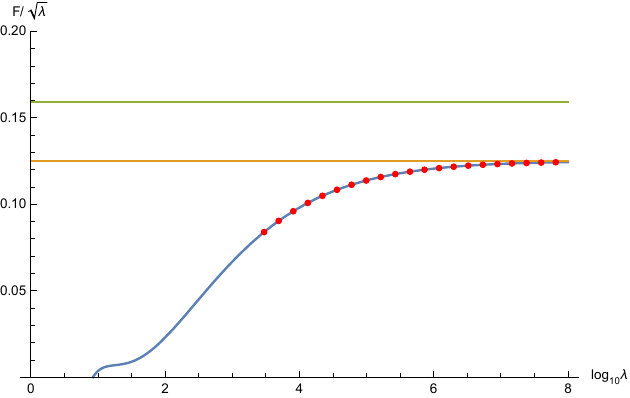}
	\caption{\rm The counterpart of Figure 4 in \cite{Tseytlin}. The green and dark yellow lines represent the constants $1/2\pi$ and $1/8$, respectively. The blue line is the analytic approximation in \eqref{FF}. The red dots are the numerical results obtained by evaluating the Fredholm determinant with the Bornemann method.}
	\label{LambdaExp}
\end{figure}

Interestingly the result in \eqref{FF} is in conflict with the leading order result presented in \cite{Tseytlin} where it was argued that $\cF \sim \frac{1}{2\pi}\sqrt{\lambda}$. As illustrated in Figure~\ref{LambdaExp} this is in clear disagreement with our numerical results. While we are not completely convinced why the result in  \cite{Tseytlin} disagrees with ours we suspect that it could be due to the assumption in \cite{Tseytlin} that the leading order behavior of $\cF$ for large $\lambda$ is controlled by the leading order behavior of $\mathsf{X}$. Namely, in \cite{Tseytlin} the authors use\footnote{The expression of matrix $\mathsf{S}$ is given in \eqref{s}.}
\begin{equation}\begin{aligned}
		\cF &= \frac{1}{2}\log \det \left(\mathbf{1} - \mathsf{X}\right) = \frac{1}{2}\log \det \left[ \mathbf{1} + \frac{\lambda}{2\pi^2} \mathsf{S} + O(\lambda^0) \right]\\
		&\quad  =  \frac{1}{2}\log \det \left( \mathbf{1} + \frac{\lambda}{2\pi^2}\mathsf{S} \right) + o(\sqrt{\lambda})\,.
	\end{aligned}
	\label{ConjectureRelation}
\end{equation}

Both analytical and numerical analysis supports the following behavior, see \cite{Tseytlin} 
\begin{equation}
	\frac{1}{2}\log \det \left( \mathbf{1} + \frac{\lambda}{2\pi^2}\mathsf{S} \right) \sim \frac{1}{2\pi} \sqrt{\lambda} 
	\label{LeadingOrderApproximation}
\end{equation}
Thus we suspect that there is something wrong in going from the first to the second line in \eqref{ConjectureRelation} which is the cause of the conflict with our numerical analysis. To illustrate this subtlety we have shown numerically that $\cF/\sqrt{\lambda}$ can also depend on the sub-leading order expansion of $\mathsf{X}$, for example, if we multiply the identity matrix in \eqref{LeadingOrderApproximation} with an arbitrary constant, then the coefficient of $\sqrt{\lambda}$ in the final result changes. As discussed in Appendix \ref{sec:term2} similar subtle issues are present in the analytic evaluation of $\Delta_k$.

\subsection{Untwisted correlators and Wilson loops in the $\mathbf{E}$ theory}\label{FSection}

The free energy in \eqref{FF} can be used to calculate the large $\lambda$ behavior of the functions $\delta_k(\lambda)$ that control the two-point functions of the untwisted sector operators  in the $\mathbf{E}$ theory, see \eqref{DeltaAnddelta}. It was argued in \cite{Billo1} that $\delta_k(\lambda)$ obey the following relation 
\begin{equation}\label{eq:deltakrelF}
	\delta_k(\lambda) = -2k\left[ (2k^2-1)\lambda\partial_\lambda \cF + (\lambda\partial_\lambda)^2\cF \right],
\end{equation}
	where $\cF$ is the free energy defined in \eqref{FreeEnergyDifference}. Using \eqref{eq:deltakrelF} and the conjectured analytic form of the large $\lambda$ behavior of $\mathcal{F}$ in \eqref{FF} we find that the large $\lambda$ behavior of $\delta_k(\lambda)$ is 
\begin{multline}
\delta_k(\lambda) =  -\frac{k(4k^2-1) }{16}\sqrt{\lambda} + \frac{3}{8}k(4k^2-2)\\- \frac{3}{8}\frac{k(4k^2-3)\log 2}{\sqrt{\lambda}} -\frac{3}{2}  \frac{k(4k^2-4) \log^22}{\lambda}+O(\lambda^{-3/2})\,.
\end{multline}
This is a new result that provides three more orders in the large $\lambda$ expansion of $\delta_k(\lambda)$ as compared to the previous literature \cite{Billo1,Tseytlin}. Note that we again find a discrepancy between our result for the coefficient of the $\sqrt{\lambda}$ term in $\delta_k(\lambda)$ above and the results of \cite{Tseytlin}. This is due to the discrepancy in $\mathcal{F}$ discussed around \eqref{ConjectureRelation}.

As a final application of our result for $\mathcal{F}$ we can calculate the first four terms in the strong coupling expansion of the function $q(\lambda)$ that controls the vacuum expectation value of a 1/2-BPS circular expectation loop in the $\mathbf{E}$ theory. As derived in \cite{Tseytlin} and discussed around equations \eqref{eq:Deltaqdef} and \eqref{eq:Deltaqres} this function can be calculated by taking a derivative of $\mathcal{F}$, $\Delta q (\lambda) = - \frac{\lambda^2}{4} \partial_{\lambda} \mathcal{F}(\lambda)$. Using our result for $\mathcal{F}$ in \eqref{FF} we find 
\begin{equation}
\Delta q (\lambda) = - \frac{1}{64} \lambda^{3/2}+\frac{3}{32} \lambda - \frac{3\log2}{32} \lambda^{1/2} - \frac{3\log^22}{8} +O(\lambda^{-1/2})\,.
\end{equation}
%

\section{Three point functions}
\label{sec:Threeptfc}

In this section, we study the strong coupling expansion of three-point functions of single trace operators in the  {\bf E} theory. As discussed in Section \ref{sec:N2summary}, $G_{m,n}$, which are defined in equation~\eqref{FTCorrelators}, can be calculated with an expectation value in the matrix model. In the planar limit this is:
\begin{equation}
		G_{m,n} =  \langle O_m(a) O_n(a) O_{m+n}(a) \rangle\\\,.
\end{equation} 
Its value in  $\cN = 4$ YM is given by, see for example \cite{Billo:2022xas},
\begin{equation}\label{GmnN=4}
	\quad G_{m,n}^{(0)} = \frac{mn(m+n)}{2}\left( \frac{N}{2}\right)^{m+n-1}\,.
\end{equation}
We will calculate $\Delta_{m,n}(\lambda)$ which is defined by
\begin{equation}
	G_{m,n} = G_{m,n}^{(0)} \left( 1 + \Delta_{m,n}(\lambda) + O(1/N)^2 \right).
\end{equation}
	There are three possibilities for the parities of $m$ and $n$. If $m$ and $n$ are both even, the three point functions belong to the un-twisted sector, and therefore in the planar limit $\Delta_{2k, 2l} = 0$. 
	For the cases where $m$ and $n$ are both odd, or one is odd and the other is even, $\Delta_{m,n}$ is generally non-zero. Their leading order behaviour at large $\lambda$ was calculated previously in \cite{Billo:2022xas}. In the next few sections we will calculate the next three terms in its expansion. The reader who is not interested in the calculation can find the result in equations~\eqref{Xuao3pt1} and~\eqref{Xuao3pt2} below. We want to emphasize that the calculation below is fully analytic, but we use the 
	conjectured form~\eqref{Dkl} for  $\mathsf{D}_{kl}$ as input.

Before calculating $\Delta_{m,n}$ we need to calculate several auxiliary objects called $T$, $C$ and $M$. The $T$'s are defined by 
the expectation values
	\begin{equation}
	T_{m}  = \langle \Omega_{m}(a) \rangle ,\quad T_{m,n}  = \langle \Omega_{m}(a) \Omega_{n}(a) \rangle ,\quad\text{etc}\,.
\end{equation} 
The mixing coefficients $C_{m,n}$ were defined previously in equation~\eqref{mixingCmn}. The matrix $M$ is given by inverting a matrix: $M = (\mathbf{1} + C)^{-1}$. We will first present the expressions for $T$, $C$, $M$ in the untwisted sector and twisted sectors. These results can be found in \cite{Billo:2022xas} but we include them here because we need them in Section~\ref{3ptFunction} where we put everything together and calculate the strong coupling expansion of $\Delta_{m,n}(\lambda)$.
	The two-point functions of operators with even dimensions belong to the un-twisted sector, therefore\footnote{All equations presented in this section are valid at the leading order in the large $N$ expansion.}:
	\begin{equation}\label{TEven}
		T_{2k,2l} = T_{2k,2l}^{(0)} = \frac{N^{k+l+2}}{2^{k+l}} \frac{(2k)!(2l)!}{k!(k+1)!l!(l+1)!} \,.
	\end{equation}
	In $\cN = 4$ YM, the mixing coefficients are
	\begin{equation}\label{CnmN=4YM}
		C_{n,m}^{(0)} = \lim\limits_{\lambda\rightarrow 0} C_{n,m} = \left( \frac{N}{2} \right)^{\frac{n-m}{2}}  \left( \begin{array}{c} n \\ \frac{n-m}{2}  \\ \end{array}\right),\quad \text{if } n > m,
	\end{equation}
and $C_{n,m}^{(0)} = 0$ if $n\le m$, thus for the un-twisted sector, we have
	\begin{equation} \label{CMatrixEven}
		C_{2k, 2l} =  C_{2k,2l}^{(0)} = \left(\frac{N}{2}\right)^{k-l}   \left( \begin{array}{c} 2k \\ k - l \\ \end{array}\right),\quad\text{if }k>l.
	\end{equation}
	We introduce the vev-less version of $\Omega_n(a)$:
	\begin{equation}
		\widehat{\Omega}_n(a) \equiv \Omega_n(a) - \langle \Omega_n(a) \rangle = \Omega_n(a) - T_n,
	\end{equation}
	and we denote the transformation coefficients from $\widehat{\Omega}_n(a)$ to $O_n(a)$ as $M_{n,m}$: 
	\begin{equation} \label{ExpandOn}
		O_n(a) = \sum_{m \le n} M_{n, m } \widehat{\Omega}_m (a).
	\end{equation}
	Using~\eqref{O2Omega} on finds: 
	\begin{equation}
		M_{n,m} = \left( \frac{1}{\mathbf{1} + C} \right)^{-1}_{n,m}\,.
	\end{equation}

The matrices $M$ and $C$ are lower-triangular, so we obtain from~\eqref{CnmN=4YM}
	\begin{equation}\label{MnmN=4YM}
		M_{n,m}^{(0)} = \left( - \frac{N}{2} \right)^{\frac{n-m}{2}} \frac{n}{m} \left( \begin{array}{c} \frac{n+m-2}{2} \\ \frac{n-m}{2} \\\end{array}\right),\quad\text{if } n\ge m\, ,
	\end{equation}
	and $M_{n,m}^{(0)} = 0$ if $n < m$. Therefore in the un-twisted sector we have 
	\begin{equation}
	M_{2k, 2l} = M_{2k, 2l}^{(0)} = \left( - \frac{N}{2} \right)^{k-l} \frac{k}{l} \left( \begin{array}{c} k+l-1\\ k-l\\\end{array}\right) ,\quad \text{if } k\ge l.
	\end{equation}
		When $\lambda = 0$, one has \cite{RodriguezGomez:2016wl}:
		\begin{equation}\label{Chebyshev}
				O_n^{(0)}(a) = {\rm tr}\, p_n(a),\quad p_n(a)\equiv 2 \left(\frac{N}{2}\right)^{\frac{n}{2}} T_n \left( \frac{a}{\sqrt{2N}} \right)+ \delta_{n,2} \mathbf{1} = a^n + \cdots\,.
		\end{equation}
		where $T_n(x)$ is the Chebyshev polynomial of the first kind. 
	
		The two-point functions $T_{2k+1, 2l+1}$ can be evaluated using $\mathsf{D}_{k,l}$. From  \eqref{MnmN=4YM} and \eqref{Chebyshev}, one finds
	\begin{equation}
		\Omega_{2k+1}(a) = \left( \frac{N}{2} \right)^{k+ \frac{1}{2}} \sum_{i=0}^{k-1} c_{k,i} \omega_{k-i}(a),\quad\text{with}\quad c_{k,i}=\left( \begin{array}{c} 2k+1 \\ i \\\end{array}\right)\sqrt{2k-2i+1},
	\end{equation}
	and therefore for $k,l \ge 1$
	\begin{equation}\label{TOdd}
		T_{2k+1, 2l+1}  =  \left( \frac{N}{2} \right)^{k+l + 1} \sum_{i=0}^{k-1}\sum_{j=0}^{l-1} c_{k,i}c_{l,j} \mathsf{D}_{k-i,l-j}\,.
	\end{equation}

	Using our conjectured expression~\eqref{Dkl} for $\mathsf{D}_{kl}$ it is then straightforward to show that
		\begin{equation} \label{TOddExp}
			 \begin{aligned}
				T_{2k+1,2l+1} =&  \left(\frac{N}{2} \right)^{k+l+1} 4\pi^2 \frac{(2k+1)!(2l+1)!}{k!(k-1)!l!(l-1)!}\left[\frac{1}{k+l} \dfrac{1}{\lambda}  - \frac{8\log 2}{\lambda^{3/2}} \right.\\
				& 	\left.  +  \frac{16\log^22}{\lambda^2} (2k+2l-1) + \frac{4\pi^2}{3\lambda^2} \frac{k(k-1)+l(l-1)}{k+l-1} \right.\\
				& \left. - \frac{32}{3} \frac{\log2}{\lambda^{5/2}}  \left(\pi ^2 (k(k-1)+l(l-1)) + 4 \log^22 (k+l-1) (2 k+2 l-1)\right) \right.\\
				& \left. + \frac{\zeta(3)}{\lambda^{5/2}} (-32 k^2 -32 l^2 + 24 k + 24 l + 32 k l) + O(\lambda)^{-3}\right]\,.
		\end{aligned}
	\end{equation}
		The mixing coefficients $C_{2k+1, 2l+1}$ for $k > l \ge 1$ can be obtained from $T_{2k+1, 2l+1}$ using the following formula \cite{Billo:2022xas}: 
	\begin{equation}\label{COddFromTOdd}
		C_{2k+1, 2l+1} = \left| \begin{array}{cccc}  T_{3,3} & T_{3,5} & \cdots & T_{3, 2k+1} \\ T_{5,3} & T_{5,5} & \cdots & T_{5,2k+1}\\ \vdots & \vdots & \ddots & \vdots \\ T_{2l+1,3} & T_{2l+1,5} & \cdots & T_{2l+1, 2k+1} \end{array} \right| \Big/		
		\left| \begin{array}{cccc}  T_{3,3} & T_{3,5} & \cdots & T_{3, 2l+1} \\ T_{5,3} & T_{5,5} & \cdots & T_{5,2l+1}\\ \vdots & \vdots & \ddots & \vdots \\ T_{2l+1,3} & T_{2l+1,5} & \cdots & T_{2l+1, 2l+1} \end{array} \right| \ .
	\end{equation}
	Here the denominator is the determinant of an $l\times l$ matrix $T_{2i+1, 2j+1}$ with $1\le i,j \le l$, and the matrix in the numerator is obtained by performing in the last column the replacement $l \rightarrow k$. 
	Substituting~\eqref{TOddExp} in~\eqref{COddFromTOdd} one obtains
	\begin{equation} \label{CMatrixOdd}
	\begin{aligned}
			C_{2k+1, 2l+1} =&  \left(\frac{N}{2}\right)^{k-l} \frac{2k+1}{2l+1}  \left( \begin{array}{c} 2k \\ k - l \\ \end{array}\right) \left[ 1 - \frac{8\log 2}{\lambda^{1/2}} (k-l) + \right.\\
			& \left. + \frac{16 \log^22 }{\lambda} (k-l) (2k-2l-1) + \frac{4\pi^2}{3\lambda}(k+l)(k-l)\right.\\
			& \left. -\frac{32\log 2}{3\lambda^{3/2}} (k-l-1) (k-l) \left(4 \log ^22 (2 k-2 l-1)+\pi ^2  (k+l)\right)\right.\\
			& \left. + \frac{\zeta (3)}{\lambda^{3/2}} (k-l) \left(-32 k^2 +16 l^2 -32 k l+24 k+24 l-1 \right) + O(\lambda)^{-2}\right]\,.
	\end{aligned} 
	\end{equation}
		Inverting $\mathbf{1} + C$ with odd indices we then find that for $ k \ge l$
	\begin{equation}\label{MOdd}
		\begin{aligned}
				M_{2k+1, 2l+1} =& \left( -\frac{N}{2}\right)^{k-l} \frac{2k+1}{2l+1} \frac{k}{l}  \left( \begin{array}{c} k+l -1\\ k - l\\\end{array}\right) \left[ 1 - \frac{8\log2}{\lambda^{1/2}} (k-l) \right.\\
				& \left. + \frac{16\log^22}{\lambda}(k-l)(2k-2l-1) + \frac{4\pi^2}{3\lambda} (k+l)(k-l) \right.\\
				& \left. -\frac{32\log 2}{3\lambda^{3/2}} (k-l-1) (k-l) \left(4 \log ^22 (2 k-2 l-1)+\pi ^2  (k+l)\right)\right.\\
				& \left. + \frac{\zeta (3)}{\lambda^{3/2}}  (k-l) \left(16 k^2 -32 l^2 -32 k l-24 k-24 l-1 \right) + O(\lambda)^{-2} 
				\right]\,.
		\end{aligned}
		\end{equation}
		
The last quantity we will need is the three-point functions of $\Omega_n(a)$. This can be found in \cite{Billo:2022xas} and reads:
		\begin{equation} \label{Tmnp}
			T_{2m, 2k+1, 2l+1} = T_{2m} \left[ 1 + \frac{m(m+1)}{N^2} \left( k+l+1 + \lambda \partial_\lambda \right) \right] T_{2k+1, 2l+1} + O(1/N)^4.
		\end{equation}

	\subsection{Three point function at large $\lambda$ } \label{3ptFunction}
	Using \eqref{ExpandOn}, we have
		\begin{equation}\label{3ptFunctionRaw}
			\begin{aligned}
				\langle O_{2m}(a) &  O_{2l+1}(a) O_{2p+1}(a) \rangle = \sum_{(n,r,s)=1}^{(m,l,p)} M_{2m, 2n}M_{2l+1,2r+1}M_{2p+1, 2s+1} \langle \widehat{\Omega}_{2n}\widehat{\Omega}_{2r+1} \widehat{\Omega}_{2s+1} \rangle \\
				&=  \sum_{(n,r,s)=1}^{(m,l,p)} M_{2m, 2n}M_{2l+1,2r+1}M_{2p+1, 2s+1}(T_{2n, 2r+1, 2s+1} - T_{2n}T_{2r+1, 2s+1})\,.\\
			\end{aligned}
		\end{equation}
			On the second line above, we substituted the definition of $\widehat{\Omega}_n$ and simplified.  Using \eqref{Tmnp}, the term in the bracket is given, in the large $N$ limit, by 
		\begin{equation}
			T_{2n, 2r+1, 2s+1} - T_{2n}T_{2r+1, 2s+1} = \frac{n(n+1)}{N^2}T_{2n} \left[ (r+s+1) T_{2r+1, 2s+1} + \lambda \partial_\lambda T_{2r+1, 2s+1}  \right].
		\end{equation}
		Using~\eqref{3ptFunctionRaw}, we find that the summation of $n$ can be performed explicitly. Namely, using the expressions for  $T_{2n}$ and $M_{2m, 2n}$ in \eqref{TEven} and \eqref{MnmN=4YM}, one has
		\begin{equation}
				\sum_{n=1}^m M_{2m, 2n} T_{2n} \frac{n(n+1)}{N^2} = m \left(\frac{N}{2} \right)^{m-1}\,.
		\end{equation}
		The sum over $r, s$ can also be done directly using \eqref{TOddExp} and \eqref{MOdd}. Putting everything together, we find
		\begin{equation}\label{3ptFunctionGeneral}
		\begin{aligned}
					\langle O_{2m} O_{2l+1} O_{2p+1} \rangle =&  \left(\frac{N}{2}\right)^{p+l+m}  \frac{16\pi^2m}{\lambda} l(2l+1)p(2p+1) \left[ 1 - \frac{4 \log 2}{\lambda^{1/2}} (2p+2l-1 )\right. \\
					&\left.\  + \frac{16\log^22}{\lambda} (2p+2l-1)(p+l-1)\right.\\
					&\left.\ - \frac{64\log^32}{3\lambda^{3/2}}(2p+2l-1)(p+l-1)(2p+2l-3) \right.\\
					&\left.\  + \frac{\zeta (3) }{2\lambda^{3/2} } (2p+2l-1) \left( 16 p^2 + 16 l^2-16 pl-4 p-4 l-3 \right)+ O(\lambda)^{-2} \right]\,.
		\end{aligned} 
		\end{equation}
		If we take $m = k$ and $p = k+l$, use \eqref{Def3ptFunctions} and the value of $G_{m,n}^{(0)}$ in \eqref{GmnN=4}, 
		we finally find 
	\begin{equation}
	\label{Xuao3pt1}		
	\begin{aligned} 
		1 + \Delta_{2k,2l+1} =&  \frac{16\pi^2}{\lambda}  l(k+l) \left[ 1 -  \frac{4\log 2 }{\lambda^{1/2}} (2k + 4l -1)  + \frac{16\log^22}{\lambda} (2k+4l-1)(k+2l-1)\right. \\
		&\left. - \frac{64\log^32}{3\lambda^{3/2}}(2k+4l-1)(k+2l-1)(2k+4l-3) \right.\\
		&\left. + \frac{\zeta (3) }{2\lambda^{3/2} } (2k+4l-1) \left( 16 k^2 +16 l^2 +16 k l-4 k-8 l-3 \right)+ O(\lambda)^{-2} \right]\\
	\end{aligned}
\end{equation}
	If we take $m = k + l + 1$ and $p = k$ instead, we find
%
%
%
	\begin{equation}
	\label{Xuao3pt2}		
	\begin{aligned} 
	1 + \Delta_{2k+1,2l+1} =&  \frac{16\pi^2}{\lambda} kl \left[ 1  -  \frac{4\log 2 }{\lambda^{1/2}} (2k+2l-1 )  + \frac{16\log^22}{\lambda} (2k+2l-1)(k+l-1 ) \right. \\
	&\left. - \frac{64\log^32}{3\lambda^{3/2}}(2k+2l-1)(k+l-1)(2k+2l-3) \right.\\
	&\left. + \frac{\zeta (3) }{2\lambda^{3/2} } (2k+2l-1) \left( 16 k^2 + 16 l^2-16 kl-4 k-4 l-3 \right)+ O(\lambda)^{-2} \right]\\
\end{aligned}
\end{equation}
The leading term in the large $\lambda$ expansion in equations~\eqref{Xuao3pt1} and~\eqref{Xuao3pt2} agrees with the result in \cite{Billo:2022xas}, the next three subleading terms are our novel results based on the conjecture for $\mathsf{D}_{kl}$ in \eqref{Dkl}.

In \cite{Beccaria:2022ypy}, which appeared while the first version of this manuscript was being finished, it was pointed out that the expression for $\Delta_k(\lambda)$ simplifies when using the variable $\lambda'$, where	
\begin{equation}	
	( \lambda')^{1/2} = \lambda^{1/2} -4\log 2\,.
\end{equation}
We observe that something similar happens for $\Delta_{k,l}(\lambda)$, namely we have
\begin{multline}
		\label{Xuao3pt3}		
		1 + \Delta_{2k,2l+1}(\lambda) = 
		 \frac{16\pi^2}{\lambda}  l(k+l)\left( \frac{\lambda'}{\lambda} \right)^{(2k+4l-1)/2}\\
		 \left[ 1  + 
		\frac{\zeta (3) }{2(\lambda')^{3/2} } (2k+4l-1) \left( 16 k^2 +16 l^2 +16 k l-4 k-8 l-3 \right) 
		+ O(\lambda')^{-2} \right]\,,
\end{multline}
and
\begin{multline}
		\label{Xuao3pt4}				
		1 + \Delta_{2k+1,2l+1}(\lambda) =  \frac{16\pi^2}{\lambda} kl \left( \frac{\lambda'}{\lambda} \right)^{(2k+2l-1)/2}\\
		\left[ 1  + \frac{\zeta (3) }{2(\lambda')^{3/2} } (2k+2l-1) \left( 16 k^2 + 16 l^2-16 kl-4 k-4 l-3 \right)
		+ O(\lambda')^{-2} \right]\,.
\end{multline}

That this simplification also occurs for $\Delta_{k,l}(\lambda)$ seems to provide additional support for the conjectured form of $\mathsf{D}_{kl}$ in \eqref{Dkl}. Also, based on the form for $\Delta_k(\lambda)$ in \cite{Beccaria:2022ypy}, it is tempting to speculate that the expressions ~\eqref{Xuao3pt3} and~\eqref{Xuao3pt4} are correct up to $O(\lambda')^{-5/2}$ instead of only up to $O(\lambda')^{-2}$ displayed above, and that the coefficient of the next non-zero term is proportional to $\zeta(5)$.

\section{Discussion}
\label{sec:Discussion}

In this work, our main focus was on developing an efficient numerical algorithm that allows for the calculations of correlation functions in the $\mathbf{E}$ theory at strong coupling. To this end, we exploited recent supersymmetric localization results that reduce the calculations of these observables to matrix models that can be analyzed with a variety of techniques. Based on our numerical studies we were able to extract analytic expression for a few of the leading terms in the strong coupling expansion of these observables. 

Parallel to our work, analytic techniques have been developed to derive these results. The authors of \cite{Beccaria:2022ypy} took advantage of techniques on Bessel kernels developed in \cite{Belitsky:2020qir} to confirm some of the results presented in our Section~\ref{sec:Deltakinteq} and Section~\ref{sec:FreeEnergy}. The results were used in \cite{Billo:2022lrv} to study the strong coupling expansion of three-point extremal correlators. 

We note that the kernel of the integral operator appearing in this chapter is {\it integrable}~\cite{Deift, Korepin}. This type of kernel appears often in the mathematical physics literature. In that context, one often translates the calculation to a Riemann-Hilbert problem and the expansion for large parameters is carried out with a non-linear steepest descent method \cite{DeiftZhou}. Perhaps these methods can also be applied to our setup to calculate the large $\lambda$ expansion of the correlators analytically.

The numerical techniques we developed can also be efficiently applied to calculate correlators at finite values of the coupling. This provides valuable lessons about the properties of large $N$ gauge theories at finite coupling. This can be used for instance to understand the convergence properties of the weak and strong coupling expansion and to gain insights into non-perturbative effects. We believe that very similar methods can be applied also to other examples of 4d $\mathcal{N}=2$ SCFTs that can be analyzed with supersymmetric localization. The Lagrangian theories discussed in Section~\ref{sec:N2summary} are natural candidates for such an analysis. Work along these lines has recently appeared in \cite{Billo:2022gmq, Billo:2022fnb,Galvagno:2020cgq,Galvagno:2021bbj} for a class of quiver gauge theories and we hope that our techniques will find an application and can be adapted to the study of these models.

The strong coupling results obtained here have connections to string theory through the AdS/CFT correspondence. It will be most interesting to understand how to calculate any of the correlators we analyzed by using world sheet methods. This problem appears to be highly non-trivial since it requires performing string world sheet calculations in the AdS$_5\times S^5/\mathbb{Z}_2$ orientifold at high order in the $\alpha'$ perturbative expansion. We hope that our results will serve as an impetus to tackle this challenging world sheet analysis. Some progress has been made in the literature reproducing the extremal correlators. The leading order analysis in supergravity \cite{Billo:2022gmq, Billo:2022fnb} reproduces the leading order of strong coupling limit in \eqref{Xuao3pt3} and \eqref{Xuao3pt4}. Partial results on higher-derivative corrections in type IIB supergravity are reported in \cite{Skrzypek:2023fkr} which tried to explain the term $\sim \zeta(3) \lambda^{-3/2}$ in our expansion \eqref{Xuao3pt3} and \eqref{Xuao3pt4}.

Throughout the chapter, we have been discussing the planar limit exclusively. It is desirable to study the $1/N$ corrections to the observables. The main difficulty is that at finite $N$, the operator mixing between $S^4$ and $\mathbb{R}^4$ will involve not only single-trace operators but also multi-trace ones, which will make the Gram-Schmidt method much more complicated. Some comments on extremal correlators are in the Appendix D of \cite{Billo:2022xas}. A novel double-scaling limit with $\lambda^{3/2}/N^2$ fixed is discussed in \cite{Beccaria:2023kbl}, which also deserves a closer study.

\cleardoublepage%

  \graphicspath{{\chaptersdir/ads8/image/}}%
  \chapter{ Partition functions on squashed seven-spheres and holography }\label{ch:ads8}

\section{Introduction}

The AdS/CFT correspondence posits dualities between large $N$ field theories and string theories in asymptotically AdS spacetimes. \cite{Maldacena:1997re, Witten:1998qj, Gubser:1998bc} In its original proposal, the conformal field theory is living on a manifold $\cM_d$ conformally equivalent to the boundary of the bulk. The coupling between the conformal field theory and the background helps us gain insight into the dynamics of the field theory. As the bulk theories typically lie in the semi-classical regime, one usually needs to exert non-perturbative tools to investigate the field theories, the well-known examples include supersymmetric localization \cite{Pestun:2016zxk}, bootstrap, and integrability. Through a simplified version of AdS/CFT correspondence \cite{Klebanov:2002ja}, the O($N$) model, which is a set of $N$ conformally coupled scalars, corresponds to higher spin gravity in the bulk, both of them can be attacked analytically. In our setup, instead, we consider a Euclidean Einstein gravity living in an asymptotically locally AdS$_8$ space without matter coupling, which does not correspond to O($N$) model, but shares some universal behaviors as dictated by the conformal symmetry. On the boundary, we consider non-supersymmetric conformal field theories.

The phase structure of asymptotically AdS spaces is an interesting topic, originating from the famous Hawking-Page phase transition \cite{Hawking:1982dh} between thermal AdS$_4$ space and Euclidean Schwarzschild black hole. Holographically, the Hawking-Page transition is conjectured to be dual to the confinement-deconfinement transition in QCD, where the two phases correspond to different behaviors of Wilson loops. \cite{Witten:1998zw} Generalizations of Hawking-Page transition has been discussed in \cite{Blackman:2011in} where the asymptotical boundary geometry is $S^2\times S^d$, and in \cite{Aharony:2019vgs} where the boundary geometry is $S^{d_1} \times S^{d_2}$. Besides, thermodynamical properties and phase transitions are also discussed for AdS-Taub-NUT spaces \cite{Khodam-Mohammadi:2008hww, Chamblin:1998pz, Awad:2000gg, Clarkson:2002uj, Astefanesei:2004kn}, where the asymptotic geometry is a U(1) fiber bundle over a K\"ahler-Einstein space $\cB$, where the metric looks like:
\be 
ds^2 = dr^2 + a^2(r) ds^2_{\cB} + b^2(r) (d\psi + A_\cB)^2,
\ee 
where $A_\cB$ is the K\"ahler potential and the K\"ahler form is given by $J_\cB = dA_\cB$. The metric on the asymptotic boundary where the conformal field theory lives is given by taking $r\rightarrow \infty$:
\be 
ds^2_{\rm bdy} = ds_\cB^2 + \lambda^2 (d\psi + A_\cB)^2,\quad \lambda \equiv \lim\limits_{r\rightarrow \infty} \frac{b(r)}{a(r)}.
\ee 
The squashing parameter $\lambda$ controls the size of the U(1) bundle, which no longer has the interpretation as the inverse temperature as in thermal field theories. Besides NUT, there exists another family of spaces, which has the same asymptotic symmetry but with a different topology, dubbed Bolt, which has a horizon where the Killing vectors degenerate. By varying the size of the U(1) bundle and comparing the gravitational free energies, one observes a first-order phase transition between Taub-NUT and Taub-Bolt. In \cite{Bobev:2016sap}, a generalization of Euclidean AdS$_4$-Taub-NUT space with two squashing parameters was studied.

In our work, we focus on a squashed asymptotically locally AdS$_8$ geometry preserving SO(5)$\times$SO(3) isometry \cite{Page:1985hg,Gibbons:1989er}\footnote{See \cite{Hiragane:2003qq} for a similar bulk setup but with positive cosmological constant. }, whose boundary geometry is a squashed seven sphere constructed by an SU(2) bundle over $\mathbb{S}^4$. By evaluating and comparing the gravitational free energy of SU(2) analog of NUT and Bolt spaces, we are able to study the phase structure of our AlAdS$_8$ geometry. One special property of our squashed $\mathbb{S}^7$ is that there exist two squashing parameters $\lambda = 1, \lambda^2 = 1/5$, for which the squashed sphere is an Einstein manifold \cite{jensen1973einstein}, i.e., $R_{\mu\nu} = k g_{\mu\nu}$ for some constant $k$. For the two special cases, we are able to solve the Einstein equations analytically and then evaluate the Euclidean free energy by using the standard holographic renormalization. \cite{Clarkson:2002uj, Emparan:1999pm, Bueno:2022log} For general $\lambda > 0$, we can evaluate the free energy numerically. Interestingly, although having the analogs of NUT and Bolt phases, there exists only one phase for a given value $\lambda$, meaning there is no Hawking-Page-like transition between the two competing phases. A similar phase structure has been found in \cite{Aharony:2019vgs}, which appears only if the the boundary geometry $S^{d_1}\times S^{d_2}$ satisfies $d_1 + d_2 \ge 9$.

The field theory contents we're interested in are Euclidean conformal field theories living on the squashed seven sphere with SO(5)$\times$SO(3) isometry, where the squashing naturally couples the field theory with the background metric. It is the stress tensor of boundary CFT that couples to the squashed metric, the dynamical properties of which are constrained largely by the conformal symmetry \cite{Osborn:1993cr,Erdmenger:1996yc}, which reflects the universal character for all CFTs. The free energy in odd-dimensional field theories is an important quantity since it's conjectured to reflect the degrees of freedom in the field theory and thus monotonically decreasing along with RG flows, which is called $F$-theorem. \cite{Jafferis:2011zi, Pufu:2016zxm} The quantity conjectured to be decreasing along the RG flow is related to the free energy $F \equiv - \log |Z_{S^d}|$ by a sign:
\be 
\widetilde F \equiv (-1)^{\frac{d-1}{2}} \log |Z_{S^d}|.
\label{FreeEnergySd}
\ee 
The $F$-theorem was proven only in 3d CFT by \cite{Casini:2012ei}, but is supported by pieces of evidence in other dimensions. \cite{Pufu:2016zxm} In our scenario, the CFT living on round $\mathbb{S}^7$ is a free UV fixed point, and introducing squashing on the metric corresponds to switching on a marginal spin-2 deformation generated by the stress tensor: \cite{Bobev:2017asb}
\be 
S[g_{\mu\nu}, \phi] = S[g^{(0)}_{\mu\nu},\phi] - \frac{\epsilon}{2} \int d^d x \sqrt{g^{(0)}} h^{\mu\nu}T_{\mu\nu},\quad T_{\mu\nu} \equiv -\frac{2}{\sqrt{g}} \frac{\delta S[g_{\mu\nu},\phi]}{\delta g^{\mu\nu}} .
\label{Deformation}
\ee 
For a conformal field theory without conformal anomaly, which is true in odd dimensions, we expect the first-order derivative $\widetilde F'(\epsilon)$ to vanish at $\epsilon = 0$, which is proportional to the one-point function of the stress tensor. And the second-order derivative $\widetilde F''(\epsilon)$ should be negative as dictated by $F$-theorem since the marginal spin-2 deformation induces an RG flow from the UV fixed point, along which the free energy must be decreasing. It can be expressed as a double integral on the round sphere:
\be
F''(\epsilon)\Big|_{\epsilon=0} = -\frac{1}{4}\int d^dx d^dy\sqrt{g^{(0)}(x)g^{(0)}(y)} h^{ab}(x)h^{cd}(y)\langle T_{ab}(x)T_{cd}(y)\rangle_{\partial \cM}.
\ee
Quite remarkably, we are able to evaluate the second-order derivatives for general CFTs living on not only squashed SO(5)$\times$SO(3) seven spheres but also another U($k+1$)$\times$U(1) family of squashed $(2k+1)$-spheres, the latter was only conjectured from results in high-derivative gravity in \cite{Bueno:2018yzo}. The result is universal and applies to all CFTs, and this is the first time to obtain them from the field theory to our knowledge.

 The above universal results only apply to cases where the squashing is small. For finite squashing, we consider two toy models: the O($N$) vector model, which is equivalent to conformally coupled scalars, and a free fermion model. The free energy of the field theories can be obtained by taking advantage of the spectrum of the Laplacian and the Dirac operators on our squashed metric \cite{Nilsson:1983ru, Eastaugh:1985ew}, where the path integral boils down to a Gaussian integral, making the evaluation of the free energy for generic squashing $\lambda$ possible. \cite{Hartnoll:2005yc, Anninos:2012ft, Anninos:2013rza} We have also investigated the behavior of free energy at small squashing analytically and at large squashing numerically. The comparison among free energies of holographic CFTs, the O($N$) model, and the free fermion model has been studied at small squashing \cite{Bobev:2017asb, Bueno:2018yzo, Bueno:2020odt} and large squashing \cite{Bobev:2017asb, Hartnoll:2005yc, Anninos:2012ft} where the boundary is a squashed three-sphere. We have extended this exploration to the seven-sphere with SU(2) bundle.

The structure of the chapter is as follows. In Section \ref{BulkTheory}, we introduce two bulk metric ansatzes which preserve SO(5)$\times$SO(3) isometry and solve the Einstein equations. We also calculate the gravitational free energy as a function of the squashing parameter $\lambda$. The field theory calculations are included in Section \ref{BoundaryTheory}, where we begin with the small-squashing behavior for generic CFT living on SU(2) squashed seven spheres and U(1) squashed $(2k+1)$-spheres. We also evaluate the free energy of both conformally-coupled scalar and free fermion fields using different methods. In section \ref{SectionofComparison}, we compare bulk and boundary free energies.

\section{Bulk story}
\label{BulkTheory}
In this section, we study the generalization of Hawking-Page transition in asymptotically locally AdS$_8$ with Euclidean signature. We begin with an example of Hawking-Page transition in asymptotically locally AdS$_4$. Then we go to 8d, introducing two ways to construct squashed sphere metrics and use them to obtain the metrics we are after. Taking the squashed seven sphere as a boundary, the bulk geometry can be obtained by numerically solving the equation of motions. In the end, we can take advantage of the bulk solution to obtain the renormalized free energy.

\subsection{The phase transition between AdS$_4$ NUT and Bolt}
	\subsubsection{The geometries}
	To illustrate better the Hawking-Page phase transition and competition between phases, we take an example of the aymptotically locally AdS$_4$ NUT and Bolt spaces. The metric is given by: \cite{Emparan:1999pm, Stephani:2003tm}
	\begin{equation} \label{eq:ads4NUT}
		ds^2 = 4n^2 V(\rho) (\sigma_3)^2 + \frac{d\rho^2}{V(\rho)} + (\rho^2 - n^2) (\sigma_1^2 + \sigma_2^2),
	\end{equation}
	where the 1-forms $\sigma_i$ are defined in \eqref{eq:1forms} below and $V$ is given by
	\begin{equation}
		V(\rho) = \frac{(\rho^2 + n^2) - 2m\rho + l^{-2} (\rho^4 - 6n^2\rho^2 - 3n^4)}{\rho^2 - n^2},
	\end{equation}
	where $n$ is the NUT charge, $m$ is the generalised mass and $l$ is the AdS length scale. The asymptotic behaviour for $\rho \rightarrow \infty$ of the metric in \eqref{eq:ads4NUT} is locally the same as the one for AdS$_4$. The only difference being that the boundary is a squashed $S^3$ with a single squashing parameter, which can be seen by taking the limit $\rho \rightarrow \infty$ on \eqref{eq:ads4NUT}:
	\begin{equation}
		ds^2 = \frac{l^2}{\rho^2}d\rho^2 + r^2 \left[ (\sigma_1^2 + \sigma_2^2) + \frac{4n^2}{l^2} (\sigma_3)^2 \right], 
	\end{equation}
	for which we can identify $n/l$ as a squashing parameter of the squashed $S^3$ on the boundary. When $n/l = 1/2$, the metric on the sphere $ds^2 = \sigma_1^2 + \sigma_2^2 + \sigma_3^2$ reduces to that of a round sphere. The asymptotic boundary is the same for all bulk phases, which are distinct in the bulk. Let us first introduce the NUT solution, which is obtained by requiring the Dirac-Misner string to be unobservable and there should be no conical defect at $\rho = n$, this gives a constraint on the mass parameter:
	\begin{equation}
		m = n - \frac{4n^3}{l^2}.
	\end{equation}
	Now we zoom in to the ``near-horizon region'' by using the new radial coordinate $R = \sqrt{8n(\rho - n)}$, the leading order expansion of the metric in terms of $R$ gives:
	\begin{equation}
		ds^2 = dR^2 + \frac{1}{4} R^2 (\sigma_1^2 + \sigma_2^2 + \sigma_3^2),
	\end{equation}
	which is the coordinate on the origin of smooth $\mathbb{R}^4$. We can see that the NUT solution reduces to AdS$_4$ when $n/l = 1/2$.
	
	When the emblackening factor $1/V(\rho)$ diverges at $\rho = \rho_b > n$, we get the second set of solutions called Bolt. Similarly requiring the absence of conical singularity at the ``horizon'', we get:
	\begin{equation} \label{eq:ads4BoltConstraints}
		V(\rho_b) = 0,\quad V'(\rho_b) = \frac{1}{2n}.
	\end{equation}
	From the first condition in \eqref{eq:ads4BoltConstraints} one finds that the mass of the Bolt should satisfy
	\begin{equation}
		m_b = \frac{\rho_b^2 + n^2}{2\rho_b} + \frac{1}{2l^2} \left( \rho_b^3 - 6n^2 \rho_b - \frac{3n^4}{\rho_b} \right).
	\end{equation}
	The second condition in \eqref{eq:ads4BoltConstraints} yields a relation between $\rho_b$ and $n$ and $l$:
	\begin{equation} \label{eq:rbBolt}
		\rho_{b\pm} = \frac{l^2}{12n} \left( 1 \pm \sqrt{ 1 - 48 \frac{n^2}{l^2} + 144 \frac{n^4}{l^4} } \right).
	\end{equation}
	There are two branches of Bolt solutions, reminiscent of the large and small Schwarzschild black holes in AdS$_4$ in the original Hawking-Page transition \cite{Hawking:1982dh}, where the small black hole never dominates. By requiring the positivity of the discriminant in \eqref{eq:rbBolt} and that $\rho_b > n$, we find the range of the squashing parameter $n/l$ for the Bolt solutions to exist:
	\begin{equation}
		0 < \frac{n^2}{l^2} = \frac{2-\sqrt{3}}{12} \approx 0.022.
	\end{equation}
	In particular, the Bolt solutions don't exist for $n/l = 1/2$ corresponding to the AdS$_4$ in the NUT solutions. Now we show that the near-horizon geometry is rather different. Take the new radial coordinate $R \equiv \sqrt{8n(\rho - \rho_b)}$ and take the limit $R \rightarrow 0$, we get:
	\begin{equation}
		ds^2 = dR^2 + \frac{R^2}{4} d\psi^2 + (\rho_b^2 - n^2) (\sigma_1^2 + \sigma_2^2),
	\end{equation}
	which has the geometry $\mathbb{R}^2 \times S^2$. $\psi$ plays the role of angular coordinate in $\mathbb{R}^2$ whose range is $0 \le \psi < 4\pi$. The two sphere has a radius $\sqrt{\rho_b^2 - n^2}$ which is positive because we require $\rho_b > n$.
	\subsubsection{The phase transition}
	Now we understand both the near-horizon geometry and asymptotic boundary for the two families of solutions, we can get down to studying the phase transition. We start from the Euclidean Einstein-Hilbert action:
	\begin{equation}
		S = - \frac{1}{16\pi G_N} \int_M d^4x \sqrt{g}(R - 2\Lambda) - \frac{1}{8\pi G_N} \int_{\partial M} d^3x \sqrt{h} K,
	\end{equation}
	where $h$ denotes the determinant of the induced metric on the boundary $\partial M$ of our choice, and $K$ is the trace of the extrinsic curvature. The integral is proportional to the volume of the asymptotically AdS space which diverges, so we need to introduce counter terms:
	\begin{equation}
		S_{\rm ct} = \frac{1}{8\pi G_N} \int_{\partial M} d^3 x \sqrt{h} \left( \frac{\cR}{2} + 2 \right),
	\end{equation}
	where $\cR$ is the scalar curvature on the boundary. A natural choice of boundary is the surface of constant $\rho$, this gives us the following renormalized action:
	\begin{equation}
	\begin{aligned}
		S_{\rm NUT} & = \frac{4\pi l^2}{G_N} \beta (1-2\beta),\quad \beta \equiv \frac{n^2}{l^2},\quad 0 < \beta < \infty,\\
		S_{\rm Bolt}^{(\pm)} &= \frac{ \pi l^2 }{216 G_N}\left[ \left( 72 - \frac{1}{\beta} \right) \pm \left( 48 - \frac{1}{\beta} - 144 \beta \right) \sqrt{ 1 + 48 \beta  (-1+3\beta)} \right],\\
		& \qquad 0 < \beta \lesssim 0.022.\\
	\end{aligned}
	\end{equation}
	\begin{figure}[t]
		\centering
		\includegraphics[width=8cm]{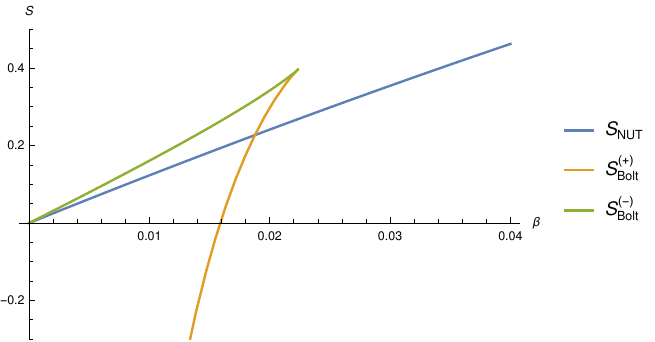}
		\caption{{\rm The Euclidean action $S$ versus the squashing parameters $\beta = n^2/l^2$ for the three families of classical asymptotically locally AdS$_4$ saddles. The actions are expressed in units of $l^2/G_N$. }}
		\label{fig:ads4NUTBolt}
	\end{figure}
	The action versus the squashing parameter $\beta = n^2/l^2$ is plotted in Fig. \ref{fig:ads4NUTBolt}. The saddle-point approximation \eqref{eq:GibbonsHawking} tells us that the the saddle with the lowest free energy dominates. As we know, $\beta = 1/4$ corresponds to the round three sphere on the boundary, so when $\beta$ is close to 0, the smaller $\beta$ is, the more squashed the sphere becomes. According to the plot, when the squashing is small, the NUT saddle dominates. In fact, NUT is the only saddle for $\beta \gtrsim 0.022$. When the sphere gets more squashed such that $$\beta < \beta_* = \frac{7-2\sqrt{10}}{36} \approx 0.019, $$ the dominant saddle becomes Bolt$^{(+)}$. The Bolt solutions with smaller radius never dominate. 
	
	The above example shows how the Hawking-Page phase transition happens when the asymptotic boundary is a squashed sphere, whose squashing parameter determines which phase dominates. The goal for our study is to test whether similar story applies to a Euclidean asymptotically locally AdS$_8$ space with SO(5)$\times$SO(3) isometry.

\subsection{A tale of two metrics}\label{ATableTwoMetrics}
In this chapter, we consider squashed seven spheres with SO(5)$\times$SO(3) isometry, whose metric can be constructed by distinguished yet equivalent ways. The first one is by embedding the sphere in a projective space admitting the standard Fubini-Study metric, and the other one is by Hopf fibration over a projective space, the fibration is non-trivial because of the existence of K\"ahler or hyper-K\"ahler potential on the projective space.

We start with the first construction following \cite{Awada:1982pk,Duff:1986hr}, where we embed $\mathbb{S}^7$ in $\mathbb{HP}^2$. The standard Fubini-Study metric on $\mathbb{HP}^2$ is:
\be 
ds^2 = (1+\bar{q}_kq_k)^{-1} d\bar{q}_i dq_i - (1+\bar{q}_kq_k)^{-2}\bar{q}_idq_id\bar{q}_jq_j,\quad q_1,q_2\in \mathbb{HP}^2,
\label{FubiniStudy}
\ee
where the repeated indices are summed over $\{1,2\}$. We introduce a parametrization of $(q_1, q_2)\in \mathbb{HP}^2$:
\be
q_1 = U\tan\chi \cos\frac{1}{2}\mu,\quad q_2 = V\tan\chi\sin\frac{1}{2}\mu,\quad 0\le \chi\le\pi/2,\quad 0\le \mu\le\pi.
\ee
Here, $U, V\in$ SU(2) can be regarded as quaternions with unit length, so we parametrize them by Euler angles $(\theta, \phi, \psi)$ and $(\Theta, \Phi, \Psi)$ respectively:
\be
U = e^{\textbf{k}\phi/2} e^{\textbf{i}\theta/2} e^{\textbf{k}\psi/2},\quad 
V = e^{\textbf{k}\Phi/2} e^{\textbf{i}\Theta/2} e^{\textbf{k}\Psi/2};\quad 0\le \theta\le \pi,\ 0\le \phi\le 2\pi,\ 0\le \psi \le 4\pi,
\label{parametrization}
\ee
where $\mathbf{i,j,k}$ are units of quaternion. The Maurer-Cartan form can be worked out directly:
\be
\begin{aligned}
	& 2U^{-1}dU = \mathbf{i}\sigma_1 + \mathbf{j}\sigma_2 + \mathbf{k}\sigma_3,\quad 2V^{-1}dV = \mathbf{i}\Sigma_1 +\mathbf{j}\Sigma_2 +\mathbf{k}\Sigma_3,	\\
\end{aligned}
\ee
where the left-invariant one-forms are:
\be \label{eq:1forms}\small \begin{aligned}
	& \sigma_1 = \cos\psi d\theta + \sin\psi\sin\theta d\phi,\quad \sigma_2 = -\sin\psi d\theta + \cos\psi\sin\theta d\phi,\quad \sigma_3 = d\psi + \cos\theta d\phi;\\
	& \Sigma_1 = \cos\Psi d\Theta + \sin\Psi\sin\Theta d\Phi,\quad \Sigma_2 = -\sin\Psi d\Theta + \cos\Psi\sin\Theta d\Phi,\quad \Sigma_3 = d\Psi + \cos\Theta d\Phi.\\
\end{aligned}\ee
Using the parametrization above, the metric (\ref{FubiniStudy}) becomes
\be\begin{aligned}
	ds^2 &= d\chi^2 + \frac{1}{4}\sin^2\chi\left[d\mu^2 + \frac{1}{4}\sin^2\mu \omega_i^2 + \frac{1}{4}\cos^2\chi(\nu_i + \cos\mu \omega_i)^2\right],\\
	& \quad \quad \nu_i \equiv \sigma_i + \Sigma_i,\quad  \omega_i \equiv \sigma_i - \Sigma_i,\quad i = 1,2,3.\\
\end{aligned}
\ee
By setting $\chi$ to be constant, we're equivalently taking the co-dimension 1 surface foliating the projective space and surrounding the original point $\chi=0$. Introducing the squashing parameter $\lambda \equiv \cos\chi$, we obtain the metric on a squashed seven sphere with unit radius \cite{Awada:1982pk}:\footnote{The scale factor $1/4$ ahead is to make sure when $\lambda = 1$, the metric goes back to unit sphere $\mathbb{S}^7$. }
\be
ds^2 = \frac{1}{4} \left[ d\mu^2 +\frac{1}{4}\sin^2\mu\omega_i^2 + \frac{1}{4}\lambda^2 (\nu_i +\cos\mu \omega_i)^2 \right].
\label{SquashedSphereMetric1}
\ee
One can introduce the following vielbein:
\be 
e^1 = \frac{1}{2}d\mu,\quad e^i = \frac{1}{4}\sin\mu\omega_{i-1},\quad e^I = \frac{\lambda}{4}(\nu_{I-4} + \cos\mu \omega_{I-4}),\quad i = 2,3,4;\ I = 5,6,7.
\ee 
Under the tetrad, the Ricci tensor is diagonal:
\be 
R_{ab} = {\rm diag} \left(\alpha,\alpha,\alpha,\alpha,\beta,\beta,\beta\right),\quad \alpha = 12-6\lambda^2,\quad \beta = 4\lambda^2 + \frac{2}{\lambda^2}.
\label{RicciTensor}
\ee 
For $\alpha = \beta$, the space becomes Einstein. There're two possibilities for that condition, $\lambda = 1$ and $\lambda_* = \frac{1}{\sqrt{5}}$. These two values will both play important roles in the gravitational free energy.

Now let's look at the other way to construct squashed seven-sphere with $SO(5)\times SO(3)$ isometry. It's motivated by $k=1$ SU(2) Yang-Mills instanton on $\mathbb{S}^4$. Considering $\mathbb{S}^7$ as $\mathbb{S}^3$ bundle over $\mathbb{S}^4$ with a gauge potential $A_i = \cos^2\frac{\tilde{\mu}}{2}\tilde{\Sigma}_i$, the metric is
\be
ds^2 = \frac{1}{4}\left[ds^2_{\mathbb{S}^4} + \lambda^2 (\tilde{\sigma}_i - A_i)^2 \right],\quad ds^2_{\mathbb{S}^4} = d\tilde{\mu}^2 + \frac{1}{4}\sin^2\tilde{\mu}\tilde{\Sigma}_i^2.
\label{SquashedSphereMetric2}
\ee
This is an ``inverse-Kaluza-Klein'' procession, where we construct a metric on $\mathbb{S}^7$ out of a four-dimensional metric plus an SU(2) bundle, and the size of the bundle is described by the squashing parameter $\lambda$. In fact, we find the metric returns to $\mathbb{S}^3 \times \mathbb{S}^4$ if we set $A^i = 0$, where $\tilde{\sigma}_i$ parametrizes $\mathbb{S}^3$ and $\tilde{\Sigma}_i$ is on $\mathbb{S}^4$. One can define the following vielbein:
\be 
e^1 = \frac{1}{2} d\tilde{\mu},\quad e^i = \frac{1}{4}\sin\tilde{\mu}\tilde{\Sigma}_{i-1},\quad e^I = \frac{\lambda}{2}\left(\tilde{\sigma}_{I-4} - A_{I-4}\right),\quad i=2,3,4;\ I=5,6,7,
\ee 
under which the Ricci tensor is the same as the before \eqref{RicciTensor}, indicating that the two metrics are closely related.

The two metrics (\ref{SquashedSphereMetric1}) and (\ref{SquashedSphereMetric2}) constructed above share some similarities, they are both characterized by one squashing parameter $\lambda$, become round when $\lambda = 1$, and are Einstein when $\lambda = 1$ or $\frac{1}{\sqrt{5}}$; they both preserve explicitly the SO(5)$\times$SO(3) isometry. Although they're different metrics, there exists a map between them\footnote{The relation in equation (8.1.32) of \cite{Duff:1986hr} is not correct, and should actually be inversed.}
\be
\tilde{\sigma}_i = \sigma_i,\quad \mathbf{i}\Sigma_1 + \mathbf{j}\Sigma_2 + \mathbf{k}\Sigma_3 = \tilde{V}(\mathbf{i}\tilde{\omega}_1  + \mathbf{j}\tilde{\omega}_2 + \mathbf{k}\tilde{\omega}_3)\tilde{V}^{-1},\quad \tilde{\mu} = \pi - \mu,
\label{relation}
\ee
where $\tilde{\omega}_i \equiv \tilde{\sigma}_i - \tilde{\Sigma}_i$, and $2\tilde{V}^{-1}d\tilde{V} = \mathbf{i}\tilde{\Sigma}_1 + \mathbf{j}\tilde{\Sigma}_2 + \mathbf{k}\tilde{\Sigma}_3$ same as defined before. The map can be written equivalently in the following more compact way:
\be 
U = \tilde{U},\quad V = \tilde{U}\tilde{V}^{-1}\quad  \Leftrightarrow \quad \tilde{U} = U,\quad \tilde{V} = V^{-1}U;\quad \tilde{\mu} = \pi - \mu,
\label{Utilde}
\ee 
recall that $U, V$ are parametrized by the Euler angles as in \eqref{parametrization} and likewise for $\tilde{U}, \tilde{V}$. The relation above is nothing but a twist performed in the definition of some angles on the $\mathbb{S}^3$ bundle between the two metrics, which is philosophically the same as the transformation between (\ref{StdMetricS3}) and (\ref{MostStdMetricS3}) for a similar construction on complex projective spaces, this explains the similarities between the two metrics.

Since we're concerning Euclidean ${\rm AlAdS_8}$ in the bulk with a squashed seven sphere at the boundary, we make the following two ansatzes:\footnote{We have replaced all tilded quantities with un-tilded ones for simplicity.}
\begin{subequations}
	\begin{align}
		\label{Ansatz1}
		ds^2 &= dr^2 + a^2(r) \left(d\mu^2 + \frac{1}{4}\sin^2\mu \,\Sigma_i^2\right) + b^2(r) (\sigma_i - A_i)^2,\quad i = 1,2,3 \\
		ds^2 &= f_1^2(r)dr^2 + f_2^2(r)\left(d\mu^2 + \frac{1}{4}\sin^2\mu \omega_i^2 \right) +\frac{ f_{i+2}^2(r)}{4}(\nu_i + \cos\mu\omega_i)^2,  \label{Ansatz2}
\end{align} \end{subequations}
where $a(r),b(r),f_1(r),...,f_5(r)$ are undetermined functions of $r$ only. In the first ansatz, we set the coefficient of $dr$ to be 1 for simplicity, which can also be kept in general. Also note that in the second ansatz, $f_3, f_4,$ and $f_5$ can be different, while in the first ansatz, Einstein equations require only one function $b(r)$ can be arbitrary. This suggests an enhanced permutation symmetry in the ansatz (\ref{Ansatz1}), and the twist performed in \eqref{relation} breaks the symmetry. In the main text of the chapter, we will focus on the first ansatz, and we put our partial results on the second ansatz in appendix \ref{solutionAnsatz1} and a more general study for future work.

\subsection{Equation of Motion}
We consider Einstein gravity in Euclidean AlAdS$_{d+1}$ space with negative cosmology constant, the equations of motion are:
\be 
G_{\mu\nu} + \Lambda g_{\mu\nu}=0 ,\quad \Lambda = -\frac{d(d-1)}{2\ell^2},
\ee 
where $\ell$ is the scale of AdS. Plugging in the ansatz, we take advantage of the vielbein formalism in GR to simplify the Einstein field equations. We take the achtbein $e^m_\mu$ such that the metric tensor $g_{\mu\nu} = \delta_{mn}e^m_\mu e^n_\nu$. In the local achtbein coordinate, the connection, Riemann tensor, Ricci tensor, scalar curvature, and Einstein equations are given by:\footnote{We follow the notations and conventions of \cite{blau2011lecture}. }
\be
\begin{aligned}
	&\Gamma_{lmn} = \frac{1}{2}(d_{lmn} - d_{lnm} + d_{mnl} - d_{mln} + d_{nml} - d_{nlm}),\quad d_{lmn} \equiv e_{l\mu} e_n^\nu \partial_\nu e_m^\mu\\
	&R_{klmn} = 2\partial_{[k} \Gamma_{|mn|l]} + 2\Gamma^a_{\ m[l}\Gamma_{|an|k]} + 2\Gamma^a_{\ [kl]}\Gamma_{mna},\quad R_{km} = R^{\ \, l}_{k\ ml},\qquad R =  R^k_{\ k}\\
	&R_{km} - \frac{1}{2}R\delta_{km} + \Lambda \delta_{km} = 0,\quad k,l,... = 1,2,3,...,8. \\
\end{aligned}
\label{TetradFormalism}
\ee
By extracting independent parts of Einstein equations, we get the following equations of motion for the first ansatz:
\be \small 
\begin{aligned}
	&	-16 a^3 b a' b'-8 a^2 b^2 a'^2-4 a^4 b'^2+28 a^4 b^2+8 a^2 b^2+a^4-2 b^4 = 0,&\\
	&	-4 a b^2 a''-12 a b a' b'-4 b^2 a'^2-4 a^2 b b''-4 a^2 b'^2+28 a^2 b^2+a^2+4 b^2 = 0,&\\
	&	16 a^3 b^2 a''+32 a^3 b a' b'+24 a^2 b^2 a'^2+8 a^4 b b''+4 a^4 b'^2-84 a^4 b^2-24 a^2 b^2-a^4+10 b^4 = 0.&\\
\end{aligned}
\label{EinsteinEquations3}
\ee
Only two of them are independent because of Bianchi identity so they are not over-constrained. An analytical study is desirable, but not yet available unfortunately for now, and partial attempts were done in \cite{Bizon:2007zf} for a similar problem with vanishing cosmological constant. In this chapter, we will solve them perturbatively at both large-$r$ and small-$r$ regimes, followed by numerical evaluation, in the same manner as \cite{Bobev:2016sap}. By numerical simulation, we are able to find relations between parameters in large-$r$ series and small-$r$ ones.

Remarkably, the equations of motion (\ref{EOMAnsatz1}) for the second metric ansatz are identical to the ones above when imposing $f_3 = f_4 = f_5$. Thus the analysis below applies identically to the two metrics. In a sense, the second metric is a generalization of the first one which we will focus on in the main text.

\subsubsection{Large radius expansion}
As a space with negative cosmological constant, it's natural to perform Fefferman-Graham expansion at large-$r$:
\be
\left\{
\begin{aligned}
	a(r) = e^r A_0 + A_1 + e^{-r}A_2 + \cdots\\
	b(r) = e^r B_0 + B_1 + e^{-r}B_2 + \cdots\\
\end{aligned}
\right. .
\label{UVExpansionAnsatz2}
\ee
Solving the equations of motion order by order, we find that a general solution can be determined by three parameters, which we choose to be $A_0, B_0, $ and $A_7$. The coefficient $A_7$ is dual to the vacuum expectation value (vev) of a corresponding operator on the boundary \cite{Bianchi:2001kw}, which is the stress tensor in our case and thus should vanish. The first several terms are:
\be
\begin{aligned}
	a(r) = A_0 e^r & + e^{-r} \left(\frac{B_0^2}{8 A_0^3}+\frac{A_0}{80 B_0^2}-\frac{1}{5 A_0}\right) \\
	& +\frac{e^{-3 r} \left(-2 A_0^6 B_0^2-39 A_0^4 B_0^4+140 A_0^2 B_0^6+A_0^8-100 B_0^8\right)}{1600A_0^7 B_0^4} +O(e^{-5r}),\\
	b(r)=B_0 e^r & + \frac{e^{-r} \left(-\frac{10 B_0^4}{A_0^4}+\frac{8 B_0^2}{A_0^2}-3\right)}{80 B_0}\\
	& -\frac{e^{-3 r} \left(-2 A_0^6 B_0^2-39 A_0^4 B_0^4+140 A_0^2 B_0^6+A_0^8-100 B_0^8\right)}{1200A_0^8 B_0^3} +O(e^{-5r}). \\
\end{aligned}
\ee
As a special analytical solution, for the round ${\rm AdS}_8$ with spherical foliation, we have $a(r) = b(r) = \sinh r = \frac{1}{2}e^r - \frac{1}{2}e^{-r}$, which corresponds to $A_0 = B_0 = \frac{1}{2},A_1 = B_1 = -\frac{1}{2}$, and all other coefficients vanish.

\subsubsection{Small radius expansion and numerics - NUT}
\begin{figure}[ht]
	\centering
	\includegraphics[width=11cm]{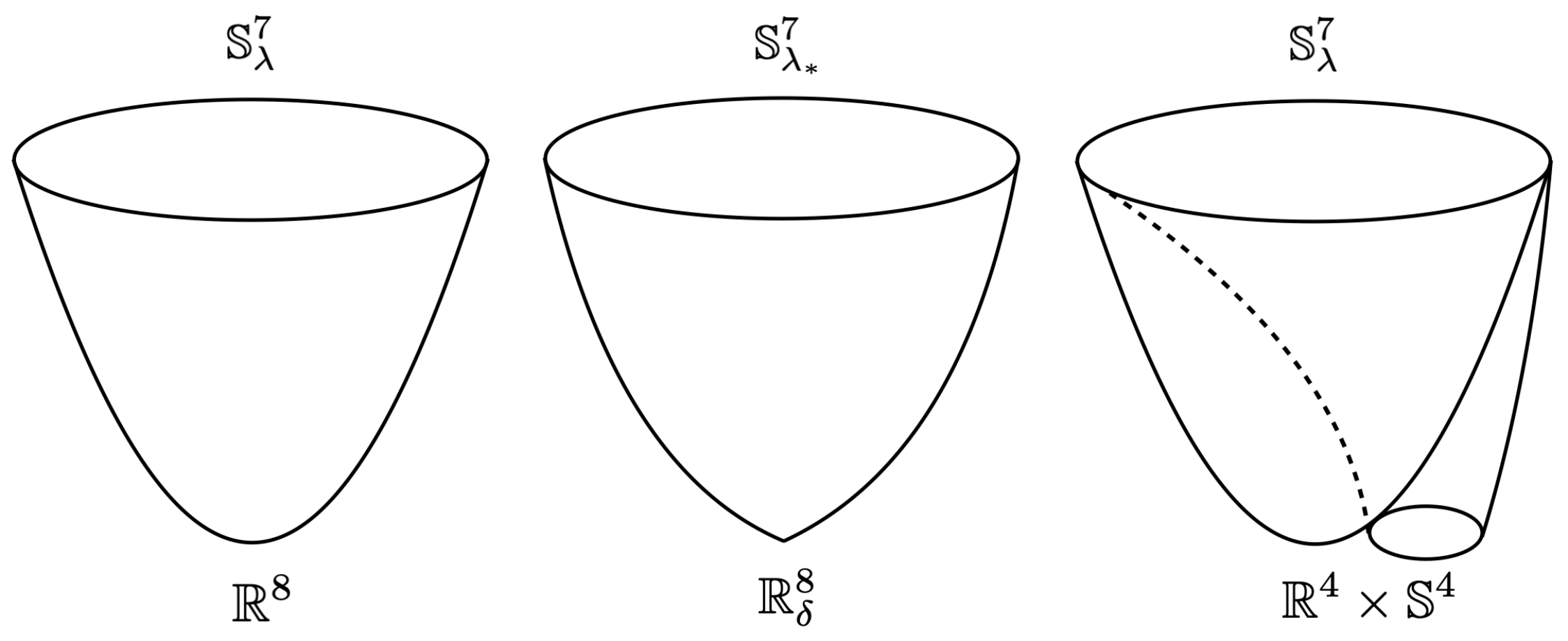}
	\caption{\rm Possible geometries in the bulk. The first one to the left is ``NUT'' geometry, which admits a squashed seven sphere at infinity and caps off smoothly at $r = r_0$. The one on the right is ``Bolt'' geometry, which admits a squashed sphere at infinity and caps off smoothly at $r = r_0$, where the geometry is $\mathbb{R}^4\times \mathbb{S}^4$. The one in the middle is a singular ``NUT'' solution which only exists for $\lambda = \lambda_* = \frac{1}{\sqrt{5}}$, it has a conifold singularity at $r = r_0$.}
	\label{Geometries}
\end{figure}

In the small-$r$ region, as $r$ decreases, both $a(r)$ and $b(r)$ decrease, and one of them will hit $0$ at some $r = r_0$, forming a black-hole horizon. Depending on how they hit 0 at $r = r_0$, there're two families of solutions, dubbed by Hawking, ``NUT'' and ``Bolt''. ``NUT'' applies to situations where $a(r)$ and $b(r)$ hits 0 at the same $r_0$, thus the near-horizon geometry looks locally like $\mathbb{R}^8$. While in ``Bolt'', $b(r)$ hits 0 at $r=r_0$ while $a(r_0)$ is finite, for which the near-horizon geometry looks like $\mathbb{R}^4\times \mathbb{S}^4$.  \footnote{On why there's no solution where $a(r)$ hits 0 first when $b(r_0)$ is finite: by expanding at small $r$ and solving equations of motion order by order, one can immediately prove the ansatz doesn't admit a solution. And on why there's no wormhole solution: we've checked it numerically using different initial conditions but failed to find a wormhole. } Besides the normal ``NUT'' and ``Bolt'' geometries, which cap off smoothly at $r = r_0$, there exist a singular ``NUT'' solution which has a conifold singularity at $r = r_0$. The geometries of the solutions are illustrated in Fig.\ref{Geometries}. The two families of solutions are analogous as in $d = 4$ AdS-Taub-NUT/Bolt solution investigated in \cite{Bobev:2016sap}.

For ``NUT'', we assume the following small-$r$ expansion:
\be
\left\{ 
\begin{aligned}
	& a(r) = a_1(r-r_0)^1 + a_2(r-r_0)^2 + ... &\\
	& b(r) = b_1(r-r_0)^1 + b_2(r-r_0)^2 + ... &\\
\end{aligned}
\right. .
\ee
The equations at leading order have two solutions:
\be
\left\{
\begin{aligned}
	&a_1 = \frac{1}{2},&\\
	&b_1 = \frac{1}{2},&
\end{aligned}
\right.\qquad {\rm or}\qquad
\left\{
\begin{aligned}
	&a_1 = \frac{3\sqrt{5}}{10},&\\
	&b_1 = \frac{3}{10}.&
\end{aligned}
\right.
\label{twoChoicesAnsatz2}
\ee
For the first choice, we solved the equations up to $O(r-r_0)^{13}$, and found the series has only one free parameter, which is $a_3$, and the first several orders are
\be \small 
\begin{aligned}
	a(r) = &\frac{\rho}{2}+a_3 \rho^3+\frac{\left(-14832 a_3^2+1932 a_3-49\right) \rho^5}{2160}+\frac{1}{816480}\left(44570304 a_3^3-7822224 a_3^2\right.\\
	&\left.+434448 a_3-7595\right)
	\rho^7 + O\left(\rho^9\right), \\
	b(r) = &\frac{\rho}{2}+\left(\frac{7}{36}-\frac{4 a_3}{3}\right) \rho^3+\frac{\left(71424 a_3^2-9744 a_3+343\right) \rho^5}{6480}+\frac{1}{816480}\left(-83054592 a_3^3\right.\\
	&\left.+15344640
	a_3^2-906192 a_3+17101\right) \rho^7+O\left(\rho^9\right), \\
\end{aligned}	
\label{SmallrExpansionNUT}
\ee
where we used the short-handed notation $\rho \equiv r-r_0$. Note that when choosing $a_3 = \frac{1}{12}$, one obtains the round ${\rm AdS}_8$ solution:
\be
a(r) = b(r) = \sinh \rho\quad \Rightarrow \quad ds^2 = d\rho^2 + \sinh^2\rho\, ds^2_{\mathbb{S}^7}.
\label{RoundSphere}
\ee 
For the second choice in (\ref{twoChoicesAnsatz2}), up to order $O(r-r_0)^{13}$, the solution is fixed, with no free parameter, whose leading order expansions can be identified with hyperbolic sine functions:
\be
\begin{aligned}
	&a(r) = \frac{3\sqrt{5}}{10}\left(\rho +\frac{\rho ^3}{6}+\frac{\rho ^5}{120}+\frac{\rho ^7}{5040}+\frac{\rho^9}{362880} + O\left(\rho ^{11}\right)\right) \rightarrow  \frac{3\sqrt{5}}{10}\sinh\rho, &\\
	&b(r)  = \frac{3}{10}\left(\rho +\frac{\rho ^3}{6}+\frac{\rho ^5}{120}+\frac{\rho ^7}{5040}+\frac{\rho ^9}{362880}+O\left(\rho^{11}\right)\right)\rightarrow \frac{3}{10}\sinh\rho. &\\	
\end{aligned}
\label{branch1ansatz2IR}
\ee
As can be checked, the asymptotic boundary of the above solution is the Einstein squashed sphere (\ref{SquashedSphereMetric2}) with $\lambda^2_* = 1/5$. This geometry is special, in one sense, it's the only singular solution that integrates to infinity, which can be shown by its diverging Kretschmann scalar at $r \rightarrow r_0$:
\be 
R = -56,\quad  R_{\mu\nu}R^{\mu\nu} = 392,\quad R_{\mu\nu\rho\sigma}R^{\mu\nu\rho\sigma} = 112 + \frac{2^{12}}{3^3\sinh^4\rho}.
\label{Curvatures}
\ee 
If we zoom in to the near-horizon limit, where $a(r) \approx \frac{3\sqrt{5}}{10}\rho$ and $b(r)\approx  \frac{3}{10}\rho$, we can evaluate:
\be 
R = 0,\quad  R_{\mu\nu}R^{\mu\nu} = 0,\quad R_{\mu\nu\rho\sigma}R^{\mu\nu\rho\sigma} = \frac{2^{12}}{3^3\rho^4 }.
\ee 
The appearance of a singular solution is not typical in Taub-NUT solution, since the introduction of nut charge actually alleviates the singular behavior in a Schwarzschild black hole. A similar singular solution is found in \cite{Aharony:2019vgs} when the boundary dimension satisfies $d \ge 9$.

Given the function values of $a(r),b(r)$ and their first-order derivatives at $\rho = 0$, we can solve the differential equations (\ref{EinsteinEquations3}) numerically. In general, for a randomly-chosen parameter $a_3$, it's not guaranteed that both $a(r)$ and $b(r)$ can be integrated up to infinity without hitting 0, otherwise one has a compact spindle-like space instead of an ${\rm AlAdS}_8$ one. In this chapter, we focus on the solutions where neither $a(r)$ nor $b(r)$ has zeros, a bulk solution with asymptotic squashed sphere boundaries as (\ref{SquashedSphereMetric2}), where the squashing parameter can be identified with the UV expansion coefficients in (\ref{UVExpansionAnsatz2}) as
\be
\lambda = \lim_{r\rightarrow\infty} \frac{b(r)}{a(r)} = \frac{B_0}{A_0} .
\ee
\begin{figure}[!t]
	\centering	\includegraphics[width=5.15cm]{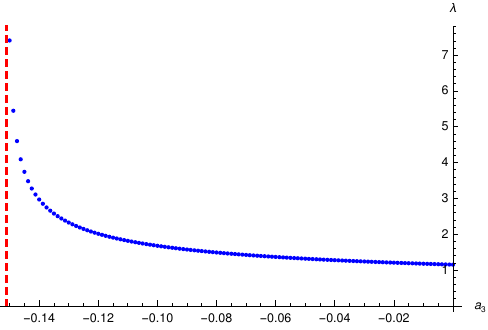}
	\qquad \qquad 
	\includegraphics[width=5.15cm]{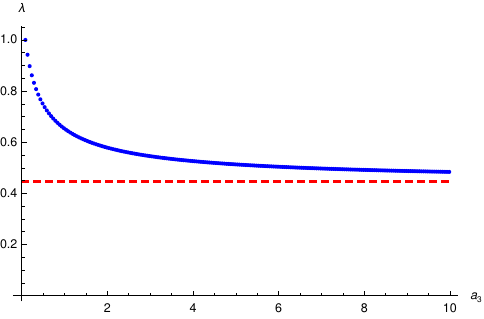}
	\hfil
	\includegraphics[width=5.15cm]{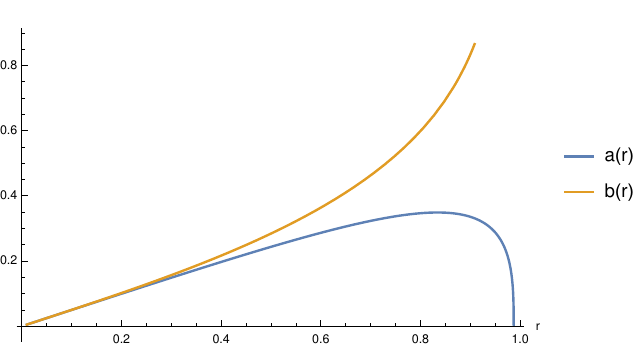}
	\qquad \qquad 
	\includegraphics[width=5.15cm]{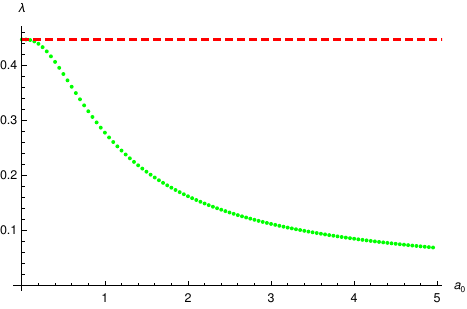}
	\caption{{\rm Upper: the relation between $\lambda$ on the boundary and $a_3$ at small-$r$, the verticle dashed line is $a_3 = a_3^*$, and the horizontal dashed line is $\lambda=\lambda_*=1/\sqrt{5}$. Downleft: numerical solution for NUT with initial value $a_3 = -0.5$. Downright: relation between $\lambda$ and $a_0$ for Bolt, the dashed line is $\lambda = \lambda_*$.}}
	\label{UVIRRelationAnsatz2}
\end{figure}
We start from $a_3 = \frac{1}{12}$ and integrate numerically to infinity, we get $\lambda = 1$, as expected. With different $a_3$, this procedure establishes a one-to-one correspondence between the squashing parameter $\lambda$ at asymptotic boundary and the initial condition $a_3$ at small-$r$, as shown in Fig.\ref{UVIRRelationAnsatz2}. Increasing the value of $a_3$ from $1/12$, the corresponding $\lambda$ decreases monotonically from $1$ to $\lambda_* = 1/\sqrt{5}\approx 0.447$; we then decrease $a_3$, and $\lambda$ increases monotonically from 1 to $\infty$ and finally stops at a special value $a_3^*\approx -0.151394$. For any $a_3<a_3^*$, $a(r)$ doesn't integrate to infinity, thus is of no interest to us. An example for $a_3 = -0.5 < a_3^*$ is shown in Fig.\ref{UVIRRelationAnsatz2}.

The NUT solution also has interesting implications in terms of holographic RG flow. It interpolates an asymptotically locally squashed AdS$_8$ at the UV region and a flat $\mathbb{R}^8$ in the IR region. The interesting fact is that from a UV theory with different squashing parameters $\lambda$, the RG flow converges to the same IR theory, which, from the leading order coefficients in (\ref{SmallrExpansionNUT}), lives in $\mathbb{R}^8$. The picture here is similar to the holographic uniformization discovered in \cite{Anderson:2011cz}, which is worth further understanding.

\subsubsection{Small radius expansion and numerics - Bolt}
The geometry of ``Bolt'' solution is supposed to be $\mathbb{R}^4\times \mathbb{S}^4$ at small-$r$ limit, where $a(r_0)>0$ is the finite radius of $\mathbb{S}^4$, and $b(r)$ goes to zero smoothly, forming $\mathbb{R}^4$ together with the radial coordinate. The geometry indicates the following small-$r$ expansion:
\be
\begin{aligned}
	& a(r) = a_0 + a_1(r-r_0)^1 + a_2(r-r_0)^2 + ... &\\
	& b(r) = b_1(r-r_0)^1 + b_2(r-r_0)^2 + ... &\\
\end{aligned} .
\ee
Repeating the procedure as above, the solutions can be obtained in any order, determined by one free parameter $a_0>0$. \footnote{Taking the limit $a_0 \rightarrow 0$, the geometry becomes closer to NUT solutions, as can be seen from the corresponding values of $\lambda$.} The leading terms are as follows

\be \small 
\begin{aligned}
	a(r) = & a_0-\frac{\left(-7 a_0^2-3\right) \rho^2}{8 a_0}-\frac{\left(49 a_0^4+98 a_0^2+39\right) \rho^4}{384 a_0^3}\\
	&\quad  -\frac{\left(-7889 a_0^6-15141
		a_0^4-10241 a_0^2-2379\right) \rho^6}{46080 a_0^5}+O\left(\rho^8\right),\\
	b(r) = & \frac{\rho}{2}-\frac{\rho^3}{12 a_0^2}-\frac{\left(-49 a_0^4-70 a_0^2-26\right) \rho^5}{480 a_0^4}\\
	&\quad  -\frac{\left(10290 a_0^6+20531 a_0^4+14000
		a_0^2+3224\right) \rho^7}{80640 a_0^6}+O\left(\rho^9\right).\\
\end{aligned}
\ee
The functions can be integrated out to infinity for any $a_0 > 0$, and the map between $a_0$ and $\lambda$ is shown in Fig.\ref{UVIRRelationAnsatz2}. As shown in the plot, as $a_0$ increases from 0 to $\infty$, $\lambda$ decreases from $\lambda_*$ to $0$ monotonically. Combine the $\lambda\sim a_0$ plot for ``Bolt'' solutions with $\lambda \sim a_3$ plot for ``NUT'' solutions, we find that for a squashed seven-sphere metric (\ref{SquashedSphereMetric2}) with any $\lambda>0$, there exists one unique bulk solution whose asymptotic boundary is squashed seven-sphere with $\lambda$. For $0<\lambda<\lambda_*$, the bulk is ``Bolt''; for $\lambda\ge \lambda_*$, the bulk is ``NUT''. The two analytical solutions, which have $\lambda = 1$ and $\lambda = \lambda_*=1/\sqrt{5}$, are both ``NUT'' spaces. Inversely speaking, the bulk metric is uniquely determined by $\lambda$, so there will not be an analog of Hawking-Page transition in our geometry. But it's still meaningful to evaluate the bulk free energy $F_{\rm  bulk}(\lambda)$ as a function of the squashing parameter, which we will do in the next section.

\subsection{Free energy}
\begin{figure}[t]
	\centering
	\includegraphics[width=7cm]{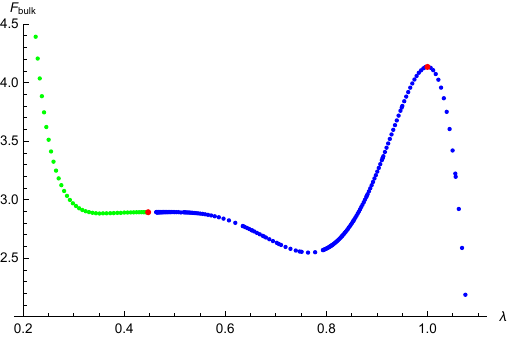}
	\caption{{\rm Numerical renormalized free energy $F_{\rm bulk}(\lambda)$ for NUT and Bolt spaces. The values of $F_{\rm bulk}$ are divided by $\frac{\ell^6}{G_N}$. Where blue points show NUT solutions, green points are Bolt solutions, and red points are the two special values $\lambda = 1, \lambda_*$ given in \eqref{AnalyticalBulkFE}. }}
	\label{FscLambda}
\end{figure}

In the semi-classical limit, the partition function localizes to solutions of Einstein equations, then the Euclidean gravitational free energy is reduced to the on-shell action:
\be 
F_{\rm bulk} = - \log \int \cD g_{\mu\nu} e^{-S_E[g_{\mu\nu}]} \xlongequal[\rm approximation]{\rm classical} - \log e^{-S_E[g_{\mu\nu}]_{\rm on-shell}} = S_E[g_{\mu\nu}]_{\rm on-shell}.
\ee 
We focus on Euclidean gravitational field without matter in the bulk, whose free energy with the GHY term on the boundary is given by \cite{Gibbons:1976ue}:
\be
S_{\rm EH} + S_{\rm GHY}= -\frac{ 1 }{16\pi G_N}\int_{\cM_8} d^{8}x\sqrt{g}\left(R -2\Lambda\right) - \frac{ 1 }{8\pi G_N}\int_{\partial\cM_8} d^7 x\sqrt{g^{(7)}}K,
\ee
where $g^{(7)}_{ab}$ is the induced metric on $\partial\cM_8$, which we choose to be a surface with constant large $r$, and $K$ is the trace of extrinsic curvature tensor on the boundary:\footnote{In our convention, Greek letters starting from $\mu$ are bulk indices, and Latin letter starting from $a$ are boundary indices. }\cite{blau2011lecture} 
\be 
g_{\mu\nu}^{(7)} \equiv g_{\mu\nu} - n_\mu n_\nu ,\quad K_{\mu\nu} \equiv (g^{(7)})_\mu^\rho (g^{(7)})_\nu^\sigma \nabla_\rho n_\sigma.
\ee 
Here $n^\mu$ is the unit normal vector of the boundary. By plugging in the metric ansatz, we obtain the action and its boundary term:
\be\begin{aligned}
	S_{\rm EH} &= \int_0^{R_{\partial}} dr\ \frac{4\pi^3 \ell^5 b}{3G_N}\left( 48 a^3 b a' b'+24 a^2 b^2 a'^2+16 a^3 b^2 a''+12 a^4 b'^2+12 a^4 b b''\right. \\
	&\left. \qquad\qquad \qquad  -84 a^4
	b^2-24 a^2 b^2-3 a^4+6 b^4 \right),\\
	S_{\rm GHY} &= - \frac{16\pi^3 \ell^6 a^3b^2}{3G_N} \left( 4 b a'+3 a b' \right).\\
\end{aligned}\label{GravAction} \ee 
For general AlAdS space, the free energy above is proportional to the volume of $\cM_8$, which is infinite, corresponding to a UV divergence. So we need to regularize the divergence by holographic renormalization. \cite{Henningson:1998gx, Henningson:1998ey, Balasubramanian:1999re} The counter-terms for $d\le 7$ are given by \cite{Clarkson:2002uj, Astefanesei:2004kn, Emparan:1999pm}\footnote{See \cite{Bueno:2022log} for a recent algorithm for deducing the counter terms up to arbitrarily large order.}
\be \small 
\begin{aligned}
	S_{\rm ct} =&  \frac{ 1 }{8\pi G_N} \int_{\partial \cM} d^d x\sqrt{g^{(d)}}\left[(d-1) + \frac{1}{2(d-2)}\cR + \frac{1}{2(d-4)(d-2)^2}\left(\cR_{ab}\cR^{ab}-\frac{d}{4(d-1)}\cR^2\right)\right.\\
	& \left.	+\frac{1}{(d-2)^3(d-4)(d-6)}\left(\frac{3d+2}{4(d-1)}\cR \cR_{ab}\cR^{ab}-\frac{d(d+2)}{16(d-1)^2}\cR^3 - 2\cR^{ab}\cR^{cd}\cR_{acbd}\right.\right.\\
	&\left.\left.-\frac{d}{4(d-1)}\nabla_a\cR\nabla^a \cR +\nabla^c\cR^{ab}\nabla_c\cR_{ab}\right)\right],
\end{aligned}
\ee
where $\cR_{abcd}, \cR_{ab}, \cR$ are the Riemann tensor, Ricci tensor, Ricci scalar of the boundary. Plugging in the metric ansatz, we get:
\be \small  \begin{aligned}
	S_{\rm ct} & = \frac{\pi^3 \ell^6 }{12000a^8b^3G_N} \left( 384000 a^{12} b^6+9600 a^{12} b^4+40 a^{12} b^2+76800 a^{10} b^6-4480 a^{10} b^4 \right. \\
	&\left.  +40 a^{10}
	b^2-19200 a^8 b^8-160 a^8 b^6+2774 a^8 b^4-6400 a^6 b^8+1504 a^6 b^6+4000 a^4
	b^{10} \right. \\
	&\left.  -13900 a^4 b^8+23200 a^2 b^{10}+7 a^{12}-11000 b^{12} \right).
\end{aligned}\ee
In the following evaluations, we set the Newton constant to be unit. Recall that there're two special values of $\lambda^2 = 1, 1/5$ for which the metric $g_{\mu\nu}$ is known analytically (\ref{RoundSphere}, \ref{branch1ansatz2IR}), the free energies can also be calculated analytically by $F_{\rm bulk}(\lambda) = S_{\rm EH} + S_{\rm GHY} + S_{\rm ct}$, which are:
\be
F_{\rm bulk}(1) = \frac{2\pi^3}{15} \frac{\ell^6}{G_N} \approx 4.134  \frac{\ell^6}{G_N}  ,\qquad F_{\rm bulk}\left(\frac{1}{\sqrt{5}}\right) = \frac{2\cdot 3^6 \pi ^3}{5^6}  \frac{\ell^6}{G_N}  \approx 2.893  \frac{\ell^6}{G_N} .
\label{AnalyticalBulkFE}
\ee
For general $\lambda$, $F_{\rm bulk}(\lambda)$ can be obtained only numerically. By large-$r$ expansion, we can check that the diverging part of $S_{\rm EH} + S_{\rm GHY}$ and $S_{\rm ct}$ exactly cancel with each other, which have the same diverging structure:
\be \small 
\begin{aligned}
	S_{\rm EH} &  + S_{\rm GHY} \sim  -S_{\rm ct} \sim \frac{B_0\pi^3 \ell^6 }{G_N}\left[-32 A_0^4 B_0^2 e^{7 r}+\frac{6}{5} e^{5 r} \left(8 A_0^2 B_0^2+A_0^4-2 B_0^4\right)\right.\\
	& +\frac{1}{600} e^{3 r} \left(\frac{100
		B_0^6}{A_0^4}-\frac{160 B_0^4}{A_0^2}+\frac{A_0^4}{B_0^2}-112 A_0^2-4 B_0^2\right)\\
	& \left.	+\frac{e^r }{9600}\left(\frac{11000
		B_0^8}{A_0^8}-\frac{23200 B_0^6}{A_0^6}+\frac{13900 B_0^4}{A_0^4}-\frac{1504 B_0^2}{A_0^2}-\frac{40
		A_0^2}{B_0^2}-\frac{7 A_0^4}{B_0^4}-2774\right)\right].
\end{aligned}
\label{IRDivergenceExpansion}
\ee
We use the notation ``$\sim$'' to denote that the equation holds up to vanishing terms at large $r$. The numerical renormalized free energy is shown in Fig.\ref{FscLambda}. As $\lambda\rightarrow 0$ or $\lambda\rightarrow +\infty$, $F_{\rm bulk}$ diverges as expected. At $\lambda = 1$, there is a local maximum, consistent with the holographic F-theorem. Up to the precision of our numerics, near $\lambda_* = \frac{1}{\sqrt{5}}$, the free energy varies quite slowly, but we are not able to determine the critical behaviors explicitly.

Since we're integrating numerically starting from $r = 0$, the numerical error keeps accumulating and explodes at around $r \approx 5$, thus one has to stop integrating somewhere at $r = R < 5$, then the action looks like:
\be 
S_{\rm EH} + S_{\rm GHY} = A e^{7R} + B e^{5R} + C e^{3R} + D e^{R} + E + O\left( e^{-R}\right).
\ee 
The diverging terms are canceled by the counter term, and what we're interested in is the finite term $E$. To extract $E$, we take a set of cutoff values $R$ and numerically fit it with the function of the form $E + F e^{-R} + G e^{-3R}+...$, until the discrepancy from the analytical results (\ref{AnalyticalBulkFE}) is small enough.

In the end, we would like to mention that the gravitational action and boundary term of the second metric ansatz with $f_3=f_4=f_5$ are identical to those posted above (\ref{GravAction}), thus the renormalized action is identical to the one we posted in Fig.\ref{FscLambda}.

\section{Field theory story}
\label{BoundaryTheory}

It's also interesting to study the free energies of quantum field theories. To look for the correct holographic correspondence of the Einstein gravity theory, one has to go to string or M-theory and identify the boundary CFT, which is typically a strongly-coupled CFT without supersymmetry. Instead, we will study the universal properties of the free energies dictated by conformal symmetry. Directly making use of the conformal symmetry, we can study universal properties shared by all CFTs living on the squashed sphere as in \cite{Bobev:2017asb}. Some of the universal properties can also be found by studying some toy models as in \cite{Hartnoll:2005yc, Anninos:2012ft, Anninos:2013rza, Bobev:2016sap, Bobev:2017asb}.

The main quantity that we are interested in is the free energy $F_{\rm CFT}(\lambda)$ of CFTs as a function of the squashing parameter that the field theory lives. In the first part of this section, we calculate $F_{\rm CFT}''(\lambda)\Big|_{\lambda = 1}$ for general CFT living on the squashed seven-sphere of the first kind (\ref{SquashedSphereMetric2}), which can be evaluated by integrating the two-point function of stress tensors on the sphere. This property is totally determined by the conformal symmetry and thus independent of the details of the theories such as couplings.

To support the study of the correspondence both around and away from $\lambda = 1$, we study in addition two free CFTs, the conformally coupled scalar, and the free fermion respectively. Using these toy models, we can evaluate $F_{\rm CFT}'''(\lambda)\Big|_{\lambda = 1}$ both analytically and numerically, which is too complicated to be evaluated as an integrated three-point function of stress tensors. Besides, we can also study the strongly deformed scaling of the field theories numerically, which can also be compared with the bulk.

Let's clarify our results for the two metric ansatzes. The result in the first subsection only works for the first ansatz which is easier to deal with, we leave a detailed study of the second ansatz for future work. But we do expect they are the same because our results for the toy models are identical for them both due to the coincidence of spectrum on the two squashed sphere metrics. What's more, the results show a very good coincidence with the bulk free energy, which is also identical for the two metrics.

\subsection{Partition function on deformed manifold}

The partition function of a general CFT on the boundary of manifold $\cM_{d+1}$ with Euclidean boundary metric $g_{ab}$ is given by:\footnote{In this section, we follow the notation of \cite{Bobev:2017asb} for the free energy. Note that the quantity $\widetilde{F}$ in $F$-theorem \cite{Pufu:2016zxm} in \eqref{FreeEnergySd} is different from the free energy by a sign, which happens to be $+1$ for $d = 7$ of our interest. } 
\be
Z = \int \cD \varphi\; e^{-S[\varphi,g_{ab}]},\quad F \equiv -\ln Z.
\ee
We can couple the field theory to the gravitational background by squashing the metric. As is mentioned in the introduction, squashing background metric is equivalent to adding a marginal deformation (\ref{Deformation}) generated by the stress tensor, where we parametrize the squashing by a parameter $\epsilon$ as follows:
\be 
g_{ab} = g_{ab}^{(0)} + \epsilon h_{ab},\quad g^{ab} = (g^{(0)})^{ab} + \epsilon h^{ab} + O(\epsilon^2) \quad \Rightarrow \quad  h^{ab} = -h_{cd}(g^{(0)})^{ac}(g^{(0)})^{bd}.
\ee 
The coupling of field theory with background metric, or equivalently the effect of the induced RG flow, is encoded in correlation functions of the stress tensor. To start with, the first-order derivative of the free energy is given by integrating the one-point function of the stress tensor: \cite{Bobev:2017asb}
\be 
F'(\epsilon)\Big|_{\epsilon = 0} = - \frac{1}{2} \int d^dx \sqrt{g^{(0)}} h^{ab}(x) \langle T_{ab}(x)\rangle_{\partial \cM}.
\ee  
In odd-dimensional CFTs, out setup included, one point function of stress tensor vanishes as is required by conformal symmetry, thus $F'(0) = 0$. The second derivative of free energy $F_{\rm CFT}(\epsilon)$ is given by the integrated two-point function of stress tensor:
\be
F''(\epsilon)\Big|_{\epsilon=0} = -\frac{1}{4}\int d^dx d^dy\sqrt{g^{(0)}(x)g^{(0)}(y)} h^{ab}(x)h^{cd}(y)\langle T_{ab}(x)T_{cd}(y)\rangle_{\partial \cM}.
\label{Integrated2ptFunc}
\ee
The integrand is evaluated on the boundary $\partial \cM$ with deformations turned off. To evaluate the two-point function on $\partial \cM$, we take a conformal map $f:\partial\cM\rightarrow \mathbb{R}^d$ which relates the line element as $$f^*(ds^2_{\mathbb{R}^d} ) = \Omega^2(x)ds^2_{\partial\cM},$$ where $f^*$ is the pullback and $\Omega^2(x)$ is the corresponding conformal factor. In odd dimensions, the stress tensors are transformed under the conformal map through \cite{Simmons-Duffin:2016gjk}
\be \label{defMab}
T_{ab}(x) = \Omega^{d-2}M^{\bar{a}\bar{b}}_{ab}T_{\bar{a}\bar{b}}(X),\quad {\rm where} \ M^{\bar{a}\bar{b}}_{ab}\equiv \frac{\partial X^{\bar{a}}}{\partial x^a}\frac{\partial X^{\bar{b}}}{\partial x^b},\quad X^{\bar{a}} \in \mathbb{R}^d,\ x^a \in \partial \cM.
\ee
Thus the stress tensor two-point function on $\mathbb{S}^7$ is related to that on $\mathbb{R}^7$ by
\be
\langle T_{ab}(x)T_{cd}(y)\rangle_{\mathbb{S}^7} =                          \Omega^5(x)\Omega^5(y)M_{ab}^{\bar{a}\bar{b}}M_{cd}^{\bar{c}\bar{d}}\langle T_{\bar{a}\bar{b}}(X)T_{\bar{c}\bar{d}}(Y)\rangle_{\mathbb{R}^7}.
\label{StressTensor2ptFunction}
\ee
And the two-point function on $\mathbb{R}^d$ is well-known in the literature by using conformal symmetry: \cite{Osborn:1993cr,Erdmenger:1996yc}
\be
\begin{aligned}
	& \langle T_{\bar{a}\bar{b}}(X)T_{\bar{c}\bar{d}}(Y)\rangle_{\mathbb{R}^d}     = C_T \frac{I_{\bar{a}\bar{b}\bar{c}\bar{d}}(X-Y)}{|X-Y|^{2d}},\quad I_{\bar{a}\bar{b}, \bar{c}\bar{d}}(X) =\cE_{\bar{e}\bar{f},\bar{c}\bar{d}}I_{\bar{a}\bar{e}}I_{\bar{b}\bar{f}}, \\
	& \cE_{\bar{a}\bar{b},\bar{c}\bar{d}} = \frac{1}{2}(\delta_{\bar{a}\bar{c}}\delta_{\bar{b}\bar{d}} + \delta_{\bar{a}\bar{d}}\delta_{\bar{b}\bar{c}}) - \frac{1}{d}\delta_{\bar{a}\bar{b}}\delta_{\bar{c}\bar{d}},\quad I_{\bar{a}\bar{b}}(X) = \delta_{\bar{a}\bar{b}} - 2 \frac{X_{\bar{a}}X_{\bar{b}}}{X^2}.\\
\end{aligned}
\label{TTCorrelation}
\ee
There exists a natural conformal map between $\mathbb{S}^d$ and $\mathbb{R}^d$: the stereographic projection, for which the conformal factor is given by:
\be \Omega = \frac{1}{2}\left( 1 + X_{\bar{a}} X^{\bar{a}} \right), 
\label{ConformalFactorOmega}
\ee 
thus one can always evaluate the reaction of the free energy under metric deformation. But for a specific metric on the sphere, the detailed construction of the global map may be different, more discussions and explicit constructions of the conformal map can be found in Appendix \ref{ConformalMapping}. Using the conformal map, one can work out the matrix $M_{ab}^{\bar{a}\bar{b}}$ defined in \eqref{defMab}, which facilitates the integral. In this section, we will investigate two different squashings, one is the SU(2) bundled seven spheres which is the theme of this chapter, and another one is the U(1) bundled squashed $\mathbb{S}^{2d+1}$ which is also of great interest.

\subsubsection{SU(2) bundle squashed $\mathbb{S}^7$}    

To apply the formalism above to our metric (\ref{SquashedSphereMetric2}) on $\mathbb{S}^7$, we can relate the two squashing parameters as follows: $$\epsilon \equiv \lambda^2 -1,$$ for which one finds the only non-vanishing components of the perturbative inverse metric $h^{ab}$ are:
\be
h^{\theta\theta} = -4,\quad h^{\phi\phi} = h^{\psi\psi} = -\frac{4}{\sin^2\theta},\quad h^{\phi\psi} = h^{\psi\phi} = \frac{4}{\tan\theta \sin\theta}.
\label{hComp}
\ee 
Now we know everything in our integral \eqref{Integrated2ptFunc}, which is the double integral on $\mathbb{S}^7$. To evaluate it, one doesn't need to perform a 14-dimensional integration, instead, one fixes one stress tensor on the south pole and integrates the other stress tensor over the sphere. \cite{Pufu:2016zxm, Cardy:1988cwa, Bobev:2017asb} What considered in the references is the correlation function of scalar operators, thus the trick is easily justified because the scalar perturbation preserves the symmetry of the sphere. However, we're dealing with the correlation function of stress tensors, whose deformation coefficients are tensors with coordinate-dependent components as shown in \eqref{hComp}, so it is not understood why the trick still works. Luckily it works well and corresponds to the other calculations that we did in this chapter. According to the trick, we put one point $y^a$ on the south pole, and only integrate on $x^a$:
\begin{equation*}
    \begin{aligned}
    F''(\epsilon)\Big|_{\epsilon=0} &= - \frac{C_T}{4}V_{\mathbb{S}^7} \int d^7x \sqrt{g^{(0)}(x)} h^{ab}(x)h^{cd}(0) \langle T_{ab}(x)T_{cd}(0)\rangle_{\mathbb{S}^7}\\
    &  = - \frac{C_T}{4}V_{\mathbb{S}^7} h^{cd}(0)\Omega^{5}(0)M_{cd}^{\bar{c}\bar{d}}(0)\int d^7x \sqrt{g^{(0)}(x)} \left[h^{ab}(x)\Omega^{5}(x)M_{ab}^{\bar{a}\bar{b}}(x)\frac{\cI_{\bar{a}\bar{b},\bar{c}\bar{d}}(X)}{|X|^{14}}\right],
\end{aligned}
\end{equation*}
where $V_{\mathbb{S}^7} = \frac{\pi^4}{3}$ is the volume of the unit seven sphere, coming from identifying the integrand with arbitrary $y^\mu \in \mathbb{S}^7$ to that with $y^\mu$ fixed at the south pole. The south pole of the sphere corresponds to the origin of $\mathbb{R}^7$, where the quantities are:
\be 
    \Omega(0) = \frac{1}{2},\quad M^{\bar{5}\bar{5}}(0) = M^{\bar{6}\bar{6}}(0) = M^{\bar{7}\bar{7}}(0) = \frac{1}{4},\quad M^{\bar{c}\bar{d}} \equiv h^{cd}M_{cd}^{\bar{c}\bar{d}}.
\ee 
Then the integrand, which we denote $I$ is given by:
\be 
    I = - \frac{C_T V_{\mathbb{S}^7}}{2^9} \sqrt{g^{(0)}(x)} \frac{(1-x^8)^9}{(x^{\bar{a}} x_{\bar{a}})^7} \sum_{\bar{c} = (5,6,7)} M^{\bar{a}\bar{b}}I_{\bar{c}\bar{c};\bar{a}\bar{b}}(X),\quad x^A \in \mathbb{R}^8, \ \bar{a} = 1,2,\cdots 7.
\ee 
    Since the integrand is rather complicated, in order to evaluate it analytically, we split the contraction into the following three terms:
\be 
    \begin{aligned}
        \sum_{\bar{c} = (5,6,7)} M^{\bar{a}\bar{b}}I_{\bar{c}\bar{c};\bar{a}\bar{b}}(X) &= \left( M^{\bar{5}\bar{5}} + M^{\bar{6}\bar{6}} + M^{\bar{7}\bar{7}} - \frac{3}{7} M^{\bar{a}\bar{a}} \right) - \frac{4(1-x^8)^2}{x^{\bar{a}} x_{\bar{a}}} \sum_{\bar{c} = 1}^7 \sum_{\bar{d} = 1}^3 M^{\bar{c}\bar{d}} X_{\bar{c}}X_{\bar{d}} \\
        & + \frac{4(1-x^8)^2}{(x^{\bar{a}} x_{\bar{a}})^2 } M^{\bar{c}\bar{d}} X_{\bar{c}}X_{\bar{d}}( x^{\bar{5}} x_{\bar{5}} + x^{\bar{6}} x_{\bar{6}} + x^{\bar{7}} x_{\bar{7}} ).
    \end{aligned}
\ee 
Thus we have $I = I_1 + I_2 + I_3$ where
\be\begin{aligned}
    I_1 &= - \frac{C_TV_{\mathbb{S}^7} }{2^9} \sqrt{g^{(0)}(x)} \frac{(1-x^8)^9}{(x^{\bar{a}} x_{\bar{a}})^7} \left( M^{\bar{5}\bar{5}} + M^{\bar{6}\bar{6}} + M^{\bar{7}\bar{7}} - \frac{3}{7} M^{\bar{a}\bar{a}} \right)  \\
    I_2 &= + \frac{C_TV_{\mathbb{S}^7}}{2^7} \sqrt{g^{(0)}(x)} \frac{(1-x^8)^{11}}{(x^{\bar{a}} x_{\bar{a}})^8} \sum_{\bar{c}=1}^7\sum_{\bar{d}=1}^3M^{\bar{c}\bar{d}} X_{\bar{c}}X_{\bar{d}} \\
    I_3 &= - \frac{C_TV_{\mathbb{S}^7}}{2^7} \sqrt{g^{(0)}(x)}  \frac{(1-x^8)^{11}}{(x^{\bar{a}} x_{\bar{a}})^{9}} M^{\bar{c}\bar{d}} X_{\bar{c}}X_{\bar{d}}( x^{\bar{5}} x_{\bar{5}} + x^{\bar{6}} x_{\bar{6}} + x^{\bar{7}} x_{\bar{7}} ). \\
\end{aligned}\label{FullIntegral}\ee
Substituting $h^{ab}$ from \eqref{hComp}, $\Omega^2$ from \eqref{ConformalFactorOmega}, and $M_{ab}^{\bar{a}\bar{b}}$ from the conformal map given by combining \eqref{StereographicProjection} and \eqref{EmbeddingMapS7}, we can express the integrand in terms of $\mathbb{S}^7$ coordinates $(\mu, \Theta, \Phi, \Psi, \theta, \phi, \psi)$. Since the three integrals above are all divergent near the south pole, the order of variables to be integrated out is important. Here we present our procedure, which might be not a unique way to get the correct answer. 

The dependence of the integrands on $(\Theta, \Phi, \Psi)$ is very simple, so we can integrate over them directly. Then we integrate out $\theta$ followed by $\mu$, both of which are convergent, and the resulting function only depends on $\chi \equiv \phi + \psi$. The final function is divergent, but the integration is ignorant of the divergence, which is similar to the example below:
\be 
    \int_{-1}^1 \frac{1}{x^2}dx = -\frac{1}{x}\Big|_{-1}^1 = -2.
\ee 
By going through the procedures above, one can obtain that
\be 
\begin{aligned}
    & F''(\epsilon)\Big|_{\epsilon=0} = \int (I_1 + I_2 + I_3) = - \frac{29\pi^8}{3780}  C_T + \frac{\pi^8}{1890} C_T + 0 =  - \frac{\pi^8}{140}C_T\\
    & \quad \Rightarrow \quad  F''(\lambda)\Big|_{\lambda = 1} = 4F''(\epsilon)\Big|_{\epsilon=0} = - \frac{\pi^8}{35} C_T.\\
\end{aligned}
\label{FppSU2Bundle}
\ee 
In the last line above, we used the relation between the squashings $\epsilon = \lambda^2 - 1$. The result is universal since it's only related to conformal symmetry and thus applies to all CFTs. It must be reproduced in free field theories as well as the holographic theory, as we will see.

\subsubsection{U(1) bundle squashed $\mathbb{S}^{2k+1}$ }    
Spheres as U(1) bundles over complex projective spaces are very interesting cases and have been investigated partially in the literature. \cite{Bobev:2017asb, Bueno:2018yzo, Bueno:2020odt} As in the last section, we can obtain the second-order derivative of free energy in terms of squashing parameter for any odd-dimensional sphere with a U(1) bundle. The special cases for $\mathbb{S}^3$ and $\mathbb{S}^5$ has been discussed in \cite{Bobev:2017asb}, and a powerful formula was conjectured for general $d = 2k+1$ from the point of view of high-derivative gravity \cite{Bueno:2018yzo}:
\be 
F''_d(\epsilon)\Big|_{\epsilon = 0} = \frac{(-1)^{\frac{d-1}{2}}\pi^{d+1}(d-1)^2 }{2d!}C_T.
\label{Prediction}
\ee 
Using the embedding map of $\mathbb{S}^7$ described in (\ref{S7Embedding}), we perform the integral as in \cite{Bobev:2017asb} for generic values of $d = 2k+1$ and find the result consistent with the prediction above, proving it from the field theory side, which is again a universal result applying for general CFTs living on squashed $(2k+1)$-spheres. This is the first time to obtain them from the field theory side up as far as we know. We put the detailed analysis of the integral in Appendix \ref{FppForGeneralk} for people interested.

Taking some special values $d = 3, 5, 7$, one gets:
\be 
F''_3(0) = - \frac{\pi^4}{3}C_T,\quad F''_5(0) = \frac{\pi^6}{15}C_T,\quad F''_7(0) = - \frac{\pi^8}{280} C_T.
\label{FppU1Bundle}
\ee 
The first two are identical to the results in \cite{Bobev:2017asb}. Comparing $F''(\epsilon)$ on the seven spheres with U(1) bundle in (\ref{FppU1Bundle}) and SU(2) bundle in (\ref{FppSU2Bundle}), we find the one with SU(2) bundle is twice larger, which means the squashing with SU(2) bundle perturbs the field theory more strongly, which is intuitively correct. 

\subsection{Conformally-coupled scalar}
The conformally coupled scalar theory, or the $O(N)$ model,\footnote{ In this chapter, whenever we refer to the $O(N)$ model, we always assume $N = 1$, which is the conformally coupled scalar model. } is holographic dual to higher spin gravity \cite{Klebanov:2002ja}, whose free energy is different from the Einstein gravity that we studied. However, by comparing the free energies of Einstein's gravity and field theories here, we can observe some universal properties shared among them. \cite{Hartnoll:2005yc,Anninos:2012ft,Anninos:2013rza,Bobev:2016sap,Bobev:2017asb}. It would also be interesting to study the high spin gravity in the future. 

The partition function of massless conformally-coupled scalars living on squashed seven-sphere is as follows\footnote{The coefficient of the coupling term for general $d$ is $\frac{d-2}{4(d-1)}$. }
\be
Z_{\rm sc} = \int D\phi e^{-S_{\rm sc}[g_{ab},\phi] },\quad S_{\rm sc} = \frac{1}{2} \int d^7x\sqrt{g}  \left((\partial\phi)^2 + \frac{5}{24}R\phi^2\right).
\ee
After a Gaussian integration, the free energy is given by
\be
F_{\rm sc} = -\log Z_{\rm sc}  =  \frac{1}{2} \log \det \left(-\nabla^2 + \frac{5}{24}R\right).
\ee
The eigenvalues of Laplacian on the squashed seven sphere (\ref{SquashedSphereMetric1}) and their degeneracies are: \cite{Nilsson:1983ru}
\be
\begin{aligned}
	& \lambda_{n,r} = n(n+6) + \frac{1-\lambda^2}{\lambda^2} (n-2r)(n-2r+2),\quad\quad n = 0,1,2,\cdots; \\
	& m_{n,r} = \frac{1}{6}(n+3)(n-r+2)(n-2r+1)^2(r+1),\quad \ r = 0, 1, \cdots , \left[\frac{n}{2}\right]. \\
\end{aligned}
\label{ScalarSpectrum}
\ee
As shown in (\ref{RicciTensor}), the Ricci scalar of the two squashed sphere metrics are identical, thus the calculation here applies in both cases. We can obtain the spectrum of conformal Laplacian on squashed seven sphere with the Ricci scalar
\be
\tilde{\lambda}_{n,r} = \lambda_{n,r} + \frac{5}{24}R,\qquad R = 6 \left(8-2 \lambda ^2+\frac{1}{\lambda ^2}\right).
\ee
From (\ref{ScalarSpectrum}), we find the eigenvalues are unbounded, thus the free energy has a UV divergence and needs to be regularized. In the following, we will use two regularization methods. The first one is heat-kernel regularization following  \cite{Anninos:2012ft,Anninos:2013rza, Bobev:2016sap, Bobev:2017asb}, for which we can calculate $F_{\rm sc}(\lambda)$ numerically for general $\lambda$. The second one is zeta-function regularization following \cite{Hartnoll:2005yc, Bueno:2020odt}, with which we can calculate the derivatives at $\lambda = 1$ analytically. Combining the numerical results for generic $\lambda$ and the analytical results at $\lambda= 1$, we can justify the precision of numerical simulation, as well as compare them to the bulk calculation from various aspects.

\subsubsection{Heat-kernel regularization}

In this section, we follow the heat-kernel technique of \cite{Hawking:1976ja}, see also \cite{Monin:2016bwf, Vassilevich:2003xt}. Consider a general spectrum $\{\lambda_i\}$ with degeneracy $\{ m_i\} $, one can define the heat-kernel function and spectral zeta-function:
\be 
K(t) \equiv \sum_{i} m_i e^{-t\lambda_i},\quad \zeta_{\Delta}(p) \equiv \sum_{i}\frac{m_i}{\lambda_i^p}.
\ee 
After evaluating the Gaussian integral in the partition function, the free energy is proportional to the determinant of the kinetic operator. For conformally coupled scalar fields, we have
\be 
F = \frac{1}{2} \log\det \Delta = \frac{1}{2} \sum_i m_i \log \lambda_i = - \frac{1}{2} \zeta_\Delta'(0).
\label{FreeEnergySC}
\ee 
The spectral zeta function is related to the heat kernel with a Mellin transformation:
\be 
G(p) \equiv \Gamma(p)\zeta_\Delta(p) = \int_0^\infty dt K(t) t^{p-1}.
\ee 
On the left hand side above, we can expand $G(p)$ at small $p$:
\be 
G(p) = \frac{\zeta_\Delta(0)}{p} - \gamma\zeta_\Delta(0) + \zeta_\Delta'(0) +O(p).
\ee 
Meanwhile, on the right hand side, the integral is divergent in the small-$t$ region because of the divergence of heat kernel at small-$t$, which in general $d$-dimensional field theories is
\be 
K(t) =\sum_{k=0}^{\frac{d+1}{2}} a_{d/2 - k}t^{-d/2 + k} + O(t).
\ee 
In fact, both the divergence of $K(t)$ at $t = 0$ on the right hand side and the pole at $p = 0$ on the left hand side reflect the UV divergence of the determinant, which requires a regularization. The essence of heat-kernel regularization is to regulate the integral on the right hand side. Practically, we divide the integral domain into $(0,1] \cup [1,\infty)$ and perform the divergent integral over $(0,1]$ as follows, this essentially eliminates the divergence by throwing off the diverging contribution from a small cutoff:
\be \begin{aligned}
	&\int_0^1 dt \ \frac{1}{t} \ a_{d/2 - k}t^{-d/2 + k} = \lim_{\epsilon\rightarrow 0} \int_\epsilon^1 dt \ \frac{1}{t} \ a_{d/2 - k}t^{-d/2 + k}\\
	&\qquad\qquad\qquad  = \frac{ a_{d/2 - k}}{k-d/2} \lim_{\epsilon\rightarrow 0} \left(1 - \frac{1}{\epsilon^{d/2-k}} \right) \quad \Rightarrow \quad  \frac{ a_{d/2 - k}}{k-d/2}.
\end{aligned}\ee
After heat-kernel regularization, $G(p)$ is reduced to the following finite value:
\be 
G(p) = \int_0^1 dt\left[ K(t) - \sum_{k=0}^{\frac{d+1}{2}} a_{d/2 - k}t^{-d/2 + k} \right] t^{p-1} + \sum_{k=0}^{\frac{d+1}{2}} \frac{ a_{d/2 - k}}{k-d/2} + \int_1^\infty dt K(t) t^{p-1}.
\ee 
The expression above has no pole at $p = 0$, which requires $\zeta_\Delta(0) = 0$, thus the free energy (\ref{FreeEnergySC}) is equal to the integral of heat-kernel, this is the central equation we're going to calculate:
\be 
F_{\rm sc} = - \frac{1}{2}\zeta_\Delta'(0) = G(0) = -\frac{1}{2}  \int_0^\infty \frac{1}{t}K(t)dt.
\ee 
From the discussion above, the heat kernel regularization is essentially also a zeta function regularization through the spectral zeta function. Since the integrals above are evaluated numerically, we don't require $\lambda$ to be special values. For the conformally coupled scalars, we obtain the following coefficients of the divergent terms:
\be \small 
\begin{aligned}
	a_{7/2} =& \frac{\sqrt{\pi } \lambda ^3}{384}, \\
	a_{5/2} =& \frac{\sqrt{\pi } \lambda  \left(2 \lambda ^4-8 \lambda ^2-1\right)}{1536}, \\
	a_{3/2} =& \frac{\sqrt{\pi } \left(4 \left(255 \lambda ^6-504 \lambda ^4+257 \lambda ^2+60\right) \lambda ^2+15\right)}{184320 \lambda }, \\
	a_{1/2} =& -\frac{\sqrt{\pi } \left(2 \left(51100 \lambda ^{10}-60240 \lambda ^8-13674 \lambda ^6+6224 \lambda ^4+11109 \lambda ^2+1260\right)
		\lambda ^2+105\right)}{15482880 \lambda ^3}, \\
	a_{-1/2} =& \frac{\sqrt{\pi }}{247726080 \lambda ^5} [ 8 \left(733586 \lambda ^{14}-2142496 \lambda ^{12}+2212060 \lambda ^{10}-905552 \lambda ^8 \right. \\
	& \left. +81835 \lambda ^6+16808 \lambda ^4+5607
	\lambda ^2+420\right) \lambda ^2+105 ]. \\
\end{aligned}
\ee
\begin{figure}[t]
	\centering
	\includegraphics[width=8cm]{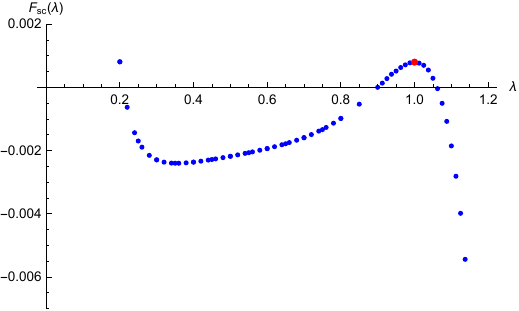}
	\caption{\rm Renormalized free energy with squashing parameter $\lambda$, the red points correspond to the free energy on the round sphere.}
	\label{Tbllambda}
\end{figure}
The calculation can theoretically be conducted using the formula listed above, but in practice, the double-summation of $n,r$ converges extremely slowly. So here we followed the technique of \cite{Anninos:2012ft}\footnote{The method has been nicely introduced in several places such as \cite{Anninos:2012ft, Bobev:2016sap, Bobev:2017asb}, thus we only present our final results here. }, using the Euler-MacLaurin formula to approximate the summation over thousands of points by integrals and function values, reducing the time-cost to an acceptable range. Using heat-kernel regularization, we numerically reproduced the round sphere free energy obtained in \cite{Klebanov:2011gs}; and the dependence of $\lambda$ is plotted in Fig.\ref{Tbllambda}. By comparing it with the bulk free energy shown in Fig.\ref{FscLambda}, we can identify similar behaviors as well as differences. The difference can mostly be attributed to the distinctions between strongly coupled holographic field theories and free ones we consider here, but also to field theory details such as correlation functions.

Using heat kernel regularization, we are able to evaluate not only the values of free energy but also derivatives with regard to the squashing $\lambda$. We studied the second and third-order derivatives at $\lambda = 1$, which provides a reference for our analytical results obtained in zeta function regularization. The procedure is nearly the same after taking the derivative of $\lambda$:
\be
\frac{d^n}{d\lambda^n} F_{\rm sc} = -\frac{1}{2}\int_0^\infty \frac{1}{t} \frac{\partial^n}{\partial \lambda^n}K(t) dt.
\ee
We also need to distill the divergent part of the integrand:
\be
\frac{\partial^n}{\partial\lambda^n}K(t) = \sum_{k=0}^{\frac{d+1}{2}} a_{d/2-k}^{(n)}t^{-d/2+k} + O(t).
\ee
By dividing the integral domain into $(0,1]$ and $[1,\infty)$, regulating the divergences, and evaluating the converging part numerically, we obtain the renormalized derivative, which is consistent with the results from zeta-function regularization which we obtain later:
\be
\begin{aligned}
	F_{\rm sc}^{(2)}(1) =& -0.289411 = 1.0009\left(-\frac{15\pi^2}{512}\right),\\
	F_{\rm sc}^{(3)}(1) =& -5.275 = 1.005\left(-\frac{62815 \pi ^2}{118272}\right).
\end{aligned}
\ee
For $F_{\rm sc}^{(2)}(1)$ and $F_{\rm sc}^{(3)}(1)$ at $\lambda = 1$, the divergent coefficients are as follows:
\be \small 
\begin{aligned}
	a_{7/2}^{(2)} =& \frac{\sqrt{\pi }}{64},\qquad a_{5/2}^{(2)} = -\frac{\sqrt{\pi }}{192},\qquad a_{3/2}^{(2)} = \frac{1453 \sqrt{\pi }}{30720}, \\
	a_{1/2}^{(2)} =& -\frac{152689 \sqrt{\pi }}{1290240},\qquad a_{-1/2}^{(2)} = \frac{2317181 \sqrt{\pi }}{41287680}. \\
	a_{7/2}^{(3)} =& \frac{\sqrt{\pi }}{64},\qquad a_{5/2}^{(3)} = \frac{3 \sqrt{\pi }}{64},\qquad a_{3/2}^{(3)} = \frac{16553 \sqrt{\pi }}{30720},\\
	a_{1/2}^{(3)} =& -\frac{2051699 \sqrt{\pi }}{1290240},\qquad a_{-1/2}^{(3)} = \frac{10868771 \sqrt{\pi }}{5898240}.\\
\end{aligned}
\ee

\subsubsection{Zeta-function regularization}
For the scalar spectrum (\ref{ScalarSpectrum}) we have
\be
F_{\rm sc}(\lambda) = \frac{1}{2} \log \det \left(-\nabla^2 + \frac{5}{24}R\right) = \frac{1}{2}\sum_{n=0}^\infty\sum_{r=0}^{\left[\frac{n}{2}\right]}m_{n,r}\log\tilde\lambda_{n,r}.
\ee
Thus, following \cite{Bueno:2020odt}, by formally commuting the infinite sum and derivative over $\lambda$, we obtain the following expressions for the $i$-th order derivative of $F_{\rm sc}$ at $\lambda = 1$:
\be
\begin{aligned}
	F^{(i)}_{\rm sc}(1) & =  \frac{1}{2}\sum_{n=0}^\infty\sum_{r=0}^{\left[\frac{n}{2}\right]}m_{n,r}\frac{d^i}{d\lambda^i}\log\tilde\lambda_{n,r}\Big|_{\lambda = 1} \\
	& = \frac{1}{2}\sum_{k=0}^\infty\sum_{r=0}^{k}m_{2k,r}\frac{d^i}{d\lambda^i}\log\tilde\lambda_{2k,r}\Big|_{\lambda = 1} + \frac{1}{2}\sum_{k=0}^\infty\sum_{r=0}^{k}m_{2k+1,r}\frac{d^i}{d\lambda^i}\log\tilde\lambda_{2k+1,r}\Big|_{\lambda = 1}. \\
\end{aligned}
\ee
For the $i = 0$ case, we met with a divergent series with logarithms, which needs to be regularized using the generalized zeta-function:
\be
\sum_{k=0}^\infty (k+ a)^n \log (k+a) = - \zeta_{a}'(-n),\quad \zeta_a(s) \equiv \sum_{m=0}^\infty \frac{1}{(a+m)^s},\quad a\in \mathbb{R}.
\label{GenerlizedZetaFunction}
\ee
The result we obtain, which is also shown in Table \ref{FpTbl}, is the same as in \cite{Klebanov:2011gs}:
\be
F_{\rm sc}(1) = \frac{60 \pi ^6 \log 2 + 82 \pi ^4 \zeta (3)-150 \pi ^2 \zeta (5)-945 \zeta (7)}{61440 \pi ^6} \approx 0.000797.
\ee
For the first-order derivative, which should vanish because of the vanishing of the conformal anomaly in odd-dimensional field theory, the expression is
\be
F_{\rm sc}'(1) = -\sum_{k=0}^\infty \frac{16}{21}\left(k + \frac{1}{2}\right) (k+1) \left(k +\frac{3}{2} \right)^2 ( k+2)\left(k+\frac{5}{2}\right).
\ee  
This expression seems to be non-zero using normal zeta-function regularization, contradicting the conformal symmetry. However, using a new parameter $k' = k + 3/2$, we get a non-trivial cancellation among generalized zeta functions:
\be
F_{\rm sc}'(1) =  -\frac{4}{21}\sum_{k'} (k'^2 - 5 k'^4 + 4 k'^6) =  -\frac{4}{21}(\zeta_{3/2}(-2) - 5\zeta_{3/2}(-4) + 4\zeta_{3/2}(-6)) = 0.
\ee
Since this kind of divergent series can't be regularized using a method that is both stable and linear, the shift of the summing parameter will bring a different answer. \cite{Monin:2016bwf} To guarantee we have the correct shift, one thing we can do is find a physical meaning of the parameter or use another summation method, such as comparing our results against those obtained by heat kernel regularization, integrated correlators, or the bulk results. Indeed, all these results justify that the universal value is obtained by omitting the term without the $\pi$ factors, which we will see below. Another justification is that although shifting variables $k' = k+a$ shifts the result, all terms with $\pi$ factors don't change.

For the second and third-order derivatives, we arrange the sum in the form where the first line is divergent and the second line is convergent:
\be \small 
\begin{aligned}
	F_{\rm sc}^{(2)} (1)& = \frac{1}{2} \sum_{k=0}^\infty  \left(\frac{544 k^6}{945}+\frac{272 k^5}{45}+\frac{23588 k^4}{945}+\frac{6928 k^3}{135}+\frac{17018 k^2}{315}+\frac{7003 k}{270}+\frac{6917}{1680} \right. \\
	&\qquad\qquad \left. -\frac{5 (64 k (2 k+7) (13 k (2 k+7)+155)+29619)}{48 (4 k+5)^2 (4 k+7)^2 (4 k+9)^2} \right) \\
	& = \frac{6499}{43200} + \frac{652-75 \pi ^2}{2560} \quad \Rightarrow \quad  -\frac{15 \pi ^2}{512}.
\end{aligned}
\ee
\begin{equation*} \small 
\begin{aligned}
	& F^{(3)}_{\rm sc}(1) \\
	&=  \frac{1}{2}\sum_{k=0}^\infty \left( -\frac{15296 k^6}{10395}-\frac{7648 k^5}{495}-\frac{234104 k^4}{3465}-\frac{233872
		k^3}{1485}-\frac{102317 k^2}{495}-\frac{143801 k}{990}-\frac{6209419}{166320} \right.\\
	&\left. -\frac{8 k (2 k+7) (8 k (2 k+7) (11960664 k (2
		k+7)+205534087)+9393298205)+142669238625}{18480 (4 k+5)^3 (4 k+7)^2 (4 k+9)^3}\right) \\
	&\qquad  =  -\frac{614077}{950400}-\frac{62815 \pi ^2}{118272} \quad \Rightarrow \quad -\frac{62815 \pi ^2}{118272}.\\
\end{aligned}\end{equation*} 
  The procedure outlined above can be directly applied to evaluate higher-order derivatives, which are less insightful from the holography point of view, thus we will not bother to put them here.
\begin{table}[!h]
	\renewcommand{\arraystretch}{1.7}
	\centering
	\begin{tabular}{cccc}
		\hline
		& $F$ & $F''$ & $F'''$ \\\hline
		Scalar  & $\frac{60 \pi ^6 \log 2 + 82 \pi ^4 \zeta (3)-150 \pi ^2 \zeta (5)-945 \zeta (7)}{61440 \pi ^6}$ & $-\frac{15 \pi ^2}{512}$ & $-\frac{62815 \pi ^2}{118272}$ \\
		Fermion & $\frac{300 \pi ^6 \log2 + 518 \pi ^4 \zeta (3)+1050 \pi ^2 \zeta (5)+945 \zeta (7)}{7680 \pi ^6}$ & $-\frac{45 \pi ^2}{64}$ & $-\frac{101335 \pi ^2}{7392}$ \\\hline
	\end{tabular}
	\caption{\rm $F^{(i)}_{\rm sc}(1)$ and $F^{(i)}_{\rm f}(1)$ for $i=0,2,3$.}
	\label{FpTbl}
\end{table}

\subsection{Free fermion}
In this section, we calculate the derivatives of free energy at $\lambda = 1$ for free fermions analytically, using zeta function regularization introduced before. The free energy is
\be
F_{\rm f}(\lambda) = - \log \det \left(-i\slashed{\nabla}\right) = -\sum_{n,r}m_{n,r}\log \tilde{\lambda}_{n,r}.
\ee
The spectrum of Dirac operator on squashed seven-sphere with SO(5)$\times$SO(3) isometry was calculated in \cite{Nilsson:1983ru}, and in \cite{Eastaugh:1985ew} by another method. The eigenstates of Dirac operator in $\mathbb{S}^7$ correspond to the SO(8) irreps $(n,0,0,1)$ and $(n,0,1,0)$, with negative and positive values respectively. \cite{Duff:1986hr} Consider the branching rules of the two SO(8) irreps under SO(5)$\times$SO(3):
\be \small 
\begin{aligned}
	(n,0,0,1) & \rightarrow \sum_{r=0}^{[n/2]} (n+1-2r,r;n+1-2r) + \sum_{r=0}^{[n/2]}(n+1-2r,r;n-1-2r)\\
	& + \sum_{r=0}^{[n/2]}(n-1-2r, r+1;n+1-2r) + \sum_{r=0}^{[n/2]} (n-1-2r,r;n-1-2r). &\\
	(n,0,1,0) & \rightarrow \sum_{r=0}^{[n/2]+1} (n-2r,r+1;n-2r) + \sum_{r=0}^{[(n-1)/2]} (n-2r,r;n-2r) \\
	& + \sum_{r=0}^{[n/2]} (n-2r,r;n-2r+2) + \sum_{r=0}^{[n/2-1]} (n-2r,r; n-2r-2). \\
\end{aligned}
\label{decomposition}
\ee 
The label $(p, q; r)$ correspond to irrep of SO(5) with Dynkin indices $(p, q)$ and irrep for Sp(1) with index $r$, whose dimensions are:
\be 
\begin{aligned}
	&m_{p,q} = \frac{1}{6}(p+1)(q+1)(p+q+2)(p+2q+3),\quad m_r = r + 1.  \\
\end{aligned}
\ee 
Thus the degeneracy of irrep $(p, q; r)$ is $m_{p,q,r} = m_{p,q}m_r$. In the branching rule, there're three classes of SO(5)$\times$SO(3) irreps, with specific relations between $r$ and $p$: (i)$(p,q;p)$, (ii)$(p,q;p+2)$, (iii)$(p,q;p-2)$. 

For case (i), the eigenvalues $\tilde\lambda$ are solutions of the following quatic equation:\footnote{Notice that the equation in \cite{Nilsson:1983ru} has a typo in the last term in the denominator.}
\be 
\begin{aligned}
	& (\tilde\lambda-\sigma_{1+})(\tilde\lambda-\sigma_{1-})(\tilde\lambda-\sigma_{2+})(\tilde\lambda-\sigma_{2-}) - \frac{\left(\lambda ^2-1\right) \left(5 \lambda ^2-1\right)}{\lambda^4}  p (p+2) = 0; \\
	&\qquad  \sigma_{1\pm} = -\frac{3}{2}\lambda \pm  \frac{1}{2\lambda}\sqrt{9\lambda^4 + 2\lambda^2(8 p q+8 p+8 q^2+24 q+9) + 4 p^2+8 p+9},\\
	&\qquad \sigma_{2\pm} = \frac{\lambda}{2} \pm \frac{1}{2\lambda}\sqrt{\lambda^4 +2\lambda^2(8 p q+8 p+8 q^2+24 q+17) + 4 p^2+8 p+1}.
\end{aligned}
\ee 
The equation becomes trivial when $\lambda^2 = 1, 1/5$. And we will name the four corresponding branches of eigenvalues $(\Sigma^{1+}, \Sigma^{2+},\Sigma^{1-},\Sigma^{2-})$, which reduce to $(\sigma^{1+}, \sigma^{2+},\sigma^{1-},\sigma^{2-})$ respectively if we take $\lambda^2 = 1, 1/5$.

For cases (ii) and (iii), the eigenvalues are denoted by $D^{2\pm}$ and $D^{3\pm}$ respectively:
\be
\begin{aligned}
	& D^{2\pm} = \frac{\lambda}{2} \pm \frac{1}{2\lambda}\sqrt{\lambda^4+2\lambda^2(8 p q+2 p+8 q^2+24 q+5)+(2p+5)^2}; \\
	& D^{3\pm} = \frac{\lambda}{2} \pm \frac{1}{2\lambda}\sqrt{\lambda^4+2\lambda^2(8 p q+14 p+8 q^2+24 q+17)+(2p-1)^2}. \\
\end{aligned}
\ee
To determine which eigenvalue to choose for the irreps in (\ref{decomposition}), we need to compare the above eigenvalues with \cite{Duff:1986hr}, which discussed the special case $\lambda^2 = 1/5$. It turns out that corresponding to the eight irreps of SO(5)$\times $SO(3) appearing in the branching rule following the order of \eqref{decomposition}, we need to specify the following eigenvalues:
\be
\Sigma^{2-},\ D^{3-},\ D^{2-},\ \Sigma^{1-};\qquad \Sigma^{1+},\ \Sigma^{2+},\ D^{2+},\ D^{3+}.
\ee
When $\lambda = 1$, the eigenvalues and degeneracies above are significantly simplified:
\be
\begin{aligned}
	D^{+} &= n + \frac{7}{2},\quad D^{-} = -n - \frac{7}{2},\quad n=0,1,2,...;\\
	m_n &= \frac{1}{90} (n+1) (n+2) (n+3) (n+4) (n+5) (n+6). \\
\end{aligned}
\ee
To calculate $F_{\rm f}(1)$, we need generalized zeta function as in (\ref{GenerlizedZetaFunction}) to regulate logarithmic terms, which is also shown in Table \ref{FpTbl}, and corresponds to the results of \cite{Klebanov:2011gs}:
\be
F_{\rm f}(1) = \frac{300 \pi ^6 \log2 + 518 \pi ^4 \zeta (3)+1050 \pi ^2 \zeta (5)+945 \zeta (7)}{7680 \pi ^6} \approx 0.0369.
\ee
The first, second, and third-order derivatives can be calculated using the ordinary zeta-function regularization, and we extract the terms with factors of $\pi$ as the physically relevant ones as argued before:
\begin{equation*}  \small 
\begin{aligned}
	F'_{\rm f}(1) = & \frac{128}{105}\sum_{k=0}^\infty (k+1)\left(k+\frac{3}{2}\right)(k+2)\left(k+\frac{5}{2}\right)(k+3)\\
	= &\frac{32}{105} \sum_{m=2}^\infty (4m^6 - 5m^4 + m^2) = \frac{32}{105}[4\zeta(-6)-5\zeta(-4)+\zeta(-2)] = 0.\\
	F''_{\rm f}(1) = & -\sum_{k=0}^\infty\left(\frac{2176 k^6}{945}+\frac{8704 k^5}{315}+\frac{125984 k^4}{945}+\frac{311552
		k^3}{945}+\frac{3100 k^2}{7}+\frac{295696 k}{945}+\frac{112481}{1260}\right.\\
	& \left.  -\frac{256}{105 (2 k+5)}+\frac{45}{8 (4 k+7)^2}+\frac{45}{8 (4 k+9)^2}+\frac{256}{105 (2
		k+3)}\right)\\
	= & -\frac{5608873}{264600} + \left( \frac{14269}{2520} - \frac{45 \pi ^2}{64}\right) \quad \Rightarrow \quad - \frac{45 \pi ^2}{64}.\\
	F'''_{\rm f}(1) =& - \sum_{k=0}^\infty\left(\frac{16 k (k+4) (8 k (k+4) (956 k (k+4)+9981)+255129)+5118283}{20790}\right.\\
	&\qquad\qquad \left. + \frac{\frac{600000}{2 k+5}-\frac{506675}{(4 k+7)^2}-\frac{506675}{(4
			k+9)^2}-\frac{248832}{(2 k+5)^3}-\frac{600000}{2 k+3}+\frac{248832}{(2 k+3)^3}}{4620}\right)\\
	=& - \frac{101335 \pi ^2}{7392} + \frac{54409409}{363825}\quad \Rightarrow \quad - \frac{101335 \pi ^2}{7392}.
\end{aligned}
\end{equation*}

\section{Comparison}
\label{SectionofComparison}
Up to now, we have studied the free energies of three theories: holographic CFT corresponding to Euclidean Einstein gravity, O($N$) vector model, and free fermion model. In this section, we will compare their free energies as functions of the squashing parameter $\lambda$. 

 The number of degrees of freedom in field theory is characterized by its central charge $C_T$ in the stress-tensor two-point functions (\ref{StressTensor2ptFunction}), thus to compare the free energy among different theories, the most convenient normalization is to divide them by the corresponding central charges. For free scalar and massless Dirac fermion, the central charges are given by, respectively: \cite{Bueno:2020odt,Osborn:1993cr,Erdmenger:1996yc,Buchel:2009sk}\footnote{In \cite{Bobev:2017asb}, the convention is different, where the factor of $V_{\mathbb{S}^{d-1}}$ is absent. }
\be
C_{T, {\rm sc}} = \frac{d}{d-1}\frac{1}{V_{\mathbb{S}^{d-1}}^2},\qquad C_{T,{\rm f}} = \frac{d}{2}2^{[d/2]}\frac{1}{V_{\mathbb{S}^{d-1}}^2},\quad V_{\mathbb{S}^{d-1}} \equiv \frac{2\pi^{d/2}}{\Gamma\left( d/2\right)} .
\ee
The two-point function of stress tensor can also be evaluated holographically from the bulk side, and the value of $C_T$ in Einstein gravity is given by \cite{Buchel:2009sk, Liu:1998bu}:
$$
C_{T, {\rm holo}} = \frac{\Gamma(d+2)}{8(d-1)\Gamma(d/2)\pi^{(d+2)/2}} \frac{\ell^{d-1} }{G_N} .
$$
Specifically for $d=7$, we have:
$$
C_{T, {\rm sc}} = \frac{525}{512\pi^6},\quad C_{T, {\rm f}} = \frac{1575}{64\pi^6}, \quad C_{T,{\rm holo}} = \frac{448}{\pi^5} \frac{\ell^6}{G_N}.
$$
\begin{figure}[t]
	\centering
	\includegraphics[width=8cm]{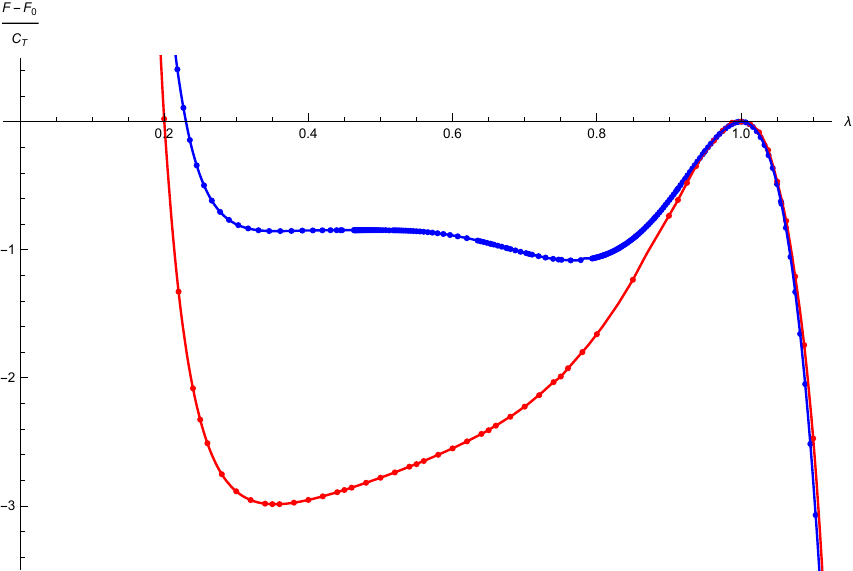}
	\caption{\rm Blue points: gravitational free energy. Red points: conformally coupled scalar action. The zero-point energy where $\lambda = 1$ is extracted, and $C_T$ is divided. The lines are interpolation lines to make the plots more clear.}
	\label{Comparison}
\end{figure}
The first comparison can be made between normalized $F_{\rm bulk}$ and $F_{\rm sc}$ for general $\lambda$ collected in Fig. \ref{Comparison}. Except for the expected correspondence at weak and strong deformation regions, they seem to behave differently, especially around $\lambda_* = 1/\sqrt{5} \approx 0.447$. There appear to be some local minima in both cases which are less understood. The gravitational free energy seems to have a platform around $\lambda_*$, which deserves further numerical study with higher precision.

In the weak-squashing regime, we collect the analytic results of second-order and third-order derivatives at $\lambda=1$ for scalar and fermion in Table \ref{CompareTbl}. In the same table, we also include the same quantities of $F_{\rm bulk}$ obtained by numerical fitting. The small discrepancy of the second-order derivative is probably due to numerical errors, which reflects our normalization by dividing $C_T$ in each theory is reasonable. The three-point function of the stress tensor depends on the details of the field theory, which is different in different theories. It was observed in \cite{Bobev:2017asb} that for some special field theories on U(1) squashed three spheres, the third-order derivatives of free energies of the squashing parameter $\epsilon$ appear to be vanishing, which is no longer true in our example. Take the holographic field theory as an example, we have:
\be 
    \frac{1}{C_T} \frac{d^3 F}{d\epsilon^3}\Big|_{\epsilon = 0} = \frac{1}{8C_T} F'''(1) - \frac{3}{8C_T} F''(1) \approx -670,
\ee 
which is different from zero. It will be interesting to explore this further.

\begin{table}[!h]
	\centering
	\renewcommand{\arraystretch}{1.5}
	\begin{tabular}{cccc}
		\hline & scalar & free fermion & holographic field \\\hline
		$ F''(1)/C_T$ &  $-\frac{\pi^8}{35}\approx -271.1$ & $-\frac{\pi^8}{35}\approx -271.1$ & $-271.8$ \\
		$ F'''(1)/C_T$ & $-\frac{12563 \pi ^8}{24255}\approx -4900$ & $-\frac{40534 \pi ^8}{72765}\approx -5300$ & $-6200$\\\hline
	\end{tabular}
	\caption{\rm Comparison of the second and third-order derivatives of free energy at $\lambda = 1$.}
	\label{CompareTbl}
\end{table}


\begin{figure}[t]
	\centering
	\includegraphics[width=5.15cm]{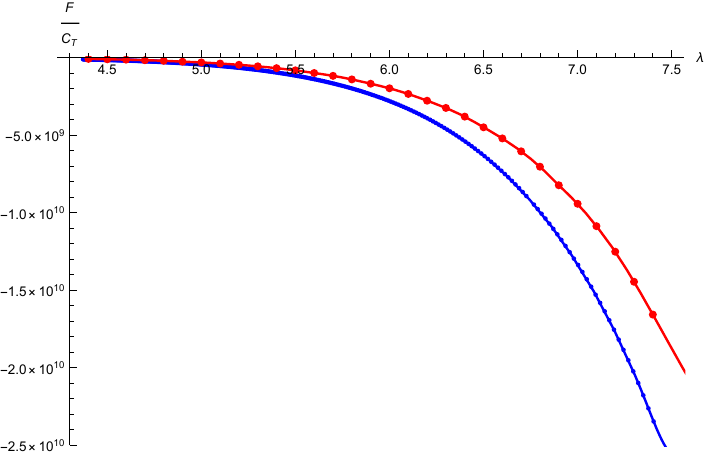}
	\qquad \qquad 
	\includegraphics[width=5.15cm]{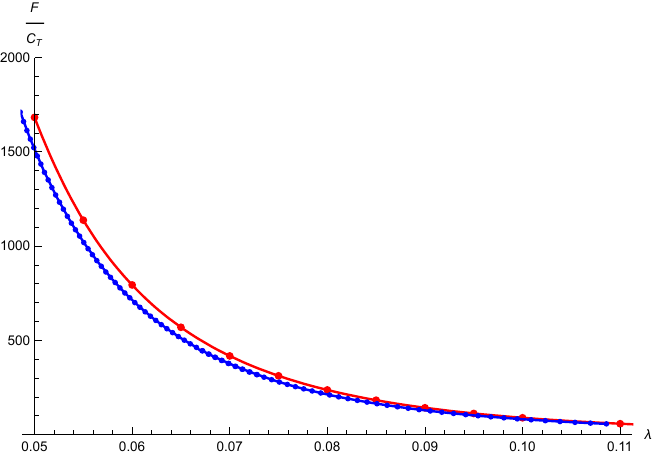}
	\hfil 
	\includegraphics[width=5.15cm]{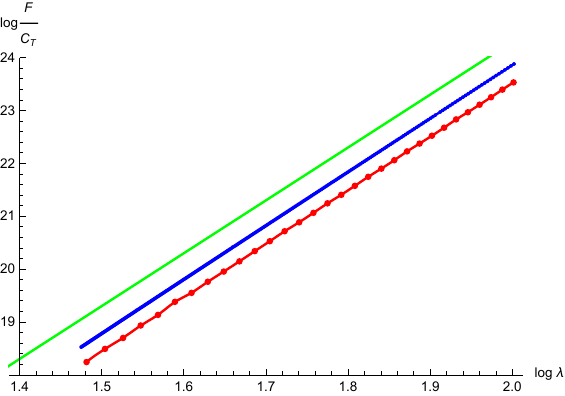}
	\qquad \qquad 
	\includegraphics[width=5.15cm]{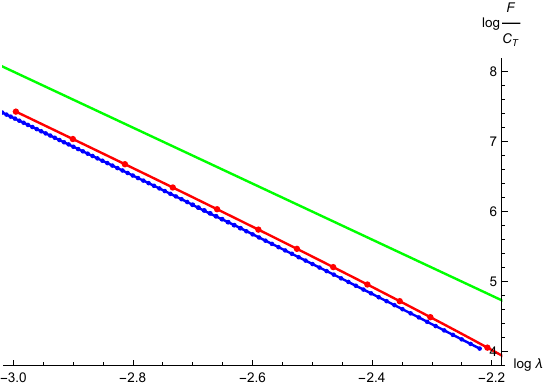}
	\caption{{\rm Free energy at large and small $\lambda$, respectively. The blue line is bulk free energy, and the red one is for free scalar. The green lines are reference lines, whose slopes are 10 and $-4$ respectively.}}
	\label{LargeSquashing}
\end{figure}
The comparison at strong-squashing regime, corresponding to big and small values of $\lambda$, has been reported in the literature. \cite{Aharony:2019vgs,Hartnoll:2005yc} Our results are collected in Fig. \ref{LargeSquashing}. Although the free energies are not exactly the same as in weak-squashing regime, they turn out to be proportional to each other and have the same scaling of $\lambda$ or $\lambda^{-1}$ respectively, which is the same as AlAdS$_4$ study in \cite{Hartnoll:2005yc}.

For small $\lambda$, the size of the SU(2) bundle can be omitted compared with the $\mathbb{S}^4$ base space, this means $b(r)$ will cap off before $a(r)$ as $r$ go from infinity, where the local geometry is given by $\mathbb{S}^4\times \mathbb{R}^4$, corresponding to a Bolt space. Following \cite{Aharony:2019vgs}, we expect the free energy of the conformal field theory to be identical to a theory living on $\mathbb{S}^4$, which is proportional to the normalized volume of $\mathbb{S}^4$ and scales like $F\sim \frac{1}{\lambda^4}$, and indeed they have been reproduced for free energies both in the bulk and boundary. 

At large $\lambda$, the $\mathbb{S}^4$ becomes tiny and the fiber effect becomes strong, and numerics suggests to us the power is $F\sim \lambda^{10}$. This power can be compared against \cite{Aharony:2019vgs}: there they consider an asymptotic boundary $\mathbb{S}^4\times\mathbb{S}^3$ there when $\mathbb{S}^3$ becomes large, the near horizon geometry becomes $\mathbb{R}^5\times \mathbb{S}^3$, corresponding to a CFT living on $\mathbb{S}^3$, thus the free energy will scale as $\lambda^3$, which is much different from our result. The difference in our case is that we don't have such a solution where $a(r)$ caps off before $b(r)$, thus the field theory argument above doesn't work anymore. Notice that $10 = 7 + 3$, which is the sum of the dimension of the full boundary and the bundle, it will be interesting to understand this scaling better.

We can also discuss the large squashing behavior for field theories living on the U(1) squashed $(2k+1)$-spheres. If we still label the size of the bundle as $\lambda$, then as $\lambda \rightarrow 0$, the free energy scales as an effective theory on $\mathbb{S}^{2k}$, thus $F\sim \lambda^{-2k}$; and as $\lambda \rightarrow \infty$, we expect a power equal to the sum of boundary dimension and the fibre, which is $\lambda^{2k+2}$. The special case $k = 1$ has been already considered in the literature. At the bulk side, it was done in \cite{Emparan:1999pm} where $O(\lambda^4)$ scaling at large $\lambda$ and $O(\lambda^{-2})$ at small $\lambda$ are obtained analytically. The $O(\lambda^{-2})$ scaling has also been reproduced for conformally coupled scalar and fermions. \cite{Hartnoll:2005yc, Bobev:2017asb, DeFrancia:2000xm}

\section{Discussion}

As the first result of this chapter, we construct a novel one-parameter family of AlAdS$_8$ solutions of Euclidean Einstein gravity preserving SO(5)$\times$SO(3) isometry with two closely related metrics. We study the renormalized gravitational free energy as a function of the squashing parameter and show its correspondence with free field theories living on a squashed seven sphere. With the solutions containing SU(2) fiber bundles, our work has extended the literature on AdS vacuum solutions. On the field theory side, we have studied the free energy for an arbitrary CFT living on a perturbed metric. Using zeta-function and heat-kernel regularization, we study the free energy of free conformal scalar and free fermion theories living on the squashed sphere. We compare free energies in the bulk and boundary both at small and large deformations.

As a second result of this chapter, we expressed the second-order derivative of the free energy with regard to the squashing parameter in terms of the central charge in general CFT. The relation is identical to the corresponding one in the bulk, which reflects the universal property of conformal symmetry and acts as a sanity check of our result. We checked this for general conformal field theories living on the U(1) squashed sphere or the SU(2) squashed seven spheres. From a renormalization group point of view, the squashing on the round sphere induces an RG flow generated by the stress tensor of the field theory, along which the $\widetilde F$ quantity related to the free energy must be decreasing, this is also confirmed in our analysis.

\subsection*{Future directions}

The story we uncovered is interesting but there are some open questions left by our analysis. Let us mention some of them.

More investigations on the field theory free energy will further support our results. It's interesting to evaluate the renormalized free energy for free fermions using heat-kernel regularization to get the strong deformation behavior, where a crucial step is to obtain the heat-kernel coefficients. One should also try to obtain the strong deformation behavior analytically, following \cite{DeFrancia:2000xm}.

Our second metric ansatz turns out to be more general than the first one and admits three squashing parameters. The discussions in the appendix \ref{solutionAnsatz1} only show the behaviors of the metric, without exploring further the free energy. It would be definitely interesting to investigate the extended phase space of the AlAdS$_8$ solution as well as its holographic partners. We conjecture that there should be an interface between the NUT and Bolt phase in the bulk configuration space parametrized by $(\lambda_1, \lambda_2, \lambda_3 ) \equiv \left(\frac{\cF_{30}}{\cF_{20}}, \frac{\cF_{40}}{\cF_{20}}, \frac{\cF_{50}}{\cF_{20}} \right) $, passing through the special point corresponding to $\lambda_* = \frac{1}{\sqrt{5}}$. On the field theory side, we need to generalize the spectrum of conformal Laplacian \cite{Nilsson:1983ru, Eastaugh:1985ew} when the symmetry group is further broken from SO(5)$\times$SO(3) to SO(5)$\times$U(1). At strong deformation regime, we expect the bulk and boundary free energy to scale consistently as a homogeneous polynomial of $(\lambda_1, \lambda_2, \lambda_3)$ whose order is dictated by our analysis which essentially corresponds to the special case $\lambda_1 = \lambda_2 = \lambda_3$.

We also need to understand better the $\lambda^{10}$ scaling of the free energies, which reflects important properties of the field theory. Notice that a similar scaling also emerges for U(1) squashed spheres, a numerical study would also be helpful for those cases. We are also looking forward to a formal justification of our steps to evaluate integrated two-point correlators on the sphere, which may also help us evaluate three-point correlators of stress tensors.

Based on our work, there are also many generalizations that are worthy of exploring.

As we mentioned in the beginning, as dictated by the Klebanov-Polyakov correspondence \cite{Klebanov:2002ja}, the O($N$) vector field and free fermion on the boundary are dual to higher spin gravity in the bulk. The higher spin modes in the latter contributes non-trivially to the free energy, and thus needs to be studied independently. There are two types of higher spin gravities, called A and B, correspond to complex scalar and free fermion fields on the boundary respectively.\footnote{See \cite{Giombi:2012ms} for a review and \cite{Giombi:2014iua,Giombi:2016pvg,Brust:2016xif} for results with AdS$_8$.} Following \cite{Giombi:2014iua, Giombi:2016pvg, Brust:2016xif} for a new solution similar to our ansatzes \eqref{Ansatz1} and \eqref{Ansatz2} will be desirable. This will provide new opportunities for quantitative checks of the higher-spin/vector model duality.

The authors of \cite{Ooguri:2016pdq} conjectured that any non-supersymmetric AdS vacua supported by fluxes must be unstable. An example with a 11D Euclidean AdS$_5 \times \mathbb{CP}^3$ backups the conjecture. \cite{Ooguri:2017njy} It would be interesting to study the stability under real-time evolution of our bulk space, following \cite{Bizon:2007zf} whose geometry is asymptotically flat.

One can further explore the integrated three-point functions of stress tensor on seven spheres, and compare them against the proposal in \cite{Bueno:2018yzo, Bueno:2020odt}. It was pointed out in \cite{Bobev:2017asb, Bueno:2018yzo} that free bosons and free fermions may play as lower and upper bounds for the free energies in general field theories. For U(1) bundled $(2k+1)$-spheres, this can be explicitly checked by numerically evaluating the renormalized Euclidean Einstein-Hilbert action as well as analytically on the field theory side using the zeta-function regularization with the known eigenvalues of Laplacian \cite{Bobev:2017asb} and Dirac operators \cite{Bar:1996wqp}.

Our metric ansatz is based on quaternionic line bundle, it might be interesting to also consider octoninic line bundle considered in \cite{Bakas:1998rt, Kanno:1999hr}. Generalizing our solutions here to include higher derivative corrections is also an interesting direction to explore. \cite{Bueno:2018uoy} A more ambitious plan is to follow the top-down approach of \cite{Bobev:2019bvq, Cordova:2018eba} to understand the AdS$_8$ vacua in string theory.

\cleardoublepage%

  \graphicspath{{\chaptersdir/conclusion/image/}}%
\chapter{This is conclusion}\label{ch:conclusion}

\ldots

\instructionsconclusions

\cleardoublepage

%


\appendix

\chapter{This is myappendix}\label{ch:myappendix}

\ldots

\instructionsappendices

\cleardoublepage


\chapter{More details on the superconformal E-theory}\label{ch:appendixetheory}

\section{Wrong analytical calculation of $\Delta_k(\lambda)$ for large $\lambda$}
\label{sec:expansionDeltak}

In this section we try to calculate analytically the first two terms in the large $\lambda$ expansion of $\Delta_k(\lambda)$. We warn the reader however that although the calculation appears to be quite straightforward, the result turns out to be wrong. 

\subsection*{Evaluation of the $\lambda^{-1}$ term in $\Delta_k(\lambda)$}\label{sec:term1}

The calculation is based on the $LDU$ decomposition of matrices and uses the following fact. If $A = L \mathbb{D} U$ where $L$ is a lower triangular matrix, $ \mathbb{D}$ is a diagonal matrix and  $U$ is an upper triangular matrix, then	\begin{equation}
\label{eq:LDUblock}
A_{(k)} = L_{(k)}  \mathbb{D}_{(k)} U_{(k)}\,,
\end{equation}
where $A_{(k)}$ is the upper left $k\times k$ block in the matrix $A$ and we have used the same notation for the other matrices.
Furthermore, if $L$ is lower uni-triangular, i.e. $L$ has ones on the diagonal, and $U$ is upper uni-triangular, then
\begin{equation}
\label{ratiodet}
\frac{ \det{A_{(k)}}}{\det{A_{(k-1)}}} = \mathbb{D}_{kk}\,,
\end{equation}
with the convention that $A_{(0)} = 1$. 

The $\lambda^{-1}$ term in the strong coupling expansion of $\Delta_k(\lambda)$ was calculated in \cite{Billo1} with a different method, we calculate it here again with formula \eqref{ratiodet}. By performing the Mellin transformation of \eqref{Xx}, one has \cite{Billo1}:
\begin{align}
\label{xs}
\mathsf{X} \underset{\lambda \to \infty}{\sim}
- \frac{\lambda}{2 \pi^2}  \, \mathsf{S} + O(\lambda)^0
\end{align}
where $\mathsf{S}$ is a tri-diagonal infinite matrix whose elements are
\begin{align}
\label{s}
\mathsf{S}_{kl}
=  \frac{1}{4} (-1)^{k+l} \sqrt{\frac{2l+1}{2k+1}} \,
\Big(\frac{\delta_{k-1,l}}{k\,(2k-1)} + \frac{\delta_{k,l}}{k\,(k+1)} +
\frac{\delta_{k+1,l}}{(k+1)\,(2k+3)}\Big) ~.
\end{align}
One can then show that
\begin{align}
\label{Dll}
\Dx \equiv  \frac{1}{1-\mathsf{X}}  =  \frac{2\pi^2}{\lambda} \mathsf{S}^{-1} + O(\lambda)^{-3/2}\,.
\end{align} 
Now let us define a lower uni-triangular matrix $L$ and a diagonal matrix $\mathbb{D}$
\begin{equation}\label{eq:LDdef}
L_{k l} =  
\begin{cases}
1 & \text{if} ~k =  l~\\[4mm]
-\dfrac{ \sqrt{2 k +1}}{\sqrt{ 2 l+1}}& \text{if} ~k= l+1\\[4mm] 
0& \text{else}
\end{cases},\qquad 	\mathbb{D}_{k l} =  
\begin{cases}
\dfrac{\lambda}{4 \pi^2}\dfrac{1}{(2 k+1) 2 k }& \text{if} ~k = l~.\\[4mm] 
0& \text{else}
\end{cases}
\end{equation}
One can check explicitly that
\begin{equation}
L^T \mathbb{D} L = \frac{\lambda}{2 \pi^2}\ \mathsf{S}
\end{equation}
which in turn leads to
\begin{align}
\label{eq1}
L^{-1} \mathbb{D}^{-1} L^{-T} = \frac{2 \pi^2}{\lambda}\ \mathsf{S}^{-1} = \mathsf{D} + O(\lambda)^{-3/2}\,.
\end{align}
Since $L^{-1}$ is again a lower triangular matrix, one can use formula \eqref{ratiodet} to get:
\begin{align}
1 + \Delta_k(\lambda) + O(\lambda)^{-3/2} =  \frac{\det \mathsf{D}_{(k)}}{\det \mathsf{D}_{(k-1)}} + O(\lambda)^{-3/2} = \mathbb{D}^{-1}_{kk}  + O(\lambda)^{-3/2}\,.
\end{align}
Substituting the value of $\mathbb{D}_{kk}^{-1}$, we get:
\begin{align}\label{eq:approx1}
1 + \Delta_k(\lambda) =  \dfrac{4 \pi^2}{\lambda}(2 k+1) 2 k +O(\lambda)^{-3/2}
\end{align}

This agrees with both the analytical evaluation in \cite{Billo1} and the numerical evaluation \eqref{NikolayConjecture} in the main text above.

\subsection*{Evaluation of the $\lambda^{-3/2}$ term in $\Delta_k(\lambda)$}\label{sec:term2}
To proceed to higher order in the $1/\lambda$ expansion we add one more term to the expansion of $\mathsf{X}$:
\begin{align}
\label{x2}
\mathsf{X}  = - \dfrac{\lambda}{2 \pi^2}\mathsf{S} + \frac{1}{3} \mathbf{1} +O(\lambda)^{-1/2}\,.
\end{align}
One then finds
\begin{equation}
\mathsf{D} = \frac{1}{\mathbf{1} - \mathsf{X}} = \left( \dfrac{\lambda}{2 \pi^2}\mathsf{S} + \frac{2}{3}\mathbf{1} \right)^{-1} + O(\lambda)^{-2}\,.
\end{equation}

We define new matrices $L$ and $\mathbb{D}$, different from the ones in \eqref{eq:LDdef}, 
\begin{equation}
L_{k l} =  
\begin{cases}
1 & \text{if} ~k =  l~\\[4mm]
-\dfrac{ E_k}{E_l}& \text{if} ~k= l+1\\[4mm] 
0& \text{else}
\end{cases},\qquad 	\mathbb{D}_{k l} =  
\begin{cases}
\dfrac{E_{k-1}}{E_k}  \dfrac{\lambda}{2 \pi^2}\dfrac{1}{4 k \sqrt{4 k^2-1} }& \text{if} ~k = l~,\\[4mm] 
0& \text{else}
\end{cases}
\end{equation}
where $E_k$ is expressed in terms of the modified Bessel function as
\begin{equation}
E_k = \sqrt{2 k+1} \frac{ I_{2 k+1} \left( \sqrt{\frac{3\lambda}{2 \pi^2}} \right)}{  I_{1} \left( \sqrt{\frac{3\lambda}{2 \pi^2}} \right)}\,.
\end{equation} 

To proceed further we recall some properties of the modified Bessel function. A somewhat nonstandard recursion relation satisfied by the modified Bessel functions reads:
\begin{align}
\label{eq:recursionBessel}
\nu I_{2 \nu +3}(z) - \left( 1 + \nu (\nu +1) \dfrac{8}{z^2} \right) (2 \nu +1) I_{2 \nu+1}(z) 
+ (\nu +1) I_{2 \nu -1}(z)=0
\end{align}
This can be proven by using the standard recursion relation 
\begin{equation}\label{eq:standardrecursion}
\dfrac{2 \nu}{z} I_{\nu} (z) = I_{\nu-1}(z) -  I_{\nu+1}(z)\,.
\end{equation}
Using \eqref{eq:standardrecursion} on both terms in the right hand side of \eqref{eq:standardrecursion} gives
\begin{equation}
\dfrac{2 \nu}{z^2 }I_{\nu} (z) = \dfrac{1}{2 (\nu-1)}I_{\nu-2}(z) -\dfrac{\nu}{\nu^2-1}I_{\nu}(z) +  \dfrac{1}{2 (\nu+1)}I_{\nu+2}(z) \,.
\end{equation}
This relation is equivalent to \eqref{eq:recursionBessel}, after renaming the order of $I_{\nu}(z)$.

Using the recursion relation of modified Bessel functions \eqref{eq:recursionBessel}, one can verify directly that
\begin{equation}
L^T \mathbb{D} L =   \dfrac{\lambda}{2 \pi^2}\mathsf{S} + \frac{2}{3}\mathbf{1}\,,
\end{equation}
and hence
\begin{align}
L^{-1} \mathbb{D}^{-1} L^{-T} =  \left( \dfrac{\lambda}{2 \pi^2}\mathsf{S} + \frac{2}{3}\mathbf{1} \right)^{-1} = \mathsf{D} + O(\lambda)^{-2}\,.
\end{align}
Applying once again formula \eqref{ratiodet} we find:
\begin{align}\label{eq:0108b}
1 + \Delta_k(\lambda) + O(\lambda)^{-2} =  \frac{\det \mathsf{D}_{(k)}}{\det \mathsf{D}_{(k-1)}} + O(\lambda)^{-2} = \mathbb{D}^{-1}_{kk}  + O(\lambda)^{-2}\,.
\end{align}

Plugging in the value of $\mathbb{D}_{kk}$ and expanding the modified Bessel functions for large $\lambda$, we obtain
\begin{equation}\label{eq:approx2}
1 + \Delta_k(\lambda) = \dfrac{8\pi^2 k (2 k+1) }{\lambda} 
- \frac{32 \sqrt{\frac{2}{3}} \pi ^3 k^2 (2 k+1)}{\lambda^{3/2}} +O(\lambda)^{-2}\,.
\end{equation}
Although the calculations above appear to be quite straightforward, it turns out that the coefficient $C_{3/2}$ of the $\lambda^{-3/2}$ term in the large $\lambda$ expansion of $\Delta_k(\lambda)$  calculated analytically here is different from the one obtained numerically, see \eqref{NikolayConjecture}:
\begin{equation}
\begin{aligned}
C_{3/2}^{\rm ana} &= - 32 \sqrt{\frac{2}{3}} \pi ^3 k^2 (2 k+1)\approx 810.1 k^2(2k+1),\\
C_{3/2}^{\rm num} &= - 128\pi^2\log 2 k^2(2k+1) \approx 875.7k^2(2k+1)\,.\\
\end{aligned}
\end{equation}
We believe that the reason for this discrepancy lies in the fact that the analytic calculation above is subtle and leads to a wrong result due to the fact that we have treated the infinite dimensional matrix $\mathsf{X} $ as a finite dimensional one. For example, first taking a large $\lambda$ expansion and then inverting an infinite matrix could potentially yield a different result than first inverting the matrix and then expanding at large $\lambda$. Also, we are calculating the large $\lambda$ expansion of the matrix $\mathsf{X}$ for each component separately. In doing this, perhaps one needs to be more careful to ensure that this expansion is uniform over all components of the matrix. While we have no rigorous proof that the analytic calculation above is wrong, due to the subtleties discussed here and the fact that the numerical results are very accurate and well-behaved we have chosen to trust the result of our numerical analysis and in particular the large $\lambda$ expansion in \eqref{NikolayConjecture}.

\section{Quadrature rules}\label{app:discretisation}

A quadrature rule for approximation to an integral is a set of points $x_i$ and weights $w_i$ such that
\begin{equation}
I \equiv \int_a^b dx f(x) \approx \sum_{i=1}^n w_i f(x_i) \equiv I_n\,.
\end{equation}
The quadrature rules we describe in this section converge exponentially fast. Namely, if $f(x)$ can be extended to an analytical function $f(z)$ in a region around the interval $\left[a,b\right]$ then 
\begin{equation}
| I - I_n| \le e^{- c n}\,,
\end{equation}
with $c$ a positive constant. Typically, the larger the region of analyticity, the larger $c$. The reason we want fast quadrature rules is that $n$ should not be too large because we solve linear algebra problems
with matrices of size $n \times n$, see equation \eqref{eq:eq12}. The memory requirement will then be of order $O(n)^2$, and the timing of order $O(n)^3$, so this scales quite badly with $n$.

\subsection*{Gauss-Laguerre}
This is a quadrature scheme for the interval $[0, +\infty[$. It is the Gaussian quadrature scheme based on Laguerre polynomials. The weights $w_i$ and points $x_i$ with $i = 1 , \ldots, n$ are such that all polynomials of degree not greater than $ 2 n -1$ are integrated exactly. More specifically, if
\begin{equation}
I \equiv \int_0^{+\infty}\! dx\ e^{-x} f(x),\qquad I_n \equiv \sum_{i=1}^n w_i f(x_i)\,,
\end{equation}
then $I = I_n$  for all polynomials $f(x)$ of degree not greater than $ 2 n -1$. Initially, this seemed to be a natural quadrature scheme to use in our setup because the function $W(t)$ in \eqref{defW} decays exponentially fast for $t \to +\infty$. Also, although the weights $w_i$ and points $t_i$  are not directly available in {\tt Mathematica}, they can be calculated efficiently with the Golub-Welsch algorithm. This quadrature scheme works quite well for $\lambda \lesssim 100$, however for larger values of $\lambda$ we need to use larger values of $n$ ($n \gtrsim 100$). For large values of $n$, it turns out that many weights are very small ($\sim 10^{-100}$), so they are in a sense wasted. This is also documented in the literature, see Sections 5 and 6 in \cite{Tref}. Therefore, it is better to use a different quadrature scheme.

\subsection*{Quadrature rules on the interval $[-1 ,1]$}

Some well-known quadrature rules are: 
\begin{description}
	\item[Gauss-Legendre] Here $x_i$ are given by the zeros of Legendre polynomials. These are actually points in the interval $[0,1]$, but a linear change of variable converts these to the interval  $[-1 ,1]$.
	\item[Clenshaw-Curtis] Here $x_k = \cos\left(\frac{k \pi}{n}\right)$ with $k=0,1, \ldots, n$.
	\item[Fej\'er type 1] Here $x_k = \cos\left(\frac{(k - \frac{1}{2}) \pi}{n}\right)$ with $k=1,2, \ldots, n$. 
\end{description}

Gauss-Legendre is perhaps theoretically nicer than 	Clenshaw-Curtis, but the advantage of Clenshaw-Curtis is that its weights and points are much easier to compute. Also, {\tt Mathematica} has a function to compute the weights and points of Clenshaw-Curtis. However, the end points $x_0=1$ and $x_n = -1$ are part of the quadrature points. In the integral equation we want to solve, we prefer not to use the left end point. Therefore, we settled on Fej\'er type 1.
\subsection*{Fej\'er type 1}
The quadrature points are:
\begin{equation}\label{eq:Fejertheta}
x_k = \cos \theta_k,\quad\text{with}\quad 	\theta_k = (2 k -1) \frac{\pi}{2 n}, \quad k = 1 , 2, \ldots, n\,,
\end{equation}
and the weights $w_k$ can for example be found in \cite{Waldvogel}\footnote{If the reader wants to check formula \eqref{eq:Fejerweights}, we found Fej\'er's own calculation very readable \cite{Fejer}.}
\begin{equation}\label{eq:Fejerweights}
w_k = \frac{2}{n} \left[ 1 - 2 \sum_{r =1}^{ \frac{n-1}{2} } \frac{\cos(2 r \theta_k)}{4 r^2 -1} \right]
\end{equation}

The points $x_k$ and weights $w_k$ are not directly available in {\tt Mathematica}. However, there is code available at Wolfram Function Repository  which can be used for their calculation \cite{WolframResource}.\footnote{The function is called \texttt{FejerQuadratureWeights} and is based on the Fast Fourier Transform. We checked for many cases that \texttt{FejerQuadratureWeights} indeed produces $x_i$ and $w_i$ that are numerically the same as the ones given by \eqref{eq:Fejertheta} and \eqref{eq:Fejerweights}.}
\subsection*{Error analysis}
To integrate numerically over the interval $[0,+\infty[$ we truncate the interval to  $[0,L]$ and then use  Fej\'er type 1 on the truncated interval.
This amounts to the approximation
\begin{equation}
\int_0^{+\infty}\!\!\! dt\ \ f(t) \approx \int_0^L \!\!\! dt\ \ f(t) \approx \sum_{i=1}^n w_i f(x_i)\,.
\end{equation}
This procedure leads to two sources of error: firstly, there is the truncation error which goes to zero if $L \to +\infty$; secondly, there is the discretization error which goes to zero if $n \to +\infty$.
One possibility is to keep $L$ and $n$ independent, and test for accuracy by increasing both $L$ and $n$ separately. Another possibility is to relate $L$ and $n$ in such a way that the truncation error is roughly equal to the discretization error. We observed that the scaling $L \sim n^{2/3}$ works well in practice. 

It is important to note that if $n \to \infty$ then $L \to \infty$, so the truncation error goes to zero. Also if $n \to \infty$, the discretization size $h = \frac{L}{n} \sim n^{-1/3} \to 0$, so the discretization error will go to zero as well.

\section{Degenerate kernels}
\label{sec:degkernels}

In this section we collect some formulae for integral operators with degenerate kernel. These formulae are well-known but included here for convenience.
The integral operator
\begin{equation}\label{eq:deg1}
f(x) \mapsto f(x) + \int_{\alpha}^{\beta}\!\!\!\! dy\ K(x,y) f(y) \,,
\end{equation}
has degenerate kernel (also known as kernel of finite rank or separable kernel) if $K(x,y)$ can be expressed as the finite sum\footnote{We do not use the summation convention for repeated indices in this section.}
\begin{equation}\label{eq:deg2}
K(x,y) = \sum_{i,j=1}^n a_i(x)\ C_{ij}\ b_j(y)\,.
\end{equation}
Integral equations with degenerate kernel can be solved in closed form as follows. Suppose the integral equation is
\begin{equation}\label{eq:deg3}
f(x) + \int_{\alpha}^{\beta}\!\!\!\! dy\ K(x,y) f(y) = g(x) \,,
\end{equation}
with $K(x,y)$ as in \eqref{eq:deg2}. Define $f_i =  \int_{\alpha}^{\beta}\!\! dy\ b_i(y) f(y)$, then equation \eqref{eq:deg3} is
\begin{equation}\label{eq:deg4}
f(x) + \sum_{i,j=1}^n a_i(x) \ C_{ij} f_j = g(x) \,.
\end{equation}
Multiplying with $b_k(x)$ and integrating over $x$ gives
\begin{equation}\label{eq:deg5}
f_k + \sum_{i,j=1}^n A_{ki} C_{ij} f_j =g_k \,,
\end{equation}
where
\begin{equation}\label{eq:deg6}
A_{ki} = \int_{\alpha}^{\beta}\!\!\!\! dx\ b_k(x) a_i(x) \,,
\end{equation}
and $g_k =  \int_{\alpha}^{\beta}\!\! dx\ b_k(x) g(x)$. 
The system of $n$ linear equations \eqref{eq:deg5} can be solved exactly for $f_k$. Inserting this solution in \eqref{eq:deg4} gives\footnote{$M_{ij}^{-1}$ is the $ij$ component of the matrix $M^{-1}$.}
\begin{equation}\label{eq:deg7}
f(x) = g(x) - \sum_{i,j,l=1}^n a_i(x) C_{ij} (1 + AC)_{jl}^{-1} g_l \,.
\end{equation}
The solution of \eqref{eq:deg3} is thus
\begin{equation}\label{eq:deg8}
f(x) = g(x) -  \int_{\alpha}^{\beta}\!\!\!\! dy\ L(x,y) g(y) \,,
\end{equation}
with 
\begin{equation}\label{eq:deg9}
L(x,y) = \sum_{i,j=1}^n a_i(x) [ C ( 1 + AC)^{-1}]_{ij} b_j(y) \,.
\end{equation}
In the literature $L(x,y)$ is called the resolvent.

The Fredholm determinant of an integral operator with degenerate kernel can also be calculated analytically. Here is a derivation of the formula.
We discretize the integral with discretisation points $x_{\mu}$ and weights $w_{\mu}$ with $\mu, \nu = 1,2, \ldots, m$. Then the Fredholm determinant is equal to
\begin{equation}
\det(1 + K) = \lim_{m \to \infty} \det_{\mu,\nu=1}^m ( \delta_{ \mu \nu} + K(x_{\mu} , x_{\nu}) w_{\nu})\,.
\end{equation}
Write
\begin{equation}
K(x_{\mu} , x_{\nu}) w_{\nu} = \sum_{i,j=1}^n a_i(x_{\mu})\ C_{ij}\ b_j( x_{\nu}) w_{\nu} = ( M C N )_{\mu\nu}\,,
\end{equation}
with $M_{\mu i} = a_i(x_{\mu})$ and $N_{j \nu} =  b_j( x_{\nu}) w_{\nu}$. Then
\begin{equation}
\det(\mathbf{1}_{m \times m } + MCN) = \det(\mathbf{1}_{n \times n } + CNM)\,.
\end{equation}
Since
\begin{equation}
( CNM)_{ij} = \sum_{k=1}^n \sum_{\nu=1}^m C_{ik} b_k(x_{\nu}) w_{\nu} a_j(x_{\nu})\,,
\end{equation}
we find that in the limit $m \to \infty$
\begin{equation}
( CNM)_{ij} = \sum_{k=1}^n C_{ik} A_{kj}\,,
\end{equation}
with $A_{kj}$ defined in \eqref{eq:deg6}.
Alltogether, one has
\begin{equation}\label{eq:deg10}
\det(\mathbf{1} + K) = \det(\mathbf{1}_{n \times n } + CA) \,.
\end{equation}
This is a closed form formula for the Fredholm determinant because the determinant on the right hand side is of a finite $n \times n$ matrix.

\section{Numerical data}\label{app:tables}

Here we provide two tables with some of our numerical data for $\Delta_k(\lambda)$ for $k=1,2,3$ as well as for the free energy $\mathcal{F}$.
\begin{table}[H]
	\small 
	$$
	\begin{array}{|c||c|c|c|}
	\hline
	\lambda  & 1+\Delta_1(\lambda) &1+\Delta_2(\lambda) &1+\Delta_3(\lambda) \\
	\hline 
	\hline
	e^3 & 0.967087942591236   (00\! \pm \! 11 ) & 0.998182 &
	0.999886471522795   (00\! \pm \! 11 ) \\
	e^4 & 0.84413354083007   (60\! \pm \! 19 ) & 0.976408560611106 
	(00\! \pm \! 22 ) & 0.996135433920974   (00\! \pm \! 11 ) \\
	e^5 & 0.591355036026568   (60\! \pm \! 22 ) & 0.867655885531129 
	(00\! \pm \! 11 ) & 0.95770388339493   (20\! \pm \! 22 ) \\
	e^6 & 0.3263339352130   (10\! \pm \! 12 ) & 0.62836903552457  
	(64\! \pm \! 11 ) & 0.80851244411776   (50\! \pm \! 30 ) \\
	e^7 & 0.1524442716515   (11\! \pm \! 10 ) & 0.36148255438002  
	(6\! \pm \! 5 ) & 0.547853721918213   (5\! \pm \! 8 ) \\
	e^8 & 0.064525566747991   (9\! \pm \! 9 ) & 0.1749433084896  
	(77\! \pm \! 15 ) & 0.2997641417256   (9\! \pm \! 5 ) \\
	e^9 & 0.02579506509172   (98\! \pm \! 16 ) & 0.07589996453754  
	(0\! \pm \! 7 ) & 0.1407940457851   (2\! \pm \! 4 ) \\
	e^{10} & 0.00997266161958   (43\! \pm \! 27 ) & 0.0308295464823 
	(8\! \pm \! 6 ) & 0.0600521108447   (98\! \pm \! 30 ) \\
	e^{11} & 0.003779867460756   (4\! \pm \! 4 ) & 0.0120383586528 
	(52\! \pm \! 14 ) & 0.0241553127444   (05\! \pm \! 15 ) \\
	e^{12} & 0.001415785217   (9\! \pm \! 4 ) & 0.00459090014  
	(08\! \pm \! 27 ) & 0.00937868572   (6\! \pm \! 8 ) \\
	e^{13} & 0.0005265339   (42\! \pm \! 19 ) & 0.001726033   (07\!
	\pm \! 12 ) & 0.003564618   (8\! \pm \! 4 ) \\
	e^{14} & 0.000194980   (46\! \pm \! 21 ) & 0.00064338   (77\!
	\pm \! 14 ) & 0.00133750   (4\! \pm \! 4 ) \\
	\hline
	\end{array}
	$$
	\caption{$\Delta_k(\lambda)$ with $k=1,2,3$ for some values of $\lambda$. These values are calculated with the Nystr\"om method explained in Section \ref{sec:Deltanum}. We have used different values of $L$ and $m$ to estimate the accuracy of the numerical values.}
\end{table}
\begin{table}[H]
	$$
	\begin{array}{|l|l||l|l|}
	\hline
	\lambda  & {\cal F} &\lambda  & {\cal F}\\
	\hline 
	\hline
	1 & 9.3017 \times 10^{-6} & 100 & 0.227879 \\
	2 & 0.0000663066 & 200 & 0.50639 \\
	3 & 0.000200935 & 300 & 0.759961 \\
	4 & 0.000430494 & 400 & 0.991783 \\
	5 & 0.000764363 & 500 & 1.20637 \\
	6 & 0.00120686 & 600 & 1.40708 \\
	7 & 0.00175905 & 700 & 1.59638 \\
	8 & 0.00241988 & 800 & 1.77606 \\
	9 & 0.00318697 & 900 & 1.94753 \\
	\hline
	10 & 0.00405711 & 1000 & 2.11184 \\
	20 & 0.0175565 & 2000 & 3.49445 \\
	30 & 0.0373082 & 3000 & 4.60102 \\
	40 & 0.060823 & 4000 & 5.55364 \\
	50 & 0.0866247 & 5000 & 6.40401 \\
	60 & 0.11382 & 6000 & 7.17993 \\
	70 & 0.141848 & 7000 & 7.89843 \\
	80 & 0.170347 & 8000 & 8.57086 \\
	90 & 0.199078 & 9000 & 9.20523 \\
	\hline
	\end{array}
	$$
	\caption{The free energy ${\cal F}$ for some values of $\lambda$. These values are calculated with the Bornemann method explained in Section \ref{sec:Fnum}. We have used appropriate settings of $L$ and $m$ to ensure that all printed digits are correct.}
\end{table}

\cleardoublepage


\chapter{More details on the AlAdS$_8$ background}\label{ch:appendixads8}

\section{Partial results with the second ansatz}
\label{solutionAnsatz1}

In this section, we collect our results for the second squashed sphere metric (\ref{Ansatz2}). We solved the Einstein equations both perturbatively and numerically, and evaluate the initial values that can be integrated up to the boundary. The results indicate a richer family of solutions worth exploring further.

\subsection{Einstein equations of motion}
\label{EinsteinEOM}

Using the tetrad method introduced in the main text (\ref{TetradFormalism}), the Einstein equations for the metric (\ref{SquashedSphereMetric1}) are as follows:
\be 
\begin{aligned}
	&\frac{ f_1^2 f_3^2}{2 f_2^4}+\frac{ f_1^2 f_4^2}{2 f_2^4}+\frac{ f_1^2 f_5^2}{2 f_2^4}-\frac{6
		f_1^2}{f_2^2}+\frac{f_1^2 f_5^2}{4  f_3^2 f_4^2}+\frac{f_1^2 f_4^2}{4  f_3^2 f_5^2}		+\frac{f_1^2 f_3^2}{4 
		f_4^2 f_5^2}-\frac{f_1^2}{2  f_3^2}-\frac{f_1^2}{2  f_4^2} &\\ 
	&	-\frac{f_1^2}{2  f_5^2}-21 f_1^2+\frac{4 f_2' f_3'}{f_2
		f_3}+\frac{4 f_2' f_4'}{f_2 f_4}+\frac{4 f_2' f_5'}{f_2 f_5}+\frac{6 f_2'^2}{f_2^2}+\frac{f_3' f_4'}{f_3
		f_4}+\frac{f_3' f_5'}{f_3 f_5}+\frac{f_4' f_5'}{f_4 f_5} = 0, &\\
	&-\frac{3 f_1' f_2'}{f_1 f_2}-\frac{f_1' f_3'}{f_1 f_3}-\frac{f_1' f_4'}{f_1 f_4}-\frac{f_1' f_5'}{f_1
		f_5}-\frac{3 f_1^2}{f_2^2}+\frac{f_1^2 f_5^2}{4  f_3^2 f_4^2}+\frac{f_1^2 f_4^2}{4  f_3^2 f_5^2}
	+\frac{f_1^2 f_3^2}{4
		f_4^2 f_5^2}-\frac{f_1^2}{2  f_3^2}-\frac{f_1^2}{2  f_4^2}-\frac{f_1^2}{2  f_5^2}&\\
	& -21 f_1^2	&\\
	&	+\frac{3
		f_2''}{f_2}+\frac{3 f_2' f_3'}{f_2 f_3}+\frac{3 f_2' f_4'}{f_2 f_4}+\frac{3 f_2' f_5'}{f_2 f_5}+\frac{3
		f_2'^2}{f_2^2}
	+\frac{f_3''}{f_3}+\frac{f_3' f_4'}{f_3 f_4}+\frac{f_3' f_5'}{f_3 f_5} +\frac{f_4''}{f_4}+\frac{f_4'
		f_5'}{f_4 f_5}+\frac{f_5''}{f_5} = 0, &\\
	&	-\frac{4 f_1' f_2'}{f_1 f_2}-\frac{f_1' f_4'}{f_1 f_4}-\frac{f_1' f_5'}{f_1 f_5}+\frac{3  f_1^2 f_3^2}{2
		f_2^4}+\frac{ f_1^2 f_4^2}{2 f_2^4}+\frac{ f_1^2 f_5^2}{2 f_2^4}-\frac{6 f_1^2}{f_2^2}-\frac{f_1^2 f_5^2}{4  f_3^2
		f_4^2}
	-\frac{f_1^2 f_4^2}{4  f_3^2 f_5^2}  +\frac{3 f_1^2 f_3^2}{4  f_4^2 f_5^2} &\\
	&	+\frac{f_1^2}{2  f_3^2}-\frac{f_1^2}{2
		f_4^2}-\frac{f_1^2}{2  f_5^2}-21 f_1^2		+\frac{4 f_2''}{f_2}+\frac{4 f_2' f_4'}{f_2 f_4}+\frac{4 f_2' f_5'}{f_2
		f_5}+\frac{6 f_2'^2}{f_2^2}+\frac{f_4''}{f_4}+\frac{f_4' f_5'}{f_4 f_5} +\frac{f_5''}{f_5} = 0, &\\
	&	-\frac{4 f_1' f_2'}{f_1 f_2}-\frac{f_1' f_3'}{f_1 f_3}-\frac{f_1' f_5'}{f_1 f_5}+\frac{ f_1^2 f_3^2}{2
		f_2^4}+\frac{3  f_1^2 f_4^2}{2 f_2^4}+\frac{ f_1^2 f_5^2}{2 f_2^4}-\frac{6 f_1^2}{f_2^2}-\frac{f_1^2 f_5^2}{4  f_3^2
		f_4^2}
	+\frac{3 f_1^2 f_4^2}{4  f_3^2 f_5^2} 	-\frac{f_1^2 f_3^2}{4  f_4^2 f_5^2} &\\
	&-\frac{f_1^2}{2  f_3^2}+\frac{f_1^2}{2
		f_4^2}-\frac{f_1^2}{2  f_5^2}-21 f_1^2+\frac{4 f_2''}{f_2}+\frac{4 f_2' f_3'}{f_2 f_3}+\frac{4 f_2' f_5'}{f_2
		f_5}+\frac{6 f_2'^2}{f_2^2}+\frac{f_3''}{f_3}+\frac{f_3' f_5'}{f_3 f_5} +\frac{f_5''}{f_5}	= 0, &\\
	&	-\frac{4 f_1' f_2'}{f_1 f_2}-\frac{f_1' f_3'}{f_1 f_3}-\frac{f_1' f_4'}{f_1 f_4}+\frac{ f_1^2 f_3^2}{2
		f_2^4}+\frac{ f_1^2 f_4^2}{2 f_2^4}+\frac{3  f_1^2 f_5^2}{2 f_2^4}-\frac{6 f_1^2}{f_2^2}+\frac{3 f_1^2 f_5^2}{4 
		f_3^2 f_4^2}	-\frac{f_1^2 f_4^2}{4  f_3^2 f_5^2}	-\frac{f_1^2 f_3^2}{4  f_4^2 f_5^2}&\\
	&-\frac{f_1^2}{2 
		f_3^2}-\frac{f_1^2}{2  f_4^2}+\frac{f_1^2}{2  f_5^2}-21 f_1^2+\frac{4 f_2''}{f_2}
	+\frac{4 f_2' f_3'}{f_2 f_3}+\frac{4
		f_2' f_4'}{f_2 f_4}+\frac{6 f_2'^2}{f_2^2}+\frac{f_3''}{f_3}+\frac{f_3' f_4'}{f_3 f_4} +\frac{f_4''}{f_4} = 0, &\\
\end{aligned}
\label{EOMAnsatz1}
\ee
where we have taken the cosmological constant explicitly to be $-21$. Only four of the equations are independent because of Bianchi Identity. If one considers the special case where: 
\be
f_1(r) \rightarrow 1,\quad f_2(r)\rightarrow a(r),\quad f_3(r) = f_4(r) = f_5(r) \rightarrow b(r),
\ee
then the last three equations become identical, and the remaining three equations become identical to the three equations for our first metric ansatz in (\ref{EinsteinEquations3}). Thus the solutions discussed in the main text belong to a subset of solutions here, which we will solve both analytically at the near horizon and near boundary regions, and numerically for general squashing parameters $\lambda_1, \lambda_2, \lambda_3$.

\subsection{Large radius expansion}
We perform the following Fefferman-Graham expansion:
\be
f_1(r) = 1,\quad f_i(r) = \sum\limits_{j = 0}^{+\infty} \cF_{ij}e^{-(j-1)r},\quad i = 2,3,4,5
\ee
We solve the expansion perturbatively up to $O(\exp(-9r))$, it turns out the solutions are determined by totally 7 free parameters which can be chosen as $\{\cF_{i0}, \cF_{j7}\}_{i = 2,3,4,5}^{j = 2,3,4}$. The first several terms up to $O(e^{-3r})$ are
\be \small 
\begin{aligned}
	f_2(r) = &\cF_{20}e^{r} + e^{-r} \left(\frac{ \cF_{30}^2+\cF_{40}^2+\cF_{50}^2}{24
		\cF_{20}^3}\right.&\\
	&\left.-\frac{\cF_{20} \left(\cF_{30}^4-2
		\left(\cF_{40}^2+\cF_{50}^2\right)
		\cF_{30}^2+\left(\cF_{40}^2-\cF_{50}^2\right)^2\right)}{240 
		\cF_{30}^2 \cF_{40}^2 \cF_{50}^2}-\frac{1}{5 \cF_{20}}\right) + O(e^{-3r}), &\\
	f_3(r) = &\cF_{30}e^{r} + \frac{e^{-r}}{240\cF_{30}}\left( -\frac{2  \left(13 \cF_{30}^2+\cF_{40}^2+\cF_{50}^2\right)
		\cF_{30}^2}{\cF_{20}^4}\right.&\\
	&\left.	+\frac{-13 \cF_{30}^4+2
		\left(\cF_{40}^2+\cF_{50}^2\right) \cF_{30}^2+11
		\left(\cF_{40}^2-\cF_{50}^2\right)^2}{ \cF_{40}^2
		\cF_{50}^2}+\frac{24 \cF_{30}^2}{\cF_{20}^2}\right) +O(e^{-3r}), &\\
	f_4(r) = &\cF_{40}e^{r} + \frac{e^{-r}}{240\cF_{40}}\left(-\frac{2  \left(\cF_{30}^2+13 \cF_{40}^2+\cF_{50}^2\right)
		\cF_{40}^2}{\cF_{20}^4}\right.&\\
	&\left.	+\frac{11 \cF_{30}^4+2 \left(\cF_{40}^2-11
		\cF_{50}^2\right) \cF_{30}^2-13 \cF_{40}^4+11 \cF_{50}^4+2
		\cF_{40}^2 \cF_{50}^2}{ \cF_{30}^2 \cF_{50}^2}+\frac{24
		\cF_{40}^2}{\cF_{20}^2}\right) + O(e^{-3r}), &\\
	f_5(r) =  &\cF_{50}e^{r} + \frac{e^{-r}}{240\cF_{50}}\left(-\frac{2  \left(\cF_{30}^2+\cF_{40}^2+13 \cF_{50}^2\right)
		\cF_{50}^2}{\cF_{20}^4}\right.&\\
	&\left.	+\frac{11 \cF_{30}^4+\left(2 \cF_{50}^2-22
		\cF_{40}^2\right) \cF_{30}^2+11 \cF_{40}^4-13 \cF_{50}^4+2
		\cF_{40}^2 \cF_{50}^2}{ \cF_{30}^2 \cF_{40}^2}+\frac{24
		\cF_{50}^2}{\cF_{20}^2}\right) + O(e^{-3r}).&\\
\end{aligned}
\ee
Same as argued after \eqref{UVExpansionAnsatz2}, we expect the coefficients $\{ \cF_{j7}\}^{j = 2,3,4}$ to vanish, thus the family of solutions have four free parameters.

\subsection{Small radius expansion and numerics - NUT}
\begin{figure}[ht]
	\includegraphics[width=7cm]{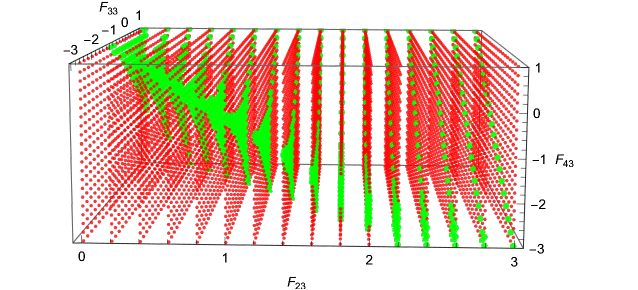}
	\includegraphics[width=5cm]{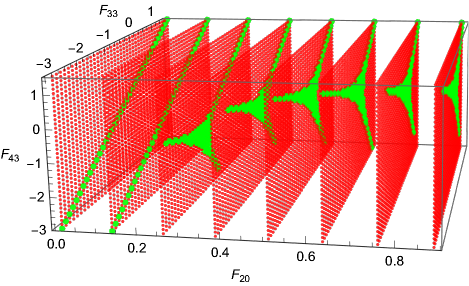}
	\caption{\rm Left: Green points correspond to initial values for NUT solution that integrate to infinity, and red points for those not. Right: Same plot for the Bolt solution.}
	\label{UVAllowedPoints}
\end{figure}

For NUT, we assume the following small-$r$ expansion:
\be
\begin{aligned}
	&f_1(r) = 1, &\\
	&f_i(r) = F_{ij}(r-r_0)^j,\quad i = 2,3,4,5,\quad j = 1,2,3,\cdots. &\\
\end{aligned}
\ee
In addition, we assume the SO(5)$\times$SO(3) symmetry to be preserved along the constant-$r$ surface, thus we require
\be
F_{31} = F_{41} = F_{51} > 0.
\ee
The equation of first order has two solutions:
\be
\left\{
\begin{aligned}
	&F_{21} = \frac{1}{2},&\\
	&F_{31} = F_{41} = F_{51} = \frac{1}{2},&
\end{aligned}
\right.\qquad {\rm or}\qquad
\left\{
\begin{aligned}
	&F_{21} = \frac{3\sqrt{5}}{10},&\\
	&F_{31} = F_{41} = F_{51} = \frac{3}{10}.&
\end{aligned}
\right.
\label{twoChoices}
\ee
For the first choice, we have solved \eqref{EOMAnsatz1} up to order $O(r-r_0)^{13}$, there're three free parameters, which we can choose to be $F_{23}, F_{33},F_{43}$, the first several terms up to $O(r-r_0)^5$ are:
\be
\begin{aligned}
	f_2(r) = &\frac{\rho }{2}+\rho ^3 F_{23}+\rho ^5 \left[-\frac{2}{5}  \left(F_{33}^2+F_{43} F_{33}+F_{43}^2\right)+\frac{7}{30}   (F_{33}+F_{43})\right.&\\
	&\left.	-\frac{1}{60} F_{23} (96  
	(F_{33}+F_{43})-91)-9 F_{23}^2-\frac{49}{720}\right]+O\left(\rho ^7\right), &\\
	f_3(r) = &\frac{\rho }{2  }+\rho ^3 F_{33}-\frac{\rho^5}{720  } \left[-144  \left(13 F_{33}^2-12 F_{43} F_{33}-12 F_{43}^2\right)+252   (F_{33}-4 F_{43})\right.&\\
	&	\left.+96 F_{23} (12   (F_{33}+6
	F_{43})-7)+49\right]+O\left(\rho ^7\right), &\\
	f_4(r) = &\frac{\rho }{2  }+\rho ^3 F_{43}-\frac{\rho^5}{720  } \left[44  \left(12 F_{33}^2+12 F_{43} F_{33}-13 F_{43}^2\right)-252   (4 F_{33}-F_{43})\right.&\\
	&\left.+96 F_{23} (12   (6 F_{33}+F_{43})-7)+49\right]+O\left(\rho ^7\right), &\\
	f_5(r) = &\frac{\rho }{2  }+\frac{\rho ^3}{12  } (-12   (F_{33}+F_{43})-48 F_{23}+7) + \frac{\rho ^5}{240  } \left[48  \left(13 F_{33}^2+38 F_{43} F_{33}+13 F_{43}^2\right)\right. \\
	&\left. -644(F_{33}+F_{43}) + 112 F_{23} (48   (F_{33}+F_{43})-23)+11520 F_{23}^2+147\right]+O\left(\rho ^7\right),&\\
\end{aligned}
\ee
where $\rho \equiv r - r_0$. When assigning $F_{23} = F_{33} = F_{43} = 1/12$, we obtain the standard ${\rm AdS_8}$ solution with $f_2 = f_3 = f_4 = f_5 = \sinh \rho$. Take the small $r$ expansion as the initial condition, we can perform numerics for general value of $\{ F_{23}, F_{33}, F_{43} \}$. As expected, not all initial values integrate to infinity: some of them vanishes at some finite $r$. By numerical simulation, we determine which initial values can be integrated to infinity and plot them in Fig.\ref{UVAllowedPoints}. There're two observations to be made. First, on each slice of the plot, the green region has the same shape as that obtained in \cite{Bobev:2016sap}. Second, initial values with $F_{33} = F_{43} = F_{53}$ are always allowed, which correspond to the first metric ansatz.

For the second choice in (\ref{twoChoices}), up to order $O(r-r_0)^{13}$, the solution is fixed, with no free parameter, whose leading order expansions can be identified with hyperbolic sine functions:
\be
\begin{aligned}
	&f_2(r) = \frac{3\sqrt{5}}{10}\left(\rho +\frac{\rho ^3}{6}+\frac{\rho ^5}{120}+\frac{\rho ^7}{5040}+\frac{\rho^9}{362880} + O\left(\rho ^{11}\right)\right) \rightarrow  \frac{3\sqrt{5}}{10}\sinh\rho, &\\
	&f_3(r)  = \frac{3}{10}\left(\rho +\frac{\rho ^3}{6}+\frac{\rho ^5}{120}+\frac{\rho ^7}{5040}+\frac{\rho ^9}{362880}+O\left(\rho^{11}\right)\right)\rightarrow \frac{3}{10}\sinh\rho, &\\	
	& f_3(r) = f_4(r) = f_5(r). &\\
\end{aligned}
\ee
With above solution, the metric has an asymptotic boundary given by squashed sphere in (\ref{SquashedSphereMetric1}) with $\lambda^2 = \frac{1}{5}$ as:
\be
ds^2 = dr^2 + \frac{9}{5} \sinh^2r ds_{\tilde{\mathbb{S}}^7}^2.
\ee
The curvatues of the geometry above is the same as the singular solution we obtained in the other metric (\ref{Curvatures}). 

\subsection{Small radius expansion and numerics - Bolt}
The ansatz for Bolt is
\be
\begin{aligned}
	&f_1(r) = 1, &\\
	&f_2(r) = F_{2j}(r-r_0)^j, j = 0,1,2,3,\cdots, &\\
	&f_i(r) = F_{ij}(r-r_0)^j, j = 1,2,3\cdots, &\\
	&F_{31} = F_{41} = F_{51} > 0,&\\
\end{aligned}
\ee
where we require $F_{20}>0$ for convenient. The leading terms are
\be
\begin{aligned}
	f_2(r) = & F_{20}-\frac{\rho ^2 \left(-7 F_{20}^2-3\right)}{8 F_{20}}-\frac{\rho ^4 \left(49 F_{20}^4+98 F_{20}^2+39\right)}{384 F_{20}^3}+\frac{\rho ^6}{46080 F_{20}^5} \\
	& \left[7 F_{20}^6 \left(1536  \left(F_{33}^2+F_{43}
	F_{33}+F_{43}^2\right)+1127\right)\right.\\
	&\quad  \left.	+3 F_{20}^4 \left(1536  \left(F_{33}^2+F_{43} F_{33}+F_{43}^2\right)+896   (F_{33}+F_{43})+5047\right)\right.\\
	&\quad \left.	+F_{20}^2 (1152  
	(F_{33}+F_{43})+10465)+2475\right]+O\left(\rho ^8\right). \\
	f_3(r) = & \frac{\rho }{2  }+\rho ^3 F_{33}+\frac{\rho ^5}{480   F_{20}^2} \left(F_{20}^2 \left(576  \left(2 F_{33}^2-3 F_{43} F_{33}-3 F_{43}^2\right)-420   F_{33}+49\right)\right.\\
	&\quad \left.	-72   (F_{33}+6
	F_{43})+35\right)+O\left(\rho ^7\right). \\
	f_4(r) = & \frac{\rho }{2  }+\rho ^3 F_{43}+\frac{\rho ^5}{480   F_{20}^2} \left(F_{20}^2 \left(576 \left(2 F_{43}^2 - 3 F_{43} F_{33}-3 F_{33}^2\right)-420   F_{43}+49\right) \right.\\
	& \quad \left. -72   (6F_{33}+F_{43})+35\right)+O\left(\rho ^7\right). \\
	f_5(r) = & \frac{\rho }{2  }+\rho ^3 \left(-\frac{1}{4   F_{20}^2}-F_{33}-F_{43}\right)\\
	&\quad +\frac{\rho ^5}{480   F_{20}^4} \left(F_{20}^4 \left(576  \left(2 F_{33}^2+7 F_{43} F_{33}+2 F_{43}^2\right)+420  
	(F_{33}+F_{43})+49\right)\right.\\
	&\quad \left.	+4 F_{20}^2 (162   (F_{33}+F_{43})+35)+90\right)+O\left(\rho ^7\right).\\
\end{aligned}
\ee
Taking the first two orders as the initial conditions, one can also solve the Einstein equations (\ref{EOMAnsatz1}) numerically. We can also obtain the set of initial values which can be integrated to infinity, as plotted in Fig.\ref{UVAllowedPoints}. The shape of the region is also similar to the one obtained in \cite{Bobev:2016sap}.


\section{Conformal mapping from sphere to plane}
\label{ConformalMapping}
In the main text, to perform the integration of the stress tensor correlation function, one needs the explicit conforming map from $\mathbb{S}^7$ to $\mathbb{R}^7$. In this section, we'll state how we get it and discuss how to construct the map in different situations.

Generally speaking, the conformal map is the composition of an embedding map from $\mathbb{S}^d$ to $\mathbb{R}^{d+1}$, together with a standard stereographic projection from $\mathbb{R}^{d+1}$ to the equatorial plane $\mathbb{R}^d$:
\be 
Y^{\bar a} = \frac{y^{\bar a} }{1-y^{d+1}},\quad \bar{a} = 1,2,...,d.\quad Y^{\bar{a}} \in \mathbb{R}^d,\quad y^A \in \mathbb{R}^{d+1}.
\label{StereographicProjection}
\ee 
The conformal factor is given by
\be 
ds^2_{\mathbb{R}^{d}} = \Omega^2 ds^2_{\mathbb{R}^{d+1}}  = \Omega^2 ds^2_{\mathbb{S}^d},\quad \Omega = \frac{1}{2}(1+Y_{\bar{a}} Y^{\bar{a}} ) = \frac{1}{1-y^{d+1}}.
\ee 
Thus the only step dependent on the specific metric is the embedding map from $\mathbb{S}^d$ to $\mathbb{R}^{d+1}$, which we'll discuss. In general, the exercise of obtaining the embedding map from the induced coordinate is a highly non-trivial task even for two-dimensional cases. The examples considered in this section are only some special cases where such a map can be constructed systematically.

\subsection{U(1) bundle}
\label{U1Bundle}

We start with an easier and more established situation, i.e., U(1) Hopf fibration. Spaces constructed by Hopf fibration with U(1) bundle are common in the literature, and closely related to generalized Taub-NUT spaces. One of them is $\mathbb{S}^{2k+1}$ constructed by U(1) bundle over $\mathbb{CP}^k$, preserving U(1)$\times$ SU$(k+1)$ symmetry. The metric can be written explicitly:
\be 
ds^2_{\mathbb{S}^d} = \frac{1}{d+1}ds^2_{\mathbb{CP}^k} + \left( d\psi + \frac{A_{\mathbb{CP}^k}}{d+1} \right)^2,\quad d = 2k+1,\ 0\le \psi\le 2\pi,
\label{U1FiberSphere}
\ee
where the Fubini-Study metric on the projective space and the K\"ahler potential $A_{\mathbb{CP}^k}$ are defined recursively:
\be \begin{aligned}
	ds^2_{\mathbb{CP}^k} &= (2k+2)\left[ d\xi_k^2 + \frac{1}{2k} \sin^2\xi_k ds^2_{\mathbb{CP}^{k-1}} +  \sin^2\xi_k \cos^2\xi_k \left( d\psi_k + \frac{1}{2k} A_{\mathbb{CP}^{k-1}} \right)^2 \right], \\
	A_{\mathbb{CP}^k} &= (2k+2)\sin^2\xi_k\left( d\psi_k + \frac{1}{2k} A_{\mathbb{CP}^{k-1}} \right),\quad 0\le \xi_i\le \frac{\pi}{2},\ 0\le\psi_i\le 2\pi.  \\
\end{aligned}
\ee 
The lowest value is $k = 1$, which reproduces the initial construction of Hopf fibration for $\mathbb{S}^3$:
\be 
ds^2_{\mathbb{CP}^1} = 4(d\xi_1^2 + \sin^2\xi_1 \cos^2\xi_1 d\psi_1^2),\quad A_{\mathbb{CP}^1} = 4\sin^2\xi_1 d\psi_1.
\ee 
The metric on $\mathbb{S}^3$ is:
\be 
ds^2 = d\xi_1^2 + \cos^2\xi_1\sin^2\xi_1 d\psi_1^2 + \left( d\psi + \sin^2\xi_1 d\psi_1 \right)^2 .
\ee 
To make the SO(4) invariance manifest, we can do the following trick: replace $d\psi^2$ by $(\cos^2\xi_1 + \sin^2\xi_1)d\psi^2$, collect terms with $\sin^2\xi_1$ and $\cos^2\xi_1$ respectively, then the metric becomes:
\be 
ds^2 = d\xi_1^2 + \cos^2\xi_1 d\psi^2 + \sin^2\xi_1 (d\psi + d\psi_1)^2 .
\label{StdMetricS3}
\ee 
This form may look more familiar, which can be obtained by the simple embedding formula, as we'll explain later:\footnote{Here we reverse the order of $x^a$ for the convenience of evaluating the integral.}
\be 
x^4 = \cos\xi_1 \cos \psi,\quad x^3 = \cos\xi_1\sin\psi,\quad x^2 = \sin\xi_1 \cos (\psi+\psi_1),\quad x^1 =  \sin\xi_1 \sin (\psi+\psi_1).
\label{EmbeddingMap}
\ee 
Now we head forward to $k=2$ for a metric on $\mathbb{S}^5$. Again, this time we replace $d\psi^2$ by $(\cos^2\xi_2 + \sin^2\xi_2)d\psi^2$, leave the former on its own, and put the latter together with the other terms:
\be 
ds^2 = d\xi_2^2 + \cos^2\xi_2 d\psi^2 + \sin^2\xi_2 \left[ d\xi_1^2 + \cos^2\xi_1 (d\psi + d\psi_2)^2 + \sin^2\xi_1 (d\psi + d\psi_1 + d\psi_2)^2 \right],
\label{StdMetricS5}
\ee 
where the term in the bracket is organized in order to produce the same form as (\ref{StdMetricS3}). So we can write down the embedding formula easily:
\be \begin{aligned}
	& x^6 = \cos\xi_2 \cos\psi,\quad x^5 = \cos\xi_2\sin\psi, \\
	& x^4 = \sin\xi_2 \cos \xi_1 \cos(\psi+\psi_2),\quad x^3 = \sin\xi_2 \cos \xi_1 \sin (\psi+\psi_2), \\
	&x^2 = \sin\xi_2 \sin \xi_1 \cos (\psi + \psi_1 + \psi_2),\quad x^1 = \sin\xi_2 \sin \xi_1 \sin (\psi + \psi_1 + \psi_2) .
\end{aligned}\ee 
Now we understand how the procedure works. Let's take one last example, the metric on $\mathbb{S}^7$ with $k=3$. Again, we replace $d\psi^2$ by $(\cos^2\xi_3 + \sin^2\xi_3)d\psi^2$ first:
\be 
ds^2 = d\xi_3^2 + \cos^2\xi_3 d\psi^2 + \sin^2 \xi_3 [\ A\ ].
\ee 
In our expectation, we need to organize $A$ in a form similar to (\ref{StdMetricS5}), where we need to depart $d\psi^2$ again by $(\cos^2\xi_2 + \sin^2\xi_2)d\psi^2$, finally we get
\be \begin{aligned}
	A & = d\xi_2^2 + \cos^2\xi_2 (d\psi + d\psi_3)^2 \\
	& + \sin^2\xi_2 \left[ d\xi_1^2 + \cos^2\xi_1 (d\psi + d\psi_2 + d\psi_3)^2 + \sin^2\xi_1 (d\psi + d\psi_1 + d\psi_2 + d\psi_3)^2 \right] .
\end{aligned}
\ee 
We can then write down the embedding map without any difficulty:
\be \begin{aligned}
	&x^8 = \cos\xi_3 \cos\psi,\quad x^7 = \cos\xi_3 \sin\psi,\quad x^6 = \sin\xi_3 \cos\xi_2 \cos(\psi + \psi_3), \\
	&x^5 = \sin\xi_3 \cos\xi_2 \sin(\psi + \psi_3),\quad x^4 = \sin\xi_3 \sin\xi_2 \cos\xi_1 \cos (\psi + \psi_2 + \psi_3), \\
	& x^3 = \sin\xi_3 \sin\xi_2 \cos\xi_1 \sin (\psi + \psi_2 + \psi_3),\quad x^2 = \sin\xi_3 \sin\xi_2 \sin\xi_1 \cos(\psi + \psi_1 + \psi_2 + \psi_3), \\
	& x^1 = \sin\xi_3 \sin\xi_2 \sin\xi_1 \sin(\psi + \psi_1 + \psi_2 + \psi_3).
\end{aligned}\label{S7Embedding}\ee 

Upon following the procedures above, we can write down the mapping for general $k$:
\be\begin{aligned}
	& x^{2k+2} = \cos\xi_k \cos\psi ,\quad x^{2k+1} = \cos\xi_k \sin\psi, \\
	& x^{2k} = \sin\xi_k \cos\xi_{k-1} \cos\theta_k,\quad x^{2k-1} = \sin\xi_k \cos\xi_{k-1} \sin\theta_k, \\
	& x^{2k-2} = \sin\xi_k \sin\xi_{k-1} \cos\xi_{k-2} \cos\theta_{k-1},\quad x^{2k-3} = \sin\xi_k \sin\xi_{k-1} \cos\xi_{k-2} \sin\theta_{k-1}, \\
	&... ... \\
	& x^4 = \prod_{i=2}^k \sin\xi_i \cos\xi_1 \cos\theta_2,\quad x^3 =  \prod_{i=2}^k \sin\xi_i \cos\xi_1 \sin\theta_2, \\
	& x^2 =  \prod_{i=1}^k \sin\xi_i \cos\theta_1,\quad x^1 =  \prod_{i=1}^k \sin\xi_i \sin\theta_1,
\end{aligned}\label{GeneralEmbedding}\ee 
where we have defined
\be 
\theta_i = \psi + \sum_{j = i}^k \psi_j,\quad i = 1,2,...,k.
\ee 
With the embedding map, one can write down the conformal map from spheres to planes, and evaluate the integrated two-point function of the stress tensor (\ref{Integrated2ptFunc}). We have evaluated this explicitly, and find it corresponds to the prediction of (\ref{Prediction}).

\subsection{SU(2) bundle}
Our metric on $\mathbb{S}^7$ that preserves explicitly SO(5)$\times $SO(3) isometry is the simplest example of Hopf fibration with SU(2) bundle, thus we expect some tricks can help us obtain the embedding map like (\ref{EmbeddingMap}), which is the simplest fibration with U(1) bundle. We discussed in section \ref{ATableTwoMetrics} that there're two equivalent ways to construct metrics on a sphere: by embedding into projective space, or Hopf fiber over projective space. As we will see, the embedding map of the first class of metrics can be obtained by the identification of projective space with Euclidean space in the near-origin limit. And the embedding map of the second class of metrics can be obtained by an explicit coordinate transformation \eqref{relation} from the first class.

Let's start with the first class of metrics. To illustrate how it works, we take an example first and consider $\mathbb{S}^3 \in \mathbb{CP}^2$, following the same steps in section \ref{ATableTwoMetrics}, we can obtain the metric on $\mathbb{S}^3$. The Fubini-Study metric is given by
\be 
ds^2 = (1+\bar{q}_kq_k)^{-1} d\bar{q}_i dq_i - (1+\bar{q}_kq_k)^{-2}\bar{q}_idq_id\bar{q}_jq_j,\quad q_1,q_2\in \mathbb{CP}^2.
\ee
We take the following parametrization on $\mathbb{CP}^2$:
\be\begin{aligned} 
	q_1 &= U\tan\chi\cos\frac{\mu}{2},\quad q_2 = V\tan\chi\sin\frac{\mu}{2},\quad U = e^{\mathbf{i}\theta/2},\quad V = e^{\mathbf{i}\Theta/2}, \\
	& \quad 0\le \mu \le \pi,\ 0\le \chi \le \frac{\pi}{2},\ 0\le \theta, \Theta \le 4\pi.
\end{aligned}\label{ParaCP2}\ee 
The Maurer-Cartan form and left-invariant one-form are
\be 
2U^{-1}dU = \mathbf{i}\sigma,\quad 2V^{-1}dV = \mathbf{i}\Sigma ,\quad \sigma = d\theta,\quad \Sigma = d\Theta.
\ee 
The Fubini-Study metric on $\mathbb{CP}^2$ is given by
\be 
\begin{aligned}
	ds^2 &= d\chi^2 + \frac{1}{4}\sin^2\chi \left[ d\mu^2 + \frac{1}{4}\sin^2\mu \omega^2 + \frac{1}{4}\cos^2\chi (\nu + \omega \cos\mu)^2 \right], \\
	& \quad \nu \equiv \sigma+\Sigma,\quad \omega \equiv \sigma - \Sigma .
\end{aligned}
\ee 
The line element in the bracket is the squashed three-sphere metric, with the squashing parameter identified with $\lambda\equiv \cos\chi$. In the limit $\chi\rightarrow 0$, we get the metric on the round sphere:
\be 
ds^2_{\mathbb{S}^3} = \frac{1}{4}\left( d\mu^2 + \cos^2 \frac{\mu}{2}d\theta^2 + \sin^2\frac{\mu}{2}d\Theta^2\right) .
\ee 
To compare with (\ref{StdMetricS3}), we can define $\tilde{\mu} \equiv \mu/2, \tilde{\theta}\equiv \theta/2,\tilde{\Theta}\equiv\Theta/2$ to get rid of the $1/4$ factor:
\be 
ds^2 = d\tilde{\mu}^2 + \cos^2\tilde{\mu}d\tilde{\theta}^2 + \sin^2\tilde{\mu}d\tilde{\Theta}^2.
\label{MostStdMetricS3}
\ee 
The metric (\ref{StdMetricS3}) is related to this one by a twist over angles:
\be 
\tilde{\mu} = \xi_1,\quad \tilde{\theta} = \psi,\quad \tilde{\Theta} = \psi + \psi_1 .
\ee 
Now we're ready to discuss the embedding map of $\mathbb{S}^3$ in $\mathbb{R}^4$. It still seems non-trivial to write down an embedding map from (\ref{MostStdMetricS3}) to $\mathbb{R}^4$, but the magic happens at $\chi\rightarrow 0$ limit of $\mathbb{CP}^2$: the space is identical to $\mathbb{R}^4$. We write down the map between $\mathbb{CP}^2$ and $\mathbb{R}^4$:
\be 
q_1 = x^1 + \mathbf{i}x^2,\quad q_2 = x^3 + \mathbf{i} x^4 .
\ee     
Compare it with (\ref{ParaCP2}), one obtains the following map between $(\mu, \theta, \Theta)$ and $(x^1,x^2,x^3,x^4)$:
\be 
x^1 = \cos\frac{\mu}{2}\cos\frac{\theta}{2},\quad x^2 = \cos\frac{\mu}{2}\sin\frac{\theta}{2},\quad x^3 = \cos\frac{\mu}{2}\cos\frac{\Theta}{2},\quad x^4 = \cos\frac{\mu}{2}\sin\frac{\Theta}{2}.
\ee 
This provides an embedding rule from the round sphere metric of the first class. However, the metrics that we're more interested in belong to the other class, for example, those we discuss in the last section: they're all obtained by Hopf fiber. For U(1) bundle, the problem is not severe as we can identify the two metrics (\ref{StdMetricS3}) and (\ref{MostStdMetricS3}) by observation. But for SU(2) bundle, it's no longer direct. Luckily, the map (\ref{relation}) between them has been worked out forty years ago. \cite{Awada:1982pk, Duff:1986hr}

The example we discuss in the main text is $\mathbb{S}^7$ embedded in $\mathbb{HP}^2$. Similarly, we identify the near-origin limit of $\mathbb{HP}^2$ as $\mathbb{R}^8$:
\be 
q_1 = x^1 + x^2 \mathbf{i} + x^3 \mathbf{j} + x^4 \mathbf{k},\quad q_2 = x^5 + x^6 \mathbf{i} + x^7 \mathbf{j} + x^8 \mathbf{k}.
\ee 
Meanwhile,
\be     
q_1 = \cos\frac{\mu}{2} \tilde{U},\quad q_2 = \sin\frac{\mu}{2} \tilde{V},
\ee 
where $\tilde{U}$ and $\tilde{V}$ are defined in terms of $(\mu, \Theta, \Phi, \Psi, \theta, \phi, \psi)$ according to (\ref{Utilde}). We get the following embedding map:
\be\small \label{EmbeddingMapS7}
\begin{aligned}
	&x^1 = \cos \frac{\mu }{2} \left(  \cos \frac{\theta }{2} \cos \frac{\Theta
	}{2} \cos \frac{1}{2} (-\Phi +\psi -\Psi +\phi )+\sin \frac{\theta }{2}
	\sin \frac{\Theta }{2} \cos \frac{1}{2} (-\Phi -\psi +\Psi +\phi ) \right), \\
	&x^2 =\cos \frac{\mu }{2} \left( \sin \frac{\theta }{2} \cos \frac{\Theta
	}{2} \cos \frac{1}{2} (\Phi -\psi +\Psi +\phi )-\cos \frac{\theta }{2}
	\sin \frac{\Theta }{2} \cos \frac{1}{2} (\Phi +\psi -\Psi +\phi )\right), \\
	&x^3 = \cos \frac{\mu }{2} \left( \sin \frac{\theta }{2} \cos \frac{\Theta
	}{2} \sin \frac{1}{2} (\Phi -\psi +\Psi +\phi )-\cos \frac{\theta }{2}
	\sin \frac{\Theta }{2} \sin \frac{1}{2} (\Phi +\psi -\Psi +\phi )\right), \\
	&x^4 = \cos \frac{\mu }{2}\left(  \sin \frac{\theta }{2} \sin \frac{\Theta
	}{2} \sin \frac{1}{2} (-\Phi -\psi +\Psi +\phi )+\cos \frac{\theta }{2}
	\cos \frac{\Theta }{2} \sin \frac{1}{2} (-\Phi +\psi -\Psi +\phi )\right), \\
	& x^5 = \cos \frac{\theta }{2} \sin \frac{\mu }{2} \cos \frac{\psi +\phi }{2}, \qquad\qquad x^6 = \sin \frac{\theta }{2} \sin \frac{\mu }{2} \cos \frac{\phi -\psi }{2}, \\
	& x^7 = \sin \frac{\theta }{2} \sin \frac{\mu }{2} \sin \frac{\phi -\psi }{2},\qquad\qquad  x^8 = \cos \frac{\theta }{2} \sin \frac{\mu }{2} \sin \frac{\psi +\phi }{2}. \\
\end{aligned}\ee 
The conformal map between $\mathbb{S}^7$ and $\mathbb{R}^7$ is obtained by combining the stereographic projection and the embedding map above, then we can evaluate the integral of the stress tensor two-point function as in the main text.

\section{Integrated correlators on general U(1) bundle spheres}
\label{FppForGeneralk}
In this section, we provide more details of our reproduction of the formula (\ref{Prediction}) for general $k$, where the integral (\ref{Integrated2ptFunc}) is performed on $\mathbb{S}^{2k+1}$, which we rewrite as below:
\be 
I = -\frac{C_T}{4}V_{2k+1} \int_{\mathbb{S}^{2k+1}} \sqrt{g_{2k+1}^{(0)}} d^{2k+1}x \Omega^{2k-1}(0)\Omega^{2k-1}(x) M^{\bar{a}\bar{b}}(0) M^{\bar{c}\bar{d}}(x) \frac{I_{\bar{a}\bar{b};\bar{c}\bar{d}}(X)}{X^{2(2k+1)}}
\ee 
The volume of $\mathbb{S}^{2k+1}$ and the square root of the metric (\ref{U1FiberSphere}) are given by:
\be 
V_{2k+1} = \frac{2\pi^{k+1}}{k!},\quad \sqrt{g_{2k+1}^{(0)}} = \prod_{i=1}^k \cos\xi_i\sin^{2i-1}\xi_i 
\ee 
The conformal factor and $X^2$ are known from the standard stereographic projection (\ref{StereographicProjection}):
\be 
\Omega(x) = \frac{1}{1-\cos\psi \cos\xi_k},\ \Omega(0) = \frac{1}{2};\quad X^{2} = \frac{1+\cos\psi\cos\xi_k}{1-\cos\psi\cos\xi_k}
\ee 
We define the new quantities:
\be 
M^{\bar{a}\bar{b}} = h^{ab}\frac{\partial X^{\bar{a}} }{\partial x^a}\frac{\partial X^{\bar{b}} }{\partial x^b} = -\frac{\partial X^{\bar{a}} }{\partial  \psi}\frac{\partial X^{\bar{b}} }{\partial \psi} \equiv -\Lambda^{\bar{a}} \Lambda^{\bar{b}},
\ee 
where we used the fact that the only non-vanishing component of $h^{ab}$ is $h^{\psi\psi} = -1$. Using another fact that the only non-zero component of $M^{\bar{a}\bar{b}}(0)$ is $M^{\bar{\psi}\bar{\psi}} = -\frac{1}{4}$, the contraction between $ M^{\bar{a}\bar{b}}M^{\bar{c}\bar{d}}$ and $I_{\bar{a}\bar{b};\bar{c}\bar{d}}$ can be simplied as:
\be \begin{aligned}
	M^{\bar{a}\bar{b}}(0) M^{\bar{c}\bar{d}}(x) I_{\bar{a}\bar{b};\bar{c}\bar{d}}(X) & = \frac{1}{4}(\Lambda^{2k+1})^2 - \frac{1}{X^2} (\Lambda^{\bar{a}} X_{\bar{a}}) (\Lambda^{2k+1}X_{2k+1})\\
	& + \frac{1}{X^4}(\Lambda^{\bar{a}} X_{\bar{a}})^2 (X^{2k+1})^2 - \frac{1}{4(2k+1)}(\Lambda^{\bar{a}}\Lambda^{\bar{a}})\\
\end{aligned}\ee 
Using our general form of coordinate transformation (\ref{GeneralEmbedding}), we can get a general form of the quantities appearing above, which, if we put together into the integrand together with what we have above, we obtain:
\be\begin{aligned} 
	F''(0) &= -\frac{C_T\pi^{k+1}}{(2k+1)2^{2k+5}k!} \int\sqrt{g}d^{2k+1}x \frac{1}{(1+\cos\psi\cos\xi_k)^{2k+3}}\left[ 8 k \cos 2 \psi \cos ^2\xi_k \right. \\
	&\left. \qquad  +4 \cos \psi[ (6 k-1) \cos \xi_k+(2 k+1) \cos
	3 \xi _k] +4 (3 k+1) \cos2 \xi _k \right. \\
	&\left. \qquad +(2 k+1) \cos 4 \xi _k+10 k-5 \right] 
\end{aligned}\label{IntegralFpp}\ee 
Note that the integral doesn't depend on $\psi_i$, which reflects the symmetry of the manifold. As a first step, we can integrate out $\prod_{i=1}^k d\psi_i$ and $\prod_{j=1}^{k-1}d\xi_j$:
\be 
\int \prod_{i=1}^k\prod_{j=1}^{k-1} d\psi_id\xi_j\sqrt{g} = \frac{2\pi^k\cos\xi_k\sin^{2k-1}\xi_k}{(k-1)!}
\ee 
Since the dependence of $\psi$ is relatively simple in (\ref{IntegralFpp}), we integrate over $\psi$ first. The trick is to identify the integral range $\psi\in [\pi,2\pi]$ to $[0,\pi]$ by the invariance of the integrand under $\psi \rightarrow -\psi$ and periodicity $\psi \sim \psi + 2\pi$, and integrate on a new variable $z = \cos\psi$. This gives us with a divergent integral of $\xi_k$, whose value, after throwing away the diverging terms, is:
\be 
F''(0) = - C_T \frac{k\pi^{2k+3/2}\Gamma\left(-k-\frac{1}{2}\right)}{4^k (k-1)!}
\ee 
After identifying $2k+1 = d$ and simplification, the result is exactly the formula (\ref{Prediction}) predicted by \cite{Bueno:2018yzo} from the analysis of high-derivative gravity.

\cleardoublepage





\bibliography{allpapers}
\bibliographystyle{JHEP}



\makebackcoverXII

\end{document}